\documentclass[aps,floats]{revtex4-2}
\usepackage{amsmath,amssymb}
\usepackage{graphicx,epsfig}
\usepackage[greek,english]{babel}
\usepackage{bbold}

\makeatletter\AtBeginDocument{\let\@elt\relax}\makeatother
 
\begin{document}
\bibliographystyle {plain}

\pdfoutput=1
\def\oppropto{\mathop{\propto}} 
\def\opsimeq{\mathop{\simeq}}
\def\opoverderline{\mathop{\overline}}
\def\operarrow{\mathop{\longrightarrow}}
\def\opsim{\mathop{\sim}}

\def\opmin{\mathop{\min}} 
\def\opmax{\mathop{\max}} 
\def\oplim{\mathop{\lim}}

\title{ Nonequilibrium diffusion processes via non-Hermitian electromagnetic quantum mechanics \\
 with application to the statistics of entropy production in the Brownian gyrator  
} 


\author{Alain Mazzolo}
\affiliation{Universit\'e Paris-Saclay, CEA, Service d'\'Etudes des R\'eacteurs et de Math\'ematiques Appliqu\'ees, 91191, Gif-sur-Yvette, France}

\author{C\'ecile Monthus}
\affiliation{Universit\'e Paris-Saclay, CNRS, CEA, Institut de Physique Th\'eorique, 91191 Gif-sur-Yvette, France}


\begin{abstract}
The nonequilibrium Fokker-Planck dynamics in an arbitrary force field $\vec f(\vec x)$ in dimension $N$ is revisited via the correspondence with the non-Hermitian quantum mechanics in a real scalar potential $V(\vec x)$ and in a purely imaginary vector potential $[-i \vec A(\vec x)] $ of real amplitude $\vec A(\vec x)$. The relevant parameters of irreversibility are then the $\frac{N(N-1)}{2}$ magnetic matrix elements $B_{nm}(\vec x )  =-B_{mn} (\vec x ) = \partial_n A_m (\vec x ) -  \partial_m A_n (\vec x )$, while it is enlightening to explore the corresponding gauge transformations of the vector potential $\vec A(\vec x) $. This quantum interpretation is even more fruitful to study the statistics of all the time-additive observables of the stochastic trajectories, since their generating functions correspond to the same quantum problem with additional scalar and/or vector potentials. Our main conclusion is that the analysis of their large deviations properties and the construction of the corresponding Doob conditioned processes can be drastically simplified via the choice of an appropriate gauge for each purpose. This general framework is then applied to the special time-additive observables of Ornstein-Uhlenbeck trajectories in dimension $N$, whose generating functions correspond to quantum propagators involving quadratic scalar potentials and linear vector potentials, i.e. to quantum harmonic oscillators in constant magnetic matrices.  As simple illustrative example, we finally focus on the Brownian gyrator in dimension $N=2$ to compute the large deviations properties of the entropy production of its stochastic trajectories and to construct the corresponding conditioned processes having a given value of the entropy production per unit time.

\end{abstract}

\maketitle

\section{ Introduction  }

\subsection{ On the various links between diffusion processes and quantum mechanics  }

\subsubsection{ Link between the Brownian motion and the Euclidean quantum mechanics for a free particle }

For the Brownian motion in dimension $N$, the probability $P(\vec x,t) $ 
to be at position $\vec x$ at time $t$ satisfies the heat equation 
\begin{eqnarray}
 \partial_t P(\vec x,t )     = \frac{1}{2} \Delta   P(\vec x,t ) \equiv \frac{1}{2} \vec \nabla^2  P(\vec x,t ) \equiv   \frac{1}{2} \sum_{n=1}^N   \partial_n^2   P(\vec x,t )
\label{heat}
\end{eqnarray}
which involves the Laplacian $\Delta=\vec \nabla^2$ 
built from the spatial derivatives $\partial_n \equiv \frac{\partial }{\partial_{x_n}}$ with respect to the $N$ coordinates $x_n$
for $n=1,2,.,N$.
The heat Eq. \ref{heat} corresponds to the Euclidean-time $t=i \theta$ version 
of the quantum mechanics for a free particle,
where the amplitude $\psi( \vec x, \theta )$ to be at position $\vec x$ at time $\theta$
satisfies the free Schr\"odinger equation that involves only the Laplacian
\begin{eqnarray}
i   \partial_{\theta} \psi( \vec x, \theta  )  = - \frac{1}{2}  \Delta \psi(\vec x, \theta )
\label{schrodingerfree}
\end{eqnarray}
This correspondence at the level of generators is of course even more powerful at the level of Feynman path-integrals for trajectories \cite{feynman}.
It is thus very natural to extend this analogy as much as possible by considering the Euclidean-time quantum mechanics for a particle in an electromagnetic potential (see the reminder in Appendix \ref{app_electromagnetic}), 
with the various special cases recalled in the next subsections.


\subsubsection{ Similarity transformation between detailed-balance diffusions and supersymmetric quantum mechanics  }

As described in textbooks \cite{gardiner,vankampen,risken,Pavliotis}, the generator of a Markov processes satisfying detailed-balance
can be transformed via a similarity transformation into an Hermitian operator, 
with the very important spectral consequences.
For the simplest example of a diffusion process of diffusion coefficient $D=1/2$ converging toward the normalizable steady state $P^*(\vec x) $
that can always be rewritten in terms of some function $\phi(\vec x)$ in the exponential
\begin{eqnarray}
  P^*(\vec x) = P^*(\vec 0) e^{- 2  \phi( \vec x) } 
 \label{steadyB}
\end{eqnarray}
the corresponding Fokker-Planck dynamics 
\begin{eqnarray}
 \partial_t P(\vec x,t )  = -  \vec \nabla . \left(   P(\vec x,t ) \vec f^{rev} (\vec x ) - \frac{1}{2}  \vec \nabla  P(\vec x,t )  \right) 
\label{FPdetailed}
\end{eqnarray}
satisfying detailed-balance involves the force
\begin{eqnarray}
\vec f^{rev} (\vec x ) = \frac{1}{2} \vec \nabla  \ln P^*(\vec x) = -  \vec \nabla  \phi( \vec x) 
\label{forceDB}
\end{eqnarray}
The similarity transformation
\begin{eqnarray}
 P(\vec x,t )  =   \sqrt{  P^*(\vec x) } \psi (\vec x,t ) = \sqrt{  P^*(\vec 0) }  \ e^{-   \phi( \vec x) }  \psi (\vec x,t )
\label{similaritydetailed}
\end{eqnarray}
transforms the Fokker-Planck Eq. \ref{FPdetailed}
into the Euclidean Schr\"odinger equation for $\psi (\vec x,t ) $
\begin{eqnarray}
 \partial_t   \psi (\vec x,t )   =  -   {\cal  H} \psi (\vec x,t ) 
\label{Euclideanpsi}
\end{eqnarray}
which involves the well-known Hermitian quantum supersymmetric Hamiltonian $  {\cal  H}$  
(see the review \cite{review_susyquantum} and references therein)
\begin{eqnarray}
 {\cal  H}  =  {\cal  H}^{\dagger} = \frac{  1 }{2 }  \left( - \vec \nabla + ( \vec \nabla\phi(\vec x) )\right) 
. \left(  \vec \nabla + ( \vec \nabla\phi(\vec x) )\right)  \equiv    -  \frac{1}{2}  \Delta    +  V(\vec x)
  \label{Hsusydetailed}
\end{eqnarray}
with the very specific form of the scalar potential 
\begin{eqnarray}
  V(\vec x) = \frac{  1 }{2 } \ \left( [ \vec \nabla  \phi (\vec x) ]^2- \Delta  \phi (\vec x)  \right) 
  \label{Vscalarsusydetailed}
\end{eqnarray}
while the quantum-normalized zero-energy ground-state reads
\begin{eqnarray}
\psi^{GS} (\vec x )  =   \sqrt{  P^*(\vec x) }  = \sqrt{  P^*(\vec 0) }  \ e^{-   \phi( \vec x) }  
\label{GSsimilaritydetailed}
\end{eqnarray}


\subsubsection{ Feynman-Kac formula to analyze the time-additive observables of the stochastic trajectories  }

The standard method to study the statistics of the time-additive observables of Markov trajectories
 is the introduction of the appropriate deformations of the Markov generator.
 This approach goes back to the famous Feynman-Kac formula \cite{feynman,kac,c_these,review_maj,Poland_Bistable}
 introduced to analyze any observable ${\cal O}[\vec x(0 \leq s \leq t) ]$ of the Brownian trajectory
 $\vec x(0 \leq \tau \leq t)$ 
that can be parametrized by some scalar field $V^{[{\cal O}]}(\vec x)$ 
and by some vector field $\vec A^{[{\cal O}]}(\vec x)$ in the Stratonovich interpretation
\begin{eqnarray}
{\cal O}[\vec x(0 \leq \tau \leq t) ]=    \int_{0}^{t} d \tau \left[   - V^{[{\cal O}]} (\vec x(\tau) )
+  \dot {\vec x} (\tau) . \vec A^{[{\cal O}]}( \vec x(\tau)) \right]
\label{additive}
\end{eqnarray}
Its generating function $Z^{[k]}(\vec x,t \vert \vec y,0)$ of parameter $k$
over the Brownian trajectories $ \vec x(0 \leq s \leq t) $ starting at $\vec x(0)=\vec y$ and ending at $\vec x(t)=\vec x$
can be written as the Feynman path-integral
\begin{eqnarray}
 Z^{[k]}(\vec x,t \vert \vec y,0) 
&& \equiv \overline{ \delta^{(N)} \left( \vec x(t) - \vec x \right)  e^{k{\cal O} [\vec x(0 \leq \tau \leq t) ]}  \delta^{(N)} \left( \vec x(0) - \vec y \right) }
 \nonumber \\
 &&
= \int_{\vec x(\tau=0)=\vec y}^{\vec x(\tau=t)=\vec x} {\cal D}   \vec x(\tau)  
 e^{ - \displaystyle 
  \int_0^{t} d\tau \frac{\dot {\vec x}^2  (\tau)}{2}   
+ k 
\int_{0}^{t} d \tau \left[   - V^{[{\cal O}]} (\vec x(\tau) )
+  \dot {\vec x} (\tau) . \vec A^{[{\cal O}]}( \vec x(\tau)) \right]
 }
  \nonumber \\
 && \equiv  \langle \vec x \vert e^{- t {\cal  H}^{[k]} } \vert \vec y  \rangle
 \label{geneBrown}
\end{eqnarray}
that corresponds to the Euclidean quantum propagator $ \langle \vec x \vert e^{- t {\cal  H}^{[k]} } \vert \vec x  \rangle $ 
associated to the Hamiltonian 
\begin{eqnarray}
 {\cal H}^{[k]}   =  -  \frac{1}{2}  \left( \vec \nabla  -k \vec A^{[{\cal O}]}( \vec x) \right)^2    + k V^{[{\cal O}]} ( \vec x) 
 =   \frac{1}{2}  \left( -i \vec \nabla  + i k \vec A^{[{\cal O}]}( \vec x) \right)^2    + k V^{[{\cal O}]} ( \vec x) 
 \label{FPhamiltonianpBrown}
\end{eqnarray}
Since one is usually interested into real observables ${\cal O}[\vec x(0 \leq \tau \leq t) ] $ 
parametrized by real fields $V^{[{\cal O}]} ( \vec x)  $ and $\vec A^{[{\cal O}]}( \vec x) $,
it is useful to distinguish the three following cases :

(i) If $\vec A^{[{\cal O}]}( \vec x) \equiv 0$ in the additive observable of Eq. \ref{additive}, 
then the Hamiltonian of Eq. \ref{FPhamiltonianpBrown} is Hermitian  
\begin{eqnarray}
 {\cal H}^{[k]} = ( {\cal H}^{[k]} )^{\dagger}   =  -  \frac{1}{2} \Delta    + k V^{[{\cal O}]} ( \vec x) 
 \label{FPhamiltonianpBrownscalar}
\end{eqnarray}
and involves only the scalar potential $(k V^{[{\cal O}]} ( \vec x)  )$.

(ii) If $V^{[{\cal O}]} ( \vec x) \equiv 0$ in the additive observable of Eq. \ref{additive}, 
then it is possible to consider instead $k=iq$ with real $q$ to transform the generating function of Eq. \ref{geneBrown} into the characteristic function of the observable ${\cal O}[\vec x(0 \leq \tau \leq t) ] $ 
for the Fourier parameter $q$. 
Then the Hamiltonian of Eq. \ref{FPhamiltonianpBrown} 
is complex but Hermitian 
\begin{eqnarray}
 {\cal H}^{[k=iq]} = ( {\cal H}^{[k=iq]} )^{\dagger}   =  -  \frac{1}{2}  \left( \vec \nabla  - i q \vec A^{[{\cal O}]}( \vec x) \right)^2    
 =   \frac{1}{2}  \left( -i \vec \nabla  -q \vec A^{[{\cal O}]}( \vec x) \right)^2    
 \label{FPhamiltonianpBrownvector}
\end{eqnarray}
and involves only the real vector potential $(q A^{[{\cal O}]} ( \vec x)  )$ from the quantum point of view.
Let us mention that while the Feynman-Kac formula is often 
described only for the scalar potential case of Eq. \ref{FPhamiltonianpBrownscalar}, 
its application for the vector potential case of Eq. \ref{FPhamiltonianpBrownvector}
plays a major role
to take into account topological constraints in the context of polymer physics \cite{edw67,wiegel}
and to analyze the winding properties of Brownian paths \cite{orsay_winding1,orsay_winding2,orsay_winding3,orsay_winding4,c_these,winding1,winding2}.

(iii) For the general case of an additive observable of Eq. \ref{additive} where both fields $V^{[{\cal O}]} ( \vec x) \ne 0 $
 and $\vec A^{[{\cal O}]}( \vec x) \ne 0 $ are nonvanishing, it is clear that 
 the Hamiltonian of Eq. \ref{FPhamiltonianpBrown} is different from its adjoint 
\begin{eqnarray}
 {\cal H}^{[k]} \ne ( {\cal H}^{[k]} )^{\dagger} && =  -  \frac{1}{2}  \left( \vec \nabla  +k \vec A^{[{\cal O}]}( \vec x) \right)^2    + k V^{[{\cal O}]} ( \vec x) 
 =   \frac{1}{2}  \left( -i \vec \nabla  - i k \vec A^{[{\cal O}]}( \vec x) \right)^2    + k V^{[{\cal O}]} ( \vec x) 
 \label{FPhamiltonianpBrownNonHermitian}
\end{eqnarray}
and corresponds to the quantum problem with the real scalar potential $k V^{[{\cal O}]} ( \vec x) $
and the purely imaginary vector potential $(-i k A^{[{\cal O}]} ( \vec x)  )$, described in Appendix \ref{app_electromagnetic} 
around Eqs \ref{hamiltonianquantumEuclideanAimaginary} and \ref{hamiltonianquantumEuclideanAimaginaryadjoint}.

It is also important to stress that in the field of large deviations discussed in the next subsection, even for the case (ii) discussed above,
it is standard to consider only the generating function $Z^{[k]}(\vec x,t \vert \vec y,0) $ for real $k$, and not the characteristic function corresponding to $k=iq$ with real $q$,  
i.e. it is usual to work with the real non-Hermitian Hamiltonian 
$  {\cal H}^{[k]} \ne ( {\cal H}^{[k]} )^{\dagger}$ instead of the complex Hermitian Hamiltonian
$ {\cal H}^{[k=iq]} = ( {\cal H}^{[k=iq]} )^{\dagger} $  of Eq. \ref{FPhamiltonianpBrownvector}.


\subsection{ Large deviations properties for trajectory observables of Markov trajectories  }

\label{subsec_largedevadditive}

The theory of large deviations (see the reviews \cite{oono,ellis,review_touchette} and references therein)
has become the unifying language in the field of nonequilibrium processes
(see the reviews with different scopes \cite{derrida-lecture,harris_Schu,searles,harris,mft,sollich_review,lazarescu_companion,lazarescu_generic,jack_review}, 
the PhD Theses \cite{fortelle_thesis,vivien_thesis,chetrite_thesis,wynants_thesis,chabane_thesis,duBuisson_thesis} 
 and the Habilitation Thesis \cite{chetrite_HDR}).
In particular, the approach based on the deformed Markov generators recalled above has been used 
to analyze the statistics of many interesting additive observables of various Markov processes over the years
 \cite{peliti,derrida-lecture,sollich_review,lazarescu_companion,lazarescu_generic,jack_review,vivien_thesis,lecomte_chaotic,lecomte_thermo,lecomte_formalism,lecomte_glass,kristina1,kristina2,jack_ensemble,simon1,simon2,tailleur,simon3,Gunter1,Gunter2,Gunter3,Gunter4,chetrite_canonical,chetrite_conditioned,chetrite_optimal,chetrite_HDR,touchette_circle,touchette_langevin,touchette_occ,touchette_occupation,garrahan_lecture,Vivo,c_ring,c_detailed,chemical,derrida-conditioned,derrida-ring,bertin-conditioned,touchette-reflected,touchette-reflectedbis,c_lyapunov,previousquantum2.5doob,quantum2.5doob,quantum2.5dooblong,c_ruelle,lapolla,c_east,chabane}. 
 While the large deviations properties of all types of Markov processes are of course interesting, the following summary is restricted to the case of Markov processes converging towards a steady state.
 
 
  \subsubsection{ Rate functions $I ( o ) $ and scaled-cumulant-generating-functions $E(k)$ of time-additive observables }
 
Since a time-additive observable ${\cal O}[\vec x(0 \leq s \leq t) ] $ of a Markov trajectory $\vec x(0 \leq s \leq t) $
is extensive with respect to the duration $t$,
it is useful to introduce its rescaled intensive counterpart
\begin{eqnarray}
o[\vec x(0 \leq s \leq t) ] \equiv \frac{ {\cal O}[\vec x(0 \leq s \leq t) ] }{t} \opsimeq_{t \to +\infty} o^*
\end{eqnarray}
which will converge for $t \to +\infty$ towards the steady value $o^*$  that can be computed from the steady state 
properties.
For large $t$, the fluctuations around this steady-state value $o^*$ are described by the following large deviations form for the probability $ P_t( o ) $ to see the intensive value $o$
over the time-window $t$
 \begin{eqnarray}
 P_t( o ) \opsimeq_{t \to +\infty} e^{- t I ( o )}
\label{level1def}
\end{eqnarray} 
The positive rate function $I(o) \geq 0 $ vanishes only for the steady value $o^*$ 
where it is minimum
 \begin{eqnarray}
 0 = I ( o^* ) = I' ( o^* )
\label{iaeqvanish}
\end{eqnarray}

For large time $t \to +\infty$, the generating function $Z^{[k]}(\vec x,t \vert \vec y,0)$ 
rewritten as Euclidean Schr\"odinger propagator associated to some Hamiltonian ${\cal  H}^{[k]} $
(as in the example of Eq. \ref{geneBrown} for Brownian trajectories)
will display the asymptotic behavior
\begin{eqnarray}
Z^{[k]}(\vec x,t \vert \vec y,0) = \langle \vec x \vert e^{- t {\cal  H}^{[k]} } \vert \vec y  \rangle\opsimeq_{t \to +\infty}
 e^{-t E(k) } r_k(\vec x) l_k(\vec y)
\label{genelargeT}
\end{eqnarray}
where $E(k)$ is the ground-state energy of the Hamiltonian ${\cal  H}^{[k]} $,
while $r_k(.)$ and $l_k(.)$
are the corresponding positive right and left eigenvectors 
\begin{eqnarray}
 E(k)  r_k( \vec x) && = {\cal H}^{[k]}  r_k( \vec x)
\nonumber \\
E(k)  l_k( \vec x) && = ({\cal H}^{[k]})^{\dagger}  l_k( \vec x)
\label{eigenright}
\end{eqnarray}
with the normalization
\begin{eqnarray}
\int d^N \vec x \   l_k( \vec x)  r_k( \vec x) =1
  \label{eigennorma}
\end{eqnarray}
For $k=0$ where the generating function reduces to the propagator $P(\vec x,t \vert \vec y,0) $
of the Markov process
that converges towards the steady state $P^*(x) $ for any initial condition $y$
\begin{eqnarray}
Z^{[k=0]}(\vec x,t \vert \vec y,0) = P(\vec x,t \vert \vec y,0)   \opsimeq_{t \to +\infty}
 e^{-t E(0) } r_{0}(\vec x) l_{0}(\vec y) = P^*(x)
\label{genelargeTkzero}
\end{eqnarray}
 the ground state energy vanishes $E(k=0)=0$,
while the right eigenvector corresponds to the steady state $r_{k=0}(\vec x)= P^*(\vec x )$ 
and the left eigenvector is trivial $  l_{k=0} (\vec y )=1$
\begin{eqnarray}
 E(0) && =0
 \nonumber \\
  r_{0}(\vec x) && = P^*(x)
   \nonumber \\
    l_{0} (\vec y )&& =1
\label{Ekeigenkzero}
\end{eqnarray}

The consistency between the asymptotic time behavior of Eq. \ref{genelargeT}
and the large deviation form of Eq. \ref{level1def}
via the saddle-point evaluation for large $t$
\begin{eqnarray}
\langle e^{k {\cal O}[\vec x(0 \leq s \leq t) ] } \rangle = \langle e^{k t o[\vec x(0 \leq s \leq t) ] } \rangle 
\equiv \int do  e^{k t o} P_t( o ) \opsimeq_{t \to +\infty} 
\int do e^{ t \left[ k o - I ( o ) \right] }\opsimeq_{t \to +\infty} e^{ -t E(k) }
\label{level1gen}
\end{eqnarray} 
yields that the ground-state energy $E(k)$ is the scaled-cumulant-generating-function and
 corresponds the Legendre transform of the rate function $ I ( o ) $
 \begin{eqnarray}
 k o - I ( o ) && = -E(k) 
 \nonumber \\
 k - I'(o) && =0
\label{legendre}
\end{eqnarray} 
So the reciprocal Legendre transform 
 \begin{eqnarray}
 k o + E(k) && = I(o) 
 \nonumber \\
 o + E'(k) && =0
\label{legendrereci}
\end{eqnarray} 
allows us to compute the rate function $I(o)$ from the knowledge of the energy $E(k)$.
In particular, the steady value $o^*$ satisfying Eq. \ref{iaeqvanish}
is conjugated to the value $k=0$ and thus corresponds to the first-order perturbation theory in $k$ of the ground-state energy $E(k)$ around $E(k=0)=0$ in Eq. \ref{legendrereci}
 \begin{eqnarray}
  o^* = - E'(k=0)
\label{linktypperturbationEp}
\end{eqnarray}
Once one has elucidated the large deviations properties of the observable
via its rate function $I(o)$,
it is often interesting to analyze the rare Markov trajectories that have been able to produce
a given anomalous value $o \ne o^*$ different from the steady value $o^*$
via the notion of canonical conditioning recalled in the next subsection.

 
 \subsubsection{ Canonical conditioning of parameter $k$ based on the generating function $Z^{[k]}(\vec x,t \vert \vec y,0)  $  }
 
 \label{subsec_remindercanonical}
 
As explained in detail in the two complementary papers \cite{chetrite_conditioned,chetrite_optimal}
and in the Habilitation thesis \cite{chetrite_HDR},
 the canonical conditioning of parameter $k$ based
  the generating function $Z^{[k]}(\vec x,t \vert \vec y,0) $ 
  that is summarized below becomes equivalent in the large-time limit $t \to + \infty$
with the microcanonical conditioning that would impose the Legendre value $o=- E'(k)$ of Eq. \ref{legendrereci} for the intensive observable $o$.
 The idea is that for each value $k$, 
 one introduces the conditional probability $ {\cal P}^{Cond[k]}(\vec z,\tau) $ 
to be at position $\vec z$ at the internal time $\tau \in ]0,t[$ 
\begin{eqnarray}
{\cal P}^{Cond[k]}(\vec z,\tau) 
=  \frac{Z^{[k]}(\vec x, t \vert \vec z, \tau) Z^{[k]}(\vec z, \tau \vert \vec y, 0)}{Z^{[k]}(\vec x, t \vert \vec y, 0)}
\label{markovcondk}
\end{eqnarray}
which is normalized over $\vec z$ at any time $\tau$
\begin{eqnarray}
\int d^N \vec z \ {\cal P}^{Cond[k]}(\vec z,\tau) 
= 1
\label{markovcondknorma}
\end{eqnarray}
and that satisfies the boundary conditions at times $\tau=0$ and $\tau=t$
\begin{eqnarray}
{\cal P}^{Cond[k]}(\vec z,\tau=0) && = \delta^{(N)} (\vec z -\vec y)
\nonumber \\
{\cal P}^{Cond[k]}(\vec z,\tau=t) && = \delta^{(N)} (\vec z -\vec x)
\label{markovcondkbound}
\end{eqnarray}

For large time $t\to +\infty$, the conditional probability ${\cal P}^{Cond[k]}(\vec z,\tau)  $ of Eq. \ref{markovcondk} at any interior time $\tau$ satisfying $0 \ll \tau \ll t$
can be evaluated from the asymptotic property of Eq. \ref{genelargeT}
for the three involved generating functions 
\begin{eqnarray}
{\cal P}^{Cond[k]}(\vec z,\tau) 
&& \opsimeq_{ 0 \ll \tau \ll t}   \frac{e^{-(t-\tau) E(k) } r_k(\vec x) l_k(\vec z) 
e^{-\tau E(k) } r_k(\vec z) l_k(\vec y)}
{e^{-t E(k) } r_k(\vec x) l_k(\vec y)} 
\nonumber \\
&& = l_k(\vec z) r_k(\vec z) 
 \equiv {\cal P}^{Cond[k]Interior}(\vec z) 
\label{markovcondktinfty}
\end{eqnarray}
to obtain that it does not depend on the interior time $\tau$ and that it reduces to the product of the left eigenvector $l_k(\vec z) $ and the right eigenvector $r_k(\vec z) $ of Eqs \ref{eigenright} and \ref{eigennorma}.
The knowledge of the left eigenvector $l_k(\vec z) $ is then necessary to construct 
the generator of the conditioned process that has Eq. \ref{markovcondktinfty} as steady state
(see subsection \ref{subsec_canonical} of the main text for more details).

This canonical conditioning of parameter $k$ in the large-time limit $t \to + \infty$
is thus a huge simplification
with respect to the finite-time Doob process conditioned to end at the given  position $x(t)=x$
and at the given value ${\cal O}[\vec x(0 \leq s \leq t) ] =O$ of the additive observable,
whose construction requires the knowledge of the finite-time joint propagator $ P(\vec x,O,t \vert \vec y,0,0)   $ and produces time-dependent generators,
as described for the various examples studied recently \cite{refMazzoloJstat,Alain_OU,refdeBruyne2021,c_microcanonical,us_LocalTime}.


\subsection{ Goals of the present paper  }

In the present paper, the main goal is to analyze the statistics of time-additive observables 
of Eq. \ref{additive} when the diffusion process $\vec x(t)$ satisfies the Langevin 
stochastic differential system involving the $N$ independent Wiener processes $w_n(t)$
\begin{eqnarray}
dx_n(t) &&=  f_n(\vec x (t) ) \ dt + dw_n(t)
\label{langevin}
\end{eqnarray}
where the space-dependent force $\vec f(\vec x)$ ensures that 
the corresponding Fokker-Planck dynamics for the propagator $P(\vec x,t \vert \vec y,0) $
[i.e. the probability distribution to be at $\vec x$ at time $t$ when starting at $\vec y$ at time $t=0$]
\begin{eqnarray}
 \partial_t P(\vec x,t \vert \vec y,0)   
 =  -  \sum_{n=1}^N \partial_n  \left( f_n( \vec x)  P(\vec x,t \vert \vec y,0)  - \frac{1}{2} 
   \partial_n   P(\vec x,t \vert \vec y,0) \right)
 \equiv - {\cal  H} P(\vec x,t \vert \vec y,0)
\label{fokkerplanck}
\end{eqnarray}
converges towards some normalized steady state $P^*(\vec x)$.
The interpretation of the Fokker-Planck Eq. \ref{fokkerplanck}
 as an Euclidean Schr\"odinger equation
involves the non-Hermitian quantum Hamiltonian ${\cal  H} \ne {\cal  H}^{\dagger}$
\begin{eqnarray}
{\cal  H}&&  =   \vec \nabla . \left(  - \frac{1}{2}  \vec \nabla +   \vec f(\vec x)\right)
=     - \frac{1}{2}  \vec \nabla^2
 +   \vec f(\vec x) . \vec \nabla
 +    [\vec \nabla . \vec f(\vec x) ]  
 \nonumber \\
 {\cal  H}^{\dagger} &&  = - \left( \frac{1}{2}  \vec \nabla +   \vec f(\vec x)\right). \vec \nabla
 =     - \frac{1}{2}  \vec \nabla^2 -   \vec f(\vec x) . \vec \nabla
\label{FPhamiltonian}
\end{eqnarray}
This Hamiltonian ${\cal  H} $ can be rewritten as an Euclidean non-Hermitian quantum Hamiltonian
involving a real scalar potential $V(\vec x) $ and a 
purely imaginary vector potential $[-i \vec A(\vec x)] $ of real amplitude $\vec A(\vec x)$,
that we have already encountered in Eq. \ref{FPhamiltonianpBrownNonHermitian} of the Introduction
  and that is discussed in details around Eq. \ref{hamiltonianquantumEuclideanAimaginary}  
  of Appendix \ref{app_electromagnetic}
\begin{eqnarray}
{\cal  H} &&   =  -  \frac{1}{2}  \left( \vec \nabla -   \vec A(\vec x) \right)^2    +  V(\vec x)
=  - \frac{1}{2}  \vec \nabla^2 +\vec A ({\vec x}). \vec \nabla
+ \frac{1}{2}   \left( \vec \nabla.\vec A ({\vec x})\right)  - \frac{1}{2}  \vec A^2 (\vec x)  + V (\vec x)
\nonumber \\
{\cal  H}^{\dagger} &&   =
-  \frac{1}{2} \left(  \vec \nabla  +  \vec A (\vec x) \right)^2  + V (\vec x) 
   =  - \frac{1}{2}  \vec \nabla^2 - \vec A ({\vec x}). \vec \nabla
- \frac{1}{2}   \left( \vec \nabla . \vec A ({\vec x})\right)  - \frac{1}{2}  \vec A^2 (\vec x)  + V (\vec x)
\label{FPhamiltonianQuantum}
\end{eqnarray}
where the vector potential $ \vec A (\vec x ) $ coincides with the force $\vec f( \vec x) $
\begin{eqnarray}
  \vec A (\vec x ) \equiv  \vec f( \vec x)
 \label{vectorpot}
\end{eqnarray}
while the scalar potential $V(\vec x) $ involves both the divergence $[\vec \nabla . \vec f(\vec x) ] $ and the square $ {\vec f \ }^2( \vec x)$ of the force
\begin{eqnarray}
V(\vec x) \equiv  \frac{  [\vec \nabla . \vec f(\vec x) ] +\vec f ^2( \vec x) }{2 }   
\label{scalarpot}
\end{eqnarray}
Equivalently, the path-integral for the propagator $P(\vec x,t \vert \vec y,0) $ associated to the Langevin system of Eq. \ref{langevin}
\begin{eqnarray}
P(\vec x,t \vert \vec y,0)
= \int_{\vec x(\tau=0)=\vec y}^{\vec x(\tau=t)=\vec x} {\cal D}   \vec x(\tau)  
 e^{ - \displaystyle 
  \int_0^{t} d\tau {\cal L} (\vec x(\tau), \dot {\vec x}(\tau) )
 } 
\label{pathintegral}
\end{eqnarray}
involves the classical Lagrangian
\begin{eqnarray}
 {\cal L} (\vec x(\tau), \dot {\vec x}(\tau))
 && \equiv \frac{1}{2} \left( \dot {\vec x } (\tau) - \vec f( \vec x(\tau)) \right)^2
 +\frac{  1 }{2 }  [\vec \nabla . \vec f(\vec x(\tau)) ]  
 \nonumber \\
 && = \frac{1}{2}  \dot {\vec x}^2  (\tau) 
 - \dot {\vec x } (\tau) . \vec f( \vec x(\tau))
+ \frac{\vec f^2( \vec x(\tau))+[\vec \nabla . \vec f(\vec x(\tau)) ] }{2}   
 \nonumber \\
 && \equiv 
 \frac{1}{2}  \dot {\vec x}^2  (\tau) 
 - \dot {\vec x } (\tau) . \vec A( \vec x(\tau))
 + V(\vec x(\tau))
\label{lagrangian}
\end{eqnarray}
The last rewriting in terms of the vector potential $\vec A ( \vec x) $ of Eq. \ref{vectorpot}
and of the scalar potential $V ( \vec x) $ of Eq. \ref{scalarpot}
corresponds as it should to the Euclidean classical electromagnetic Lagrangian of Eq. \ref{lagrangianEuclideanimaginary}
discussed in detail in Appendix \ref{app_electromagnetic}.
Please note that from now on, we will use the shorter name "the vector potential $\vec A ( \vec x) $ " and not 
"a purely imaginary vector potential $[-i \vec A(\vec x)] $ of real amplitude $\vec A(\vec x)$" anymore.

In the present paper, our main goal is to describe all the advantages of this non-Hermitian electromagnetic quantum interpretation, 
both for the diffusion process $\vec x(t)$ itself to characterize its irreversibility,
and for all its time-additive observables of Eq. \ref{additive}.  
This general framework will be then applied to Ornstein-Uhlenbeck processes 
that are well-known for their explicit Gaussian finite-time propagators and steady states
described in textbooks \cite{gardiner,vankampen,risken}, but that  
have nevertheless been reconsidered over the years to 
study specific problems like diffusion in a real magnetic field \cite{Poland_BrownianB},
as well as to address the new issues raised by 
the progresses in the field of nonequilibrium.
It will be thus interesting to consider first the general case in arbitrary dimension $N$ studied in \cite{Thouless,CGetJML} 
with various applications to spin models and electrical arrays described in \cite{CGetJML},
and then the specific model of the Brownian gyrator in dimension $N=2$ that has attracted a lot of interest recently  \cite{Gyr_vanWijland,Gyr_2013,Gyr_elec,Gyr_exp,Gyr_Tryphon_Harvesting,Gyr_Tryphon_Engine,Gyr_Tryphon_Geometry,Gyr_Inferring,Gyr_Inference,cerasoli} as the simplest model that is maintained out-of-equilibrium by two reservoirs at two different temperatures.


\subsection{ Organization of the paper  }

The paper is organized as follows.
In section \ref{sec_mappingQ}, we stress that the relevant irreversibility parameters of the diffusion are the corresponding magnetic matrix elements $B_{nm}(\vec x )  =-B_{mn} (\vec x ) = \partial_n A_m (\vec x ) -  \partial_m A_n (\vec x )$ and we describe the gauge transformation of the vector potential $\vec A(\vec x) $
based on the decomposition of the force $\vec f(\vec x) $ into its reversible and irreversible contributions..
In section \ref{sec_additive}, the generating function of an arbitrary time-additive observable of the stochastic trajectories
is studied via the corresponding quantum problem with additional scalar and/or vector potentials, to show that the analysis of its large deviations properties and of the corresponding Doob conditioned process can be drastically simplified via the choice of the appropriate gauge.
This general framework is then applied to Ornstein-Uhlenbeck-processes in dimension $N$  in section \ref{sec_OU}
and to their quadratic trajectory observables in section \ref{sec_OUadditive}, where the corresponding quantum problems involve quadratic scalar potentials and linear vector potentials, i.e. to quantum harmonic oscillators in constant magnetic matrices.
 Finally in section \ref{sec_Gyr}, we consider
  the Brownian gyrator in dimension $N=2$ to characterize the statistics of the entropy production of its stochastic trajectories.
Our conclusions are summarized in section \ref{sec_conclusion}.
Various appendices contain complementary material or more technical computations.


\section{ Magnetic matrix and appropriate gauge for the irreversibility}

\label{sec_mappingQ}

In this section, we describe the advantages of the interpretation 
of the Fokker-Planck generator of Eq. \ref{fokkerplanck}
as the non-Hermitian quantum Hamiltonian of Eq. \ref{FPhamiltonianQuantum}
to characterize the irreversibility of the diffusion.


\subsection{ The antisymmetric magnetic matrix $ B_{nm}(\vec x )  =-B_{mn} (\vec x ) $
 as the relevant parameters of irreversibility } 

In arbitrary dimension $N$, the magnetic 'field' corresponds to the 
the antisymmetric matrix $B_{nm}(\vec x )  =-B_{mn} (\vec x ) $
that can be computed from the vector potential $\vec A ( \vec x) $ of Eq. \ref{vectorpot}
 via the formula generalizing the three-dimensional curl
\begin{eqnarray}
 B_{nm}(\vec x )  =-B_{mn} (\vec x )  
 \equiv \partial_n A_m (\vec x ) -  \partial_m A_n (\vec x ) 
  =\partial_n f_m (\vec x ) -  \partial_m f_n (\vec x ) 
 \label{magneticN}
\end{eqnarray}
and that is invariant under gauge transformations of the vector potential $ \vec A(\vec x ) $.
It is then essential to distinguish two cases :

(i) When the magnetic matrix identically vanishes ${\bold B}(\vec x )=0 $, 
then the vector potential $ \vec A(\vec x ) =\vec f(\vec x)$ of Eq. \ref{vectorpot} corresponding 
to the force $\vec f(\vec x) $ can be written as the gradient of some function $\phi(\vec x) $
\begin{eqnarray}
{\rm When } \ {\bold B}(\vec x )=0 : \ \ \ \   \vec A(\vec x ) =\vec f(\vec x)= -  \vec \nabla  \phi( \vec x)
  \label{magneticNzero}
\end{eqnarray}
and one recovers the detailed-balance dynamics of Eq. \ref{FPdetailed}.
Then the standard similarity transformation of Eq. \ref{similaritydetailed}
that transforms the Fokker-Planck generator into the
 the Hermitian quantum Hamiltonian of Eq. \ref{Hsusydetailed}
 without any vector potential can be reinterpreted, in  
 the non-Hermitian electromagnetic quantum langage,
 as the gauge transformation from the initial vector potential 
$\vec A(\vec x ) = -  \vec \nabla  \phi( \vec x)$
 into the new vector potential that identically vanishes.

(ii) When the magnetic matrix does not vanish ${\bold B}(\vec x ) \ne 0 $,
then the corresponding vector potential $\vec A(\vec x) $ cannot be completely
eliminated by a gauge transformation anymore.
To see more clearly that the $\frac{N(N-1)}{2}$ independent matrix elements $B_{nm}(\vec x )  $
are then the relevant intrinsic parameters of the model that characterize the irreversibility of the nonequilibrium Langevin dynamics,
 it is useful to consider the gauge transformation from the initial vector potential 
$\vec A(\vec x ) $
to the new vector potential that will only contains the irreversible contribution of the force.


\subsection{ Decomposition of the force $\vec f(\vec x) = \vec f^{rev} (\vec x)+\vec f^{irr} (\vec x)$ 
into its reversible and irreversible contributions  }

For irreversible diffusions, it is convenient to continue to parametrize the steady state $ P^*(\vec x)$
by the function $\phi(\vec x)$ in the exponential as in the reversible case of Eq. \ref{steadyB}
\begin{eqnarray}
  P^*(\vec x) = P^*(\vec 0) e^{- 2  \phi( \vec x) } 
 \label{steadyirr}
\end{eqnarray}
The stationary Fokker-Planck Eq. \ref{fokkerplanck}
\begin{eqnarray}
0  =  - \vec \nabla . \left(  P^*(\vec x ) \vec f(\vec x) - \frac{1}{2}  \vec \nabla  P^*(\vec x )  \right) 
\equiv    - \vec \nabla .  {\vec J}^*(\vec x)
\label{divjsteady}
\end{eqnarray}
means that the steady current ${\vec J}^*(\vec x) $ associated to the steady state $ P^*(\vec x )$
\begin{eqnarray}
 {\vec J}^*(\vec x) \equiv  P^*(\vec x ) \vec f( \vec x) - \frac{1}{2} \vec \nabla  P^*(\vec x) 
 = P^*(\vec x ) \left( \vec f( \vec x) - \frac{1}{2} \vec \nabla  \ln P^*(\vec x) \right)
 = P^*(\vec x ) \left( \vec f( \vec x) + \vec \nabla  \phi(\vec x) \right)
\label{jsteady}
\end{eqnarray}
should be divergenceless.
In the field of nonequilibrium, it is then standard 
to decompose the force $\vec f(\vec x) = \vec f^{rev} (\vec x)+\vec f^{irr} (\vec x)$ 
into its reversible and irreversible contributions as follows :

(i) the reversible contribution $\vec f^{rev} ( \vec x) $ is the force already written in 
Eq. \ref{forceDB} that satisfies the detailed-balance condition
\begin{eqnarray}
\vec f^{rev} (\vec x ) = \frac{1}{2} \vec \nabla  P^*(\vec x) = -  \vec \nabla  \phi( \vec x) 
\label{forceRev}
\end{eqnarray}
i.e. that would produce a vanishing steady current in Eq. \ref{jsteady}
\begin{eqnarray}
\vec  J^{*rev} (\vec x) = \vec 0
 \label{jsteadyrev}
\end{eqnarray}

(ii) the remaining irreversible contribution of the force
\begin{eqnarray}
\vec f^{irr} (\vec x) \equiv \vec f(\vec x)  - \vec f^{rev} (\vec x)
\label{forceIrrev}
\end{eqnarray}
 is then directly responsible for the steady current of Eq. \ref{jsteady} via
\begin{eqnarray}
  \vec J^*(\vec x)  =  P^*(\vec x ) \vec f^{irr}( \vec x)  
 \label{jsteadyirrev}
\end{eqnarray}
The vanishing divergence of Eq. \ref{divjsteady} for the steady current $\vec J^*(\vec x)$
\begin{eqnarray}
0 =  \vec \nabla .  {\vec J}^*(\vec x) = \vec \nabla .   \left( P^*(\vec x ) \vec f^{irr}( \vec x)  \right)
= \vec f^{irr}( \vec x) . \vec \nabla P^*(\vec x ) +P^*(\vec x )    \left( \vec \nabla . \vec f^{irr}( \vec x)  \right)
 \label{divjsteadyirrev}
\end{eqnarray}
yields that the divergence of the irreversible force $\vec f^{irr}( \vec x) $ should satisfy
\begin{eqnarray}
 \vec \nabla .  \vec f^{irr}( \vec x)  
  = -  \vec f^{irr}( \vec x) .   \vec \nabla \ln   P^*(\vec x )
  = 2 \vec f^{irr} ( \vec x)  . \vec \nabla \phi(\vec x )
 \label{divforceirrev}
\end{eqnarray}

Another way to take into account the vanishing divergence of the steady current $\vec J^*(\vec x)$
amounts to introduce some antisymmetric stream matrix $\psi_{nm}(\vec x)= - \psi_{nm}(\vec x)$ to write the current components as
\begin{eqnarray}
   J^*_n(\vec x)  =  \sum_m \partial_m \psi_{nm}(\vec x)
 \label{jsteadystream}
\end{eqnarray}
This generalizes the standard property that a divergence-less three-dimensional vector can be written as a curl.
It is actually more convenient to use instead the antisymmetric matrix obtained after the rescaling by the steady state $P^*(\vec x)$
\begin{eqnarray}
  \Omega_{nm} (\vec x) \equiv \frac{ \psi_{nm}(\vec x) }{P^*(\vec x)} = -  \Omega_{mn} (\vec x)
 \label{omeganm}
\end{eqnarray}
to rewrite the irreversible force of Eq. \ref{jsteadyirrev}
as
\begin{eqnarray}
   f^{irr}_n( \vec x) && = \frac{  J^*_n(\vec x) }{ P^*(\vec x ) }
   =\frac{  1 }{ P^*(\vec x ) }  \sum_m \partial_m \left[ P^*(\vec x) \Omega_{nm} (\vec x)\right]
   =\frac{  1 }{ P^*(\vec x ) }  \sum_m  \Omega_{nm} (\vec x)  \partial_m P^*(\vec x) 
   +   \sum_m \partial_m  \Omega_{nm} (\vec x)
   \nonumber \\
   && =   \sum_m \Omega_{nm} (\vec x) \partial_m \ln P^*(\vec x) 
   +   \sum_m \partial_m  \Omega_{nm} (\vec x)
   = -2 \sum_m \Omega_{nm} (\vec x) \partial_m \phi(\vec x) 
   +   \sum_m \partial_m  \Omega_{nm} (\vec x)
 \label{Firrevomega}
\end{eqnarray}


\subsection{ Gauge transformation towards the vector potential $ {\vec A}^{irr} (\vec x ) =\vec f^{irr}( \vec x) $ associated to the irreversible force } 

\label{subsec_gaugeirr}

The decomposition of the force $\vec f( \vec x)  $ into its reversible and irreversible components
described in the previous subsection
translates into the following gauge transformation 
for the vector potential of Eq. \ref{vectorpot}
\begin{eqnarray}
  \vec A (\vec x ) && = \vec f( \vec x)=  \vec f^{rev}( \vec x) + \vec f^{irr}( \vec x) 
= - \vec \nabla \phi(\vec x) + \vec f^{irr}( \vec x) 
   \equiv - \vec \nabla \phi(\vec x) +  \vec A^{irr} (\vec x )
 \label{vectorpotgaugeirr}
\end{eqnarray}
where the new vector potential 
\begin{eqnarray}
  \vec A^{irr} (\vec x ) \equiv  \vec f^{irr}( \vec x)
    \label{vectorpotirr}
\end{eqnarray}
only involves the irreversible force $\vec f^{irr}( \vec x) $
instead of the total force for the initial vector potential $ \vec A (\vec x ) = \vec f( \vec x)$.
The magnetic matrix of Eq. \ref{magneticN} associated to the total vector potential of Eq. \ref{vectorpotgaugeirr}
can be then rewritten in terms of the irreversible force $\vec f^{irr}( \vec x) $ only
\begin{eqnarray}
 B_{nm}(\vec x )  =-B_{mn} (\vec x )  
 = \partial_n A^{irr}_m (\vec x ) -  \partial_m A^{irr}_n (\vec x ) 
  =\partial_n f^{irr}_m (\vec x ) -  \partial_m f^{irr}_n (\vec x ) 
 \label{magneticNirr}
\end{eqnarray}
The advantage of these magnetic matrix elements $B_{nm}(\vec x ) $ is that they can be directly obtained via Eq. \ref{magneticN}
from the force $\vec f (\vec x)$ that appears in the Langevin system of Eq. \ref{langevin} defining the model,
while the computation of the irreversible force $ \vec f^{irr}( \vec x) $ of Eq. \ref{forceIrrev} requires the knowledge of the steady state $P^*(\vec x)  $.


\subsection{ Effect of the gauge transformation
on the path-integral representation of the propagator} 

The effect of the gauge transformation of Eq. \ref{vectorpotgaugeirr}
on the path-integral of Eq. \ref{pathintegral} for the propagator 
can be analyzed from the integral over time of 
the term involving the vector potential in the Lagrangian of Eq. \ref{lagrangian} 
\begin{eqnarray}
  \int_0^{t} d\tau \dot {\vec x} (\tau).   \vec A( \vec x(\tau)) 
 && =  \int_0^{t} d\tau \dot {\vec x} (\tau). 
 \left[ - \vec \nabla \phi(\vec x(\tau)) +  \vec A^{irr} (\vec {\vec x}(\tau)) \right]
 = - \int_0^{t} d\tau \partial_{\tau}  \phi (\vec x(\tau))
 +  \int_0^{t} d\tau \dot {\vec x} (\tau). \vec A^{irr}( \vec x(\tau)) 
 \nonumber \\
 && = \phi(\vec x(0)) -  \phi(\vec x(t)) 
 +  \int_0^{t} d\tau \dot {\vec x} (\tau).  \vec A^{irr} ( \vec x(\tau)) 
\label{pathintegrallast}
\end{eqnarray}
So the appropriate change of variable for the propagator of Eq. \ref{pathintegral}
coincides with the standard similarity transformation recalled in Eq. \ref{similaritydetailed} of the Introduction
\begin{eqnarray}
P(\vec x,t \vert \vec y,0)
 = e^{ \displaystyle \phi ( \vec y) - \phi ( \vec x)} {\hat P}(\vec x,t \vert \vec y,0)
\label{changetowardshat}
\end{eqnarray}
 and
yields that the new propagator
\begin{eqnarray}
{\hat P}(\vec x,t \vert \vec y,0)
 = \int_{\vec x(\tau=0)=\vec y}^{\vec x(\tau=t)=\vec x} {\cal D}   \vec x(\tau)  
 e^{ - \displaystyle 
  \int_0^{t} d\tau {\hat {\cal L}}  (\vec x(\tau), \dot {\vec x}(\tau) )
 } 
\label{pathintegralhat}
\end{eqnarray}
is governed by the new Lagrangian
\begin{eqnarray}
{\hat {\cal L}} (\vec x(\tau), \dot {\vec x}(\tau))
  \equiv 
 \frac{1}{2}  \dot {\vec x}^2  (\tau) 
 - \dot {\vec x } (\tau) . \vec A^{irr} ( \vec x(\tau))
 + V(\vec x)
\label{lagrangianhat}
\end{eqnarray}
that involves the same scalar potential $V( \vec x) $ of Eq. \ref{scalarpot}
but the new vector potential ${\vec A}^{irr}(\vec x )  $.

The property of Eq. \ref{divforceirrev}
allows to simplify the scalar potential of Eq. \ref{scalarpot}
using the decomposition of the force $\vec f( \vec x) $ into its reversible and irreversible contributions
\begin{eqnarray}
V(\vec x) && = 
 \frac{  1 }{2 } \vec \nabla . [ {\vec f \ }^{rev}(\vec x) + {\vec f \ }^{irr}(\vec x) ] 
+  \frac{  1 }{2 } \left[ {\vec f \ }^{rev}( \vec x) + {\vec f \ }^{irr}( \vec x)\right]^2
\nonumber \\
&& = \frac{  1 }{2 } \vec \nabla . [ - \vec \nabla  \phi (\vec x) + {\vec f \ }^{irr}(\vec x) ] 
+  \frac{  1 }{2 } \left[  - \vec \nabla  \phi (\vec x) + {\vec f \ }^{irr}( \vec x)\right]^2
\nonumber \\
&& = \frac{  1 }{2 } \ \left( [ \vec \nabla  \phi (\vec x) ]^2- \Delta  \phi (\vec x)  \right)
+  \frac{  1 }{2 } \left[ {\vec f \ }^{irr}( \vec x)\right]^2
\label{scalarpotexpanded}
\end{eqnarray}

Since ${\vec A \ }^{irr}( \vec x)={\vec f \ }^{irr}( \vec x) $, the Lagrangian of Eq. \ref{lagrangianhat}
can be refactorized into
\begin{eqnarray}
{\hat {\cal L}} (\vec x(\tau), \dot {\vec x}(\tau))
  \equiv 
 \frac{1}{2} \left( \dot {\vec x}  (\tau) -    \vec f^{irr} ( \vec x(\tau)) \right)^2
 + \frac{  1 }{2 } \ \left( [ \vec \nabla  \phi (\vec x(\tau)) ]^2- \Delta  \phi (\vec x(\tau))  \right)
\label{lagrangianhatfactor}
\end{eqnarray}


\subsection{ Effect of the gauge transformation
on the non-Hermitian quantum Hamiltonian ${\cal  H}$}

The Euclidean Schr\"odinger equation of Eq. \ref{fokkerplanck}
for the initial propagator $P(\vec x,t \vert \vec y,0)   $
translates via the change of variable of Eq. \ref{changetowardshat}
into the following Euclidean Schr\"odinger equation for the new propagator ${\hat P}(\vec x,t \vert \vec y,0) $
\begin{eqnarray}
 \partial_t {\hat P}(\vec x,t \vert \vec y,0)   
 =  -  {\hat {\cal  H}} {\hat P}(\vec x,t \vert \vec y,0)
\label{Euclideanhat}
\end{eqnarray}
where the corresponding Hamiltonian $ {\hat {\cal  H}}$ obtained from the initial Hamiltonian ${\cal  H}$ of Eq. \ref{FPhamiltonianQuantum}
via 
the similarity transformation 
\begin{eqnarray}
 {\hat {\cal  H}}  = e^{ \displaystyle  \phi ( \vec x)}{\cal  H} e^{ \displaystyle - \phi ( \vec x) }
  && =  e^{ \displaystyle  \phi ( \vec x)} \left[
   -  \frac{1}{2}  \left( \vec \nabla -   \vec A(\vec x) \right)^2    +  V(\vec x)
 \right]   e^{ \displaystyle - \phi ( \vec x) }
\nonumber \\
&&    =    -  \frac{1}{2}  \left( \vec \nabla -   \vec A^{irr}(\vec x) \right)^2    +  V(\vec x)
  \label{hamiltonianhat}
\end{eqnarray}
involves the same scalar potential $V (\vec x) $ and the new vector potential $  {\vec A}^{irr}(\vec x ) $ as a consequence of the conjugation property
\begin{eqnarray}
 e^{ \displaystyle  \phi ( \vec x)} \vec \nabla   e^{ \displaystyle - \phi ( \vec x) }
 &&  =  \vec \nabla - [\vec \nabla \phi (\vec x)  ]
    \label{conjugderi}
\end{eqnarray}
Using Eq. \ref{divforceirrev},
the expansion of Eq. \ref{hamiltonianhat} with the vector potential $ \vec A^{irr}(\vec x)=\vec f^{irr}(\vec x)$
and with the scalar potential of Eq. \ref{scalarpotexpanded} 
yields that the Hamiltonian $ {\hat {\cal  H}} $ 
\begin{eqnarray}
 {\hat {\cal  H}}  &&    =    -  \frac{1}{2}  \left( \vec \nabla -   \vec f^{irr}(\vec x) \right)^2    +  V(\vec x)
 =  -  \frac{1}{2}  \Delta 
 +   \frac{  1 }{2 } \ \left( [ \vec \nabla  \phi (\vec x) ]^2- \Delta  \phi (\vec x)  \right) 
 + \vec f^{irr}(\vec x) .  \vec \nabla 
  +  \vec f^{irr}(\vec x) . \vec \nabla \phi(\vec x)
  \nonumber \\
 && \equiv   {\hat {\cal  H}}_{rev} + {\hat {\cal  H}}_{irr} 
  \label{hamiltoniantildeexpanded}
\end{eqnarray}
can be decomposed into the following reversible and irreversible contributions.

(i)  The reversible contribution ${\hat {\cal  H}}_{rev} $ 
associated to the reversible dynamics when the irreversible component of the force vanishes 
 $ \vec f^{irr}( \vec x)=0 $ in Eq. \ref{hamiltoniantildeexpanded}
\begin{eqnarray}
  {\hat {\cal  H}}_{rev}    && = 
 -  \frac{1}{2}  \Delta 
 +   \frac{  1 }{2 } \ \left( [ \vec \nabla  \phi (\vec x) ]^2- \Delta  \phi (\vec x)  \right) 
  \label{hamiltonianhateqsusy}
\end{eqnarray}
corresponds to the 
well-known Hermitian supersymmetric quantum Hamiltonian recalled in Eqs \ref{Hsusydetailed}
and \ref{Vscalarsusydetailed} of the Introduction
\begin{eqnarray}
  {\hat {\cal  H}}_{rev}    = {\hat {\cal  H}}_{rev}^{\dagger}
 =   \frac{  1 }{2 }  \left( - \vec \nabla + ( \vec \nabla\phi(\vec x) )\right) 
. \left(  \vec \nabla + ( \vec \nabla\phi(\vec x) )\right) 
 =   \frac{  1 }{2 }  \sum_n  Q_n^{\dagger} Q_n 
 \label{hamiltonianhateqsusyQ}
\end{eqnarray}
that can be factorized in terms of the $N$ first-order differential operators $Q_n $ and their adjoints $Q_n^{\dagger} $
 \begin{eqnarray}
Q_n && \equiv \partial_n + (\partial_n \phi(\vec x) ) 
\nonumber \\
Q_n^{\dagger} && \equiv - \partial_n + (\partial_n \phi(\vec x) )
\label{Qnsusy}
\end{eqnarray}

The product
 \begin{eqnarray}
Q_n Q_m && = \bigg( \partial_n + (\partial_n \phi(\vec x) ) \bigg)  \bigg( \partial_m + (\partial_m \phi(\vec x) )\bigg)
\nonumber \\
&& = \partial_n \partial_m + (\partial_m \phi(\vec x) ) \partial_n + (\partial_n\partial_m \phi(\vec x) )
+ (\partial_n \phi(\vec x) )\partial_m + (\partial_n \phi(\vec x)) (\partial_m \phi(\vec x) )
= Q_m Q_n
\label{QnQm}
\end{eqnarray}
yields that two annihilation operators $Q_n$ and $Q_m$ commute 
 \begin{eqnarray}
[Q_n , Q_m ] && = 0 = [Q_n^{\dagger} , Q_m^{\dagger} ]
\label{Qnsusycommut0}
\end{eqnarray}
and equivalently two creation operators $Q_n^{\dagger}$ and $Q_m^{\dagger}$ commute.
The comparison of the two products involving one creation operator $Q_n^{\dagger}$
and one annihilation operator $Q_m $
 \begin{eqnarray}
Q_n^{\dagger} Q_m && = \bigg( - \partial_n + (\partial_n \phi(\vec x) ) \bigg)  \bigg( \partial_m + (\partial_m \phi(\vec x) )\bigg)
\nonumber \\
&&=  - \partial_n \partial_m - (\partial_m \phi(\vec x) ) \partial_n - (\partial_n\partial_m \phi(\vec x) )
+ (\partial_n \phi(\vec x) )\partial_m + (\partial_n \phi(\vec x) )(\partial_m \phi(\vec x) )
\nonumber \\
Q_m Q_n^{\dagger}  && 
= \bigg( \partial_m + (\partial_m \phi(\vec x) )\bigg)\bigg( - \partial_n + (\partial_n \phi(\vec x) ) \bigg)  
\nonumber \\
&&=  - \partial_m \partial_n + (\partial_n \phi(\vec x) ) \partial_m  + ( \partial_m \partial_n \phi(\vec x) )
 -  (\partial_m \phi(\vec x) ) \partial_n + (\partial_m \phi(\vec x) )(\partial_n \phi(\vec x) ) 
\label{Qnsusyprod}
\end{eqnarray}
yields that the commutators involve the second derivatives of the function $ \phi(\vec x)$ 
\begin{eqnarray}
[  Q_m, Q_n^{\dagger}]  \equiv Q_m Q_n^{\dagger}- Q_n^{\dagger} Q_m =  2 (\partial_n\partial_m \phi(\vec x) ) 
\label{Qnsusycommut}
\end{eqnarray}

(ii)  The irreversible contribution in Eq. \ref{hamiltoniantildeexpanded}
can be rewritten in terms of the annihilation operators $Q_n$ of Eq. \ref{Qnsusy}
as
\begin{eqnarray}
 {\hat {\cal  H}}_{irr} = \vec f^{irr}(\vec x) .  \left(  \vec \nabla 
  +  ( \vec \nabla \phi(\vec x) ) \right)
  = \sum_{n=1}^N f^{irr}_n ( \vec x)  Q_n
  \label{Hirr}
\end{eqnarray}
The adjoint operator $ {\hat {\cal  H}}^{\dagger}_{irr}  $ reads using Eq. \ref{divforceirrev} 
\begin{eqnarray}
 {\hat {\cal  H}}^{\dagger}_{irr} 
 && = \sum_{n=1}^N Q_n^{\dagger} f^{irr}_n ( \vec x) 
=  \sum_{n=1}^N \left(   - \partial_n + (\partial_n \phi(\vec x) )  \right) f^{irr}_n ( \vec x) 
= \sum_{n=1}^N \left(   - f^{irr}_n ( \vec x) \partial_n - ( \partial_n f^{irr}_n ( \vec x)) + f^{irr}_n (\partial_n \phi(\vec x) )  \right)  
 \nonumber \\
&& =\sum_{n=1}^N \left(   - f^{irr}_n ( \vec x) \partial_n - f^{irr}_n (\partial_n \phi(\vec x) )  \right)
 = -   \sum_{n=1}^N f^{irr}_n( \vec x)  Q_n   
  \label{Hirrdagger}
\end{eqnarray}
so that the irreversible contribution $ {\hat {\cal  H}}_{irr}$ is antiHermitian
\begin{eqnarray}
 {\hat {\cal  H}}^{\dagger}_{irr}    = -  {\hat {\cal  H}}_{irr}
  \label{Hirrantih}
\end{eqnarray}
The change of propagators of Eq. \ref{changetowardshat}
corresponds to the following changes for the eigenvectors of Eqs \ref{Ekeigenkzero}
\begin{eqnarray}
 {\hat r}_0(\vec x) &&  = e^{   \phi( \vec x)} r_0(\vec x) = e^{   \phi( \vec x)} P^*(\vec x )  = P^*(\vec 0 )  e^{-  \phi(\vec x) } 
\nonumber \\
 {\hat l}_0(\vec x) &&  = e^{   - \phi( \vec x)} l_0(\vec x) =  e^{   - \phi( \vec x)}
\label{changetowardshateigen}
\end{eqnarray}
The factorized forms of Eqs \ref{hamiltonianhateqsusyQ}, \ref{Hirr}
and \ref{Hirrdagger}
show that the new right eigenvector $ {\hat r}_0(\vec x) $
and the new left eigenvector $ {\hat l}_0(\vec x) $ of the Hamiltonian $ {\hat {\cal  H}}$
associated to zero-energy
are annihilated by the $N$ commuting annihilation operators $Q_n$ of Eq. \ref{Qnsusy}
 \begin{eqnarray}
Q_n e^{   - \phi( \vec x)} && = \bigg( \partial_n + (\partial_n \phi(\vec x) ) \bigg) e^{   - \phi( \vec x)} =0
\label{Qnsusyannihil}
\end{eqnarray}
and are thus annihilated by both the reversible and irreversible contribution to the Hamiltonian
\begin{eqnarray}
0 && = {\hat {\cal  H}} {\hat r}_0(\vec x) =  {\hat {\cal  H}}_{rev}  e^{   - \phi( \vec x)}   =  {\hat {\cal  H}}_{irr}  e^{   - \phi( \vec x)}
\nonumber \\
0 && = {\hat {\cal  H}}^{\dagger} {\hat l}_0(\vec x) =  {\hat {\cal  H}}^{\dagger}_{rev}  e^{   - \phi( \vec x)}    =  {\hat {\cal  H}}^{\dagger}_{irr} e^{   - \phi( \vec x)}
  \label{annihilationboth}
\end{eqnarray}



 \section{ Statistics of time-additive observables of stochastic trajectories } 

\label{sec_additive}

In this section, we describe the how the interpretation 
of the Fokker-Planck generator of Eq. \ref{fokkerplanck}
as the non-Hermitian quantum Hamiltonian of Eq. \ref{FPhamiltonianQuantum}
 is even
more useful to analyze the statistics of the time-additive observables of stochastic trajectories.


 \subsection{ Stochastic differential equations and Fokker-Planck dynamics for the joint propagator} 

As recalled in the Introduction, 
a time-additive observable ${\cal O}[\vec x(0 \leq s \leq t) ]$  
can be written as Eq. \ref{additive} in terms of some scalar field $V^{[{\cal O}]}(\vec x)$ 
and some vector field $\vec A^{[{\cal O}]}(\vec x)$
in the Stratonovich interpretation.
Here it is important to stress that for the stochastic differential equations (SDE) of Eq. \ref{langevin}
where there is no space-dependent function in factor of the Wiener processes increments $dw_n(t)$, 
the Ito and the Stratonovich interpretations coincide. 
However for the observable of Eq. \ref{additive},
the elementary increment $dO(t) $ between $t$ and $(t+dt)$ 
reads in terms of the $N$ Langevin increments $ d  x_n(t)$ of Eq. \ref{langevin}
\begin{eqnarray}
dO(t) && =  - V^{[{\cal O}]} (\vec x(t) ) dt
+ \sum_{n=1}^N  A^{[{\cal O}]}_n( \vec x(t)) dx_n(t)
\nonumber \\
&& = \left[   - V^{[{\cal O}]} (\vec x(t) ) +  \sum_{n=1}^N  A^{[{\cal O}]}_n( \vec x(t)) 
f_n(\vec x (t) )\right] dt 
+ \sum_{n=1}^N  A^{[{\cal O}]}_n( \vec x(t)) dw_n(t)
\nonumber \\
&& \equiv f_{Strato}^{[{\cal O}]}(\vec x(t) ) dt 
+ \sum_{n=1}^N  A^{[{\cal O}]}_n( \vec x(t)) dw_n(t) \ \ \ \ \ \ \ \ \ \ \ \ \ \ [{\rm Stratonovich \ Interpretation}]
\label{additivediffelementary}
\end{eqnarray}
So the $N$ Wiener processes increments $dw_n(t)$ are multiplied by the space-dependent functions $ A^{[{\cal O}]}_n( \vec x(t))$ and one needs to specify that the force appearing in the
stochastic differential equation of Eq. \ref{additivediffelementary} 
corresponds to the Stratonovich interpretation
\begin{eqnarray}
 f_{Strato}^{[{\cal O}]}(\vec x ) \equiv   - V^{[{\cal O}]} (\vec x ) +  \sum_{n=1}^N  A^{[{\cal O}]}_n( \vec x) 
f_n(\vec x  )
\label{forcestrato}
\end{eqnarray}
The Stratonovich SDE of Eq. \ref{additivediffelementary} can be translated into the following Ito SDE
\begin{eqnarray}
dO(t) && = f_{Ito}^{[{\cal O}]}(\vec x(t) ) dt 
+ \sum_{n=1}^N  A^{[{\cal O}]}_n( \vec x(t)) dw_n(t) \ \ \ \ \ \ \ \ \ \ \ \ \ \ [{\rm Ito \ Interpretation}]
\label{ItoSDE}
\end{eqnarray}
with the modified force
\begin{eqnarray}
 f_{Ito}^{[{\cal O}]}(\vec x ) = f_{Strato}^{[{\cal O}]}(\vec x )  + \frac{1}{2} 
  \sum_{n=1}^N  \partial_n A^{[{\cal O}]}_n( \vec x) 
  =  f_{Strato}^{[{\cal O}]}(\vec x )  + \frac{1}{2}  \vec \nabla . \vec A^{[{\cal O}]}( \vec x) 
 \label{forceIto}
\end{eqnarray}
So the Ito and the Stratonovich interpretations will actually differ only for the observables ${\cal O}[\vec x(0 \leq \tau \leq t) ] $
that are characterized by a nonvanishing divergence $\vec \nabla . \vec A^{[{\cal O}]}( \vec x) \ne 0 $ of 
the vector field $\vec  A^{[{\cal O}]}$ appearing in their parametrization of Eq. \ref{additive}.

Since the observable $O(t)$ can be considered as a supplementary $(N+1)$ coordinate
 for the Stratonivich SDE system of Eq. \ref{langevin}, one can write the corresponding Fokker-Planck dynamics
 for the joint propagator $ P(\vec x,O,t \vert \vec y,0,0)   $
 \begin{eqnarray}
  \partial_t P(\vec x,O,t \vert \vec y,0,0) && = 
  V^{[{\cal O}]} (\vec x ) \partial_O P(\vec x,O,t \vert \vec y,0,0)  
  -  \sum_{n=1}^N \left( \partial_{x_n}  +  A^{[{\cal O}]}_n [ \vec x] \partial_O \right)   \bigg[  f_n[ \vec x ] 
    P(\vec x,O,t \vert \vec y,0,0) \bigg]
 \nonumber \\
 &&  + \frac{1}{2}  \sum_{n=1}^N \left( \partial_{x_n}  +  A^{[{\cal O}]}_n [ \vec x] \partial_O \right)^2
    P(\vec x,O,t \vert \vec y,0,0) 
\label{fokkerplanckO}
\end{eqnarray}
It is thus simpler to analyze the statistics of the observable ${\cal O}[\vec x(0 \leq \tau \leq t) ] $ via its generating function 
as mentioned in the Introduction and as described in the next subsection.


 \subsection{ Generating function $Z^{[k]}(\vec x,t \vert \vec y,0)$ via the quantum problem with deformed scalar and vector potentials}

Since Feynman path-integrals are written in the Stratonovich interpretation,
the generating function $Z^{[k]}(\vec x,t \vert \vec y,0)$ of parameter $k$
of the observable ${\cal O}[\vec x(0 \leq \tau \leq t) ] $ of Eq. \ref{additive}
over the stochastic trajectories $ \vec x(0 \leq s \leq t) $ starting at $\vec x(0)=\vec y$ and ending at $\vec x(t)=\vec x$
can be directly written via the path-integral of Eq. \ref{pathintegral} based on the Lagrangian of Eq. \ref{lagrangian}
as
\begin{eqnarray}
 Z^{[k]}(\vec x,t \vert \vec y,0) 
&& \equiv \overline{ \delta^{(N)} \left( \vec x(t) - \vec x \right)  e^{k{\cal O} [\vec x(0 \leq \tau \leq t) ]}  \delta^{(N)} \left( \vec x(0) - \vec y \right) }
 \nonumber \\
 &&
= \int_{\vec x(\tau=0)=\vec y}^{\vec x(\tau=t)=\vec x} {\cal D}   \vec x(\tau)  
 e^{ - \displaystyle 
  \int_0^{t} d\tau {\cal L} (\vec x(\tau), \dot {\vec x}(\tau) )
+ k 
\int_{0}^{t} d \tau \left[   - V^{[{\cal O}]} (\vec x(\tau) )
+  \dot {\vec x} (\tau) . \vec A^{[{\cal O}]}( \vec x(\tau)) \right]
 }
  \nonumber \\
 &&
 \equiv
 \int_{\vec x(\tau=0)=\vec y}^{\vec x(\tau=t)=\vec x} {\cal D}   \vec x(\tau)  
 e^{ - \displaystyle \int_{0}^{t} d \tau
{\cal L}^{[k]} (\vec x(\tau), \dot {\vec x}(\tau) )
 }
 \label{genedef}
\end{eqnarray}
This path-integral involves the $k$-deformed classical Lagrangian ${\cal L}^{[k]} (\vec x(\tau), \dot {\vec x}(\tau) )  $
with respect to the initial Lagrangian ${\cal L} (\vec x(\tau), \dot {\vec x}(\tau) ) $ of Eq. \ref{lagrangian}
\begin{eqnarray}
{\cal L}^{[k]} (\vec x(\tau), \dot {\vec x}(\tau) ) 
&& \equiv  {\cal L} (\vec x(\tau), \dot {\vec x}(\tau) ) + k V^{[{\cal O}]}( \vec x(\tau) - k  \dot {\vec x} (\tau) . \vec A^{[{\cal O}]}( \vec x(\tau))
\nonumber \\
&&
\nonumber \\
&& = \frac{1}{2}  \dot {\vec x}^2  (\tau) 
 - \dot {\vec x } (\tau) . \vec A^{[k]}( \vec x(\tau))
 + V^{[k]}(\vec x(\tau))
\label{lagrangiank}
\end{eqnarray}
with the
$k$-deformed vector potential  $\vec A^{[k]} (\vec x ) $ with respect to the initial vector potential  $ A_n( \vec x) $ of Eq. \ref{vectorpot}
\begin{eqnarray}
 \vec A^{[k]} (\vec x ) \equiv \vec A( \vec x) +k \vec A^{[{\cal O}]}( \vec x)
 = \vec f( \vec x)+k \vec A^{[{\cal O}]}( \vec x)
 \label{vectorpotp}
\end{eqnarray}
and with the $k$-deformed scalar potential $V^{[k]}(\vec x) $
with respect to the initial scalar potential $V ( \vec x)  $ of Eq. \ref{scalarpot}
\begin{eqnarray}
V^{[k]}(\vec x) \equiv V ( \vec x) + k V^{[{\cal O}]} ( \vec x)
=  \frac{  [\vec \nabla . \vec f(\vec x) ] +\vec f ^2( \vec x) }{2 } + k V^{[{\cal O}]} ( \vec x)
\label{scalarpotp}
\end{eqnarray}
The magnetic antisymmetric matrix of Eq. \ref{magneticN}
associated to the $k$-deformed vector potential 
$ \vec A^{[k]}( \vec x)$ of Eq. \ref{vectorpotp}
\begin{eqnarray}
 B^{[k]}_{nm}(\vec x )  =-B^{[k]}_{mn} (\vec x )  \equiv \partial_n A^{[k]}_m (\vec x ) -  \partial_m A^{[k]}_n (\vec x ) 
  =B_{nm}(\vec x ) + k B^{[{\cal O}]}_{nm}(\vec x )
 \label{magneticNp}
\end{eqnarray}
contains the supplementary magnetic antisymmetric matrix $ B^{[{\cal O}]}_{nm}(\vec x ) $ associated to the supplementary vector potential $A^{[{\cal O}]}_n (\vec x ) $
\begin{eqnarray}
 B^{[{\cal O}]}_{nm}(\vec x )  =- B^{[{\cal O}]}_{mn} (\vec x ) 
  \equiv \partial_n A^{[{\cal O}]}_m (\vec x ) -  \partial_m A^{[{\cal O}]}_n (\vec x ) 
 \label{magneticNpsmall}
\end{eqnarray}

So the generating function $Z^{[k]}(\vec x,t \vert \vec y,0) $ satisfies the Euclidean Schr\"odinger equation analogous to Eq. \ref{fokkerplanck}
\begin{eqnarray}
 \partial_t Z^{[k]}(\vec x,t \vert \vec y,0)  
 =  -  {\cal H}^{[k]} Z^{[k]}(\vec x,t \vert \vec y,0)
\label{EuclideanZ}
\end{eqnarray}
where the $k$-deformed quantum Hamiltonian ${\cal  H}^{[k]} $ involves the deformed scalar and vector potentials 
\begin{eqnarray}
 {\cal H}^{[k]}   && =     -  \frac{1}{2}  \left( \vec \nabla -   \vec A^{[k]}(\vec x) \right)^2    +  V^{[k]}(\vec x)
 \label{FPhamiltonianp}
\end{eqnarray}
Using Eqs \ref{vectorpotp} and \ref{scalarpotp}, this Hamiltonian can be expanded
to see more clearly the $k-$deformation with respect to the initial Hamiltonian $ {\cal H}= {\cal H}^{[k=0]} $
of Eqs \ref{FPhamiltonian} involving the force $\vec f(\vec x)$
\begin{eqnarray}
 {\cal H}^{[k]}   &&  =   -  \frac{1}{2}  \left( \vec \nabla  -k \vec A^{[{\cal O}]}( \vec x) -   \vec f( \vec x)\right)^2    +   \left[\frac{  [\vec \nabla . \vec f(\vec x) ] +\vec f ^2( \vec x) }{2 } + k V^{[{\cal O}]} ( \vec x) \right]
 \nonumber \\
 && =  -  \frac{1}{2}  \left( \vec \nabla  -k \vec A^{[{\cal O}]}( \vec x) \right)^2  
 +    \left( \vec \nabla  -k \vec A^{[{\cal O}]}( \vec x)    \right) \vec f( \vec x)
 - \frac{[\vec \nabla . \vec f(\vec x) ]}{2}  
 -  \frac{[ \vec f( \vec x)]^2 }{2}  
   +   \left[\frac{  [\vec \nabla . \vec f(\vec x) ] +\vec f ^2( \vec x) }{2 } + k V^{[{\cal O}]} ( \vec x) \right]
    \nonumber \\
 && =  -  \frac{1}{2}  \left( \vec \nabla  -k \vec A^{[{\cal O}]}( \vec x) \right)^2  
 +    \left( \vec \nabla  -k \vec A^{[{\cal O}]}( \vec x)    \right) . \vec f( \vec x)
  + k V^{[{\cal O}]} ( \vec x) 
 \label{FPhamiltonianpexpanded}
\end{eqnarray}

Since the 
generating function $Z^{[k]}(\vec x,t \vert \vec y,0)$ 
can be alternatively computed from the joint propagator $P(\vec x,O,t \vert \vec y,0,0) $ introduced in Eq. \ref{fokkerplanckO}
\begin{eqnarray}
 Z^{[k]}(\vec x,t \vert \vec y,0) 
&& \equiv \int_{-\infty}^{+\infty} dO   e^{k O}  P(\vec x,O,t \vert \vec y,0,0)
 \label{genejointpropa}
\end{eqnarray}
 the Euclidean Schr\"odinger equation of Eq. \ref{EuclideanZ} with the Hamiltonian of Eq. \ref{FPhamiltonianpexpanded}
can be also derived from the Fokker-Planck Eq. \ref{fokkerplanckO} satisfied by the 
joint propagator $P(\vec x,O,t \vert \vec y,0,0) $ using integration by parts with respect to $O$
 \begin{eqnarray}
 \partial_t  Z^{[k]}(\vec x,t \vert \vec y,0)  &&  = \int_{-\infty}^{+\infty} dO   e^{k O}   \partial_t  P(\vec x,O,t \vert \vec y,0,0)
   \nonumber \\
  && = \int_{-\infty}^{+\infty} dO   e^{k O} 
  \left[
   V^{[{\cal O}]} (\vec x ) \partial_O P(\vec x,O,t \vert \vec y,0,0)  
  -  \sum_{n=1}^N \left( \partial_{x_n}  +  A^{[{\cal O}]}_n [ \vec x] \partial_O \right)   \bigg[  f_n[ \vec x ] 
    P(\vec x,O,t \vert \vec y,0,0) \bigg] \right. \nonumber \\
  && ~~ \left. + \frac{1}{2}  \sum_{n=1}^N \left( \partial_{x_n}  +  A^{[{\cal O}]}_n [ \vec x] \partial_O \right)^2
    P(\vec x,O,t \vert \vec y,0,0) 
  \right]
  \nonumber \\
  && = 
  - k V^{[{\cal O}]} (\vec x )  Z^{[k]}(\vec x,t \vert \vec y,0) 
  -  \sum_{n=1}^N \left( \partial_{x_n}  - k  A^{[{\cal O}]}_n ( \vec x)  \right)   \bigg[  f_n[ \vec x ] 
     Z^{[k]}(\vec x,t \vert \vec y,0)  \bigg]
  \nonumber \\
  && ~~
  + \frac{1}{2}  \sum_{n=1}^N \left( \partial_{x_n}  - k  A^{[{\cal O}]}_n ( \vec x)  \right)^2
     Z^{[k]}(\vec x,t \vert \vec y,0) 
\label{fokkerplanckOforZ}
\end{eqnarray}

Let us stress that in the present paper, we only consider the case of dimension $N \geq 2$
to have at least one off-diagonal element in the antisymmetric magnetic matrix of Eq.  \ref{magneticN}.
But the analogous description for additive observables of diffusion processes on a one-dimensional ring, 
where the only magnetic parameter reduces to the total magnetic flux through the ring 
and corresponds to the famous Aharonov-Bohm effect is discussed in detail in \cite{c_lyapunov}
with the corresponding gauge transformations of the vector potential along the ring.

As recalled in the subsection \ref{subsec_largedevadditive} of the Introduction,
the asymptotic behavior of Eq. \ref{genelargeT}
of the generating function $Z^{[k]}(\vec x,t \vert \vec y,0)$ allows to compute the rate function $I(o)$ from
 the ground-state energy $E(k)$ of the $k$-deformed Hamiltonian ${\cal  H}^{[k]} $.
From the point of view of gauge-transformations, it is important to stress here that 
the corresponding similarity of Hamiltonians 
changes only the eigenvectors,
while the energy spectrum is invariant, so that the ground-state energy $E(k)$ of ${\cal  H}^{[k]} $
can be also obtained as the ground-state energy of any gauge-transformed Hamiltonian of ${\cal  H}^{[k]} $.
To be more concrete, let us now describe an important example of time-additive observable.


\subsection{ Example of time-additive observable : the entropy production $\Sigma [\vec x(0 \leq \tau \leq t)]  $ of trajectories $\vec x(0 \leq \tau \leq t) $ }

\label{subsec_entropyprod}

The entropy production $\Sigma [\vec x(0 \leq \tau \leq t)]  $
characterizes the irreversibility of the Langevin dynamics at the level of the stochastic trajectories
and is thus nowadays one of the most studied time-additive observables for diffusion processes  \cite{jarzynski,brazil,entropyProd_OUdimN,gawedzki,seifert,engel,saito,ito,dechant_wasserstein,renaud,dechant,sasa,dechant_ito,ito2022}.

\subsubsection{ Reminder on definition of the entropy production $\Sigma [\vec x(0 \leq \tau \leq t)]  $ 
associated to the stochastic trajectory $ \vec x(0 \leq \tau \leq t)  $ }

In the steady state,
the probability ${\cal P}[\vec x(0 \leq \tau \leq t)] $ of the stochastic trajectory $ \vec x(0 \leq \tau \leq t)$  involves the inside of the path-integral of Eq. \ref{pathintegral},
while the initial condition $\vec x(0) $ is drawn with the steady state distribution $P^*(\vec x(0)) $
\begin{eqnarray}
  {\cal P}[\vec x(0 \leq \tau \leq t)]
&& =  P^*(\vec x(0))  
 e^{ - \displaystyle 
  \int_0^{t} d\tau  \left[ 
  \frac{1}{2} \left( \dot {\vec x } (\tau) - \vec f( \vec x(\tau)) \right)^2
 +\frac{  1 }{2 }  [\vec \nabla . \vec f(\vec x(\tau)) ] 
\right] }
\label{ptraj}
\end{eqnarray}
The probability of the corresponding time-reversed trajectory 
\begin{eqnarray}
x_n^R(s)  \equiv x_n(\tau=t-s)
\label{Rtraj}
\end{eqnarray}
reads 
\begin{eqnarray}
  {\cal P}[x^R(0 \leq s \leq t)] && = 
  P^*(\vec x^R(0))
  e^{ - \displaystyle 
  \int_0^{t} ds \left[ \frac{1}{2} \left( \dot {\vec x}^R  (s) - \vec f( \vec x^R(s)) \right)^2
 +\frac{  1 }{2 }  [\vec \nabla . \vec f(\vec x^R(s)) ] \right] }
  \nonumber \\
&&
= P^*(\vec x(t))
 e^{ - \displaystyle 
  \int_0^{t} d\tau  \left[ 
    \frac{1}{2} \left( \dot {\vec x } (\tau) + \vec f( \vec x(\tau)) \right)^2
 +\frac{  1 }{2 }  [\vec \nabla . \vec f(\vec x(\tau)) ] 
\right] }
\label{pathFPR}
\end{eqnarray}
The entropy production $\Sigma [\vec x(0 \leq \tau \leq t)]  $ associated to the trajectory $\vec x(0 \leq \tau \leq t) $ is defined as the logarithm of the ratio between the probability of Eq \ref{ptraj} for this trajectory $\vec x(0 \leq \tau \leq t) $
and the probability of Eq. \ref{pathFPR} for the corresponding reversed trajectory $x^R(0 \leq s \leq t) $
\begin{eqnarray}
\Sigma [\vec x(0 \leq \tau \leq t)] && \equiv  
 \ln \left( \frac{ {\cal P}[\vec x(0 \leq \tau \leq t)] }{ {\cal P}[\vec x^R(0 \leq s \leq t)] } \right)
 =   \ln \left( \frac{  P^*(\vec x(0)) } {  P^*(\vec x(t))  } \right)
 +  \int_0^{t} d\tau  \  2
\dot {\vec x}(\tau) . \vec f(\vec x(\tau))      
  \nonumber \\
  && = -   \int_0^{t} d\tau  \partial_{\tau}  \ln \left( P^*(\vec x(\tau)) \right)
+  \int_0^{t} d\tau  \  2
\dot {\vec x}(\tau) . \vec f(\vec x(\tau))     
  \nonumber \\
  && =   \int_0^{t} d\tau  \  \dot {\vec x}(\tau) . \left[ 
  - \vec \nabla \ln \left( P^*(\vec x(\tau)) \right)
+  2 \vec f (\vec x(\tau))      
  \right]
\label{proddifftrajdef}
\end{eqnarray}
Using the decomposition of the force $\vec f (\vec x)$ into its reversible contribution $ \vec f^{rev} (\vec x) $ of Eq. \ref{forceRev}
and its irreversible contribution $ \vec f^{irr} (\vec x) $ of Eq. \ref{forceIrrev}, one obtains 
that the trajectory entropy production
\begin{eqnarray}
\Sigma [\vec x(0 \leq \tau \leq t)] && =   \int_0^{t} d\tau  \  \dot {\vec x}(\tau) . \left[ 
  -  2 \vec f^{rev} (\vec x(\tau))  
+  2 \vec f (\vec x(\tau))      
  \right]
\nonumber \\
&& = \int_0^{t} d\tau  \  \dot {\vec x}(\tau) . \left[    2 \vec f^{irr} (\vec x(\tau))  
  \right]
  \equiv  \int_0^{t} d\tau  \ \dot {\vec x}(\tau) . \vec A^{[{ \Sigma}]}(\vec x(\tau))
\label{proddifftraj}
\end{eqnarray}
is a time-additive observable of the form of Eq. \ref{additive} with parameters
\begin{eqnarray}
V^{[{ \Sigma}]} (\vec x ) && =0
\nonumber \\
 \vec A^{[{ \Sigma}]}( \vec x) && =   2 \vec f^{irr}(\vec x)   
\label{additivesigma}
\end{eqnarray}
i.e. there is no additional scalar potential $V^{[{ \Sigma}]} (\vec x )=0 $,
while the additional vector potential $ \vec A^{[{ \Sigma}]}( \vec x)  $ 
only involves the irreversible force $\vec f^{irr}(\vec x) $.


\subsubsection{ Generating function $ Z^{[k]}(\vec x,t \vert \vec y,0) $ of the entropy production $\Sigma [\vec x(0 \leq \tau \leq t)]  $  }

 As a consequence, the generating function of Eq. \ref{genedef}
 for the entropy production $\Sigma [\vec x(0 \leq \tau \leq t)]  $
 \begin{eqnarray}
 Z^{[k]}(\vec x,t \vert \vec y,0) \equiv \langle \delta^{(N)} \left( \vec x(t) - \vec x \right) e^{k \Sigma [\vec x(0 \leq s \leq t) ]}   \delta^{(N)} \left( \vec x(0) - \vec y \right)\rangle
 \label{geneentropy}
\end{eqnarray}
 is given by the path-integral
 of Eq. \ref{genedef} that involves the unchanged scalar potential of Eq. \ref{scalarpotp}
\begin{eqnarray}
V^{[k]}(\vec x) = V ( \vec x) 
\label{scalarpotpentropy}
\end{eqnarray}
while the $k$-deformed vector potential of Eq. \ref{vectorpotp} reads in terms of the reversible and irreversible force
\begin{eqnarray}
\vec  A^{[k]} (\vec x ) && = \vec A( \vec x) +k \vec A^{[{ \Sigma}]}( \vec x)
  =\vec f^{rev}( \vec x) + (1+2 k)  \vec f^{irr}(\vec x)   
  \nonumber \\
&& =   - \vec \nabla \phi(\vec x)
  + \left( 1+2 k \right)  \vec f^{irr}(\vec x)   
  \equiv  - \vec \nabla \phi(\vec x) + {\vec {\hat A}^{[k]}} (\vec x )
 \label{vectorpotpentropy}
\end{eqnarray}
This suggests to make the gauge transformation analog to Eq. \ref{vectorpotgaugeirr}
with the new vector potential 
\begin{eqnarray}
   {\vec {\hat A}^{[k]}} (\vec x ) \equiv (1+2 k)  \vec f^{irr}(\vec x) 
    \label{vectorpotirrentropy}
\end{eqnarray}
that only involves the irreversible force $\vec f^{irr}( \vec x) $.
The corresponding change for the generating function of Eq. \ref{geneentropy}
 analogous to Eq. \ref{changetowardshat}
\begin{eqnarray}
 Z^{[k]}(\vec x,t \vert \vec y,0)
 = e^{ \displaystyle \phi( \vec y) - \phi( \vec x)} {\hat Z}^{[k]}(\vec x,t \vert \vec y,0)
\label{changetowardstildeZ}
\end{eqnarray}
yields that the new function $ {\hat Z}^{[k]}(\vec x,t \vert \vec y,0) $
corresponds to the new vector potential of Eq. \ref{vectorpotirrentropy},
with its path-integral representation analogous to Eq. \ref{pathintegralhat}
 \begin{eqnarray}
{\hat Z}^{[k]}(\vec x,t \vert \vec y,0)
= \int_{\vec x(\tau=0)=\vec y}^{\vec x(\tau=T)=\vec x} {\cal D}   \vec x(\tau)  
 e^{ - \displaystyle 
  \int_0^{t} d\tau \left[ 
   \frac{1}{2}  \dot {\vec x}^2  (\tau) 
  + V(\vec x)
\right]
+  \int_0^{t} d\tau  \dot {\vec x} (\tau).  
\vec {\hat A}^{[k]} (\vec x )
 }
\label{pathintegraltildeZ}
\end{eqnarray}
The corresponding Hamiltonian is given by
\begin{eqnarray}
 {\hat {\cal  H}}^{[k]}     && 
 =  -  \frac{1}{2}  \left( \vec \nabla -   \vec {\hat A}^{[k]} (\vec x ) \right)^2    +  V(\vec x)
 \nonumber \\
 &&
= -  \frac{1}{2} \Delta   
  +   \vec {\hat A}^{[k]} (\vec x ) . \vec \nabla 
  +  \frac{1}{2}   [  \vec \nabla . \vec {\hat A}^{[k]} (\vec x )    ] 
   -   \frac{1}{2} [\vec {\hat A}^{[k]} (\vec x )  ]^2 
   +  V(\vec x)    
  \label{hamiltoniantildeZ}
\end{eqnarray}
Using the vector potential of Eq. \ref{vectorpotirrentropy}
and the scalar potential $V(\vec x) $ of Eq. \ref{scalarpotexpanded},
 this Hamiltonian finally reads
\begin{eqnarray}
 {\hat {\cal  H}}^{[k]}    && =  
  -  \frac{1}{2} \Delta   
  + (1+2 k)  \vec f^{irr} (\vec x ) . \vec \nabla 
  +  \frac{(1+2 k)}{2}   [  \vec \nabla . \vec f^{irr} (\vec x )    ] 
   -   \frac{(1+2 k)^2}{2} [\vec f^{irr} (\vec x )  ]^2 
 \nonumber \\
 && 
+\frac{  1 }{2 } \ \left( [ \vec \nabla  \phi (\vec x) ]^2- \Delta  \phi (\vec x)  \right)
+  \frac{  1 }{2 } \left[ {\vec f \ }^{irr}( \vec x)\right]^2
 \nonumber \\
 && =  {\hat {\cal  H}}_{rev} +  (1+2k) {\hat {\cal  H}}_{irr} 
 -  2 \left( k+k^2 \right) [\vec f^{irr} (\vec x )  ]^2 
    \label{hamiltoniantildeexpandedZ}
\end{eqnarray}
where one recognizes the Hermitian reversible contribution ${\hat {\cal  H}}_{rev}  $ of Eq. \ref{hamiltonianhateqsusy}
and the antiHermitian irreversible contribution ${\hat {\cal  H}}_{irr}  $ of Eq. \ref{Hirr} with the modified prefactor $(1+2k)$,
while the third term corresponds to a new contribution involving 
the square $[\vec f^{irr} (\vec x )  ]^2 $ of the irreversible force.


\subsubsection{ Large deviations of the intensive entropy production $\sigma[\vec x(0 \leq s \leq t) ] $ for large time $t$  }

The general framework for large deviations described in subsection \ref{subsec_largedevadditive} of the Introduction
can be applied to the special case of the intensive entropy production
\begin{eqnarray}
\sigma[\vec x(0 \leq s \leq t) ] \equiv \frac{ \Sigma[\vec x(0 \leq s \leq t) ] }{t} =  
\frac{1}{t}   \int_{0}^{t} d \tau    \   \dot {\vec x}(\tau) 
. \left[  2 \vec f^{irr}(\vec x(\tau))        \right]
\label{additiveIntensiveEntropy}
\end{eqnarray}
as follows.
In the large time limit $t \to +\infty$, the convergence will be towards the steady value $\sigma^*$ 
that can be computed  from the corresponding steady current $\vec J^*(\vec x) $
\begin{eqnarray}
\sigma[\vec x(0 \leq s \leq t) ] \opsimeq_{t \to \infty} \sigma^* 
\equiv \int d^N \vec x  \  \vec J^*(\vec x)  
. \left[  2 \vec f^{irr}(\vec x)        \right]
\label{g1Entropy}
\end{eqnarray}
One can use the relation $\vec J^*(\vec x)  =  P^*(\vec x ) \vec f^{irr}( \vec x)$ of Eq. \ref{jsteadyirrev}
to rewrite the steady value $\sigma^*$ in terms of the steady current $\vec J^*(\vec x)$ 
to recover the standard formula involving the square of the steady current 
\begin{eqnarray}
 \sigma^* = 2 \int d^N \vec x  \frac{ [ \vec J^*(\vec x) ]^2}{P^*(\vec x )} 
 \label{g1EntropyJ}
\end{eqnarray}
which shows that $\sigma^*$ is strictly positive for any nonvanishing steady current.
One can instead rewrite the steady value $\sigma^*$
 in terms of the irreversible force $\vec f^{irr}( \vec x)$ to obtain the alternative form
\begin{eqnarray}
 \sigma^* = 2 \int d^N \vec x  \ P^*(\vec x )  \left[   \vec f^{irr}(\vec x)        \right]^2
\label{g1EntropyFirr}
\end{eqnarray}
that allows to see the link of Eq. \ref{linktypperturbationEp}
with the first-order perturbation theory for the energy $E(k)$ of the Hamiltonian ${\cal  H}^{[k]}$ of Eq. \ref{hamiltoniantildeexpandedZ}
using the unperturbed eigenvectors of Eqs \ref{Ekeigenkzero}
\begin{eqnarray}
  \sigma^* = - E'(k=0)
  = \langle l_0 \vert \left( - 2 {\hat {\cal  H}}_{irr} +
  2  \left[   \vec f^{irr}(\vec x)        \right]^2
  \right) \vert r_0 \rangle
 = 2 \int d^N \vec x  \ P^*(\vec x )  \left[   \vec f^{irr}(\vec x)        \right]^2
\label{linktypperturbationEpEntropy}
\end{eqnarray}

It is now interesting to compare the Hamiltonian $ {\hat {\cal  H}}^{[k]}   $ of Eq. \ref{hamiltoniantildeexpandedZ}
with the Hamiltonian $ {\hat {\cal  H}}^{[k'=-1-k]}   $ associated to the parameter $k'=-1-k$ satisfying
\begin{eqnarray}
 k' && =   - 1-k
 \nonumber \\
  1+2k' && =   - (1+2k) 
 \nonumber \\
k' (1+k') && =   k (1+k)
   \label{prime}
\end{eqnarray}
As a consequence, the term involving $ {\hat {\cal  H}}_{irr}$
displays an opposite sign, while all the other terms are unchanged
\begin{eqnarray}
 {\hat {\cal  H}}^{[k'=-1-k]}  
&& =  {\hat {\cal  H}}_{rev} +  (1+2k') {\hat {\cal  H}}_{irr} 
 -  2 k' (1+k') \left[   \vec f^{irr}(\vec x)        \right]^2  
\nonumber \\
&&
   =  {\hat {\cal  H}}_{rev} -  (1+2k) {\hat {\cal  H}}_{irr} 
 - 2  k (1+k) 
  \left[   \vec f^{irr}(\vec x)        \right]^2  
   \label{hamiltoniantildeexpandedZGC}
\end{eqnarray}
The physical interpretation is that the irreversible force 
is changed into its opposite $f^{irr}_n( \vec x) \to - f^{irr}_n( \vec x)$,
while the steady state $P^*(\vec x)$ is unchanged,
so that the steady current of Eq. \ref{jsteadyirrev} is changed into its opposite
$\vec J^*( \vec x) \to - \vec J^*( \vec x)$.
The corresponding invariance of the ground-state energy $E(k)$ of the Hamiltonian $ {\hat {\cal  H}}^{[k]}    $
\begin{eqnarray}
E(k)   = E(k'=-1-k)
\label{GCp}
\end{eqnarray}
translates via the Legendre transform of Eq. \ref{legendre} into 
the famous Gallavotti-Cohen symmetry \cite{galla,kurchan_langevin,Leb_spo,maes1999,jepps,kurchan,zia,maes2009,maes2017}
 for the rate function $I(\sigma)$ of opposite arguments $(\pm \sigma)$
\begin{eqnarray}
 I(\sigma) = I (-\sigma) - \sigma 
    \label{GCrate}
\end{eqnarray}


\subsection{ Revisiting the canonical conditioning based on $Z^{[k]}(\vec x,t \vert \vec y,0)  $ via some appropriate gauge transformation   } 

\label{subsec_canonical}


\subsubsection{ Reminder on the canonical conditioned process of parameter $k$ for finite time $t$    }

The notion of canonical conditioning based on the generating function $Z^{[k]}(\vec x,t \vert \vec y,0)  $has been recalled in subsection \ref{subsec_remindercanonical} of the Introduction.
As described in \cite{c_microcanonical}, the conditional probability $ {\cal P}^{Cond[k]}(\vec z,\tau) $ 
of Eq. \ref{markovcondk}
satisfies the forward Fokker-Planck dynamics
 that reads in the present setting
\begin{eqnarray}
 \partial_{\tau}   {\cal P}^{Cond[k]}(\vec z,\tau)
 =
    -  \sum_{n=1}^N  \partial_{z_n} 
\bigg[  \big( f_n( \vec z )  + f_n^{Cond[k]}(\vec z,\tau) \big) {\cal P}^{Cond[k]}(\vec z,\tau)  \bigg]
   + \frac{1}{2}  \sum_{n=1}^N   \partial^2_{z_n} 
   {\cal P}^{Cond[k]}(\vec z,\tau)  
\label{markovconddiffderiforwardk}
\end{eqnarray}
where the only change with respect to the unconditioned Fokker-Planck dynamics of Eq. \ref{fokkerplanck}
satisfied by the unconditioned propagator $ P(\vec z,\tau \vert \vec y, 0)$
is the additional force $f_n^{Cond[k]}(\vec z,\tau) $
that
involves the vector potential $A^{[{\cal O}]}_n(\vec x) $ appearing in the definition of the additive observable of Eq. \ref{additive}
and the spatial derivatives of the logarithm of the generating function $Z^{[k]}(\vec x, t \vert \vec z, \tau)  $ 
\begin{eqnarray}
  f_n^{Cond[k]}(\vec z,\tau)  \equiv  
     k  A^{[{\cal O}]}_n(\vec z)          + \partial_{z_n}    \ln \left[ Z^{[k]}(\vec x, t \vert \vec z, \tau) \right]
\label{forcesupforwardk}
\end{eqnarray}


\subsubsection{ Reminder on the canonical conditioned process of parameter $k$ for large time $t \to +\infty$    }

For large time $t\to +\infty$, the conditional probability ${\cal P}^{Cond[k]}(\vec z,\tau)  $ of Eq. \ref{markovcondk} at any interior time $\tau$ satisfying $0 \ll \tau \ll t$
is given by Eq. \ref{markovcondktinfty}
that does not depend on the interior time $\tau$ and that reduces to the product of the left eigenvector $l_k(\vec z) $ and the right eigenvector $r_k(\vec z) $ of Eqs \ref{eigenright} and \ref{eigennorma}.
The corresponding additional force $f_n^{Cond[k]}(\vec z,\tau) $ of Eq. \ref{forcesupforwardk}
can be evaluated from the asymptotic property of Eq. \ref{genelargeT}
for the generating function $Z^{[k]}(\vec x, t \vert \vec z, \tau)  $ 
\begin{eqnarray}
  f_n^{Cond[k]}(\vec z,\tau)   &&  \opsimeq_{ 0 \ll \tau \ll t}  
     k  A^{[{\cal O}]}_n(\vec z)          + \partial_{z_n}    \ln \left[ e^{-(t-\tau) E(k) } r_k(\vec x) l_k(\vec z)  \right]
  \nonumber \\
  &&    =  k  A^{[{\cal O}]}_n(\vec z)          + \partial_{z_n}    \ln \left[  l_k(\vec z)  \right]
     \equiv  f_n^{Cond[k]Interior}(\vec z) 
\label{forcesupforwardktinfty}
\end{eqnarray}
to obtain that it does not depend on the interior time $\tau$ and that 
the second term only involves the spatial derivative of the logarithm of the left eigenvector $l_k(\vec z) $.
The conditioned Fokker-Planck dynamics of Eq. \ref{markovconddiffderiforwardk} 
is now governed by the corresponding time-independent generator
\begin{eqnarray}
 \partial_{\tau}   {\cal P}^{Cond[k]}(\vec z,\tau)
 =
    -  \sum_{n=1}^N  \partial_{z_n} 
\bigg[  \big( f_n( \vec z )  + f_n^{Cond[k]}(\vec z) \big) {\cal P}^{Cond[k]}(\vec z,\tau)  \bigg]
   + \frac{1}{2}  \sum_{n=1}^N   \partial^2_{z_n} 
   {\cal P}^{Cond[k]}(\vec z,\tau)  
\label{markovconddiffderiforwardktinfty}
\end{eqnarray}


\subsubsection{ Reinterpretation via a gauge transformation for the generating function $Z^{[k]}(\vec x,t \vert \vec y,0) $   } 

\label{subsec_canogauge}

The analysis recalled in the previous subsection suggests to make the following change of variables for the generating function $Z^{[k]}(\vec x,t \vert \vec y,0)  $
that involves the left eigenvector $l_k(.) $
\begin{eqnarray}
Z^{[k]}(\vec x,t \vert \vec y,0) =  \frac{l_k(\vec y)}{ l_k(\vec x) } {\mathring Z}^{[k]}(\vec x,t \vert \vec y, 0) 
\equiv  e^{ {\mathring \nu}^{[k]}(\vec y) - {\mathring \nu}^{[k]}(\vec x) } {\mathring Z}^{[k]}(\vec x,t \vert \vec y, 0) 
\label{changeZk}
\end{eqnarray}
which we will now re-interpret as a gauge transformation involving the function
\begin{eqnarray}
 {\mathring \nu}^{[k]}(\vec x)\equiv   \ln \left[  l_k(\vec x)  \right]
\label{changeZknu}
\end{eqnarray}
The initial vector potential $\vec A^{[k]} (\vec x ) \equiv \vec f( \vec x)+k \vec A^{[{\cal O}]}( \vec x) $ of Eq. \ref{vectorpotp}
is then transformed using the gradient of the function ${\mathring \nu}^{[k]}(\vec x) $ of Eq. \ref{changeZknu}
\begin{eqnarray}
 A_n^{[k]} (\vec x ) \equiv   f_n( \vec x)+k  A_n^{[{\cal O}]}( \vec x) = - \partial_n  {\mathring \nu}^{[k]}(\vec x) + 
 {\mathring A}_n^{[k]} (\vec x )
 \label{vectorpotptoring}
\end{eqnarray}
into the new vector potential 
\begin{eqnarray}
{\mathring A}_n^{[k]} (\vec x ) =  A_n^{[k]} (\vec x ) + \partial_n  {\mathring \nu}^{[k]}(\vec x)
=   f_n( \vec x)+k  A_n^{[{\cal O}]}( \vec x) + \partial_n  {\mathring \nu}^{[k]}(\vec x)   
=  f_n( \vec x) +  f_n^{Cond[k]Interior}(\vec z) 
 \label{vectorpotpring}
\end{eqnarray}
where one recognizes the additional force $f_n^{Cond[k]Interior}(\vec z) $ of Eq. \ref{forcesupforwardktinfty}.
As a consequence, the new vector potential ${\mathring A}_n^{[k]} (\vec x ) $ coincides with the total force
that appears in the conditioned Fokker-Planck dynamics of Eq. \ref{markovconddiffderiforwardktinfty}
\begin{eqnarray}
 \partial_{\tau}   {\cal P}^{Cond[k]}(\vec z,\tau)
 =   -  \sum_{n=1}^N  \partial_{z_n} 
\bigg[  {\mathring A}_n^{[k]} (\vec z )  {\cal P}^{Cond[k]}(\vec z,\tau)  \bigg]
   + \frac{1}{2}  \sum_{n=1}^N   \partial^2_{z_n} 
   {\cal P}^{Cond[k]}(\vec z,\tau)  \equiv - {\cal  H}^{Cond[k]} {\cal P}^{Cond[k]}(\vec z,\tau)
\label{markovconddiffderiforwardktinftygauge}
\end{eqnarray}
that involves the quantum Hamiltonian
\begin{eqnarray}
{\cal  H}^{Cond[k]} = \vec \nabla . \left(  - \frac{1}{2}  \vec \nabla + {\vec  {\mathring A}^{[k]}}(\vec x)\right)
=     - \frac{1}{2}  \vec \nabla^2
 +  {\vec  {\mathring A}^{[k]}}(\vec x). \vec \nabla
 +    [\vec \nabla . {\vec  {\mathring A}^{[k]}}(\vec x) ]  
\label{Hcondk}
\end{eqnarray}

However, the gauge transformation of Eq. \ref{changeZk}
means that the new function $ {\mathring Z}^{[k]}(\vec x,t \vert \vec y, 0) $
\begin{eqnarray}
{\mathring Z}^{[k]}(\vec x,t \vert \vec y, 0)   = \langle \vec x \vert e^{- t {\mathring{\cal H}}^{[k]}} \vert \vec y \rangle
\label{changeZkring}
\end{eqnarray}
corresponds to the quantum Hamiltonian $ {\mathring{\cal H}}^{[k]}$ with the new vector potential ${\mathring A}_n^{[k]} (\vec z ) $
but with the same scalar potential $V^{[k]}(\vec x) $ of Eq. \ref{scalarpotp} as the initial generating function $ Z^{[k]}(\vec x,t \vert \vec y,0)  $.
Hence the Hamiltonian of Eq. \ref{FPhamiltonianp}
is changed into the new Hamiltonian
\begin{eqnarray}
{\mathring{\cal H}}^{[k]}   && =     -  \frac{1}{2}  \left( \vec \nabla -   \vec {\mathring A}^{[k]}( \vec x) \right)^2    +  V^{[k]}(\vec x)
\nonumber \\
&& =  - \frac{1}{2}  \vec \nabla^2 +\vec {\mathring A}^{[k]} ({\vec x}). \vec \nabla
+ \frac{1}{2}   \left( \vec \nabla.\vec {\mathring A}^{[k]} ({\vec x})\right)  - \frac{1}{2} [ \vec {\mathring A}^{[k]} (\vec x) ]^2
  +  V^{[k]}(\vec x)
 \nonumber \\
 && = {\cal  H}^{Cond[k]}
 +  V^{[k]}(\vec x) - \frac{   [\vec \nabla . \vec {\mathring A}^{[k]}(\vec x) ] + [\vec {\mathring A}^{[k]}(\vec x) ]^2 }{2 } 
 \label{FPhamiltonianpring}
\end{eqnarray}
where the last line allows to compare with the Hamiltonian $ {\cal  H}^{Cond[k]} $ of Eq. \ref{Hcondk}.

It is now useful to translate the eigenvalue Eq \ref{eigenright} for the left eigenvector $l_k( \vec x) $ of the Hamiltonian ${\cal H}^{[k]} $ of Eq. \ref{FPhamiltonianp} 
\begin{eqnarray}
E(k)  l_k( \vec x) && = ({\cal H}^{[k]})^{\dagger}  l_k( \vec x)
  =   -  \frac{1}{2}  \left( \vec \nabla +   \vec A^{[k]}(\vec x) \right)^2   l_k( \vec x)  + 
V^{[k]}(\vec x)  l_k( \vec x)
\label{eigenleft}
\end{eqnarray}
 using the replacement  $l_k( \vec x) =e^{  {\mathring \nu}^{[k]}(\vec x)}  $ of Eq. \ref{changeZknu}
and the replacement  $\vec A^{[k]} (\vec x )  = - \vec \nabla  {\mathring \nu}^{[k]}(\vec x) + 
 {\vec {\mathring A}^{[k]} }(\vec x ) $ of Eq. \ref{vectorpotptoring}
\begin{eqnarray}
E(k) - V^{[k]}(\vec x)
&& = - \frac{1}{2} e^{ - {\mathring \nu}^{[k]}(\vec x)}   
  \left( \vec \nabla - \vec \nabla  {\mathring \nu}^{[k]}(\vec x) +  {\vec {\mathring A}^{[k]} }(\vec x )  \right)  
    \left( \vec \nabla - \vec \nabla  {\mathring \nu}^{[k]}(\vec x) +  {\vec {\mathring A}^{[k]} }(\vec x )  \right) e^{  {\mathring \nu}^{[k]}(\vec x)} 
 \nonumber \\
 && =  - \frac{1}{2} e^{ - {\mathring \nu}^{[k]}(\vec x)}   
  \left( \vec \nabla - \vec \nabla  {\mathring \nu}^{[k]}(\vec x) +  {\vec {\mathring A}^{[k]} }(\vec x )  \right)  
    . \left(   {\vec {\mathring A}^{[k]} }(\vec x )  e^{  {\mathring \nu}^{[k]}(\vec x)} \right)
    \nonumber \\
 && =  - \frac{[ \vec \nabla . {\vec {\mathring A}^{[k]} }(\vec x )] + [{\vec {\mathring A}^{[k]} }(\vec x ) ]^2}{2}   
 \label{eigenleftnuA}
\end{eqnarray}
As a consequence, the difference between the two Hamiltonians of Eq. \ref{FPhamiltonianpring} reduces to the constant $ E(k)$
\begin{eqnarray}
{\mathring{\cal H}}^{[k]} -  {\cal  H}^{Cond[k]}  =   E(k)
 \label{FPhamiltonianpringdiff}
\end{eqnarray}
So the function ${\mathring Z}^{[k]}(\vec x,t \vert \vec y, 0) $ of Eq. \ref{changeZkring} becomes
\begin{eqnarray}
{\mathring Z}^{[k]}(\vec x,t \vert \vec y, 0)   = \langle \vec x \vert e^{- t {\mathring{\cal H}}^{[k]}} \vert \vec y \rangle
= e^{-t E(k) } \langle \vec x \vert e^{- t {\cal  H}^{Cond[k]} } \vert \vec y \rangle
\equiv e^{-t E(k) }{\mathring P}^{[k]}(\vec x,t \vert \vec y, 0) 
\label{changeZkringPring}
\end{eqnarray}
where 
\begin{eqnarray}
{\mathring P}^{[k]}(\vec x,t \vert \vec y, 0) = \langle \vec x \vert e^{- t {\cal  H}^{Cond[k]} } \vert \vec y \rangle
\label{changeZkringPringdef}
\end{eqnarray}
is the Fokker-Planck propagator associated to the force ${\vec  {\mathring A}^{[k]}}(\vec x) $
and to the generator of Eq. \ref{Hcondk}.

In summary, 
the initial generating function $Z^{[k]}(\vec x,t \vert \vec y,0) $ has been rewritten 
using Eqs \ref{changeZk} and \ref{changeZkringPring} 
as
\begin{eqnarray}
Z^{[k]}(\vec x,t \vert \vec y,0)  
=  e^{-t E(k) + {\mathring \nu}^{[k]}(\vec y) - {\mathring \nu}^{[k]}(\vec x) } {\mathring P}^{[k]}(\vec x,t \vert \vec y, 0)
\label{changeZkPring}
\end{eqnarray}
As a consequence, the finite-time
canonical conditional probability ${\cal P}^{Cond[k]}(\vec z,\tau)  $ of Eq. \ref{markovcondk} 
only involves the Fokker-Planck propagator ${\mathring P}^{[k]}(\vec x,t \vert \vec y,0) $
via the standard Doob bridge formula
\begin{eqnarray}
{\cal P}^{Cond[k]}(\vec z,\tau) 
=  \frac{Z^{[k]}(\vec x, t \vert \vec z, \tau) Z^{[k]}(\vec z, \tau \vert \vec y, 0)}{Z^{[k]}(\vec x, t \vert \vec y, 0)}
=  \frac{{\mathring P}^{[k]}(\vec x, t \vert \vec z, \tau) {\mathring P}^{[k]}(\vec z, \tau \vert \vec y, 0)}{{\mathring P}^{[k]}(\vec x, t \vert \vec y, 0)}
\label{markovcondkPring}
\end{eqnarray}
while its behavior at any interior time $\tau$ satisfying $0 \ll \tau \ll t$
for large time $t\to +\infty$ of Eq. \ref{markovcondktinfty}
reduces to the steady state ${\mathring P}^{[k]*}(\vec z) $
\begin{eqnarray}
{\cal P}^{Cond[k]}(\vec z,\tau) 
&& \opsimeq_{ 0 \ll \tau \ll t}   {\mathring P}^{[k]*}(\vec z)
 \equiv {\cal P}^{Cond[k]Interior}(\vec z) 
\label{markovcondktinftyPring}
\end{eqnarray}


 \subsubsection{ Summary of the properties of the new gauge ${\vec {\mathring A}^{[k]} }(\vec x ) $ that simplifies the analysis
 of the generating function $Z^{[k]}(\vec x,t \vert \vec y,0)  $   } 
 
  \label{subsec_3conditions}
 
 In conclusion, the analysis of the generating function $Z^{[k]}(\vec x,t \vert \vec y,0)   $ 
 will be drastically simplified when one is able to determine the new gauge ${\vec {\mathring A}^{[k]} }(\vec x )  $ satisfying
 the three conditions :
 
 (i) the gauge transformation of Eq. \ref{vectorpotptoring}. Here it is important to stress 
 that it fixes the corresponding antisymmetric magnetic matrix elements
 to the values $B^{[k]}_{nm}(\vec x )  $ of Eq. \ref{magneticNp} involving the initial magnetic elements $B_{nm}(\vec x ) $ and the supplementary magnetic elements $B^{[{\cal O}]}_{nm}(\vec x ) $
 \begin{eqnarray}
   \partial_n {\mathring A}_m^{[k]} (\vec x )  -  \partial_m {\mathring A}_n^{[k]} (\vec x ) 
   =  B^{[k]}_{nm}(\vec x ) 
  =B_{nm}(\vec x ) + k B^{[{\cal O}]}_{nm}(\vec x )
 \label{vectorpotptoringB}
\end{eqnarray}
 
 (ii) the condition of Eq. \ref{eigenleftnuA}. This condition that is equivalent to the eigenvalue equation of Eq. \ref{eigenleft} for the left eigenvector
  fixes, up to some constant $E(k)$, the sum of the divergence $[\vec \nabla . {\vec {\mathring A}^{[k]} }(\vec x )] $ and the square $ [{\vec {\mathring A}^{[k]} }(\vec x ) ]^2$ in terms of the potential $V^{[k]}(\vec x) $ of Eq. \ref{scalarpotp} that involves the same combination for the initial force $\vec f(\vec x) $ and the supplementary scalar potential $V^{[{\cal O}]} ( \vec x) $
 \begin{eqnarray}
\frac{[ \vec \nabla . {\vec {\mathring A}^{[k]} }(\vec x )] + [{\vec {\mathring A}^{[k]} }(\vec x ) ]^2}{2}   +E(k) = V^{[k]}(\vec x)  
= \frac{  [\vec \nabla . \vec f(\vec x) ] +\vec f ^2( \vec x) }{2 } + k V^{[{\cal O}]} ( \vec x)
 \label{eigenleftnu}
\end{eqnarray}
This equation can be considered as the generalization in dimension $N>1$
of the one-dimensional Riccati equation derived in \cite{c_lyapunov} for the analysis
of additive observables concerning diffusion processes on a one-dimensional ring.

(iii) the Fokker-Planck equation of Eq. \ref{markovconddiffderiforwardktinftygauge} associated to the force ${\vec {\mathring A}^{[k]} } $ should converge towards a steady state ${\mathring P}^{[k]*}(\vec z) $ to be able to write Eq. \ref{markovcondktinftyPring}.

In section \ref{sec_OUadditive}, we will see how this framework allows to simplify the analysis of quadratic observables of
Ornstein-Uhlenbeck processes.


 \section{ Simplifications for Ornstein-Uhlenbeck processes in dimension $N$  } 
 
 \label{sec_OU}

Ornstein-Uhlenbeck processes in dimension $N$ (see the textbooks \cite{gardiner,vankampen,risken} and the more recent works \cite{Thouless,CGetJML} as well as the recent PhD thesis \cite{duBuisson_thesis}) correspond to the case where the force $\vec f(\vec x) $ of Eq. \ref{langevin} is linear with respect to the position $\vec x$
and is thus parametrized by some $N \times N$ matrix ${\bold \Gamma} =[\Gamma_{nm}] $ 
 \begin{eqnarray}
f_n(\vec x) = - \sum_{m=1}^N \Gamma_{nm} x_m
 \label{forcelinear}
\end{eqnarray}
The condition for the convergence towards a steady state is then simply that the real parts
of the $N$ eigenvalues $ \gamma_{\alpha=1,..,N} $ of the matrix ${\bold \Gamma} $ should be strictly positive 
 \begin{eqnarray}
 {\rm Re} ( \gamma_{\alpha} ) >0
 \label{conditionRealPartsGamma}
\end{eqnarray}


 \subsection{ Reminder on the matrix formulation of the Langevin system and its direct integration    }


 \subsubsection{ Direct integration of the Langevin system via its matrix formulation    }

 It is convenient to introduce the ket-notations
\begin{eqnarray}
\vert \vec x(t) \rangle \equiv \sum_{n=1}^N x_n(t) \vert n \rangle
= \begin{pmatrix} 
 x_1(t) 
 \\  ..
 \\ x_N(t) 
  \end{pmatrix}
  \ \ \ { \rm and } \ \ \ 
\vert d \vec w(t) \rangle \equiv  \sum_{n=1}^N dw_n(t) \vert n \rangle
= \begin{pmatrix} 
dw_1(s) 
 \\ ..
 \\ dw_N(s) 
  \end{pmatrix}  
\label{ket}
\end{eqnarray}
as well as the corresponding bra-notations
\begin{eqnarray}
\langle \vec x(t) \vert \equiv 
\begin{pmatrix} 
 x_1(t) 
 &  ..
 & x_N(t) 
  \end{pmatrix}
  \ \ \ { \rm and } \ \ \  
  \langle d \vec w(t) \vert \equiv 
   \begin{pmatrix} 
dw_1(s) 
 & ..
 & dw_N(s) 
  \end{pmatrix}  
\label{bra}
\end{eqnarray}

The Langevin system of Eq. \ref{langevin} then reads in the matrix form
\begin{eqnarray}
d \vert \vec x(t) \rangle &&=  -  dt \ {\bold \Gamma}  \vert \vec x(t) \rangle + \vert d \vec w(t) \rangle
\label{langevinmatrix}
\end{eqnarray}
and can be integrated to obtain the solution in terms of the initial condition $\vert \vec x(0) \rangle=\vert \vec y \rangle $
\begin{eqnarray}
\vert \vec x(t) \rangle
  = e^{ -  {\bold \Gamma} t } \vert \vec y \rangle
  +\int_0^t  e^{ - {\bold \Gamma} (t-s) }  \vert d \vec w(s) \rangle
\label{langevinmatrixsol}
\end{eqnarray}


 \subsubsection{ Average values $\mu_n(t)=\overline{x_n(t)} $ 
 and connected correlation matrix  ${\bold  C}(t) $ }

The average values $\mu_n(t) \equiv \overline{x_n(t)}$ can be computed in terms of the initial conditions $  \vec y$ via
\begin{eqnarray}
\vert \vec \mu(t) \rangle  = \begin{pmatrix} 
 \mu_1(t) 
 \\ ..
 \\  \mu_N(t) 
  \end{pmatrix}
= \overline{\vert \vec x(t) \rangle }
  =  e^{ - {\bold \Gamma} t } \vert \vec y \rangle
  \label{langevinmumatrix}
\end{eqnarray}

The symmetric connected correlation matrix 
 \begin{eqnarray}
 C_{nm}(t) = C_{mn}(t)  \equiv  \overline{ x_n(t) x_m(t) } - \mu_n(t) \mu_m(t) 
 = \overline{ [ x_n(t) - \mu_n(t)] [ x_m(t)- \mu_m(t) ]  }
 \label{mucorredef}
\end{eqnarray}
can be then computed from the ket-solution
\begin{eqnarray}
\vert \vec x(t) \rangle- \vert \vec \mu(t)\rangle
&&  = \int_0^t  e^{ - {\bold \Gamma} (t-s) }  \vert d \vec w(s) \rangle
\label{langevinmatrixsolket}
\end{eqnarray}
and the corresponding bra-solution involving the transposed matrix ${\bold \Gamma}^T $
\begin{eqnarray}
\langle \vec x(t) \vert - \langle \vec \mu(t)\vert
&&  = \int_0^t  \langle d \vec w(s') \vert e^{ - {\bold \Gamma}^T (t-s') }  
\label{langevinmatrixsolbra}
\end{eqnarray}
using the statistics of the Wiener processes increments $dw_n(s)$
\begin{eqnarray}
\overline{  \vert d \vec w(s) \rangle \langle d \vec w(s') \vert}
&& = \sum_{n=1}^N \sum_{m=1}^N \vert n \rangle \langle m \vert
\overline{ dw_n(s)dw_m(s') }
= \sum_{n=1}^N \sum_{m=1}^N \vert n \rangle \langle m \vert \delta_{n,m} ds \delta(s-s') 
\nonumber \\
&& = ds \delta(s-s') \sum_{n=1}^N  \vert n \rangle \langle n \vert 
 = ds \delta(s-s') {\bold 1}
\label{WW}
\end{eqnarray}
to obtain
 \begin{eqnarray}
{\bold  C}(t) &&  = 
\overline{ \left( \vert \vec x(t) \rangle- \vert \vec \mu(t)\rangle \right)
\left( \langle \vec x(t) \vert - \langle \vec \mu(t)\vert\right)
 }
\nonumber \\
&&  = \int_0^t ds e^{ - {\bold \Gamma} (t-s) }
   e^{ - {\bold \Gamma}^T (t-s) } 
 \label{corresol}
\end{eqnarray}
The product of the two exponentials cannot be simplified in the generic case where the matrix ${\bold \Gamma}$ does not commute with its transpose ${\bold \Gamma}^T $.
The correlation of Eq. \ref{corresol}
satisfies the dynamics
\begin{eqnarray}
\frac{d  {\bold C}(t) }{dt} = - {\bold \Gamma}{\bold C}(t) - {\bold C}(t){\bold \Gamma}^T + \mathbf{1}
\label{langevinC}
\end{eqnarray}


 \subsubsection{ Gaussian finite-time propagator $P(\vec x,t \vert \vec y,0) $  and Gaussian steady state $P^*(\vec x) $  } 

The finite-time propagator is Gaussian and can be constructed from the average values of Eq. \ref{langevinmumatrix}
and the symmetric connected correlation matrix of Eq. \ref{corresol}
\begin{eqnarray}
 P ( \vec x , t \vert \vec y,0)
&& = \frac{1}{  \sqrt{(2\pi)^N \det({\bold C}(t) ) } } 
 e^{ \displaystyle - \frac{1}{2} 
 \left( \langle \vec x \vert - \langle \vec \mu(t)\vert\right)
 {\bold \Phi} (t) 
\left( \vert \vec x \rangle- \vert \vec \mu(t)\rangle \right)
 }
 \nonumber \\
&& = \frac{1}{  \sqrt{(2\pi)^N \det({\bold C}(t) ) } } 
 e^{ \displaystyle - \frac{1}{2} 
 \left( \langle \vec x \vert - \langle \vec y\vert e^{ - {\bold \Gamma}^T t }\right)
 {\bold \Phi} (t) 
\left( \vert \vec x \rangle- e^{ - {\bold \Gamma} t } \vert \vec y \rangle \right)
 } 
\label{gaussinverse}
\end{eqnarray}
where we have introduced the notation $ {\bold \Phi} (t) $ to denote the inverse of the matrix $ {\bold C}(t)$
\begin{eqnarray}
 {\bold \Phi} (t) \equiv [{\bold C}(t)]^{-1}
\label{PhiCinverset}
\end{eqnarray}

For $t \to +\infty$, the average values of Eq. \ref{langevinmumatrix}
converge towards zero as a consequence of Eq. \ref{conditionRealPartsGamma}
 \begin{eqnarray}
\mu_n(t=+\infty)  =0
 \label{muinfty}
\end{eqnarray}
while the connected correlation matrix converge toward the finite value
$ {\bold C} (t=+\infty)= {\bold C}$ satisfying the steady version of Eq. \ref{langevinC}
  \begin{eqnarray}
{\bold \Gamma } {\bold C}+ {\bold C} {\bold \Gamma }^T=  \mathbf{1}
 \label{eqCrecover}
\end{eqnarray}
The propagator of Eq. \ref{gaussinverse}
thus converges towards the Gaussian steady state
  \begin{eqnarray}
 P^* ( \vec x )     =   
 \frac{1}{  \sqrt{ (2\pi)^N \det({\mathbf C} ) } } 
 e^{\displaystyle - \frac{1}{2} 
 \langle \vec x \vert
  {\bold \Phi} 
 \vert \vec x \rangle
 }
\label{steadyOU}
\end{eqnarray}
involving the matrix
\begin{eqnarray}
 {\bold \Phi} = {\bold \Phi} (t=+\infty)  =  {\bold C}^{-1}
\label{PhiCinverse}
\end{eqnarray}

So the function $\phi( \vec x) $ of Eq. \ref{steadyirr}
 is quadratic and involves the symmetric matrix $ {\bold \Phi} $
 \begin{eqnarray}
 \phi( \vec x) \equiv - \frac{1 }{2} \ln \left( \frac{P^*(\vec x)}{P^*(\vec 0)} \right) 
 = \frac{1}{4} 
 \langle \vec x \vert
  {\bold \Phi} 
 \vert \vec x \rangle
 =\sum_{n=1}^N \sum_{m=1}^N \frac{ \Phi_{nm}}{4} x_n x_m
\label{PhiOU}
\end{eqnarray}


 \subsubsection{ Discussion    } 

When one wishes to obtain explicit results in specific models,
the matrix solutions of Eqs \ref{langevinmumatrix} and \ref{corresol}
 should be analyzed in terms of the spectral decomposition of the matrix ${\bold \Gamma} $ \cite{Thouless,CGetJML},
 as recalled in Appendix \ref{App_explicitvisdiagoGamma}.
 
Now that we have recalled the standard solution via the direct integration of the Langevin stochatic equations,
 it is interesting to see in the following sections what insights can be gained via the quantum non-Hermitian interpretation described in section \ref{sec_mappingQ}.


 \subsection{ Corresponding non-Hermitian quantum problem :  harmonic oscillator in constant  magnetic matrix   } 
 
For the linear force of Eq. \ref{forcelinear},
the vector potential of Eq. \ref{vectorpot} is linear
\begin{eqnarray}
  A_n (\vec x ) && = f_n(\vec x) = - \sum_{m=1}^N \Gamma_{nm} x_m
   \label{vectorpotlinearOU}
\end{eqnarray}
so that the antisymmetric magnetic matrix of Eq. \ref{magneticN} is space-independent
\begin{eqnarray}
 B_{nm}  =-B_{mn}   
 \equiv \partial_n A_m (\vec x ) -  \partial_m A_n (\vec x ) 
  = \Gamma_{nm}-\Gamma_{mn}
 \label{magneticNcst}
\end{eqnarray}
and corresponds to the antisymmetric part of the matrix ${\bold \Gamma}$.

The scalar potential of Eq. \ref{scalarpot} is quadratic
\begin{eqnarray}
V(\vec x) && = \frac{  1 }{2 } \sum_{n=1}^N \partial_n f_n( \vec x)
+  \frac{  1 }{2 } \sum_{n=1}^N f_n^2( \vec x)
= - \frac{  1 }{2 } \sum_{n=1}^N \Gamma_{nn}
+  \frac{  1 }{2 } \sum_{n=1}^N  
\left( - \sum_{m=1}^N \Gamma_{nm} x_m\right) 
\left( - \sum_{j=1}^N \Gamma_{nj} x_j\right)
\nonumber \\
&& \equiv  V_0
+\sum_{m=1}^N   \sum_{j=1}^N  \frac{W_{mj}}{2} x_m  x_j
\label{scalarpotquadraticOU}
\end{eqnarray}
where the constant contribution $V_0 $ involves the trace of the matrix ${\bold \Gamma}$
\begin{eqnarray}
 V_0  \equiv - \frac{ 1  }{2} \sum_{n=1}^N\Gamma_{nn} =  - \frac{ 1  }{2} {\rm tr}({\bold \Gamma}) 
\label{W0def}
\end{eqnarray}
while the symmetric matrix ${\bold W} $ involves the matrix ${\bold \Gamma}$ and its transpose  ${\bold \Gamma}^{T}$ 
\begin{eqnarray}
 W_{mj}  \equiv    \sum_{n=1}^N   \Gamma_{nm}  \Gamma_{nj} 
=   \left({\bold \Gamma}^{T} {\bold \Gamma}\right)_{mj}
\label{Wdef}
\end{eqnarray}

So the non-Hermitian quantum problem corresponds to a $N$-dimensional harmonic oscillator in a constant magnetic matrix,
with the following technical simplifications :

(i) The Lagrangian $ {\cal L} (\vec x(\tau), \dot {\vec x}(\tau))$ of Eq. \ref{lagrangian}
is quadratic with respect to the positions $x_n (\tau) $ and the velocities $\dot x_n (\tau) $ 
\begin{eqnarray}
 {\cal L} (\vec x(\tau), \dot {\vec x}(\tau)
 && \equiv \frac{1}{2} \sum_{n=1}^N \left( \dot x_n (\tau) + \sum_{m=1}^N \Gamma_{nm} x_m(\tau) ) \right)^2
 - \frac{  1 }{2 }  \sum_{n=1}^N\Gamma_{nn} 
 \nonumber \\
 && \equiv 
 \frac{1}{2} \sum_{n=1}^N \dot x_n (\tau)
 + \sum_{n=1}^N \sum_{m=1}^N \dot x_n (\tau) \Gamma_{nm} x_m(\tau)
 +  \sum_{n=1}^N \sum_{m=1}^N \frac{ \left({\bold \Gamma}^{T} {\bold \Gamma}\right)_{nm}}{2} x_n (\tau)  x_m(\tau)
 - \frac{  1 }{2 }  \sum_{n=1}^N\Gamma_{nn} 
\label{lagrangianOU}
\end{eqnarray}

(ii) The quantum Hamiltonian of Eqs \ref{FPhamiltonian} \ref{FPhamiltonianQuantum}
is quadratic with respect to the positions $x_n$ and the momenta operators $(-i \partial_n)$
with the various forms
\begin{eqnarray}
{\cal  H}&&  =   \sum_{n=1}^N \partial_n  \left(  - \frac{1}{2}  \partial_n - \sum_{m=1}^N \Gamma_{nm} x_m\right)
\nonumber \\
&& =     - \frac{1}{2} \sum_{n=1}^N \partial_n^2
- \sum_{n=1}^N \sum_{m=1}^N \Gamma_{nm} x_m \partial_n
  -   \sum_{n=1}^N\Gamma_{nn}  
\nonumber \\
 &&   =  -  \frac{1}{2} \sum_{n=1}^N \left( \partial_n + \sum_{m=1}^N \Gamma_{nm} x_m \right)^2   
  +  \sum_{n=1}^N \sum_{m=1}^N \frac{ \left({\bold \Gamma}^{T} {\bold \Gamma}\right)_{nm}}{2} x_n  x_m
 - \frac{  1 }{2 }  \sum_{n=1}^N\Gamma_{nn} 
\label{FPhamiltonianQuantumOU}
\end{eqnarray}

It is thus instructive to see how one can recover the Fokker-Planck propagator via this quantum perspective.


 \subsection{ Recovering the Fokker-Planck propagator $P(\vec x,t \vert \vec y,0) $ from the quantum perspective } 
 
 \label{subsec_PropagatorClassicalAction}

As recalled in Appendix \ref{app_quadratic}, when the Lagrangian ${\cal L} (\vec x(\tau), \dot {\vec x}(\tau) )  $ is quadratic as in Eq. \ref{lagrangianOU}, the path-integral of Eq. \ref{pathintegral} for the quantum propagator
can be evaluated in terms of the classical action $S_{cl}(\vec x,t \vert \vec y ,0) $ alone 
\begin{eqnarray}
P(\vec x,t \vert \vec y,0)
= \sqrt{  \frac{ \det \left( - \frac{ \partial^2 S_{cl}(\vec x,t \vert \vec y, 0)}{\partial_{x_n} \partial_{y_m}}\right)}{(2 \pi)^N}}  e^{ - S_{cl}(\vec x,t \vert \vec y, 0)  } 
\label{pathintegralclassOU}
\end{eqnarray}

The application of Appendix \ref{app_quadratic} to the present case 
corresponding to the matrices ${\bold \Lambda}={\bold \Gamma} $ 
and ${\bold W} ={\bold \Gamma}^{T} {\bold \Gamma}$ yields that
the matrix ${\cal M } $ of size $(2N)\times (2N)$ in Eq. \ref{HamiltonMmatrix} reduces to
  \begin{eqnarray}
{\cal M }  = \begin{pmatrix} 
- {\bold \Gamma}   & {\bold 1}  \\
0&   {\bold \Gamma}^T
 \end{pmatrix}    
 \label{HamiltonMmatrixSimple}
\end{eqnarray}
The corresponding Hamilton's classical equations of motion of Eq. \ref{Hclassicalmotion}
\begin{eqnarray}
   \vert \dot {\vec x}(\tau)\rangle   && =   \vert  {\vec p} (\tau)\rangle  - {\bold \Gamma}  \vert  {\vec x}(\tau)\rangle
  \nonumber \\
  \vert \dot {\vec p} (\tau)\rangle    && =   {\bold \Gamma}^T\vert  {\vec p} (\tau)\rangle
\label{HclassicalmotionOU}
\end{eqnarray}
can be integrated as follows.
The second equation corresponds to a closed dynamics for the classical momentum 
$\vec p(\tau)$. Its solution in terms of the initial momentum $\vert  \vec p (0)\rangle $
\begin{eqnarray}
  \vert \vec p (\tau)\rangle     =  e^{  {\bold \Gamma}^T \tau } \vert  \vec p (0)\rangle
\label{HclassicalmotionOUp}
\end{eqnarray}
can be plugged into the first Eq. \ref{HclassicalmotionOU} to obtain the following dynamics for the position ${\vec x}(\tau) $
\begin{eqnarray}
   \vert \dot {\vec x}(\tau)\rangle    =   -  {\bold \Gamma}  \vert  {\vec x}(\tau)\rangle
   + e^{  {\bold \Gamma}^T \tau } \vert  \vec p (0)\rangle
\label{HclassicalmotionOUx}
\end{eqnarray}
The solution can be thus written in terms of the initial position $\vert   \vec y \rangle  = \vert   \vec x (\tau=0)\rangle  $
and initial momentum $\vert  \vec p (0)\rangle $ as
\begin{eqnarray}
   \vert  {\vec x}(\tau)\rangle  &&   =  e^{ - {\bold \Gamma} \tau} \bigg(
    \vert  {\vec y}\rangle + \int_0^{\tau} ds
   e^{  {\bold \Gamma} s } e^{  {\bold \Gamma}^T s } \vert  \vec p (0)\rangle
   \bigg)
  =  e^{ - {\bold \Gamma} \tau}
    \vert  {\vec y}\rangle + 
     \int_0^{\tau} ds   e^{ - {\bold \Gamma} (\tau-s)  } e^{  {\bold \Gamma}^T s }
     \vert \vec p (0)\rangle
     \nonumber \\
   &&   =e^{ - {\bold \Gamma} \tau}
    \vert  {\vec y}\rangle + 
     {\bold C}(\tau) e^{  {\bold \Gamma}^T \tau }
     \vert \vec p (0)\rangle
\label{HclassicalmotionOUxsol}
\end{eqnarray}
where one recognizes the matrix 
\begin{eqnarray}
{\bold C} (\tau) = \int_0^{\tau} ds
   e^{ - {\bold \Gamma} (\tau-s) } e^{ - {\bold \Gamma}^T (\tau-s) }
\label{ctauagain}
\end{eqnarray}
already discussed in Eq. \ref{corresol}.
 One then needs to impose 
the final position $\vert   \vec x \rangle $ at $\tau=t$ for the solution of Eq. \ref{HclassicalmotionOUxsol}
\begin{eqnarray}
\vert   \vec x \rangle && = \vert   \vec x (\tau=t)\rangle   =  
 e^{ - {\bold \Gamma} t}    \vert  {\vec y}\rangle 
 +     {\bold C}(t) e^{  {\bold \Gamma}^T t }     \vert \vec p (0)\rangle
\label{classicalmotioneigeniCLBCfin}
\end{eqnarray}
to determine the appropriate initial momentum $\vert  \vec p (0)\rangle $
using the inverse matrix ${\bold \Phi} (t) \equiv [{\bold C}(t)]^{-1}$ of Eq. \ref{PhiCinverset}
\begin{eqnarray}
\vert  \vec p (0)\rangle = e^{ - {\bold \Gamma}^T t }
{\bold \Phi} (t)\bigg(\vert   \vec x \rangle - e^{ - {\bold \Gamma} t}    \vert  {\vec y}\rangle\bigg)
\label{impulsionpzero}
\end{eqnarray}
The classical action $S_{cl}(\vec x,t \vert \vec y, 0)  $ of Eq. \ref{actionresHres} 
can be now computed from 
the solution of Eqs \ref{HclassicalmotionOUx}
and \ref{impulsionpzero} and from the constant $V_0 $ of Eq. \ref{W0def}
\begin{eqnarray}
S_{cl}(\vec x,t \vert \vec y, 0) 
&&      =    \frac{\langle \vec x   \vert  {\vec p} (t)\rangle - \langle \vec y   \vert  {\vec p} (0)\rangle  }{ 2 } 
+ t V_0
=  \frac{\langle \vec x   \vert e^{  {\bold \Gamma}^T t }
 \vert {\vec p} (0)\rangle - \langle \vec y   \vert  {\vec p} (0)\rangle  }{ 2 } - \frac{ t  }{2} {\rm tr}({\bold \Gamma}) 
  = \frac{1}{2} \bigg (\langle \vec x   \vert e^{  {\bold \Gamma}^T t } - \langle \vec y \vert  \bigg) \vert  \vec p (0)\rangle
  - \frac{ t  }{2} {\rm tr}({\bold \Gamma}) 
\nonumber \\
 &&  = \frac{1}{2} \bigg (\langle \vec x   \vert e^{  {\bold \Gamma}^T t } - \langle \vec y \vert  \bigg)
 e^{ - {\bold \Gamma}^T t }
{\bold \Phi} (t)\bigg(\vert   \vec x \rangle - e^{ - {\bold \Gamma} t}    \vert  {\vec y}\rangle\bigg)
- \frac{ t  }{2} {\rm tr}({\bold \Gamma}) 
\nonumber \\
 &&  = \frac{1}{2} \bigg (\langle \vec x   \vert  - \langle \vec y \vert e^{ - {\bold \Gamma}^T t } \bigg)
{\bold \Phi} (t)\bigg(\vert   \vec x \rangle - e^{ - {\bold \Gamma} t}    \vert  {\vec y}\rangle\bigg)
- \frac{ t  }{2} {\rm tr}({\bold \Gamma}) 
 \label{actionresHresOU}
\end{eqnarray}
To evaluate the prefactor of Eq. \ref{pathintegralclassOU}, one needs to compute the Van Vleck matrix
of the double-derivatives of the action $S_{cl}(\vec x,t \vert \vec y, 0) $ 
with respect to the final components $x_n$ and with respect to the initial components $y_m$
\begin{eqnarray}
-  \frac{ \partial^2 S_{cl}(\vec x,t \vert \vec y, 0)}{\partial_{x_n} \partial_{y_m}} =
 - \frac{ \partial}{ \partial_{y_m}}  \sum_{l=1}^N \Phi_{nl}(t)
 \left[x_l - \sum_{j=1}^N\left(e^{ - {\bold \Gamma} t} \right)_{lj} y_j \right]
  =  \sum_{l=1}^N \Phi_{nl}(t) \left(e^{ - {\bold \Gamma} t} \right)_{lm} 
  =  \left({\bold \Phi}(t) e^{ - {\bold \Gamma} t} \right)_{nm} 
\label{matrixderiSclassical}
\end{eqnarray}
As a consequence its determinant reduces to the product to two determinants
\begin{eqnarray}
 \det \left( - \frac{ \partial^2 S_{cl}(\vec x,t \vert \vec y, 0)}{\partial_{x_n} \partial_{y_m}}\right)  =
  \det \left({\bold \Phi}(t) \right) \times \det \left(e^{ - {\bold \Gamma} t} \right)
  = \frac{1}{\det \left({\bold C}(t) \right) } \times e^{ -  t [{\rm tr}({\bold \Gamma}) ]} 
\label{detmatrixderiSclassical}
\end{eqnarray}
Plugging Eqs \ref{actionresHresOU} and \ref{detmatrixderiSclassical}
into Eq. \ref{pathintegralclassOU}
\begin{eqnarray}
P(\vec x,t \vert \vec y,0)
&& = \sqrt{  \frac{e^{ -  t [{\rm tr}({\bold \Gamma}) ]} }{(2 \pi)^N \det \left({\bold C}(t) \right) }}  
e^{ - 
\frac{1}{2} \bigg (\langle \vec x   \vert  - \langle \vec y \vert e^{ - {\bold \Gamma}^T t } \bigg)
{\bold \Phi} (t)\bigg(\vert   \vec x \rangle - e^{ - {\bold \Gamma} t}    \vert  {\vec y}\rangle\bigg)
+ \frac{ t  }{2} {\rm tr}({\bold \Gamma}) 
  } 
  \nonumber \\
  && =  \frac{1 }{\sqrt{  (2 \pi)^N \det \left({\bold C}(t) \right) }}  
e^{ - 
\frac{1}{2} \bigg (\langle \vec x   \vert  - \langle \vec y \vert e^{ - {\bold \Gamma}^T t } \bigg)
{\bold \Phi} (t)\bigg(\vert   \vec x \rangle - e^{ - {\bold \Gamma} t}    \vert  {\vec y}\rangle\bigg) 
  } 
\label{pathintegralclassOUfinal}
\end{eqnarray}
allows us to recover the propagator of Eq \ref{gaussinverse} as it should for consistency between the two perspectives.


 \subsection{ Gauge transformations of the vector potential $\vec A(\vec x) $ that preserve its linearity } 
    
 \label{subsec_Gaugetransformations}
 
 Let us now consider the gauge transformation from the linear vector potential $A_n(\vec x) $ of Eq. \ref{vectorpotlinearOU}
towards some new linear vector potential $ {\tilde A}_n (\vec x )  $ parametrized by the new matrix ${\tilde {\bold \Gamma}}$ instead of $\bold \Gamma$
\begin{eqnarray}
  {\tilde A}_n (\vec x )  = -  \sum_{m=1}^N {\tilde \Gamma}_{nm} x_m 
   \label{newgaugelinear}
\end{eqnarray}  
The antisymmetric part of the matrix ${\bold {\tilde \Gamma}} $ is fixed by the magnetic matrix $ {\bold B} $ of Eq. \ref{magneticNcst}
 \begin{eqnarray}
 {\bold {\tilde \Gamma}}-{\bold {\tilde \Gamma}}^T =  {\bold B}   = {\bold \Gamma}-{\bold \Gamma}^T
  \label{magneticNcstmatrix}
\end{eqnarray}  
while its symmetric part $( {\bold {\tilde \Gamma}}+{\bold {\tilde \Gamma}}^T)$ remains arbitrary.
The corresponding gauge transformation
\begin{eqnarray}
 A_n (\vec x )  = -\partial_n {\tilde \nu}(\vec x) +  {\tilde A}_n (\vec x ) 
 \label{gaugetransfolinear}
\end{eqnarray}
involves the quadratic function
\begin{eqnarray}
 {\tilde \nu}(\vec x) && 
 =  \sum_{n=1}^N \sum_{m=1}^N  
 \frac{(\Gamma_{nm}+\Gamma_{mn}) -  ({\tilde \Gamma}_{nm}+{\tilde \Gamma}_{mn})  }{4} x_n x_m = \langle \vec x \vert \bigg(  \frac{({\bold \Gamma}+{\bold \Gamma}^T )
 - ({\bold {\tilde \Gamma}}+{\bold {\tilde \Gamma}}^T)
  }{4} \bigg)  \vert \vec x \rangle
 \label{muquadratic}
\end{eqnarray}
where one recognizes the difference between the symmetric parts $({\bold \Gamma}+{\bold \Gamma}^T ) $ 
and $({\bold {\tilde \Gamma}}+{\bold {\tilde \Gamma}}^T) $ of the two matrices.

The corresponding change for the propagator is the analog of Eq. \ref{changetowardshat}
\begin{eqnarray}
P(\vec x,t \vert \vec y,0)
 = e^{ \displaystyle {\tilde \nu} ( \vec y) - {\tilde \nu} ( \vec x)} {\tilde P}(\vec x,t \vert \vec y,0)
\label{changetowardstilde}
\end{eqnarray}
where the new propagator
\begin{eqnarray}
{\tilde P}(\vec x,t \vert \vec y,0)
 = \int_{\vec x(\tau=0)=\vec y}^{\vec x(\tau=t)=\vec x} {\cal D}   \vec x(\tau)  
 e^{ - \displaystyle 
  \int_0^{t} d\tau {\tilde {\cal L}}  (\vec x(\tau), \dot {\vec x}(\tau) )
 } 
\label{pathintegraltilde}
\end{eqnarray}
is governed by the new quadratic Lagrangian
that involves the same scalar potential $V( \vec x) $ of Eq. \ref{scalarpotquadraticOU}
but the new vector potential ${\vec {\tilde A}}(\vec x )  $ of Eq. \ref{newgaugelinear}
\begin{eqnarray}
{\tilde {\cal L}} (\vec x(\tau), \dot {\vec x}(\tau)
&& = 
 \frac{1}{2}  \dot {\vec x}^2  (\tau) 
 - \dot {\vec x } (\tau) . \vec {\tilde A} ( \vec x(\tau))
 + V(\vec x)
 \nonumber \\
&& = \frac{1}{2} \sum_{n=1}^N \dot x_n^2  (\tau) 
 + \sum_{n=1}^N \sum_{m=1}^N \dot x_n (\tau)  {\tilde \Gamma} _{nm} x_m(\tau)
+\sum_{n=1}^N   \sum_{m=1}^N  \frac{\left({\bold \Gamma}^{T} {\bold \Gamma}\right)_{nm}}{2} x_n  x_m 
- \frac{ 1  }{2} \sum_{n=1}^N\Gamma_{nn}
\label{lagrangiantilde}
\end{eqnarray}

The Euclidean Schr\"odinger equation for the new propagator ${\tilde P}(\vec x,t \vert \vec y,0) $
\begin{eqnarray}
 \partial_t {\tilde P}(\vec x,t \vert \vec y,0)   
 =  -  {\tilde {\cal  H}} {\tilde P}(\vec x,t \vert \vec y,0)
\label{Euclideantilde}
\end{eqnarray}
is governed by the new Hamiltonian $ {\hat {\cal  H}}$ analogous to Eq. \ref{hamiltonianhat}
that involves the same scalar potential $V( \vec x) $ of Eq. \ref{scalarpotquadraticOU}
but the new vector potential ${\vec {\tilde A}}(\vec x )  $ of Eq. \ref{newgaugelinear}
\begin{eqnarray}
 {\tilde {\cal  H}}  && = e^{ \displaystyle  {\tilde \nu} ( \vec x)}{\cal  H} e^{ \displaystyle - {\tilde \nu} ( \vec x) }
   =  e^{ \displaystyle {\tilde \nu} ( \vec x)} \left[
   -  \frac{1}{2}  \left( \vec \nabla -   \vec A(\vec x) \right)^2    +  V(\vec x)
 \right]   e^{ \displaystyle - {\tilde \nu} ( \vec x) }
  =    -  \frac{1}{2}  \left( \vec \nabla -   {\vec {\tilde A}}(\vec x )   \right)^2    +  V(\vec x)
\nonumber \\
&&   
=  -  \frac{1}{2}  \sum_{n=1}^N \left( \partial_n + \sum_{m=1}^N   {\tilde \Gamma} _{nm} x_m    \right)^2   
 +\sum_{n=1}^N   \sum_{m=1}^N  \frac{\left({\bold \Gamma}^{T} {\bold \Gamma}\right)_{nm}}{2} x_n  x_m 
- \frac{ 1  }{2} \sum_{n=1}^N\Gamma_{nn}
\nonumber \\
&&   
=     
 - \frac{1}{2} \sum_{n=1}^N \partial_n^2
- \sum_{n=1}^N\sum_{m=1}^N {\tilde \Gamma}_{nm} x_m \partial_n
 +\sum_{n=1}^N   \sum_{m=1}^N  \frac{\left({\bold \Gamma}^{T} {\bold \Gamma}\right)_{nm}
 - \left({\bold {\tilde \Gamma}}^{T} {\bold {\tilde  \Gamma}}\right)_{nm} }{2} x_n  x_m 
- \frac{ 1  }{2} \sum_{n=1}^N (\Gamma_{nn}+ {\tilde \Gamma}_{nn})
  \label{hamiltoniantilde}
\end{eqnarray}
which can be compared with the various forms of Eq. \ref{FPhamiltonianQuantumOU} for the initial Hamiltonian ${\cal H}$.

Let us discuss two interesting examples of such gauge transformations
that preserve the linearity of the vector potential in the next two subsections.


 \subsection{ Properties of the Lagrangian and of the Hamiltonian in the Coulomb gauge $\vec A^{Coulomb}  (\vec x ) $ }

The simplest choice of gauge transformation of the form of Eq. \ref{newgaugelinear}
is when the symmetric part vanishes $({\bold {\tilde \Gamma}}+{\bold {\tilde \Gamma}}^T) =0$.
Then Eq. \ref{magneticNcstmatrix} yields that one recovers the standard Coulomb gauge
\begin{eqnarray}
 {\bold \Gamma}^{Coulomb} = \frac{\bold B}{2}
 \label{coulombmatrix}
\end{eqnarray}
where the vector potential only involves the magnetic matrix elements $B_{nm}$
\begin{eqnarray}
A^{Coulomb}_n  (\vec x )  = - \frac{1}{2} \sum_m B_{nm} x_m 
 \label{coulomb}
\end{eqnarray}
The corresponding gauge transformation of Eq. \ref{gaugetransfolinear}
\begin{eqnarray}
 A_n (\vec x )  = -\partial_n \nu^{Coulomb}(\vec x) +  A^{Coulomb}_n (\vec x ) 
 \label{gaugetransfolinearcoulomb}
\end{eqnarray}
involves the quadratic function of Eq. \ref{muquadratic}
\begin{eqnarray}
\nu^{Coulomb}(\vec x) && 
 =  \sum_{n=1}^N \sum_{m=1}^N  \frac{   (\Gamma_{nm}+\Gamma_{mn}) }{4} x_n x_m 
 =  \langle \vec x \vert \bigg(  \frac{ 
 {\bold \Gamma}+{\bold \Gamma}^T  }{4} \bigg)  \vert \vec x \rangle
 \label{muquadraticcoulomb}
\end{eqnarray}
that will appear in the change of propagator of Eq. \ref{changetowardstilde}
\begin{eqnarray}
P(\vec x,t \vert \vec y,0)
 = e^{  \left[ \nu^{Coulomb} ( \vec y) - \nu^{Coulomb} ( \vec x)\right]} P^{Coulomb}(\vec x,t \vert \vec y,0)
\label{changetowardscoulomb}
\end{eqnarray}


 \subsubsection{ Lagrangian in the Coulomb gauge : stochastic areas as conjugated variables to the magnetic matrix elements}

  In the path-integral representation of the propagator
of Eq. \ref{pathintegraltilde}
\begin{eqnarray}
P^{Coulomb}(\vec x,t \vert \vec y,0)
&& = \int_{\vec x(\tau=0)=\vec y}^{\vec x(\tau=t)=\vec x} {\cal D}   \vec x(\tau)  
 e^{ - \displaystyle 
  \int_0^{t} d\tau {\tilde {\cal L}}^{Coulomb}  (\vec x(\tau), \dot {\vec x}(\tau) )
 } 
 \nonumber \\
 && = \int_{\vec x(\tau=0)=\vec y}^{\vec x(\tau=t)=\vec x} {\cal D}   \vec x(\tau)  
 e^{ - \displaystyle 
  \int_0^{t} d\tau \left[ \frac{1}{2}  \dot {\vec x}^2  (\tau) 
 + V(\vec x)
 \right]
 + \int_0^{t} d\tau \dot {\vec x } (\tau) . \vec  A^{Coulomb} ( \vec x(\tau))
 } 
\label{pathintegralcoulomb}
\end{eqnarray}
the last term involving the vector potential $\vec  A^{Coulomb} ( \vec x) $ can be rewritten using the antisymmetry of the magnetic matrix ${\bold B}$
\begin{eqnarray}
 \int_0^{t} d\tau \dot {\vec x} (\tau).    {\vec A}^{Coulomb} ( \vec x(\tau))  
&& =  \int_0^{t} d\tau \sum_n  \dot x_n(\tau)    A^{Coulomb}_n ( \vec x(\tau))   
= \sum_n \sum_m B_{mn}   \frac{1}{2} \int_0^{t} d\tau   x_m (\tau) \dot x_n(\tau) 
\nonumber \\
&& = \sum_n \sum_{m<n} B_{mn}   \frac{1}{2} \int_0^{t} d\tau \bigg(  x_m (\tau) \dot x_n(\tau) - x_n (\tau) \dot x_m(\tau)\bigg)
 \nonumber \\
&& 
 \equiv \sum_n \sum_{m<n} B_{mn}   {\cal A}_{mn}[\vec x(0 \leq s \leq t) ]
\label{AantiAreamn}
\end{eqnarray}
as a linear combination of the $\frac{N (N-1)}{2} $ magnetic matrix elements $B_{nm}$,
where the conjugated variables ${\cal A}_{mn}[\vec x(0 \leq s \leq t) ] $ are the projected stochastic areas  
swept by the trajectory $[\vec x(0 \leq s \leq t)] $
in the various planes $(m,n)$
\begin{eqnarray}
{\cal A}_{mn}[\vec x(0 \leq s \leq t) ] \equiv    \frac{1}{2} \int_0^{t} d\tau 
 \bigg(  x_m (\tau) \dot x_n(\tau) - x_n (\tau) \dot x_m(\tau)\bigg) = - {\cal A}_{nm}[\vec x(0 \leq s \leq t) ]
\label{areamn}
\end{eqnarray}
Since the magnetic matrix elements $B_{mn}$ of Eq. \ref{magneticNcst}
are the only relevant parameters that characterize the irreversibility of the Ornstein-Uhlenbeck process,
these projected stochastic areas ${\cal A}_{mn}[\vec x(0 \leq s \leq t) ] $ 
can be considered as the basic observables that characterize the irreversibility of the stochastic trajectories.
The relevance of these stochastic areas to characterize the irreversibility
has been already stressed in \cite{area2017,area2019,area2021}.


 \subsubsection{ Hamiltonian in the Coulomb gauge : angular momentum as conjugated operator to the magnetic matrix } 
 
 In the Coulomb gauge of Eq. \ref{coulomb} that involves only 
 the antisymmetric magnetic matrix ${\bold B}^T=- {\bold B} $, the Hamiltonian 
 of Eq. \ref{hamiltoniantilde}
 reads
\begin{eqnarray}
 {\cal  H}^{Coulomb}  &&   =    -  \frac{1}{2}  \left( \vec \nabla -   {\vec A}^{Coulomb} (\vec x )   \right)^2    +  V(\vec x)
\nonumber \\
&&   
=  -  \frac{1}{2}  \sum_{n=1}^N \left( \partial_n + \sum_{m=1}^N   \frac{ B_{nm}}{2} x_m    \right)^2   
 +\sum_{n=1}^N   \sum_{m=1}^N  \frac{\left({\bold \Gamma}^{T} {\bold \Gamma}\right)_{nm}}{2} x_n  x_m 
- \frac{ 1  }{2} \sum_{n=1}^N\Gamma_{nn}
\nonumber \\
&&   
=     
 - \frac{1}{2} \sum_{n=1}^N \partial_n^2
- \sum_{n=1}^N \sum_{m=1}^N \frac{ B_{nm}}{2} x_m \partial_n
 +\sum_{n=1}^N   \sum_{m=1}^N  \frac{\left({\bold \Gamma}^{T} {\bold \Gamma}\right)_{nm}
 + \frac{\left({\bold B}^2\right)_{nm}}{4} }{2} x_n  x_m 
- \frac{ 1  }{2} \sum_{n=1}^N \Gamma_{nn}
  \label{hamiltoniancoulomb}
\end{eqnarray}

Using again the antisymmetry of the magnetic matrix ${\bold B}^T=- {\bold B} $,
the second term can be rewritten as
\begin{eqnarray}
- \sum_{n=1}^N \sum_{m=1}^N \frac{ B_{nm}}{2} x_m \partial_n
=  \sum_{n=1}^N \sum_{m<n} \frac{ B_{nm}}{2} \left( x_n \partial_m - x_m \partial_n  \right)
= i \sum_{n=1}^N \sum_{m<n} \frac{ B_{nm}}{2} L_{nm}
  \label{hamiltoniancoulombBL}
\end{eqnarray}
where one recognizes the components $L_{nm}$ of the Hermitian angular momentum operator in dimension $N$.
\begin{eqnarray}
 L_{nm} \equiv -i \left( x_n \partial_m - x_m \partial_n  \right)
 = x_n (-i \partial_m) - x_m (-i \partial_n) = x_n p_m - x_m p_n = L_{nm}^{\dagger}
  \label{deflmn}
\end{eqnarray}


 \subsection{ Irreversible gauge ${\vec A}^{irr}(\vec x ) $ based on the irreversible force $ \vec f^{irr}( \vec x)$ } 
 
 The gauge ${\vec A}^{irr}(\vec x )  $ based on the irreversible force $ \vec f^{irr}( \vec x)$
 has been already described in subsection \ref{subsec_gaugeirr} for the general case of arbitrary forces,
 so the goal of the present subsection is to stress the simplifications that occur for the special case of linear forces.
 

\subsubsection{ Decomposition of the linear force $\vec f(\vec x) $ into its reversible and irreversible linear contributions }

For the Gaussian steady state $P^*(\vec x)$ of Eq. \ref{steadyOU}, the reversible force of Eq. \ref{forceRev} 
\begin{eqnarray}
f^{rev}_n ( \vec x) = \frac{1}{2} \vec \nabla \ln P^*(\vec x)
= - \sum_{m=1}^N   \frac{ \Phi_{nm}}{2} x_m  
\label{forceRevOU}
\end{eqnarray}
 is linear as the initial force $\vec f(\vec x) $ of Eq. \ref{forcelinear}, 
 therefore the irreversible force $\vec f^{irr}(\vec x) = \vec f(\vec x)  - \vec f^{rev} (\vec x)$ of Eq. \ref{forceIrrev}
 is also linear.
 In Eq. \ref{divforceirrev}, the divergence $\vec \nabla .  \vec f^{irr}( \vec x) $ on the left handside is then constant,
 while the scalar product $ \vec f^{irr}( \vec x) .   \vec \nabla \ln   P^*(\vec x ) $ on the right handside is homogeneous
 of order two with respect to the coordinates $x_n$. As a consequence, both should vanish
 \begin{eqnarray}
 \vec \nabla .  \vec f^{irr}( \vec x) && = 0
 \nonumber \\
  \vec f^{irr}( \vec x) .   \vec \nabla \phi(\vec x ) && =0
 \label{divforceirrevOU}
\end{eqnarray}
So the irreversible force $\vec f^{irr}( \vec x)  $ is divergence-less 
and orthogonal to the gradient $ \vec \nabla \phi(\vec x ) $ of the function $\phi(\vec x)$.
For the parametrization of Eq. \ref{Firrevomega}, these properties mean that
the antisymmetric matrix $\Omega_{nm} (\vec x) $ is independent of the position $\vec x$
\begin{eqnarray}
  \Omega_{nm}   = -  \Omega_{mn} 
 \label{omeganmOU}
\end{eqnarray}
so that Eq. \ref{Firrevomega} reduces to
 \begin{eqnarray}
   f^{irr}_n( \vec x)  =   \sum_m \Omega_{nm}  \partial_m \ln P^*(\vec x) 
   = -2  \sum_m \Omega_{nm}  \partial_m \phi(\vec x) 
     = -  \sum_l   \left( \sum_m \Omega_{nm}   \Phi_{ml} \right) x_l
     = -  \sum_l   \left( {\bold \Omega}  {\bold \Phi}   \right)_{nl} x_l
 \label{FirrevomegaOU}
\end{eqnarray}
 In summary, the linear force $\vec f(\vec x)$ of Eq. \ref{forcelinear} involving the $N^2$ matrix elements $\Gamma_{nm} $
 has been decomposed into its reversible and irreversible components parametrized
 by the $\frac{N(N+1)}{2}$ elements of the symmetric matrix $\Phi_{nm}=\Phi_{mn}  $
 and by the $\frac{N(N-1)}{2}$ elements of the antisymmetric matrix $\Omega_{nm} =-  \Omega_{mn} $.
 
 Putting Eq \ref{forceRevOU} and \ref{FirrevomegaOU} together, the total force reads
 \begin{eqnarray}
f_n (\vec x) =f^{rev}_n ( \vec x) +   f^{irr}_n( \vec x) = 
-  \sum_m   \left( 
 \frac{1}{2} {\bold \Phi}
+ {\bold \Omega}  {\bold \Phi}
  \right)_{nm} x_m
\label{forcetotOU}
\end{eqnarray}
 so the identification with the initial form of Eq. \ref{forcelinear}
 yields the identity at the matrix level 
  \begin{eqnarray}
{\bold \Gamma }  =  \frac{1}{2} {\bold \Phi}+ {\bold \Omega}  {\bold \Phi} 
 \label{gammamatrix}
\end{eqnarray}
The transposition of this equation reads using the symmetry of the matrix $ {\bold \Phi }^T=  {\bold \Phi }$ 
and the antisymmetry of the matrix ${\bold \Omega }^T=- {\bold \Omega } $ 
  \begin{eqnarray}
{\bold \Gamma }^{T}  =   \frac{1}{2} {\bold \Phi }^T   +{\bold \Phi }^T{\bold \Omega }^{T} 
= \frac{1}{2} {\bold \Phi }   - {\bold \Phi }{\bold \Omega }
 \label{gammamatrixtrans}
\end{eqnarray}
As a consequence, the antisymmetric magnetic matrix $ {\bold B } $ of Eq. \ref{magneticNcst} corresponds to 
\begin{eqnarray}
 {\bold B }  = {\bold \Gamma } - {\bold \Gamma }^{T}
 = {\bold \Omega }  {\bold \Phi } + {\bold \Phi }{\bold \Omega }
 \label{magneticNOuomega}
\end{eqnarray}

It is now useful to rewrite Eq. \ref{gammamatrix} as
  \begin{eqnarray}
\frac{1}{2} {\bold 1 }    +{\bold \Omega }  = {\bold \Gamma } {\bold \Phi }^{-1} =   {\bold \Gamma } {\bold C}
 \label{diffplusomega}
\end{eqnarray}
in terms of the symmetric matrix $ {\bold C}= {\bold \Phi }^{-1} $.
The transpose of Eq. \ref{diffplusomega} reads using the symmetry of the matrix $ {\bold C } $ 
and the antisymmetry of the matrix ${\bold \Omega }  $ 
  \begin{eqnarray}
  \frac{1}{2} {\bold 1 }   -{\bold \Omega } = {\bold C }{\bold \Gamma }^T  
 \label{diffplusomegatrans}
\end{eqnarray}
The sum of Eqs \ref{diffplusomega} and \ref{diffplusomegatrans} allows to eliminate the matrix ${\bold \Omega } $
and to recover
Eq. \ref{eqCrecover} satisfied by the correlation matrix ${\bold C}$
  \begin{eqnarray}
  {\bold 1 } =  {\bold \Gamma } {\bold C } + {\bold C }{\bold \Gamma }^{T}  
 \label{feqlinearomegaphisum}
\end{eqnarray}
The difference between Eqs \ref{diffplusomega} and \ref{diffplusomegatrans} leads to the antisymmetric matrix ${\bold \Omega } $
  \begin{eqnarray}
{\bold \Omega } = \frac{1}{2} \left( {\bold \Gamma } {\bold C } - {\bold C }{\bold \Gamma }^T \right)   
 \label{omegafromphic}
\end{eqnarray}


\subsubsection{ Quantum Hamiltonian ${\hat {\cal  H}} $ in the irreversible gauge } 

\label{subsec_HirrOU}

 The properties of Eq. \ref{divforceirrevOU}
yields that the irreversible anti-Hermitian Hamiltonian of Eq. \ref{Hirr} and \ref{Hirrantih} reduce to
\begin{eqnarray}
 {\hat {\cal  H}}_{irr}  =     \vec f^{irr}( \vec x) . \vec \nabla 
   \label{HirrOU}
\end{eqnarray}
The parametrization of Eq. \ref{FirrevomegaOU} for the irreversible force $\vec f^{irr}( \vec x) $ yields using the antisymmetry $\Omega_{nm}=- \Omega_{mn} $
 \begin{eqnarray}
 {\hat {\cal  H}}_{irr}  && =  \sum_n  f^{irr}_n( \vec x) \partial_n  
 =  -2 \sum_n    \sum_m \Omega_{nm}  [\partial_m \phi(\vec x) ] \partial_n  
\nonumber \\
&& = - \sum_n    \sum_m \Omega_{nm}  [\partial_m \phi(\vec x) ] \partial_n 
 - \sum_m    \sum_n \Omega_{mn}  [\partial_n \phi(\vec x) ] \partial_m  
 \nonumber \\
&& =  \sum_n    \sum_m \Omega_{nm} \bigg( [\partial_n \phi(\vec x) ] \partial_m-  [\partial_m \phi(\vec x) ] \partial_n \bigg)    
  \label{HirrOUomega}
\end{eqnarray}

The difference computed from Eqs \ref{Qnsusyprod}
\begin{eqnarray}
 Q_n^{\dagger} Q_m-  Q_m^{\dagger} Q_n
  =
  - 2 (\partial_m \phi(\vec x) ) \partial_n 
+ 2 (\partial_n \phi(\vec x) )\partial_m 
\label{Qnsusycommutdiff}
\end{eqnarray}
allows to rewrite the irreversible Hamiltonian of Eq. \ref{HirrOUomega} using the antisymmetry of ${\bold \Omega}$ 
as
 \begin{eqnarray}
 {\hat {\cal  H}}_{irr} 
&&  =  \sum_n    \sum_m \frac{\Omega_{nm} }{2}   \bigg( Q_n^{\dagger} Q_m-  Q_m^{\dagger} Q_n \bigg)   
  =  \sum_n    \sum_m \Omega_{nm}   Q_n^{\dagger} Q_m
=  \begin{pmatrix} 
Q_1^{\dagger}  & ... & Q_N^{\dagger} 
 \end{pmatrix}  
{\mathbf \Omega} \begin{pmatrix} 
Q_1  \\
..
\\
Q_N 
 \end{pmatrix}  
  \label{HirrOUomegaq}
\end{eqnarray}
while the reversible Hamiltonian $ {\hat {\cal  H}}_{rev} $ is given by the diagonal form of Eq. \ref{hamiltonianhateqsusyQ}
\begin{eqnarray}
  {\hat {\cal  H}}_{rev} = \frac{  1 }{2 }  \sum_n  Q_n^{\dagger} Q_n   = \frac{  1 }{2 } \begin{pmatrix} 
Q_1^{\dagger}  & ... & Q_N^{\dagger} 
 \end{pmatrix}  
{\mathbf 1} \begin{pmatrix} 
Q_1  \\
..
\\
Q_N 
 \end{pmatrix} 
 \label{hamiltonianhateqsusyQOU}
\end{eqnarray}

Finally, the commutators of Eq. \ref{Qnsusycommut} are now constant and involve
the symmetric matrix $ \Phi_{nm} $ of Eq. \ref{PhiOU}
\begin{eqnarray}
[  Q_m, Q_n^{\dagger}]  =   2 (\partial_n\partial_m \phi(\vec x) ) =   \Phi_{nm}
\label{QnsusycommutPhi}
\end{eqnarray}

The diagonalization of the Hamiltonian ${\hat {\cal  H}}  $ in terms of canonical annihilation and creation operators
can be found in Appendix \ref{app_diagoHirrcanonique} to obtain the full relaxation spectrum of the Fokker-Planck generator.


\section{Quadratic trajectory observables of Ornstein-Uhlenbeck processes  }

\label{sec_OUadditive}

In this section, we focus on the special time-additive observables ${\cal O}[\vec x(0 \leq s \leq t) ] $ of 
Ornstein-Uhlenbeck processes, whose 
 generating functions $Z^{[k]}(\vec x,t \vert \vec y,0) $
are governed by quadratic Lagrangians ${\cal L}^{[k]} (\vec x(t), \dot {\vec x}(t) )  $ in Eq. \ref{lagrangiank}
and by quadratic Hamiltonians $ {\cal H}^{[k]} $ in Eq. \ref{FPhamiltonianp}.
Their large deviations properties are also discussed in detail in the recent PhD thesis \cite{duBuisson_thesis}.


\subsection{ Generating function $Z^{[k]}(\vec x,t \vert \vec y,0) $ associated to a quadratic Lagrangian and a quadratic Hamiltonian   } 
 
The generating function $Z^{[k]}(\vec x,t \vert \vec y,0) $
will be governed by a quadratic Lagrangian ${\cal L}^{[k]} (\vec x(t), \dot {\vec x}(t) )  $ in Eq. \ref{lagrangiank} and a quadratic Hamiltonian $ {\cal H}^{[k]} $ in Eq. \ref{FPhamiltonianp}
when the additive observable ${\cal O}[\vec x(0 \leq s \leq t) ] $ of Eq. \ref{additive}
satisfies the following two conditions :

(i) when the additional scalar potential $ V^{[\cal O]} (\vec x ) $ in the $k$-deformed scalar potential $V^{[k]}(\vec x) $ of Eq. \ref{scalarpotp} is also quadratic and parametrized by some symmetric matrix ${\bold W}^{[\cal O]} $
\begin{eqnarray}
V^{[\cal O]}(\vec x)  = \sum_{n=1}^N   \sum_{m=1}^N  \frac{W^{[\cal O]}_{nm}}{2} x_n  x_m
\label{scalarpotquadraticO}
\end{eqnarray}
Then the $k$-deformed scalar potential $V^{[k]}(\vec x) $ of Eq. \ref{scalarpotp} 
with respect to the potential $V ( \vec x) $ of Eq. \ref{scalarpotquadraticOU}
reads
\begin{eqnarray}
V^{[k]}(\vec x) \equiv V ( \vec x) + k V^{[{\cal O}]} ( \vec x)
&& =   - \frac{ 1  }{2} {\rm tr}({\bold \Gamma}) 
+\sum_{m=1}^N   \sum_{j=1}^N  \frac{\left({\bold \Gamma}^{T} {\bold \Gamma}\right)_{nm}+ k W^{[\cal O]}_{nm}}{2} x_m  x_j 
\nonumber \\
&& = V_0
+\sum_{m=1}^N   \sum_{j=1}^N  \frac{ W^{[k]}_{nm}}{2} x_m  x_j 
\label{scalarpotpO}
\end{eqnarray}
with the deformed matrix
\begin{eqnarray}
{\bold W}^{[k]} \equiv {\bold \Gamma}^{T} {\bold \Gamma} +k {\bold W}^{[\cal O]} 
\label{Wkdef}
\end{eqnarray}

(ii) when the additional vector potential $ \vec A^{[\cal O]}( \vec x)$ in the $k$-deformed scalar potential $\vec A^{[k]}(\vec x) $ of Eq. \ref{vectorpotp}
 is also linear and parametrized by some matrix ${\bold \Lambda}^{[\cal O]} $
  \begin{eqnarray}
  A_n^{[\cal O]} (\vec x )  = - \sum_{m=1}^N  \Lambda^{[\cal O]}_{nm} x_m
 \label{vectorpotlinearO}
\end{eqnarray}
Then the $k$-deformed vector potential  $\vec A^{[k]} (\vec x ) $ of Eq. \ref{vectorpotp}
with respect to the potential $\vec A ( \vec x) $ of Eq. \ref{vectorpotlinearOU}
reads
\begin{eqnarray}
 \vec A^{[k]} (\vec x ) 
 = \vec A( \vec x)+k \vec A^{[{\cal O}]}( \vec x)
 && = - \sum_{m=1}^N \left( \Gamma_{nm} + k \Lambda^{[\cal O]}_{nm}\right) x_m
 \nonumber \\
  && = - \sum_{m=1}^N  {\bold \Lambda}^{[k]}_{nm} x_m
 \label{vectorpotpO}
\end{eqnarray}
with the deformed matrix
\begin{eqnarray}
{\bold \Lambda}^{[k]} = {\bold \Gamma} +k {\bold \Lambda}^{[\cal O]} 
\label{Lambdakdef}
\end{eqnarray}
The corresponding supplementary magnetic antisymmetric matrix of Eq. \ref{magneticNpsmall} is space-independent
\begin{eqnarray}
 B^{[{\cal O}]}_{nm}  =- B^{[{\cal O}]}_{mn} 
  =  \Lambda^{[\cal O]}_{nm} - \Lambda^{[\cal O]}_{mn}
 \label{OUmagneticCalO}
\end{eqnarray}
 
In Appendix \ref{app_quadratic}, we describe the general framework
to compute the finite-time propagator associated to a quadratic Lagrangian
involving arbitrary matrices ${\bold W}$ and $\bold \Lambda$ in Eqs \ref{scalarpotquadratic} and \ref{vectorpotlinear}.
In particular, we mention how the analysis can be simplified by choosing a new gauge satisfying 
Eq. \ref{Leftlower}. In this section, the goal is thus to find the simplest gauge
that will simplify the study of the generating function $Z^{[k]}(\vec x,t \vert \vec y,0) $.


\subsection{ Gauge transformation between linear vector potentials for the generating function $Z^{[k]}(\vec x,t \vert \vec y,0) $    } 

Let us consider the gauge transformation from the initial linear vector potential $ A^{[k]}_n (\vec x ) $ of Eq. \ref{vectorpotpO}
\begin{eqnarray}
 A^{[k]}_n (\vec x )  = -\partial_n {\mathring \nu}^{[k]}(\vec x) +  {\mathring A}^{[k]}_n (\vec x ) 
 \label{gaugetransfolinearups}
\end{eqnarray}
towards the new linear vector potential ${\mathring A}^{[k]}_n (\vec x )  $ 
\begin{eqnarray}
  {\mathring A}^{[k]}_n (\vec x )  = -  \sum_{m=1}^N {\mathring {\bold \Lambda} }^{[k]}_{nm} x_m 
   \label{newgaugelinearupsk}
\end{eqnarray} 
parametrized by the new matrix ${\mathring {\bold \Lambda} }^{[k]} $, whose antisymmetric part 
 is fixed by the antisymmetric magnetic matrix ${\bold B}^{[k]}$ 
as computed from the antisymmetric part of the initial matrix ${\bold \Lambda}{[k]} $ of Eq. \ref{Lambdakdef}
  \begin{eqnarray}
{\mathring {\bold \Lambda} }^{[k]}    - ({\mathring {\bold \Lambda} }^{[k]})^T={\bold B}^{[k]} 
= {\bold \Lambda}^{[k]}    -  ({\bold \Lambda}^{[k]})^T
= {\bold \Gamma} - {\bold \Gamma}^T
+k \left( {\bold \Lambda}^{[\cal O]} - ({\bold \Lambda}^{[\cal O]})^T\right)
 \label{upsulontildeBtildek}
\end{eqnarray}
The quadratic function  ${\mathring \nu}^{[k]}(\vec x)$ obtained from Eq. \ref{gaugetransfolinearups}
\begin{eqnarray}
 {\mathring \nu}^{[k]}(\vec x) && 
 =  \sum_{n=1}^N \sum_{m=1}^N  
 \frac{(\Lambda^{[k]}_{nm}+\Lambda^{[k]}_{mn}) -  ({\mathring \Lambda}^{[k]}_{nm}+{\mathring \Lambda}^{[k]}_{mn})  }{4} x_n x_m = \langle \vec x \vert \bigg(  \frac{[{\bold \Lambda}^{[k]}+({\bold \Lambda}^{[k]})^T ]
 - [{\bold {\mathring \Lambda}}^{[k]}+({\bold {\mathring \Lambda}}^{[k]})^T]
  }{4} \bigg)  \vert \vec x \rangle
 \label{muquadraticupsk}
\end{eqnarray}
appears in the change of variables of the generating function 
\begin{eqnarray}
Z^{[k]}(\vec x,t \vert \vec y,0)
 = e^{ \displaystyle {\mathring \nu}^{[k]} ( \vec y) - {\mathring \nu}^{[k]} ( \vec x)} {\mathring Z}^{[k]}(\vec x,t \vert \vec y,0)
\label{changetowardsringk}
\end{eqnarray}
The new propagator ${\mathring Z}^{[k]}(\vec x,t \vert \vec y,0) $
\begin{eqnarray}
{\mathring Z}^{[k]}(\vec x,t \vert \vec y,0)
 = \int_{\vec x(\tau=0)=\vec y}^{\vec x(\tau=t)=\vec x} {\cal D}   \vec x(\tau)  
 e^{ - \displaystyle 
  \int_0^{t} d\tau {\mathring {\cal L}}^{[k]}  (\vec x(\tau), \dot {\vec x}(\tau) )
 } =  \langle \vec x \vert e^{-t {\mathring {\cal L}}^{[k]} } \vert \vec y \rangle
\label{pathintegralring}
\end{eqnarray}
is associated to the quantum mechanics in the same scalar potential $V^{[k]}(\vec x) $ of Eq. \ref{scalarpotpO}
but in the new vector potential ${\mathring A}^{[k]}_n (\vec x )  $,
with the corresponding new quadratic Lagrangian 
\begin{eqnarray}
&& {\mathring {\cal L}}^{[k]}  (\vec x(\tau), \dot {\vec x}(\tau) )
 =    \frac{1}{2}  \sum_{n=1}^N \dot  x_n^2  (\tau) 
 + \sum_{n=1}^N  \sum_{m=1}^N \dot  x_n  (\tau)     {\mathring \Lambda}_{nm}^{[k]} x_m  (\tau)
 + \sum_{n=1}^N   \sum_{m=1}^N  \frac{W_{nm}^{[k]}}{2} x_n  (\tau) x_m (\tau) +V_0
 \nonumber \\
&& =   \frac{1}{2}  \sum_{n=1}^N \left( \dot  x_n  (\tau) +  \sum_{m=1}^N      {\mathring \Lambda}^{[k]}_{nm} x_m  (\tau) \right)^2 - \frac{ 1  }{2} \sum_{n=1}^N{\mathring \Lambda}^{[k]}_{nn}
+ \sum_{n=1}^N   \sum_{m=1}^N  \frac{ \left( 
 {\bold W}^{[k]} - ({\mathring {\bold \Lambda} }^{[k]})^T {\mathring {\bold \Lambda} }^{[k]}   
\right)_{nm} }{2} x_n  (\tau) x_m (\tau)
  +\left[ V_0 + \frac{ 1  }{2} \sum_{n=1}^N{\mathring \Lambda}^{[k]}_{nn}\right]
  \nonumber \\
\label{lagrangianRingO}
\end{eqnarray}
and the corresponding new quadratic quantum Hamiltonian
\begin{eqnarray}
{\mathring {\cal H}}^{[k]}   &&  =
 -  \frac{1}{2}  \sum_{n=1}^N \left( \partial_n +   \sum_{m=1}^N {\mathring \Lambda}^{[k]}_{nm} x_m\right)^2    
+\sum_{n=1}^N   \sum_{n=1}^N  \frac{W_{nm}^{[k]}}{2} x_n  x_m + V_0
\nonumber \\
&&  = -    \sum_{n=1}^N \partial_n \left[ \frac{1}{2} \partial_n
+ \sum_{m=1}^N {\mathring \Lambda}^{[k]}_{nm} x_m \right]
+ \sum_{n=1}^N   \sum_{m=1}^N  \frac{ \left( 
 {\bold W}^{[k]} - ({\mathring {\bold \Lambda} }^{[k]})^T {\mathring {\bold \Lambda} }^{[k]}   
\right)_{nm} }{2} x_n  (\tau) x_m (\tau)
+ \left[ V_0 + \frac{ 1  }{2} \sum_{n=1}^N{\mathring \Lambda}^{[k]}_{nn} \right]
\label{FPhamiltonianQuantumZO}
\end{eqnarray}


\subsection{ Choice of the simplest gauge ${\mathring {\bold \Lambda} }^{[k]} $ to simplify the analysis of the generating function $Z^{[k]}(\vec x,t \vert \vec y,0) $    } 

\label{subsec_simplestgaugering}

The Lagrangian of Eq. \ref{lagrangianRingO}
and the hamiltonian of Eq. \ref{FPhamiltonianQuantumZO}
will be simpler if the symmetric part of the matrix ${\mathring {\bold \Lambda} }^{[k]} $ 
(since the antisymmetric part is fixed by Eq. \ref{upsulontildeBtildek})
can be chosen to satisfy the following equation involving the symmetric matrix $ {\bold W}^{[k]} $ of Eq. \ref{Wkdef}
 \begin{eqnarray}
    ({\mathring {\bold \Lambda} }^{[k]})^T {\mathring {\bold \Lambda} }^{[k]}  =  {\bold W}^{[k]}
    = {\bold \Gamma}^{T} {\bold \Gamma} +k {\bold W}^{[\cal O]} 
   \label{Leftlowerk}
\end{eqnarray} 
The discussion regarding the solution of this equation with the appropriate supplementary conditions is postponed to subsection \ref{subsec_SolvingSquareW}. Here, we assume that the matrix ${\mathring {\bold \Lambda} }^{[k]} $ satisfying Eq. \ref{Leftlowerk}
has been found and we discuss the consequences.
 
Using Eq. \ref{Leftlowerk}, the Lagrangian of Eq. \ref{lagrangianRingO} simplifies into
\begin{eqnarray}
&& {\mathring {\cal L}}^{[k]}  (\vec x(\tau), \dot {\vec x}(\tau) )
 =   \frac{1}{2}  \sum_{n=1}^N \left( \dot  x_n  (\tau) +  \sum_{m=1}^N      {\mathring \Lambda}^{[k]}_{nm} x_m  (\tau) \right)^2 - \frac{ 1  }{2} \sum_{n=1}^N{\mathring \Lambda}^{[k]}_{nn}
  +\left[ V_0 + \frac{ 1  }{2} \sum_{n=1}^N{\mathring \Lambda}^{[k]}_{nn}\right]
  \nonumber \\
\label{lagrangianRingOsimpli}
\end{eqnarray}
where up to the constant $\left[ V_0 + \frac{ 1  }{2} \sum_{n=1}^N{\mathring \Lambda}^{[k]}_{nn}\right] $, 
one recognizes the Lagrangian of Eq. \ref{lagrangianOU}
concerning the Fokker-Planck dynamics 
in the force parametrized by $ {\mathring {\bold \Lambda}}^{[k]}$ instead of ${\bold \Gamma}$.

Likewise, the quantum Hamiltonian of Eq. \ref{FPhamiltonianQuantumZO} reduces to
\begin{eqnarray}
{\mathring {\cal H}}^{[k]}      = -    \sum_{n=1}^N \partial_n \left[ \frac{1}{2} \partial_n
+ \sum_{m=1}^N {\mathring \Lambda}_{nm} x_m \right]
+ \left[ V_0 + \frac{ 1  }{2} \sum_{n=1}^N{\mathring \Lambda}_{nn} \right]
\label{FPhamiltonianQuantumZOsimpli}
\end{eqnarray}
where up to the constant $\left[ V_0 + \frac{ 1  }{2} \sum_{n=1}^N{\mathring \Lambda}^{[k]}_{nn}\right] $, one recognizes the Hamiltonian of Eq. \ref{FPhamiltonianQuantumOU}
concerning the Fokker-Planck dynamics in the force parametrized by $ {\mathring {\bold \Lambda}}^{[k]}$ instead of ${\bold \Gamma}$.

As a consequence, the propagator ${\mathring Z}^{[k]}(\vec x,t \vert \vec y,0) $ of Eq. \ref{pathintegralring}
can be rewritten with an exponential term involving this constant $\left[ V_0 + \frac{ 1  }{2} \sum_{n=1}^N{\mathring \Lambda}^{[k]}_{nn}\right] $ as
\begin{eqnarray}
{\mathring Z}^{[k]}(\vec x,t \vert \vec y,0)
 = 
 e^{ - \displaystyle   t \left[ V_0 + \frac{ 1  }{2} \sum_{n=1}^N{\mathring \Lambda}^{[k]}_{nn}\right] } 
 {\mathring P}(\vec x,t \vert \vec y,0)
\label{RingZtoP}
\end{eqnarray}
where ${\mathring P}^{[k]}(\vec x,t \vert \vec y,0) $
is the Fokker-Planck propagator in the force parametrized by $ {\mathring {\bold \Lambda}}^{[k]}$
\begin{eqnarray}
{\mathring P}^{[k]}(\vec x,t \vert \vec y,0)
 = 
  \int_{\vec x(\tau=0)=\vec y}^{\vec x(\tau=t)=\vec x} {\cal D}   \vec x(\tau)  
 e^{ - \displaystyle 
  \int_0^{t} d\tau \left[  \frac{1}{2}  \sum_{n=1}^N \left( \dot  x_n  (\tau) +  \sum_{m=1}^N      {\mathring \Lambda}^{[k]}_{nm} x_m  (\tau) \right)^2 - \frac{ 1  }{2} \sum_{n=1}^N{\mathring \Lambda}^{[k]}_{nn} \right]
 } 
\label{Pring}
\end{eqnarray}
So its analytical expression is given by the analog of Eq. \ref{pathintegralclassOUfinal}
\begin{eqnarray}
{\mathring P}^{[k]}(\vec x,t \vert \vec y,0)
 =  \frac{1 }{\sqrt{  (2 \pi)^N \det \left( \mathring {\bold C}^{[k]} (t) \right) }}  
e^{ - 
\frac{1}{2} \bigg (\langle \vec x   \vert  - \langle \vec y \vert e^{ - {\mathring ({\bold \Lambda} }^{[k]})^T t } \bigg)
\mathring {\bold \Phi}^{[k]} (t)
\bigg(\vert   \vec x \rangle - e^{ - {\mathring {\bold \Lambda}^{[k]} } t}    \vert  {\vec y}\rangle\bigg) 
  } 
\label{propagatorRing}
\end{eqnarray}
with the following notations.
The matrix 
\begin{eqnarray}
\mathring {\bold C}^{[k]} (t) \equiv \int_0^{t} ds
   e^{ - {\mathring {\bold \Lambda}^{[k]} } (t-s) } e^{ - ({\mathring {\bold \Lambda} }^{[k]})^T (t-s) }
\label{matrixCring}
\end{eqnarray}
satisfies the dynamics
\begin{eqnarray}
\frac{d \mathring {\bold C}^{[k]}(t) }{dt} = 
- {\mathring {\bold \Lambda} }^{[k]} {\mathring{\bold C}}^{[k]}(t) 
- {\mathring {\bold C}}^{[k]}(t)({\mathring {\bold \Lambda} }^{[k]})^T + \mathbf{1}
\label{langevinCring}
\end{eqnarray}
and has for inverse the matrix denoted by
\begin{eqnarray}
\mathring {\bold \Phi}^{[k]} (t) \equiv [\mathring {\bold C}^{[k]} (t)]^{-1}
\label{matrixPhiring}
\end{eqnarray}

In summary, putting together the gauge transformation of Eq. \ref{changetowardsringk}
and the change of variables of Eq. \ref{RingZtoP}, the initial generating function $Z^{[k]}(\vec x,t \vert \vec y,0) $
has been rewritten in terms of the Fokker-Planck propagator ${\mathring P}^{[k]}(\vec x,t \vert \vec y,0) $ of Eq. \ref{propagatorRing}
as
\begin{eqnarray}
Z^{[k]}(\vec x,t \vert \vec y,0) 
 = e^{ \displaystyle {\mathring \nu}^{[k]} ( \vec y) - {\mathring \nu}^{[k]} ( \vec x) -   t \left[ V_0 + \frac{ 1  }{2} \sum_{n=1}^N{\mathring \Lambda}^{[k]}_{nn}\right] } 
 {\mathring P}^{[k]}(\vec x,t \vert \vec y,0)
\label{ZtoPRing}
\end{eqnarray}


\subsection{ Supplementary condition on the gauge to simplify the large-time behavior of the generating function    }

To have good relaxation properties for Fokker-Planck propagator ${\mathring P}^{[k]}(\vec x,t \vert \vec y,0) $,
it is convenient to choose among the various solutions of Eq. \ref{Leftlowerk}, 
the solution ${\mathring {\bold \Lambda} }^{[k]} $
 whose $N$ eigenvalues $ {\mathring {\bold \lambda} }^{[k]}_{\alpha} $ have 
 strictly positive real parts as in Eq. \ref{conditionRealPartsGamma}
 \begin{eqnarray}
 {\rm Re} ( {\mathring {\bold \lambda} }^{[k]}_{\alpha} ) >0
 \label{conditionRealParts}
\end{eqnarray}
Then, for large time $t \to +\infty$, the propagator ${\mathring P}^{[k]}(\vec x,t \vert \vec y,0) $ of Eq. \ref{propagatorRing}
will converge 
 towards its steady state analog of Eq. \ref{steadyOU}
 \begin{eqnarray}
{\mathring P}^{[k]}(\vec x,t \vert \vec y,0) \opsimeq_{t \to +\infty} {\mathring P}^{[k]*}(\vec x)
 =  \frac{1 }{\sqrt{  (2 \pi)^N \det \left( \mathring {\bold C}^{[k]}  \right) }}  
e^{ - 
\frac{1}{2} \langle \vec x   \vert 
\mathring {\bold \Phi}^{[k]} 
\vert   \vec x \rangle 
  } 
\label{propagatorRingtinfty}
\end{eqnarray}
 in terms of the limiting matrix
$\mathring {\bold C}^{[k]}= \mathring {\bold C}^{[k]}(t=+\infty)$ satisfying
the time-independent version of Eq. \ref{langevinCring}
\begin{eqnarray}
0 = 
- {\mathring {\bold \Lambda} }^{[k]} {\mathring{\bold C}}^{[k]} 
- {\mathring {\bold C}}^{[k]}({\mathring {\bold \Lambda} }^{[k]})^T + \mathbf{1}
\label{langevinCringinfty}
\end{eqnarray}
and of its inverse $\mathring {\bold \Phi}^{[k]} =[\mathring {\bold C^{[k]}}]^{-1}$.

The convergence of the propagator in Eq. \ref{propagatorRingtinfty}
allows us to obtain the leading asymptotic behavior of Eq. \ref{changetowardsringP}
 for large time $t \to + \infty$
\begin{eqnarray}
Z^{[k]}(\vec x,t \vert \vec y,0) 
  \opsimeq_{t \to +\infty} e^{ \displaystyle {\mathring \nu}^{[k]} ( \vec y) - {\mathring \nu}^{[k]} ( \vec x) -   t \left[ V_0 + \frac{ 1  }{2} \sum_{n=1}^N{\mathring \Lambda}^{[k]}_{nn}\right] } 
{\mathring P}^{[k]*}(\vec x)
\label{changetowardsringP}
\end{eqnarray}
The identification with the asymptotic behavior 
written in terms of the ground-state energy $E(k)$ of the initial Hamiltonian ${\cal H} ^{[k]} $
\begin{eqnarray}
 Z^{[k]}(\vec x,t \vert \vec y,0)  = \langle \vec x \vert e^{-t {\cal H} ^{[k]} }\vert \vec y \rangle
\opsimeq_{t \to + \infty}  e^{-t E(k) }  r_k( \vec x ) l_k( \vec y )
 \label{psilimitappH}
\end{eqnarray}
yields the following consequences :

(i)  The ground-state energy $E(k)$ of the initial Hamiltonian ${\cal H} ^{[k]} $
 only involves the trace of the matrix ${\mathring {\bold \Lambda}}^{[k]} $
\begin{eqnarray}
  E(k)  = V_0 + \frac{ 1  }{2} \sum_{n=1}^N{\mathring \Lambda}^{[k]}_{nn}
  =   -\frac{ 1  }{2} {\rm tr } ( {\bold \Gamma} ) + \frac{ 1  }{2} {\rm tr } ({\mathring {\bold \Lambda}}^{[k]} )
 \label{EGSappH}
\end{eqnarray}
The rate function $I(o)$ governing the large deviations of the observable ${\cal O}[\vec x(0 \leq s \leq t) ] $
can be then obtained via the Legendre transform of Eq. \ref{legendrereci}.

(ii)  The corresponding positive left eigenvector $l_k( \vec y ) $
involves only the quadratic function ${\mathring \nu}^{[k]}( .) $ of Eq. \ref{muquadraticupsk}
which involves only the symmetric part of the matrix ${\mathring {\bold \Lambda}}^{[k]} $
\begin{eqnarray}
   l_k( \vec y ) && =e^{ \displaystyle {\mathring \nu}^{[k]} ( \vec y)  } 
 \label{lknu}
\end{eqnarray}
while the corresponding positive right eigenvector
\begin{eqnarray}
  r_k  ( \vec x ) && =e^{ \displaystyle  - {\mathring \nu}^{[k]} ( \vec x)  } 
 {\mathring P}^{[k]*}(\vec x)
 \label{rknu}
\end{eqnarray}
also involves the steady state ${\mathring P}^{[k]*}(\vec x) $ of Eq. \ref{propagatorRingtinfty}.


\subsection{ Link with subsection \ref{subsec_canonical} and summary of the conditions to determine the simplest gauge ${\mathring {\bold \Lambda} }^{[k]} $   } 

\label{subsec_SolvingSquareW}

In conclusion, for this example of quadratic observables of Ornstein-Uhlenbeck processes,
one recovers exactly a special case of the general formalism described in subsection \ref{subsec_canogauge}
in relation with the canonical conditioning of parameter $k$ described in subsection \ref{subsec_canonical}.
Let us summarize how the $N \times N$ real matrix ${\mathring {\bold \Lambda} }^{[k]} $
should be determined in practice and mention the link with the three conditions of subsection \ref{subsec_3conditions} :

(i) The condition of Eq. \ref{vectorpotptoringB} concerning the fixed magnetic matrix ${\mathring B}_{nm}^{[k]} (\vec x ) $
reduces in the present case to the condition of Eq. \ref{upsulontildeBtildek} fixing 
the antisymmetric part of the matrix ${\mathring {\bold \Upsilon} }^{[k]}$ 
with its $\frac{N(N-1)}{2} $ elements.
As a consequence, it is convenient to introduce the symmetric part ${\mathring {\bold \Upsilon} }^{[k]}=({\mathring {\bold \Upsilon} }^{[k]})^T $ with its $\frac{N(N+1)}{2} $ independent elements
 \begin{eqnarray}
{\mathring {\bold \Upsilon} }^{[k]} \equiv \frac{{\mathring {\bold \Lambda} }^{[k]}    + ({\mathring {\bold \Lambda} }^{[k]})^T}{2}
 \label{upsilonsym}
\end{eqnarray}
to rewrite the matrix ${\mathring {\bold \Lambda} }^{[k]}    $ and its transpose $({\mathring {\bold \Lambda} }^{[k]}   )^T $ as
 \begin{eqnarray}
{\mathring {\bold \Lambda} }^{[k]}   
&& =  {\mathring {\bold \Upsilon} }^{[k]}+ \frac{{\bold B}^{[k]}}{2}
\nonumber \\
 ({\mathring {\bold \Lambda} }^{[k]})^T
&& =   {\mathring {\bold \Upsilon} }^{[k]}- \frac{{\bold B}^{[k]}}{2}
  \label{symtrans}
\end{eqnarray}

(ii) The condition of Eq. \ref{eigenleftnu} reads for the present special case
concerning the linear vector potential $\vec {\mathring A}^{[k]}(\vec x) $ of Eq. \ref{newgaugelinearupsk}
and the quadratic scalar potential $V^{[k]}(\vec x) $ of Eq. \ref{scalarpotpO}
\begin{eqnarray}
E(k) && =  V^{[k]}(\vec x) - \frac{   [\vec \nabla . \vec {\mathring A}^{[k]}(\vec x) ] + [\vec {\mathring A}^{[k]}(\vec x) ]^2 }{2 } 
\nonumber \\
&& =\frac{ 1  }{2} {\rm tr}({\mathring {\bold \Lambda} }^{[k]})  - \frac{ 1  }{2} {\rm tr}({\bold \Gamma}) 
+\sum_{n=1}^N   \sum_{m=1}^N  \frac{ ({\bold W}^{[k]} - ({\mathring {\bold \Lambda} }^{[k]})^T {\mathring {\bold \Lambda} }^{[k]}  )_{nm} }{2} x_n  x_m 
 \label{eigenleftnuOU}
\end{eqnarray}
So the vanishing of the quadratic terms allows to recover the condition of Eq. \ref{Leftlowerk} satisfied by the simplest gauge ${\mathring {\bold \Lambda} }^{[k]} $, while the vanishing of the remaining terms corresponds to Eq. \ref{EGSappH}
for the ground-state energy $E(k)$.
Plugging the parametrization of Eq. \ref{symtrans}
into the condition of Eq. \ref{Leftlowerk}
 \begin{eqnarray}
   {\bold W}^{[k]} \equiv {\bold \Gamma}^{T} {\bold \Gamma} +k {\bold W}^{[\cal O]} 
 &&  =    ({\mathring {\bold \Lambda} }^{[k]})^T {\mathring {\bold \Lambda} }^{[k]}  
 = \left({\mathring {\bold \Upsilon} }^{[k]}- \frac{{\bold B}^{[k]}}{2} \right) \left({\mathring {\bold \Upsilon} }^{[k]}+ \frac{{\bold B}^{[k]}}{2} \right) 
  \nonumber \\
 && = \left({\mathring {\bold \Upsilon} }^{[k]} \right)^2  
 + \frac{{\mathring {\bold \Upsilon} }^{[k]}{\bold B}^{[k]}- {\bold B}^{[k]}{\mathring {\bold \Upsilon} }^{[k]}}{2}
 +  \frac{\left({\bold B}^{[k]}\right)^2}{4} 
   \label{Leftlowerksymtrans}
\end{eqnarray} 
yields $\frac{N(N+1)}{2} $ quadratic equations for the $\frac{N(N+1)}{2} $ elements of the symmetric matrix
${\mathring {\bold \Upsilon} }^{[k]} $ that should be computed in terms of 
the given symmetric matrix ${\bold W}^{[k]} $ and in terms of the given antisymmetric matrix ${\bold B}^{[k]} $.

(iii) Here the condition to have the convergence towards a steady state
means that among the various solutions of (ii), one should choose the matrix ${\mathring {\bold \Upsilon} }^{[k]} $
such that the corresponding matrix ${\mathring {\bold \Lambda} }^{[k]}   =  {\mathring {\bold \Upsilon} }^{[k]}+ \frac{{\bold B}^{[k]}}{2} $ of Eq. \ref{symtrans} displays $N$ eigenvalues with positive real parts, as discussed around Eq. \ref{conditionRealParts}.

The determination of the simplest gauge ${\mathring {\bold \Lambda} }^{[k]}  $ via these three conditions
 is described explicitly in the smallest dimension $N=2$
 for the general case in subsection \ref{subsec_ringN2} of Appendix \ref{app_quadratic}, 
 while the application to the Brownian gyrator model will be given in section \ref{sec_Gyr} of the main text.
 In the next subsection, we describe how the solution for $ {\mathring {\bold \Lambda} }^{[k]} $
 can be computed via a perturbative expansion in the parameter $k$ around $k=0$ in arbitrary dimension $N$.


\subsection{ Perturbative expansion in $k$ of the simplest gauge ${\mathring {\bold \Lambda} }^{[k]} $ 
to obtain the first cumulants    }

\subsubsection{ General form of the perturbative expansion in $k$ of the simplest gauge ${\mathring {\bold \Lambda} }^{[k]} $     }

For $k=0$, the solution of Eq. \ref{Leftlowerksymtrans} reduces to ${\mathring {\bold \Lambda} }^{[k=0]} = {\bold \Gamma} $.
The perturbative expansion of the symmetric matrix ${\mathring {\bold \Upsilon} }^{[k]} $
of Eq. \ref{upsilonsym} in the parameter $k$
involves corrections of all orders $k^q$ involving unknown symmetric matrices ${\mathring {\bold \Upsilon} }^{(q)} $
 \begin{eqnarray}
{\mathring {\bold \Upsilon} }^{[k]} \equiv \frac{{\mathring {\bold \Lambda} }^{[k]}    + ({\mathring {\bold \Lambda} }^{[k]})^T}{2}
=  \frac{ {\bold \Gamma} + {\bold \Gamma}^T}{2} 
+ k {\mathring {\bold \Upsilon} }^{(1)}
+ k^2 {\mathring {\bold \Upsilon} }^{(2)}+ ... + k^q {\mathring {\bold \Upsilon} }^{(q)} +  o(k^q)
 \label{symper}
\end{eqnarray}
while Eq. \ref{upsulontildeBtildek} fixes the antisymmetric part and thus its perturbative expansion in $k$ 
that only contains terms of order zero and one
  \begin{eqnarray}
\frac{{\mathring {\bold \Lambda} }^{[k]}    - ({\mathring {\bold \Lambda} }^{[k]})^T}{2}=\frac{{\bold B}^{[k]} }{2}
= \frac{ {\bold \Gamma} - {\bold \Gamma}^T}{2}+k \frac{{\bold B}^{[\cal O]} }{2}
 \label{upsulontildeBtildekper}
\end{eqnarray}
The corresponding perturbative expansions of the matrix ${\mathring {\bold \Lambda} }^{[k]} $ and of its transpose $({\mathring {\bold \Lambda} }^{[k]})^T $ read
 \begin{eqnarray}
{\mathring {\bold \Lambda} }^{[k]}   
&& =   {\bold \Gamma}  + k \left( {\mathring {\bold \Upsilon} }^{(1)} + \frac{{\bold B}^{[\cal O]} }{2}\right) + k^2 {\mathring {\bold \Upsilon} }^{(2)} + ... + k^q {\mathring {\bold \Upsilon} }^{(q)} +  o(k^q)
\nonumber \\
 ({\mathring {\bold \Lambda} }^{[k]})^T
&& =   {\bold \Gamma}^T + k \left( {\mathring {\bold \Upsilon} }^{(1)}-\frac{{\bold B}^{[\cal O]} }{2} \right) + k^2 {\mathring {\bold \Upsilon} }^{(2)} + ... + k^q {\mathring {\bold \Upsilon} }^{(q)} +  o(k^q)
  \label{sympertrans}
\end{eqnarray}


\subsubsection{ Solving the perturbative expansion up to second order     } 

\label{subsec_perturbation}

Let us now plug the perturbative expansions of Eq. \ref{sympertrans} at order $q=2$ 
 into Eq. \ref{Leftlowerksymtrans}
 \begin{eqnarray}
 && {\bold \Gamma}^{T} {\bold \Gamma} +k {\bold W}^{[\cal O]}  
  =   ({\mathring {\bold \Lambda} }^{[k]})^T {\mathring {\bold \Lambda} }^{[k]}  
  = \left[ {\bold \Gamma}^T + k \left( {\mathring {\bold \Upsilon} }^{(1)}-\frac{{\bold B}^{[\cal O]} }{2} \right) + k^2 {\mathring {\bold \Upsilon} }^{(2)} + o(k^2) \right] 
 \left[{\bold \Gamma}  + k \left( {\mathring {\bold \Upsilon} }^{(1)} + \frac{{\bold B}^{[\cal O]} }{2}\right) + k^2 {\mathring {\bold \Upsilon} }^{(2)} + o(k^2) \right] 
 \nonumber \\
 && =   {\bold \Gamma}^T{\bold \Gamma} 
  + k \left[  {\bold \Gamma}^T\left( {\mathring {\bold \Upsilon} }^{(1)} + \frac{{\bold B}^{[\cal O]} }{2}\right) 
  + \left( {\mathring {\bold \Upsilon} }^{(1)}-\frac{{\bold B}^{[\cal O]} }{2} \right) {\bold \Gamma} \right]
  + k^2 \left[ {\bold \Gamma}^T {\mathring {\bold \Upsilon} }^{(2)}  +  {\mathring {\bold \Upsilon} }^{(2)} {\bold \Gamma}
+ \left( {\mathring {\bold \Upsilon} }^{(1)}-\frac{{\bold B}^{[\cal O]} }{2} \right)  \left( {\mathring {\bold \Upsilon} }^{(1)} + \frac{{\bold B}^{[\cal O]} }{2}\right) 
  \right]  
 + o(k^2)  
  \nonumber \\
 && =   {\bold \Gamma}^T{\bold \Gamma} 
  + k \left[  {\bold \Gamma}^T {\mathring {\bold \Upsilon} }^{(1)} +  {\mathring {\bold \Upsilon} }^{(1)}{\bold \Gamma}
  + \frac{{\bold \Gamma}^T{\bold B}^{[\cal O]} - {\bold B}^{[\cal O]}{\bold \Gamma} }{2}  \right]
  + k^2 \left[ {\bold \Gamma}^T {\mathring {\bold \Upsilon} }^{(2)}  +  {\mathring {\bold \Upsilon} }^{(2)} {\bold \Gamma}
+\left( {\mathring {\bold \Upsilon} }^{(1)}\right)^2 - \frac{({\bold B}^{[\cal O]})^2 }{4}
+ \frac{{\mathring {\bold \Upsilon} }^{(1)}{\bold B}^{[\cal O]}-{\bold B}^{[\cal O]}{\mathring {\bold \Upsilon} }^{(1)} }{2}  
  \right]  
   \nonumber \\
 &&+ o(k^2)  
   \label{Leftlowerkper}
\end{eqnarray} 
At order $k$, this yields
the following linear equation for the first-order-correction matrix ${\mathring {\bold \Upsilon} }^{(1)} $
 \begin{eqnarray}
 {\bold \Gamma}^T {\mathring {\bold \Upsilon} }^{(1)} +  {\mathring {\bold \Upsilon} }^{(1)}{\bold \Gamma}
   =  {\bold W}^{[\cal O]}   + \frac{{\bold B}^{[\cal O]}{\bold \Gamma} - {\bold \Gamma}^T{\bold B}^{[\cal O]}  }{2}
 \label{eqk1}
\end{eqnarray}
Once the solution for ${\mathring {\bold \Upsilon} }^{(1)}  $ has been found, it can be plugged into 
Eq. \ref{Leftlowerkper}
 at order $k^2$ to obtain the following linear equation for the second-order-correction matrix ${\mathring {\bold \Upsilon} }^{(2)} $
 \begin{eqnarray}
{\bold \Gamma}^T {\mathring {\bold \Upsilon} }^{(2)}  +  {\mathring {\bold \Upsilon} }^{(2)} {\bold \Gamma}
= \frac{({\bold B}^{[\cal O]})^2 }{4}
+ \frac{{\bold B}^{[\cal O]}{\mathring {\bold \Upsilon} }^{(1)}
-{\mathring {\bold \Upsilon} }^{(1)}{\bold B}^{[\cal O]} }{2}  -\left( {\mathring {\bold \Upsilon} }^{(1)}\right)^2 
 \label{eqk2}
\end{eqnarray}

In practice, these equations for $  {\mathring {\bold \Upsilon} }^{(1)} $ and $ {\mathring {\bold \Upsilon} }^{(2)} $ can 
be solved in terms of the spectral decomposition of the matrix ${\bold \Gamma}$, as described in subsection \ref{subsec_upsilonper}
of Appendix \ref{App_explicitvisdiagoGamma}.


\subsubsection{ Consequences for the generating function $E(k)$ of the scaled cumulants   } 

\label{subsub_EkOUper}

As recalled around Eq. \ref{level1gen}, the energy $E(k)$ is the scaled cumulant generating function
corresponding to the Legendre transform of the rate function $ I ( o ) $.
Plugging the perturbative expansion for ${\mathring {\bold \Lambda} }^{[k]}   $ into Eq. \ref{EGSappH}
and using that the antisymmetric matrix ${\bold B}^{[\cal O]} $ has a vanishing trace
\begin{eqnarray}
  E(k) && = \frac{ 1  }{2} {\rm tr } ({\mathring {\bold \Lambda}}^{[k]}-{\bold \Gamma} )
 = \frac{ 1  }{2} {\rm tr } \left[  k \left( {\mathring {\bold \Upsilon} }^{(1)} + \frac{{\bold B}^{[\cal O]} }{2}\right) + k^2 {\mathring {\bold \Upsilon} }^{(2)} + ... + k^q {\mathring {\bold \Upsilon} }^{(q)} +  o(k^q)\right]
 \nonumber \\
&& = \frac{ k  }{2} {\rm tr }  \left( {\mathring {\bold \Upsilon} }^{(1)} \right)
+ \frac{ k^2  }{2} {\rm tr }  \left(  {\mathring {\bold \Upsilon} }^{(2)} \right)
 +...+ \frac{ k^q  }{2} {\rm tr } \left( {\mathring {\bold \Upsilon} }^{(q)} \right)+  o(k^q)
 \equiv k E^{(1)}+k^2 E^{(2)} +... +k^q E^{(q)}+  o(k^q)
 \label{EGSappHper}
\end{eqnarray}
one obtains that the coefficients $E^{(q)}$ of the perturbative expansion in $k$ of the energy $E(k)$ 
can be directly obtained from the traces of the corrections ${\mathring {\bold \Upsilon} }^{(q)} $ of Eq. \ref{symper}
\begin{eqnarray}
E^{(q)}  = \frac{ 1  }{2} {\rm tr } \left( {\mathring {\bold \Upsilon} }^{(q)} \right)
 \label{EGSappHperq}
\end{eqnarray}

To evaluate the first two traces for $q=1$ and $q=2$, it is convenient to multiply Eqs \ref{eqk1} and \ref{eqk2} by the matrix ${\bold C}$
and to compute the trace : one
can then use the cyclic property of the trace, and use Eq. \ref{eqCrecover} satisfied by the matrix ${\bold C}$
to obtain
 \begin{eqnarray}
 {\rm tr} \left[ {\bold C} \left(  {\bold W}^{[\cal O]}   + \frac{{\bold B}^{[\cal O]}{\bold \Gamma} - {\bold \Gamma}^T{\bold B}^{[\cal O]}  }{2} \right) \right]
&& ={\rm tr} \left[ {\bold C} \left(   {\bold \Gamma}^T {\mathring {\bold \Upsilon} }^{(1)} +  {\mathring {\bold \Upsilon} }^{(1)}{\bold \Gamma}\right) \right]
   ={\rm tr} \left[   {\bold C} {\bold \Gamma}^T {\mathring {\bold \Upsilon} }^{(1)} + {\bold C}  {\mathring {\bold \Upsilon} }^{(1)}{\bold \Gamma} \right]
   \nonumber \\
   && ={\rm tr} \left[  \left( {\bold C} {\bold \Gamma}^T +{\bold \Gamma} {\bold C} \right) {\mathring {\bold \Upsilon} }^{(1)} \right] = {\rm tr} \left[   {\mathring {\bold \Upsilon} }^{(1)} \right] = 2 E^{(1)} 
 \label{eqk1Ctr}
\end{eqnarray}
and similarly
 \begin{eqnarray}
 {\rm tr} \left[ {\bold C} \left(  \frac{({\bold B}^{[\cal O]})^2 }{4}
+ \frac{{\bold B}^{[\cal O]}{\mathring {\bold \Upsilon} }^{(1)}
-{\mathring {\bold \Upsilon} }^{(1)}{\bold B}^{[\cal O]} }{2}  -\left( {\mathring {\bold \Upsilon} }^{(1)}\right)^2 
\right) \right]
&& =  {\rm tr} \left[ {\bold C} \left( {\bold \Gamma}^T {\mathring {\bold \Upsilon} }^{(2)}  +  {\mathring {\bold \Upsilon} }^{(2)} {\bold \Gamma}\right) \right]
 \nonumber \\
   && ={\rm tr} \left[  \left( {\bold C} {\bold \Gamma}^T +{\bold \Gamma} {\bold C} \right) {\mathring {\bold \Upsilon} }^{(2)} \right] = {\rm tr} \left[   {\mathring {\bold \Upsilon} }^{(2)} \right] = 2 E^{(2)} 
 \label{eqk2Ctr}
\end{eqnarray}

For the second term on the left-hand side of Eq. \ref{eqk1Ctr}, one can use again 
the cyclic property of the trace, and use the Eq. \ref{omegafromphic} satisfied by the matrix ${\bold \Omega}$
to obtain
 \begin{eqnarray}
 {\rm tr} \left[ {\bold C}  \frac{{\bold B}^{[\cal O]}{\bold \Gamma} - {\bold \Gamma}^T{\bold B}^{[\cal O]}  }{2}  \right]
&& = \frac{1}{2} {\rm tr} \left[   {\bold C} {\bold B}^{[\cal O]}{\bold \Gamma} - {\bold C}{\bold \Gamma}^T{\bold B}^{[\cal O]}    \right]= \frac{1}{2} {\rm tr} \left[ \left( {\bold \Gamma} {\bold C} - {\bold C}{\bold \Gamma}^T \right) {\bold B}^{[\cal O]}    \right]
= {\rm tr} \left[ {\bold \Omega} {\bold B}^{[\cal O]}    \right]
 \label{eqk1Ctrleft}
\end{eqnarray}
Plugging this result into Eq. \ref{eqk1Ctr} yields that the first correction $E^{(1)}  $ reads
 \begin{eqnarray}
E^{(1)} 
= \frac{1}{2} {\rm tr} \left[ {\bold C}  {\bold W}^{[\cal O]}   +{\bold \Omega} {\bold B}^{[\cal O]}   \right]
 \label{eqk1Ctrfin}
\end{eqnarray}

Let us now check the consistency via $o^* = - E'(k=0) = - E^{(1)} $
of Eq. \ref{linktypperturbationEp}
with the steady value $o^*$ 
associated to the quadratic potential $V^{[{\cal O}]}(\vec x) $ of Eq. \ref{scalarpotquadraticO}
and the linear vector potential $ \vec A^{[\cal O]}( \vec x)$ of Eq. \ref{vectorpotlinearO}
\begin{eqnarray}
 o^* && = \int d^N \vec x  \left[ - P^*(\vec x)   V^{[{\cal O}]}(\vec x)  + \vec J^*(\vec x) . \vec A^{[{\cal O}]}( \vec x) \right]
 \nonumber \\
 && =   \int d^N \vec x  \left[  - P^*(\vec x)   \sum_{n=1}^N   \sum_{m=1}^N  \frac{W^{[\cal O]}_{nm}}{2} x_n  x_m 
 -  \sum_{n=1}^N J^*_n(\vec x)  \sum_{m=1}^N  \Lambda^{[\cal O]}_{nm} x_m 
  \right]
\label{g1OU}
\end{eqnarray}
The computation of these integrals with the Gaussian steady-state $P^*(\vec x) $ of Eq. \ref{steadyOU}
and with the steady-current $\vec J^*(\vec x) $ of Eq. \ref{jsteadyirrev}
that reads using the irreversible force of Eq. \ref{FirrevomegaOU}
\begin{eqnarray}
   J^*_n(\vec x)  =  P^*(\vec x )  f^{irr}_n( \vec x)  = - P^*(\vec x ) \sum_{l=1}^N   \left( {\bold \Omega}  {\bold \Phi}   \right)_{nl} x_l
 \label{jsteadyirrevOU}
\end{eqnarray}
yields that Eq. \ref{g1OU}
becomes
\begin{eqnarray}
 o^* && =  - \sum_{n=1}^N   \sum_{m=1}^N  \frac{W^{[\cal O]}_{nm}}{2} \int d^N \vec x  \left[   P^*(\vec x)   x_n  x_m \right]
 +\sum_{n=1}^N  \sum_{m=1}^N  \Lambda^{[\cal O]}_{nm} \sum_{l=1}^N  \left( {\bold \Omega}  {\bold \Phi}   \right)_{nl}
 \int d^N \vec x  \left[   
  P^*(\vec x )   x_l x_m
  \right]
  \nonumber \\
  && =  - \sum_{n=1}^N   \sum_{m=1}^N  \frac{W^{[\cal O]}_{nm}}{2} C_{mn}
 +\sum_{n=1}^N  \sum_{m=1}^N  \Lambda^{[\cal O]}_{nm} \sum_{l=1}^N  \left( {\bold \Omega}  {\bold \Phi}   \right)_{nl}
C_{lm}
  \nonumber \\
  && =  -  \frac{1}{2} {\rm tr} \left[  {\bold W}^{[\cal O]}   {\bold C}   \right]
   +\sum_{n=1}^N  \sum_{m=1}^N  \Lambda^{[\cal O]}_{nm}   \left( {\bold \Omega}  {\bold \Phi} {\bold C}  \right)_{nm}
\label{g1OUfin}
\end{eqnarray}
Using ${\bold \Phi}   =  {\bold C}^{-1} $ of Eq. \ref{PhiCinverse}  
and the antisymmetry of the matrix ${\bold \Omega}$,
the second term can be rewritten in terms of the magnetic matrix $B^{[{\cal O}]}_{nm}  
  =  \Lambda^{[\cal O]}_{nm} - \Lambda^{[\cal O]}_{mn} $ of Eq. \ref{OUmagneticCalO}
\begin{eqnarray}
 \sum_{n=1}^N  \sum_{m=1}^N  \Lambda^{[\cal O]}_{nm}   \left( {\bold \Omega}  {\bold \Phi} {\bold C}  \right)_{nm}
&& = \sum_{n=1}^N  \sum_{m=1}^N  \Lambda^{[\cal O]}_{nm}   {\bold \Omega}_{nm}
 =  \frac{1}{2}\sum_{n=1}^N  \sum_{m=1}^N  \left[ \Lambda^{[\cal O]}_{nm}   {\bold \Omega}_{nm}
 + \Lambda^{[\cal O]}_{mn}   {\bold \Omega}_{mn}
 \right]
  = - \frac{1}{2}\sum_{n=1}^N  \sum_{m=1}^N  \left[ \Lambda^{[\cal O]}_{nm}  
 - \Lambda^{[\cal O]}_{mn}  
 \right] {\bold \Omega}_{mn}
 \nonumber \\
 && = - \frac{1}{2}\sum_{n=1}^N  \sum_{m=1}^N  B^{[{\cal O}]}_{nm}  {\bold \Omega}_{mn}
=  \frac{1}{2} {\rm tr} \left[{\bold B}^{[\cal O]}  {\bold \Omega}   \right]
\label{g1OUfin2}
\end{eqnarray}
so that Eq. \ref{g1OUfin} is the opposite of $ E^{(1)} $ of Eq. \ref{eqk1Ctrfin} as it should for the consistency 
via Eq. \ref{linktypperturbationEp}
\begin{eqnarray}
 o^* && =  -  \frac{1}{2} {\rm tr} \left[  {\bold W}^{[\cal O]}   {\bold C}   \right]
  - \frac{1}{2} {\rm tr} \left[{\bold B}^{[\cal O]}  {\bold \Omega}   \right] = -  E^{(1)} 
\label{g1OUfinal}
\end{eqnarray}

To obtain the second cumulant via Eq. \ref{eqk2Ctr}
 \begin{eqnarray}
 E^{(2)} = \frac{1}{2} {\rm tr} \left[ {\bold C} \left(  \frac{({\bold B}^{[\cal O]})^2 }{4}
+ \frac{{\bold B}^{[\cal O]}{\mathring {\bold \Upsilon} }^{(1)}
-{\mathring {\bold \Upsilon} }^{(1)}{\bold B}^{[\cal O]} }{2}  -\left( {\mathring {\bold \Upsilon} }^{(1)}\right)^2 
\right) \right]
 \label{eqk2Ctrfin}
\end{eqnarray}
one needs to compute explicitly the correction ${\mathring {\bold \Upsilon} }^{(1)} $
via the method described in subsection \ref{subsec_upsilonper}
of Appendix \ref{App_explicitvisdiagoGamma}.


 \section{ Application to the Brownian gyrator in dimension $N=2$ }

 \label{sec_Gyr}
 
 The general framework described in the two previous sections for Ornstein-Uhlenbeck processes
  in arbitrary dimension $N$ is applied in the present section to the smallest dimension $N=2$,
 where the antisymmetric magnetic matrix $B_{nm}$ of Eq. \ref{magneticNcst}
 reduces to the single off-diagonal element $B_{12}=-B_{21}$.
 To be concrete, we focus on the illustrative example of the Brownian gyrator 
 that has attracted a lot of interest recently (see \cite{Gyr_vanWijland,Gyr_2013,Gyr_elec,Gyr_exp,Gyr_Tryphon_Harvesting,Gyr_Tryphon_Engine,Gyr_Tryphon_Geometry,Gyr_Inferring,Gyr_Inference,cerasoli} and references therein),
  to analyze some quadratic trajectory observables that have interesting physical meanings
  from the point of view of the stochastic thermodynamics of this nonequilibrium model.
Large deviations properties of many quadratic observables 
of various two-dimensional Ornstein-Uhlenbeck processes
   are studied in detail in the recent PhD thesis \cite{duBuisson_thesis}.


 \subsection{ Brownian gyrator in the initial coordinates $(X_1,X_2)$ and some interesting time-additive observables }

Let us consider the Brownian gyrator model based on the quadratic energy
\begin{eqnarray}
U( \vec X ) = \frac{X_1^2}{2} + \frac{X_2^2}{2} + u X_1 X_2
\label{energy}
\end{eqnarray}
where $u \ne 0$ parametrizes the coupling between the two degrees of freedom.
The Langevin system involves
 the linear forces corresponding to the derivatives of this quadratic potential $U (\vec X)$
\begin{eqnarray}
  F_1( \vec X ) = -\partial_1 U ( \vec X )= -  X_1- u X_2
  \nonumber \\
  F_2( \vec X ) = -\partial_2 U ( \vec x )= - u X_1- X_2
\label{forcelinear2}
\end{eqnarray}
as well as two different diffusion coefficients $D_1 \ne D_2$
representing two different temperatures that maintain the system out-of-equilibrium
\begin{eqnarray}
dX_1(t) && =    -  X_1(t) dt - u X_2(t) dt + \sqrt{2 D_1}  \ dw_1(t)
 \nonumber \\
dX_2(t) && =  -  u X_1(t) dt -  X_2(t) dt + \sqrt{2 D_2}  \ dw_2(t)
\label{langevinOU}
\end{eqnarray}

The goal is to analyze some time-additive observables that will characterize the irreversibility of the model.
For the trajectory $[\vec X(0 \leq s \leq t) ] $, the change of the energy $U( \vec X ) $ of Eq. \ref{energy}
between the initial position $\vec X (0) $ and the final position $\vec X (t) $ can be decomposed
\begin{eqnarray}
U( \vec X (t)) - U( \vec X (0)) &&= \int_0^t d\tau \partial_{\tau} U( {\vec X}(\tau))
 = \int_0^t d\tau {\dot {\vec X}} (\tau) . \vec \nabla U( {\vec X}(\tau))
=  \int_0^t d\tau {\dot {\vec X}} (\tau) . \left[ - \vec F( {\vec X}(\tau)) \right]
\nonumber \\
&& = {\cal Q}_1[\vec x(0 \leq s \leq t) ] + {\cal Q}_2[\vec x(0 \leq s \leq t) ]
\label{energychange}
\end{eqnarray}
as the sum of the two heats ${\cal Q}_n[\vec x(0 \leq s \leq t) ] $ received by the two degrees of freedom respectively
\begin{eqnarray}
{\cal Q}^{[X]}_1[\vec X(0 \leq s \leq t) ]&& =   \int_0^t d\tau {\dot  X}_1 (\tau) .\left[ - F_1( {\vec X}(\tau)) \right]
= \int_0^t d\tau {\dot  X}_1 (\tau) \left[ X_1(\tau) + u X_2(\tau)\right]
\nonumber \\
{\cal Q}^{[X]}_2[\vec X(0 \leq s \leq t) ]&& =   \int_0^t d\tau {\dot  X}_2 (\tau) .\left[ - F_2( {\vec X}(\tau)) \right]
= \int_0^t d\tau {\dot  X}_2 (\tau) \left[ u X_1(\tau) +  X_2(\tau)\right]
\label{heatsX}
\end{eqnarray}

As recalled in subsection \ref{subsec_entropyprod},
one of the most important time-additive observable is the entropy production of Eq. \ref{proddifftrajdef}
that characterizes the irreversibility at the level of stochastic trajectorires
\begin{eqnarray}
\Sigma^{[X]} [\vec X(0 \leq \tau \leq t)]  \equiv  
 \ln \left( \frac{ {\cal P}[\vec X(0 \leq \tau \leq t)] }{ {\cal P}[\vec X^R(0 \leq s \leq t)] } \right)
\label{proddifftrajdefX}
\end{eqnarray}
This definition, based of the logarithm of two probability trajectories, 
yields that it will remained unchanged after the rescaling described in the next subsection.


 \subsection{ Brownian gyrator in the rescaled coordinates $(x_1,x_2)$  }

\begin{figure}[h]
\centering
\includegraphics[width=6.5in,height=5.5in]{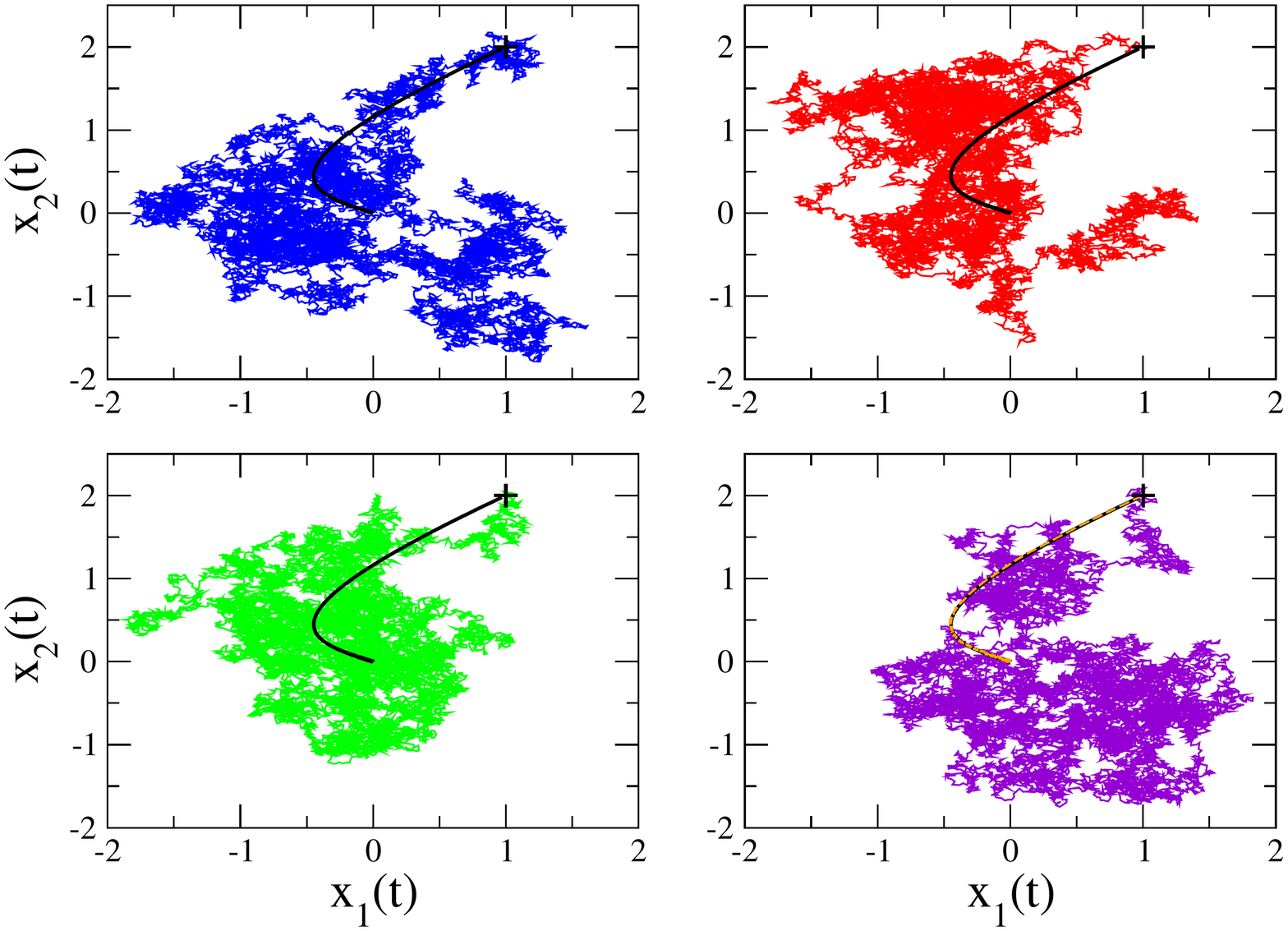}
\setlength{\abovecaptionskip}{15pt}  
\caption{Sample of diffusions satisfying the stochastic differential system Eq. \ref{langevinOUrx} on the interval $t\in[0,10]$ with the parameters $u=1/2$ and $\rho=2$. Each color corresponds to the realization of one  process. The thick black curve is the average value of the stochastic process as given by equation \ref{munclGYR} while the dashed orange curve (only shown in the lower right figure) corresponds to an average made on 10000 realizations. All processes start at $(y_1=1, y_2=2)$ (indicated by the black cross). The time step used in the discretization is $dt = 10^{-4}$. }
\label{fig1}
\end{figure}

The change of variables of Eq. \ref{bigXsmallx} described in Appendix \ref{app_rescaling}
corresponds to the simple rescaling involving the two diffusion coefficients $D_1 \ne D_2$
\begin{eqnarray}
x_1 (t) && = \frac{X_1(t)}{ \sqrt{ 2 D_1} }
\nonumber \\
x_2 (t) && = \frac{X_2(t)}{ \sqrt{ 2 D_2} }
 \label{Gyrrescal}
\end{eqnarray}
that will transform the Langevin system of Eq. \ref{langevinOU} into a system of the form of Eq. \ref{langevin}
\begin{eqnarray}
dx_1 (t) && =    -  x_1 (t) dt - u \sqrt{ \frac{D_2}{D_1} }  x_2(t) dt +   \ dw_1(t) = f_1(x_1(t),x_2(t)) dt +  dw_1(t) 
 \nonumber \\
dx_2 (t) && =  -  u \sqrt{ \frac{D_1}{D_2} } x_1 (t) dt -   x_2 (t) dt +   \ dw_2(t)=f_2(x_1(t),x_2(t)) dt +  dw_2(t) 
\label{langevinOUrx}
\end{eqnarray} 
The linear forces 
\begin{eqnarray}
 f_1(x_1,x_2)  && = - x_1- u \sqrt{ \frac{D_1}{D_2} }  x_2 = - x_1- u \rho  x_2
 \nonumber \\
f_2(x_1,x_2)  &&= -  \sqrt{ \frac{D_1}{D_2} } x_1-   x_2= -  \frac{u}{\rho} x_1-   x_2
\label{forcegyr}
\end{eqnarray}   
only involve the interaction $u$ and the new dimensionless parameter
 \begin{eqnarray}
\rho \equiv \sqrt{ \frac{D_2}{D_1} }
 \label{defrho}
\end{eqnarray}
The $2 \times 2$ matrix $\Gamma$ of Eq. \ref{forcelinear} 
  \begin{eqnarray}
 {\bold \Gamma }  
 \equiv \begin{pmatrix} 
1  & u \rho \\
\frac{u}{\rho} & 1
 \end{pmatrix}  
 \label{gammamatrixGYR}
\end{eqnarray}
yields that the constant magnetic field of Eq. \ref{magneticNcst}
\begin{eqnarray}
B= B_{12} =- B_{21}
    =  \Gamma_{12}-\Gamma_{21}
  = u \left( \rho - \frac{1}{\rho}\right)  
  = u \left( \sqrt{ \frac{D_2}{D_1} } - \sqrt{ \frac{D_1}{D_2} }\right)
  \label{BGyr}
\end{eqnarray}
is the relevant parameter for the irreversibility of the model : as expected,
it vanishes only if there is no interaction $u=0$ or if the two temperatures
given by the two diffusion coefficients are equal $D_1=D_2$. Figure \ref{fig1} shows some realizations of the stochastic process in the plane.

The explicit Gaussian forms for the finite-time propagator $ P ( \vec x , t \vert \vec y,0)$ 
and for the steady state $ P^* ( \vec x ) $ that exists for the interaction parameter $u$ in the interval $u \in ]-1,+1[$
are given in Eqs \ref{gaussinversegyr}
and \ref{steadygyr}
respectively.
Let us now focus on the time-additive observables that characterize the irreversibility.

\subsubsection{ Heats in the rescaled coordinates $(x_1,x_2)$ in terms of the stochastic area ${\cal A}[\vec x(0 \leq s \leq t) ] $ } 

The two heats of Eq. \ref{heatsX}
translate into the following time-additive observables for the rescaled trajectory $ \vec x(0 \leq s \leq t)$
\begin{eqnarray}
{\cal Q}_1[\vec x(0 \leq s \leq t) ]&& 
= 2 D_1 \int_0^t d\tau {\dot  x}_1 (\tau) \left[ x_1(\tau) + u \rho x_2(\tau)\right]
= 2 D_1 \int_0^t d\tau {\dot  x}_1 (\tau) \left[ - f_1( {\vec x}(\tau)) \right]
\nonumber \\
{\cal Q}_2[\vec x(0 \leq s \leq t) ]&& 
= 2 D_2 \int_0^t d\tau {\dot  x}_2 (\tau) \left[ \frac{u}{\rho} x_1(\tau) +  x_2(\tau)\right]
= 2 D_2 \int_0^t d\tau {\dot  x}_2 (\tau) \left[ - f_2( {\vec x}(\tau))\right]
\label{heats}
\end{eqnarray}
where one recognizes the force $\vec f(\vec x)$ of the rescaled Langevin system of Eq. \ref{langevinOUrx}.
They can be both rewritten in terms of 
the single stochastic area of Eq. \ref{areamn} existing in dimension $N=2$
\begin{eqnarray}
{\cal A}[\vec x(0 \leq s \leq t) ] && \equiv   {\cal A}_{12}[\vec x(0 \leq s \leq t) ]
 = - {\cal A}_{21}[\vec x(0 \leq s \leq t) ]
\nonumber \\
&&  =   \frac{1}{2} \int_0^{t} d\tau \left[ 
 x_1 (\tau) \dot x_2(\tau)   - \dot x_1(\tau)  x_2 (\tau)      \right] 
\label{area}
\end{eqnarray}
up to boundary terms as
\begin{eqnarray}
{\cal Q}_1[\vec x(0 \leq s \leq t) ]&& 
=  \left[ D_1 x_1^2(\tau) + D_1 u \rho x_1(\tau)x_2(\tau) \right]_{\tau=0}^{\tau=t}
- 2 D_1 u \rho {\cal A}[\vec x(0 \leq s \leq t) ]
\nonumber \\
{\cal Q}_2[\vec x(0 \leq s \leq t) ]&& 
=   \left[ D_2 x_2^2(\tau) + D_2 \frac{u}{\rho} x_1(\tau)x_2(\tau)\right]_{\tau=0}^{\tau=t}
+ 2 D_2 \frac{u}{\rho} {\cal A}[\vec x(0 \leq s \leq t) ]
\label{heatsarea}
\end{eqnarray}


\subsubsection{ Entropy production in terms of the heats ${\cal Q}_n[\vec x(0 \leq s \leq t) ] $ and in terms of the stochastic area ${\cal A}[\vec x(0 \leq s \leq t) ] $ }

The entropy production of Eq. \ref{proddifftrajdefX} translates into the entropy production
discussed around Eq. \ref{proddifftrajdef} and \ref{proddifftraj}
\begin{eqnarray}
\Sigma [\vec x(0 \leq \tau \leq t)] && \equiv  
 \ln \left( \frac{ {\cal P}[\vec x(0 \leq \tau \leq t)] }{ {\cal P}[\vec x^R(0 \leq s \leq t)] } \right)
 =  \int_0^{t} d\tau  \  \dot {\vec x}(\tau) . \left[    2 \vec f^{irr} (\vec x(\tau))    \right]
\label{SigmaGyr}
\end{eqnarray}
The link with the heats of Eqs \ref{area} can be obtained using the decomposition of the force into
its reversible and irreversible contributions of Eqs \ref{forceRev} \ref{forceIrrev}
\begin{eqnarray}
\Sigma [\vec x(0 \leq \tau \leq t)] && 
 =  \int_0^{t} d\tau  \  \dot {\vec x}(\tau) . \left[    2 \vec f (\vec x(\tau))  - 2 \vec f^{rev} (\vec x(\tau))  \right]
\nonumber \\
&&  =  \int_0^{t} d\tau  \  \dot x_1(\tau) . \left[    2  f_1 (\vec x(\tau))    \right]
+  \int_0^{t} d\tau  \  \dot x_2(\tau) . \left[    2  f_2 (\vec x(\tau))    \right]
 +  \int_0^{t} d\tau  \  \dot {\vec x}(\tau) . \left[      - \vec \nabla \ln P^*  (\vec x(\tau))  \right]
 \nonumber \\
&&  = - \frac{{\cal Q}_1[\vec x(0 \leq s \leq t) ]}{D_1}
- \frac{{\cal Q}_2[\vec x(0 \leq s \leq t) ]}{D_2}
 - \left[     \ln P^*  (\vec x(\tau))  \right]_{\tau=0}^{\tau=t}
\label{SigmaGyrheats}
\end{eqnarray}
As a consequence, the entropy production $\Sigma [\vec x(0 \leq \tau \leq t)] $ can also be rewritten in terms 
of the stochastic area of Eq. \ref{area}
up to boundary terms using Eqs \ref{heatsarea}
\begin{eqnarray}
\Sigma [\vec x(0 \leq \tau \leq t)] &&  = 
- \left[  x_1^2(\tau) + x_2^2(\tau) +  u \left( \rho + \frac{1}{\rho} \right) x_1(\tau)x_2(\tau) + \ln P^*  (\vec x(\tau)) \right]_{\tau=0}^{\tau=t}
 \nonumber \\ &&
 + 2  B {\cal A}[\vec x(0 \leq s \leq t) ]
\label{SigmaGyrarea}
\end{eqnarray}
where the prefactor $(2B)$ in front of the stochastic area involves the magnetic field of Eq. \ref{BGyr}, i.e. the irreversibility parameter of the model.


\subsubsection{ Intensive entropy production $\sigma[\vec x(0 \leq s \leq t) ] $ }

In the intensive entropy production of Eq. \ref{additiveIntensiveEntropy},
the boundary terms of the extensive entropy production of Eq. \ref{SigmaGyrarea} are of order $O\left(\frac{1}{t} \right)$ and can be neglected 
\begin{eqnarray}
\sigma[\vec x(0 \leq s \leq t) ] && \equiv \frac{ \Sigma[\vec x(0 \leq s \leq t) ] }{t} 
 =  2B \frac{{\cal A}[\vec x(0 \leq s \leq t) ]}{t}  + O\left(\frac{1}{t} \right)
\label{additiveIntensiveEntropyGyr}
\end{eqnarray}
So the study of the large deviations of the intensive entropy production $\sigma[\vec x(0 \leq s \leq t) ]$
is equivalent up to the constant prefactor $2 B$ to the study of large deviations of the intensive stochastic area
\begin{eqnarray}
a[\vec x(0 \leq s \leq t) ]  \equiv  \frac{{\cal A}[\vec x(0 \leq s \leq t) ]}{t}  
=  \frac{1}{2t } \int_0^{t} d\tau \left[ 
 x_1 (\tau) \dot x_2(\tau)   - \dot x_1(\tau)  x_2 (\tau)      \right] 
\label{intensiveareaGYR}
\end{eqnarray}

Similarly, Eq. \ref{heatsarea}
yields that the two intensive heats received by the two degrees of freedom respectively
are also directly proportional to the intensive area $a[\vec x(0 \leq s \leq t) ] $ using Eq. \ref{defrho}
to rewrite the prefactors
\begin{eqnarray}
q_1[\vec x(0 \leq s \leq t) ] \equiv \frac{{\cal Q}_1[\vec x(0 \leq s \leq t) ]}{t} && =   - 2 D_1 u \rho a[\vec x(0 \leq s \leq t) ] + O\left(\frac{1}{t} \right)
=  - 2  u \sqrt{D_1 D_2}  a[\vec x(0 \leq s \leq t) ] + O\left(\frac{1}{t} \right)
\nonumber \\
q_2[\vec x(0 \leq s \leq t) ] \equiv \frac{{\cal Q}_2[\vec x(0 \leq s \leq t) ]}{t}&& =  2 D_2 \frac{u}{\rho}  a[\vec x(0 \leq s \leq t) ] + O\left(\frac{1}{t} \right)
= 2  u \sqrt{D_1 D_2}  a[\vec x(0 \leq s \leq t) ] + O\left(\frac{1}{t} \right)
\label{intensiveheatsarea}
\end{eqnarray}
Therefore they are opposite as they should, 
since the energy cannot accumulate in the system during the long time interval $[0,t]$
\begin{eqnarray}
q_1[\vec x(0 \leq s \leq t) ]+q_2[\vec x(0 \leq s \leq t) ]  =  O\left(\frac{1}{t} \right)
\label{noaccumulation}
\end{eqnarray}
Consequently, Eq. \ref{SigmaGyrheats} yields that the intensive entropy production 
$\sigma [\vec x(0 \leq \tau \leq t)] $
is proportional to the intensive heat $q_1[\vec x(0 \leq s \leq t) ] $
\begin{eqnarray}
\sigma [\vec x(0 \leq \tau \leq t)] && 
  = - \frac{q_1[\vec x(0 \leq s \leq t) ]}{D_1}
- \frac{q_2[\vec x(0 \leq s \leq t) ]}{D_2}
 + O\left(\frac{1}{t} \right)
 = \left( \frac{1 }{D_2}- \frac{1 }{D_1}\right) q_1[\vec x(0 \leq s \leq t) + O\left(\frac{1}{t} \right)
\label{SigmaGyrheatsintense}
\end{eqnarray}
in agreement with the standard thermodynamic formula,
where the prefactor involves the difference  $ \left( \frac{1 }{D_2}- \frac{1 }{D_1}\right)$ of the inverses 
of the temperatures of the two reservoirs.

 
 \subsubsection{ Discussion } 

In the following, we will focus on the statistics of the stochastic area ${\cal A}[\vec x(0 \leq s \leq t) ] $,
since the two heats ${\cal Q}_{n=1,2}[\vec x(0 \leq s \leq t) ] $ and the entropy production ${ \Sigma}[\vec x(0 \leq s \leq t) ] $ can be then obtained via Eqs \ref{heatsarea} and \ref{SigmaGyrarea}.
Let us mention that the alternative method based on the Fourier-Matsubara decomposition 
to evaluate Gaussian functional integrals
has been used previously to compute the large deviations properties
via an integral over the frequency of the logarithm of the ratio of two determinants :

(i) for the entropy production for Ornstein-Uhlenbeck processes 
in dimension $N$ \cite{entropyProd_OUdimN};

(ii) for the heat ${\cal Q}_{n=2}[\vec x(0 \leq s \leq t) ] $  in the Brownian gyrator in dimension $N=2$ \cite{Gyr_vanWijland} for the special interaction parameter $u=-1$ in Eq. \ref{langevinOU}.


 \subsection{ Statistical properties of the stochastic area ${\cal A}[\vec x(0 \leq s \leq t) ] $    }

 
 \subsubsection{ Properties of the stochastic area ${\cal A}[\vec x(0 \leq s \leq t) ] $  as a time-additive observable  }

\begin{figure}[h]
\centering
\includegraphics[width=6.5in,height=5.5in]{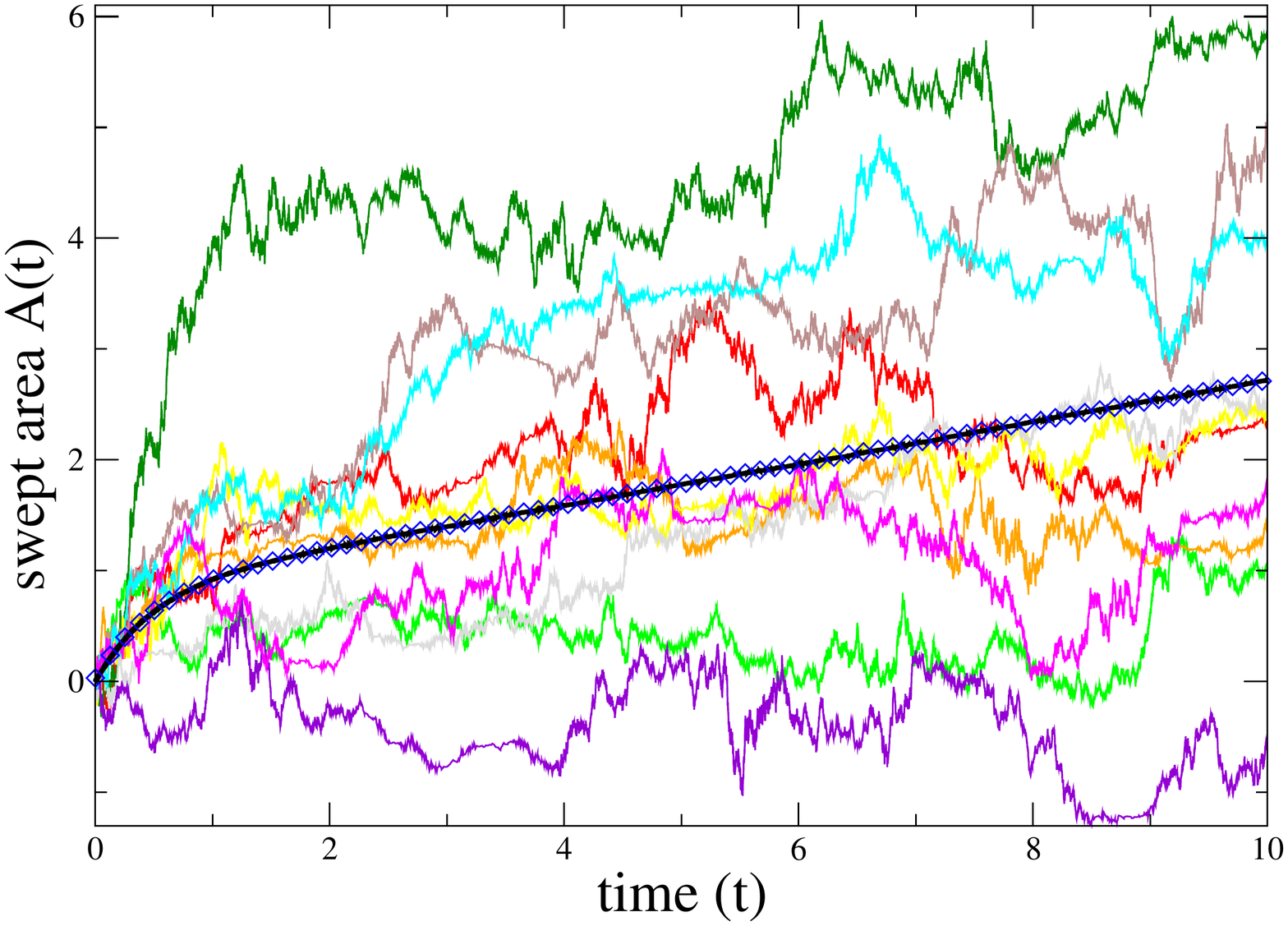}
\setlength{\abovecaptionskip}{15pt}  
\caption{Ten realizations of the stochastic area satisfying the stochastic differential equation Eq. \ref{additivediffelementaryarea} on the interval $t\in[0,10]$ with the parameters $u=1/2$ and $\rho=2$. Each color corresponds to the realization of one  process. The blue diamonds represent the average profile of the stochastic area as given by equation \ref{averageAreafinal}. The thick black curve which coincides with the theoretical average curve, corresponds to an average made on 10000 realizations. The time step used in the discretization is $dt = 10^{-4}$.}
\label{fig2}
\end{figure}

 The framework discussed in section \ref{sec_additive} can be applied to analyze the statistics of the 
 stochastic area ${\cal A}[\vec x(0 \leq s \leq t) ] $ of Eq. \ref{area}
 \begin{eqnarray}
{\cal A}[\vec x(0 \leq s \leq t) ] &&   =   \frac{1}{2} \int_0^{t} d\tau \left[ 
 x_1 (\tau) \dot x_2(\tau)   - \dot x_1(\tau)  x_2 (\tau)      \right] 
 =  \int_{0}^{t} d \tau \left[    \dot {\vec x} (\tau) . \vec A^{[{\cal A}]}( \vec x(\tau)) \right]
\label{areagyr}
\end{eqnarray}
which corresponds to the time-additive observable of the form of Eq. \ref{additive} with
\begin{eqnarray}
  A_1^{[{\cal A}]} (\vec x ) &&  = - \frac{x_2}{2}
  \nonumber \\
  A_2^{[{\cal A}]} (\vec x ) &&  =  \frac{x_1}{2}
   \label{vectorpotareagyr}
\end{eqnarray}
Since the divergence of $\vec A^{[{\cal A}]}( \vec x)  $ vanishes  $\vec \nabla . \vec A^{[{\cal A}]}( \vec x) = 0 $, the Ito and the Stratonovich forces coincide (Eq. \ref{forceIto}) in the Langevin stochastic differential equation of Eq. \ref{additivediffelementary}
that reads using the Langevin system of Eq. \ref{langevinOUrx}
\begin{eqnarray}
d{\cal A}(t) && =   \frac{x_1 (t)}{2}  dx_2(t) -  \frac{x_2 (t)}{2}  dx_1(t)
\nonumber \\
&&  = \frac{x_1 (t)}{2} \left[ - \left( \frac{u}{\rho} x_1(t)+   x_2(t) \right) dt  +dw_2(t) \right] -  \frac{x_2 (t)}{2} \left[ - \left( x_1(t)+ u \rho  x_2(t) \right) dt + dw_1(t) \right]
\nonumber \\
&& = \frac{u}{2} \left[  \rho  x_2^2(t)   -    \frac{x_1^2(t) }{\rho} \right] dt 
+ \frac{x_1 (t)}{2} dw_2(t) -  \frac{x_2 (t)}{2} dw_1(t)
\label{additivediffelementaryarea}
\end{eqnarray}
The solution of this stochastic differential equation is
\begin{eqnarray}
{\cal A}(t) = \frac{u}{2} \int_0^t \left[  \rho  x_2^2(s)   -    \frac{x_1^2(s) }{\rho} \right] ds 
+ \int_0^t \frac{x_1 (s)}{2} dw_2(s) -  \int_0^t  \frac{x_2 (s)}{2} dw_1(s)
\label{solutionadditivediffelementaryarea}
\end{eqnarray}
where $x_1(t)$ and $x_2(t)$ are given by Eq. \ref{langevinmatrixsol}. By plugging the expression of ${\bold \Gamma}$ for the Brownian gyrator, given by Eq. \ref{gammamatrixGYR}, in Eq. \ref{langevinmatrixsol}, we get the analytical expressions
\begin{eqnarray}
 x_1 (t) && =  e^{-t} \left[ \cosh(u t) y_1 - \rho \sinh(u t) y_2 \right] + \int_0^t e^{-(t-s)} \left[ \cosh(u (t-s)) dw_1(s) - \rho \sinh(u (t-s)) dw_2(s) \right]
 \nonumber \\
 x_2 (t) && =  e^{-t} \left[ \frac{-\sinh(u t)}{\rho} y_1 + \cosh(u t) y_2 \right] + \int_0^t e^{-(t-s)} \left[ \frac{-\sinh(u (t-s))}{\rho} dw_1(s) + \cosh(u (t-s)) dw_2(s) \right] 
\label{langevinOUrxsolution}
\end{eqnarray}
which can be used to compute the average value of the stochastic area ${\cal A}(t)$ of 
Eq. \ref{solutionadditivediffelementaryarea} as follows. 
First observe that

\begin{eqnarray}
  \overline{\int_0^t x_1 (s) dw_2(s)} &&  = \overline{\int_0^t   e^{-s} \left[ \cosh(u s) y_1 - \rho \sinh(u s) y_2 \right]dw_2(s)} \nonumber \\ 
  && + \overline{ \int_0^s e^{-(s-v)} \left[ \cosh(u (s-v)) dw_1(v) - \rho \sinh(u (s-v)) dw_2(v) \right] dw_2(s)  } \nonumber \\ 
  && = \int_0^t   e^{-s} \left[ \cosh(u s) y_1 - \rho \sinh(u s) y_2 \right]\overline{dw_2(s)} \nonumber \\ 
  && +  \int_0^s e^{-(s-v)} \cosh(u (s-v)) \overline{dw_1(v)dw_2(s)} - \int_0^s e^{-(s-v)}\rho \sinh(u (s-v)) \overline{dw_2(v)dw_2(s)}
  \label{calcul3integrals}
\end{eqnarray}
Since $w_1(t)$ and $w_2(t)$ are two independent Wiener processes, one can use $\overline{dw_2(t)} = 0=\overline{dw_1(s)dw_2(t)} $
 and $\overline{dw_2(s)dw_2(t)} = \delta(s-t) dt$ to obtain that the three contributions of Eq. \ref{calcul3integrals} vanish 
\begin{eqnarray}
  \overline{\int_0^t x_1 (s) dw_2(s)} = 0
\end{eqnarray}
Likewise the following integral vanishes 
  \begin{eqnarray}
  \overline{\int_0^t x_2 (s) dw_1(s)} = 0
\end{eqnarray}
so that the average value of the stochastic area ${\cal A}(t)$ of 
Eq. \ref{solutionadditivediffelementaryarea} reduces to
\begin{eqnarray}
 \overline{ {\cal A}(t)} = \frac{u}{2} \int_0^t \left[  \rho  \overline{ x_2^2(s)}   -    \frac{ \overline{x_1^2(s)} }{\rho} \right] ds 
\label{averageAreabegin}
\end{eqnarray}
The probability density function of ${(x_1,x_2)}$ is the Gaussian propagator given by Eq. \ref{gaussinversegyr}. So the values of $\overline{ x_1^2(t)}$ and $\overline{ x_2^2(t)}$ can be easily obtained. The calculations are tedious but straightforward, 
and the last integration over $s$ leads to the final expression
\begin{eqnarray}
 \overline{ {\cal A}(t)} = \frac{u}{8 \rho} e^{-2 t} \left[ -1 + 2 y_1^2 + (1 - 2 y_2^2) \rho^2 \right] + \frac{u}{8 \rho} \left[1 - 2 y_1^2 +( 2 y_2^2 -1) \rho^2 \right] + \frac{u}{4 \rho} (\rho^2 -1) t
\label{averageAreafinal}
\end{eqnarray}
Recall that $\rho = \sqrt{ \frac{D_2}{D_1} }$, so for large time $t \to +\infty$, the mean swept area increases linearly with $t$ when $D_2 > D_1$ and decreases linearly with t when $D_2 < D_1$. Figure \ref{fig2} shows some realizations of the stochastic swept area as well as its mean value.

The corresponding Fokker-Planck equation of Eq. \ref{fokkerplanckO}
 for the joint propagator $ P(\vec x,{\cal A},t \vert \vec y,0,0)   $  reads
 \begin{eqnarray}
  \partial_t P(\vec x,{\cal A},t \vert \vec y,0,0) && = 
      \left( \partial_{x_1}   - \frac{x_2}{2} \partial_{\cal A} \right)   \bigg[ \left(x_1+ u \rho  x_2 \right)
    P(\vec x,{\cal A},t \vert \vec y,0,0) \bigg]    
    +   \left( \partial_{x_2}  +  \frac{x_1}{2} \partial_{\cal A} \right)   \bigg[  \left( \frac{u}{\rho} x_1+   x_2\right) 
    P(\vec x,{\cal A},t \vert \vec y,0,0) \bigg]    
 \nonumber \\
 &&  + \frac{1}{2}   \left( \partial_{x_1}   - \frac{x_2}{2} \partial_{\cal A}  \right)^2
    P(\vec x,{\cal A},t \vert \vec y,0,0)     
+   \frac{1}{2}   \left( \partial_{x_2}  +  \frac{x_1}{2} \partial_{\cal A}  \right)^2
    P(\vec x,{\cal A},t \vert \vec y,0,0) 
\label{fokkerplanckjointpropaArea}
\end{eqnarray}
It is thus simpler to analyze the statistics of the stochastic area via its generating function as described below.

 
 \subsubsection{ Generating function $Z^{[k]}(\vec x,t \vert \vec y,0)  $ : Correspondence with a $2D$ quantum harmonic oscillator in a constant magnetic field    }

The generating function $Z^{[k]}(\vec x,t \vert \vec y,0)$ of Eq. \ref{genedef}
\begin{eqnarray}
 Z^{[k]}(\vec x,t \vert \vec y,0) 
&& \equiv \overline{ \delta^{(N)} \left( \vec x(t) - \vec x \right)  e^{k{\cal A} [\vec x(0 \leq \tau \leq t) ]}  \delta^{(N)} \left( \vec x(0) - \vec y \right) }
=
 \int_{\vec x(\tau=0)=\vec y}^{\vec x(\tau=t)=\vec x} {\cal D}   \vec x(\tau)  
 e^{ - \displaystyle 
{\cal L}^{[k]} (\vec x(t), \dot {\vec x}(t) )
 }
 \label{genearea}
\end{eqnarray}
 involves the $k$-deformed classical Lagrangian of Eq. \ref{lagrangiank}
\begin{eqnarray}
{\cal L}^{[k]} (\vec x(t), \dot {\vec x}(t) ) 
 = \frac{1}{2}  \dot {\vec x}^2  (t) 
 - \dot {\vec x } (t) . \vec A^{[k]}( \vec x(t))
 + V(\vec x)
\label{lagrangiankarea}
\end{eqnarray}
with the following vector and scalar potentials.

(i) The $k$-deformed vector potential  $\vec A^{[k]} (\vec x ) $ of Eq. \ref{vectorpotp} 
\begin{eqnarray}
  A^{[k]}_1 (\vec x ) && = f_1( \vec x) +k A_1^{[{\cal A}]}( \vec x) 
  = - x_1- \left( u \rho + \frac{k}{2} \right)x_2 
  \equiv - \Lambda_{11} x_1  - \Lambda_{12} x_2
    \nonumber \\
  A^{[k]}_2 (\vec x ) && = f_2( \vec x) +k A_2^{[{\cal A}]}( \vec x) =
  \left(\frac{k}{2} -  \frac{u}{\rho} \right) x_1-   x_2 
   \equiv - \Lambda_{21} x_1  - \Lambda_{22} x_2
 \label{vectorpotk}
\end{eqnarray}
is linear and parametrized by the matrix 
\begin{eqnarray}
{\bold \Lambda}^{[k]}  
 = \begin{pmatrix} 
1  & \left( u \rho + \frac{k}{2} \right) \\
\left(  \frac{u}{\rho} - \frac{k}{2} \right) & 1
 \end{pmatrix}   = {\bold \Gamma} +  \frac{k}{2}
 \begin{pmatrix} 
0  & 1 \\
-1 & 0
\end{pmatrix} 
 \label{upsilonarea}
\end{eqnarray}
The corresponding deformed magnetic field of Eq. \ref{magneticNp} with respect to 
the magnetic field $B$ of Eq. \ref{BGyr} reduces to
\begin{eqnarray}
 B^{[k]} = B^{[k]}_{12}  =-B^{[k]}_{21} = \partial_1 A^{[k]}_2 (\vec x ) -  \partial_2 A^{[k]}_1 (\vec x ) 
  =B + k 
 \label{magneticNpgyr}
\end{eqnarray}

(ii) The quadratic scalar potential of Eq. \ref{scalarpotquadraticOU} reads
\begin{eqnarray}
V(\vec x) &&  =  V_0+ \frac{W_{11}}{2} x_1^2+\frac{W_{22}}{2} x_2^2 + W_{12} x_1 x_2
\label{scalarpotquadraticOUgyr}
\end{eqnarray}
where the constant of Eq. \ref{W0def} reduces to
\begin{eqnarray}
 V_0  \equiv - \frac{ 1  }{2} (\Gamma_{11} +\Gamma_{22} ) =-1
\label{V0gyr}
\end{eqnarray}
while the symmetric matrix $\bold W$ reads
\begin{eqnarray}
{\bold W}   && =  {\bold \Gamma}^{T} {\bold \Gamma}
 = \begin{pmatrix} 
1  & \frac{u}{\rho} \\
u \rho & 1
 \end{pmatrix}   
 \begin{pmatrix} 
1  & u \rho \\
\frac{u}{\rho} & 1
 \end{pmatrix}  
 =  \begin{pmatrix} 
1 + \frac{u^2}{\rho^2} & u \left( \rho+\frac{1}{\rho} \right) \\
u \left( \rho+\frac{1}{\rho} \right) & 1 +u^2 \rho^2
 \end{pmatrix}  
\label{Vgyr}
\end{eqnarray}


 \subsubsection{ Choice of the simplest gauge ${\mathring {\bold \Lambda} }^{[k]} $ to simplify the analysis of the generating function $Z^{[k]}(\vec x,t \vert \vec y,0) $  }

In dimension $N=2$, the conditions to determine the simplest gauge ${\mathring {\bold \Lambda} }^{[k]} $
summarized in subsection \ref{subsec_SolvingSquareW} can be analyzed in the Pauli basis 
for $2 \times 2$ matrices as follows.

(i) The antisymmetric part of Eq. \ref{upsulontildeBtildek}
that only involves the magnetic field  $B^{[k]} =B^{[k]}_{12}  =-B^{[k]}_{21} =B + k $ of Eq. \ref{magneticNpgyr}
can be rewritten in terms of the Pauli matrix $\sigma_y$
  \begin{eqnarray}
{\mathring {\bold \Lambda} }^{[k]}    - ({\mathring {\bold \Lambda} }^{[k]})^T={\bold B}^{[k]} 
= \begin{pmatrix} 
0 & (B+k) \\
-(B+k) & 0
 \end{pmatrix} =i (B+k)  \begin{pmatrix} 
0 & -i \\
i & 0
 \end{pmatrix}  = i (B+k ) \sigma_y
 \label{upsulontildeBtildekgyr}
\end{eqnarray}

(ii)  The symmetric part of Eq. \ref{upsilonsym} can be decomposed in terms of the identity and the two Pauli matrices $(\sigma_x,\sigma_z)$
with three real coefficients $(g_0,g_x,g_z)$
\begin{eqnarray}
{\mathring {\bold \Upsilon} }^{[k]} \equiv \frac{{\mathring {\bold \Lambda} }^{[k]}    + ({\mathring {\bold \Lambda} }^{[k]})^T}{2}
= g_0 {\bold 1} + g_x  \sigma_x  + g_z  \sigma_z
 =  \begin{pmatrix} 
g_0+g_z  & g_x \\
g_x & g_0-g_z
 \end{pmatrix} 
 \label{upsilonsymgyr}
\end{eqnarray}

Using the notation
\begin{eqnarray}
g_y=\frac{B+k}{2}
 \label{gyk}
\end{eqnarray} 
one can put together (i) and (ii) 
to obtain the parametrization of Eq \ref{basePauli}
\begin{eqnarray}
 {\mathring {\bold \Lambda} } && = g_0 {\bold 1} + g_x  \sigma_x + g_y (i  \sigma_y) + g_z  \sigma_z
 =  \begin{pmatrix} 
g_0+g_z  & g_x+g_y \\
g_x-g_y & g_0-g_z
 \end{pmatrix}   
 \nonumber \\
 {\mathring {\bold \Lambda} }^T && = g_0 {\bold 1} + g_x \sigma_x - g_y (i \sigma_y) + g_z \sigma_z
 =  \begin{pmatrix} 
g_0+g_z  & g_x-g_y \\
g_x+g_y & g_0-g_z
 \end{pmatrix}    
 \label{basePaulitext}
\end{eqnarray}
One can now apply the analysis of subsection \ref{subsec_ringN2} of Appendix \ref{app_quadratic}
to the present case, where the matrix ${\bold W}$ is given by Eq. \ref{Vgyr},
with the following trace and determinant
\begin{eqnarray}
{\rm tr}({\bold W} )  && = 
2 + u^2 \left( \rho^2+\frac{1}{\rho^2} \right) 
\nonumber \\
 \det ({\bold W} ) && = [ \det ({\bold \Gamma} ) ]^2= (1-u^2)^2
\label{Vgyrtrdet}
\end{eqnarray}

The solution for the coefficient $g_0$ of Eq. \ref{coefidentityfactor2dordersolchoice} reads
 \begin{eqnarray}
g_0 = \sqrt{  \frac{  \frac{ {\rm tr}(\bold W) }{2} + \sqrt{\det (\bold W)  } }{2} - g_y^2 }
= \sqrt{\frac{  1+\frac{u^2}{2} \left( \rho^2+ \frac{1}{\rho^2} \right) + (1-u^2) }{2} - \frac{(B+k)^2}{4} }
 = \sqrt{ 1- \frac{Bk}{2}-\frac{k^2}{4}}
 \label{coefidentityfactor2dordersolchoicegyr}
\end{eqnarray}
with the corresponding solutions of Eq. \ref{basePaulieqWxzini} for $g_x$ and $g_z$ 
  \begin{eqnarray}
 g_x && = \frac{ g_0 W_{12}   -g_y \frac{W_{11}  - W_{22}}{2}  }{ \frac{ {\rm tr}(\bold W) }{2} \pm \sqrt{\det (\bold W)  } }
 = \frac{ u\left( \rho + \frac{1}{\rho}\right) \left[4  \sqrt{ 1- \frac{Bk}{2}-\frac{k^2}{4}}+ B(B+k)  \right]   }{ 2(4+B^2)}
\nonumber \\
 g_z && = \frac{g_0 \frac{W_{11}  - W_{22}}{2} + g_y W_{12}        }{ \frac{ {\rm tr}(\bold W) }{2} \pm \sqrt{\det (\bold W)  } }
 = \frac{u\left( \rho + \frac{1}{\rho}\right) \left[ B+k - B   \sqrt{ 1- \frac{Bk}{2}-\frac{k^2}{4}}
  \right]  }{4+B^2}
 \label{basePaulieqWxzinigyr}
\end{eqnarray}
The two eigenvalues of $ {\mathring {\bold \Lambda} }^{[k]} $ are given by Eq. \ref{eigenlambdaring}
  \begin{eqnarray}
{\mathring  \lambda_{\pm} }  = g_0 \pm \sqrt{ g_0^2- \sqrt{\det (\bold W)}}
 =  \sqrt{ 1- \frac{Bk}{2}-\frac{k^2}{4}}
\pm \sqrt{ u^2- \frac{Bk}{2}-\frac{k^2}{4} }
 \label{eigenlambdaringgyr}
\end{eqnarray}

Finally, the quadratic function ${\mathring \nu}^{[k]}(\vec x)$ of Eq. \ref{muquadraticupsk}
that allows us to perform the gauge transformation reads 
using ${\bold \Lambda}^{[k]}    $ of Eq. \ref{upsilonarea}
and ${\mathring {\bold \Upsilon} }^{[k]} $ of Eq. \ref{upsilonsym}
\begin{eqnarray}
 {\mathring \nu}^{[k]}(\vec x) && 
 =\frac{1}{2}  \langle \vec x \vert \bigg(  \frac{[{\bold \Lambda}^{[k]}+({\bold \Lambda}^{[k]})^T ]}{2}
- {\mathring {\bold \Upsilon} }^{[k]}   \bigg)  \vert \vec x \rangle
\nonumber \\
&& = \frac{1}{2}  \langle \vec x \vert 
 \begin{pmatrix} 
1-g_0-g_z  & \frac{u}{2}\left( \rho + \frac{1}{\rho}\right) -g_x \\
\frac{u}{2}\left( \rho + \frac{1}{\rho}\right)-g_x & 1-g_0+g_z
 \end{pmatrix} 
  \vert \vec x \rangle
  \nonumber \\
&& = (1-g_0-g_z ) \frac{x_1^2}{2} +(1-g_0+g_z ) \frac{x_2^2}{2} + \left[ \frac{u}{2}\left( \rho + \frac{1}{\rho}\right) -g_x\right] x_1 x_2
  \label{muquadraticupskgyr}
\end{eqnarray}

Let us now describe some consequences that can be derived from this explicit form of the simplest gauge ${\mathring {\bold \Lambda} }^{[k]} $.


\subsubsection{ Scaled cumulant generating function $E(k)$ and rate function $I(a)$ of the intensive stochastic area $a[\vec x(0 \leq s \leq t) ] $   } 

 The ground-state energy $E(k)$ of Eq. \ref{EGSappH}
 which only involves the trace of the matrix ${\mathring {\bold \Lambda}}^{[k]} $ of Eq. \ref{basePaulitext}
 can be directly obtained from the coefficient $g_0$ of Eq. \ref{coefidentityfactor2dordersolchoicegyr}
\begin{eqnarray}
  E(k)   =   -\frac{ 1  }{2} {\rm tr } ( {\bold \Gamma} ) + \frac{ 1  }{2} {\rm tr } ({\mathring {\bold \Lambda}}^{[k]} )
  = -1 + g_0 = -1 + \sqrt{ 1- \frac{Bk}{2}-\frac{k^2}{4}}
 \label{EGSappHgyr}
\end{eqnarray}
The rate function $I(a)$ governing the large deviations of the intensive stochastic area $a[\vec x(0 \leq s \leq t) ] $ 
of Eq. \ref{intensiveareaGYR}
can be then obtained via the Legendre transform of Eq. \ref{legendrereci}
 \begin{eqnarray}
  a  && =- E'(k) = \frac{ B+k}{ 4 \sqrt{ 1- \frac{Bk}{2}-\frac{k^2}{4}} }
 \nonumber \\
I(a) && = k a + E(k)  = k a -1 + \sqrt{ 1- \frac{Bk}{2}-\frac{k^2}{4}}
= k a -1 +\frac{ B+k}{ 4 a }
\label{legendrerecia}
\end{eqnarray} 
The first equation gives $a$ as a function of $k$,
and for $k=0$, one obtains the steady value $a^*$ of Eq. \ref{linktypperturbationEp} 
 \begin{eqnarray}
  a^*   =- E'(k=0) = \frac{ B}{ 4  }
\label{steadyagyr}
\end{eqnarray} 
Instead of calculating $k$ as a function of $a$, 
one can use the first equation to obtain the following second-order equation for $k$
 \begin{eqnarray}
0 && =\left( B+k   \right)^2 - \left(  4 a \sqrt{ 1- \frac{Bk}{2}-\frac{k^2}{4}} \right)^2  
\nonumber \\
&& = \left( 1+4 a^2\right) \left[  k^2 + 2B k + \frac{  B^2 - 16 a^2 }{1+4 a^2} \right]
\label{invertka}
\end{eqnarray} 
with the two solutions
 \begin{eqnarray}
k_{\pm} = - B \pm 2a \sqrt{ \frac{B^2+4}{1+4 a^2} }
\label{invertkpma}
\end{eqnarray} 
The first equation of Eq. \ref{legendrerecia} means that $a$ and $(B+k)$ should have the same sign,
so the appropriate solution is $k_+$
 \begin{eqnarray}
k(a)= k_+ = - B + 2a \sqrt{ \frac{B^2+4}{1+4 a^2} }
\label{invertkpa}
\end{eqnarray} 
which can now be plugged into the second equation of Eq. \ref{legendrerecia}
to obtain the explicit rate function
 \begin{eqnarray}
I(a) && = k a -1 +\frac{ B+k}{ 4 a }
=    - Ba + 2a^2 \sqrt{ \frac{B^2+4}{1+4 a^2} }
-1 +\frac{1}{2}  \sqrt{ \frac{B^2+4}{1+4 a^2} }
 =   - Ba -1  +\frac{1}{2} (1+4 a^2)  \sqrt{ \frac{B^2+4}{1+4 a^2} }
\nonumber \\
&&=  \frac{\sqrt{ (B^2+4) (1+4 a^2 ) } - 2 (Ba+1)}{2}  
=  \frac{ (B-4a)^2}{2 \left[ \sqrt{ (B^2+4) (1+4 a^2 ) } + 2 (Ba+1)\right] }
\label{rateagyra}
\end{eqnarray} 
which vanishes and is minimum at the steady value $a^*=\frac{B}{4}$ of Eq. \ref{steadyagyr}
as it should (Eq. \ref{iaeqvanish}).

For large $a \to \pm \infty$, the corresponding values of $k(a)$ in Eq. \ref{invertkpa} remain finite
 \begin{eqnarray}
k(a=\pm \infty)=  -B \pm \sqrt{B^2+4}
\label{invertkpainfty}
\end{eqnarray} 
while the rate function $I(a)$ of Eq. \ref{rateagyra} displays the linear asymptotic behaviors
 \begin{eqnarray}
I(a) && \opsimeq_{a \to + \infty} a k(+\infty) = a \left[ \sqrt{B^2+4} - B\right]
\nonumber \\
I(a) && \opsimeq_{a \to - \infty} a k(-\infty) = (-a)  \left[ \sqrt{B^2+4} + B\right]
\label{legendrereciainfty}
\end{eqnarray} 
in agreement with the fact that the scaled cumulant generating function of Eq. \ref{EGSappHgyr}
is real only on the interval $k \in [k(-\infty),k(+\infty)]$
\begin{eqnarray}
  E(k)  = -1 +\frac{1}{2}  \sqrt{ 4 - 2 Bk- k^2} = -1 +\frac{1}{2}  \sqrt{ [k(+\infty) -k ] [k-k(-\infty) ] } 
 \label{EGSappHgyrinterval}
\end{eqnarray}

\begin{figure}[h]
\centering
\includegraphics[width=6.2in,height=3.2in]{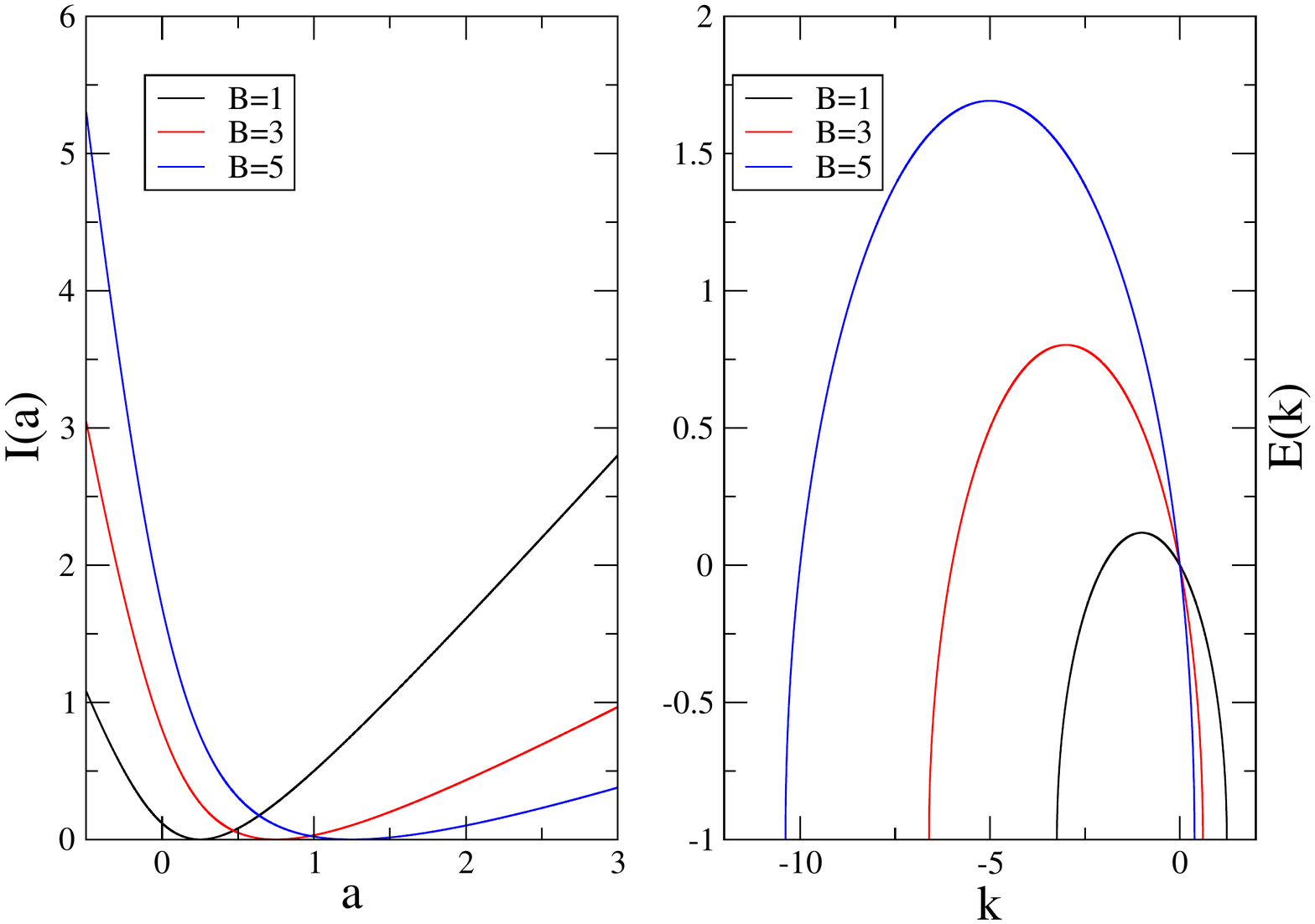}
\setlength{\abovecaptionskip}{15pt}  
\caption{Rate function $I(a)$ given by Eq. \ref{rateagyra} and scaled cumulant generating function 
$E(k)$ given by Eq. \ref{EGSappHgyrinterval} of the Brownian gyrator for various values of the magnetic field $B$.}
\label{fig3}
\end{figure}

Finally, let us mention the appropriate translation 
of the Gallavotti-Cohen symmetry of Eq. \ref{GCrate}
via $\sigma=2B a$ of Eq. \ref{additiveIntensiveEntropyGyr} : 
the difference between the rate function $I(a)$ of Eq. \ref{legendrerecia}
at opposite arguments $(\pm a)$ reduces to
 \begin{eqnarray}
I(a) -I(-a)  =   - 2 Ba 
\label{legendrereciaGC}
\end{eqnarray} 
while the corresponding invariance of the scaled cumulant generating function 
$E(k)$ of Eq. \ref{EGSappHgyrinterval}
is 
\begin{eqnarray}
E(k)   = E(k'=-2B-k)
\label{GCpa}
\end{eqnarray}
Figure \ref{fig3} shows the rate function and the scaled cumulant generating function for several values of the magnetic field $B$.

\subsubsection{ Canonical conditioning for large time $t$   }

As explained around Eq. \ref{forcesupforwardktinfty}, the 
additional force $f_n^{Cond[k]Interior}(\vec z) $ of the canonical conditioning of parameter $k$
involves the vector potential $ A^{[{\cal A}]}_n(\vec x)$ of Eq. \ref{vectorpotareagyr}
and the derivatives of the quadratic function ${\mathring \nu}^{[k]} ( \vec x) = \ln \left[  l_k(\vec z)  \right]$ of Eq. \ref{muquadraticupskgyr}
\begin{eqnarray}
f_1^{Cond[k]Interior}(\vec x) && =  k  A^{[{\cal A}]}_1(\vec x)            + \partial_{x_1}{\mathring \nu}^{[k]} ( \vec x)
=    \left[ 1-g_0-g_z \right] x_1 + \left[ \frac{u}{2}\left( \rho + \frac{1}{\rho}\right) -g_x -  \frac{k}{2} \right]  x_2
\nonumber \\
f_2^{Cond[k]Interior}(\vec x) && =  k  A^{[{\cal A}]}_2(\vec x)            + \partial_{x_2}{\mathring \nu}^{[k]} ( \vec x)
=            \left[ \frac{u}{2}\left( \rho + \frac{1}{\rho}\right) -g_x + \frac{k}{2}\right] x_1 + \left[1-g_0+g_z \right] x_2 
\label{forcesupforwardkringinteriorarea}
\end{eqnarray}
where one can plug the explicit expressions of the coefficients $(g_0,g_x,g_z)$ of Eqs \ref{coefidentityfactor2dordersolchoicegyr} and \ref{basePaulieqWxzinigyr} to obtain
the final result

\begin{eqnarray}
f_1^{Cond[k]Interior}(\vec x) && 
=    \left[ 1-\sqrt{ 1- \frac{Bk}{2}-\frac{k^2}{4}}-\frac{u\left( \rho + \frac{1}{\rho}\right) \left[ B+k - B   \sqrt{ 1- \frac{Bk}{2}-\frac{k^2}{4}}  \right]  }{4+B^2} \right] x_1
 \nonumber \\ && 
   + \left[ \frac{u}{2}\left( \rho + \frac{1}{\rho}\right) - \frac{ u\left( \rho + \frac{1}{\rho}\right) \left[4  \sqrt{ 1- \frac{Bk}{2}-\frac{k^2}{4}}+ B(B+k)  \right]   }{ 2(4+B^2)} -  \frac{k}{2} \right]  x_2
\nonumber \\
f_2^{Cond[k]Interior}(\vec x) && 
=            \left[ \frac{u}{2}\left( \rho + \frac{1}{\rho}\right) - \frac{ u\left( \rho + \frac{1}{\rho}\right) \left[4  \sqrt{ 1- \frac{Bk}{2}-\frac{k^2}{4}}+ B(B+k)  \right]   }{ 2(4+B^2)} + \frac{k}{2}\right] x_1 
 \nonumber \\ &&
 + \left[1-\sqrt{ 1- \frac{Bk}{2}-\frac{k^2}{4}}+\frac{u\left( \rho + \frac{1}{\rho}\right) \left[ B+k - B   \sqrt{ 1- \frac{Bk}{2}-\frac{k^2}{4}}  \right]  }{4+B^2} \right] x_2 
\label{forcesupforwardkringinteriorareaexpli}
\end{eqnarray}

As recalled after Eq. \ref{markovconddiffderiforwardktinfty},
this canonical conditioning of parameter $k$
is asymptotically equivalent for large time $t \to + \infty$ to the microcanonical conditioning 
on the intensive stochastic area $a[\vec x(0 \leq s \leq t) ]$ of Eq. \ref{intensiveareaGYR}
at the corresponding Legendre value $a=- E'(k)$ of Eq. \ref{legendrerecia}.


\section{ Conclusion  }

\label{sec_conclusion}

In summary, we have revisited the nonequilibrium diffusion processes in dimension $N$ via the correspondence with the non-Hermitian quantum mechanics in a real scalar potential $V(\vec x)$ and in a purely imaginary vector potential of real amplitude $\vec A(\vec x)$, to get a more intuitive understanding of nonequilibrium diffusion processes, 
as well as more efficient computation tools to analyze the properties of their trajectory observables.

We have first stressed the role of the $\frac{N(N-1)}{2}$ magnetic matrix elements $B_{nm}(\vec x )  =-B_{mn} (\vec x ) = \partial_n A_m (\vec x ) -  \partial_m A_n (\vec x )$ as the relevant parameters of irreversibility and we have emphasized the advantages of making the gauge transformation
 of the vector potential $\vec A(\vec x) $ based on the decomposition of the force into its reversible and irreversible components.
 
We have then explained why this perspective is even more useful to analyze the generating functions of all the time-additive observables of stochastic trajectories. Our main conclusion is that the computation of large deviations rate functions and the construction of the corresponding Doob conditioned processes can be drastically simplified via the choice of an appropriate gauge for each purpose.

We have then applied this general framework to study the special time-additive observables of Ornstein-Uhlenbeck trajectories, whose generating functions correspond to quantum propagators involving quadratic scalar potential and linear vector potentials, i.e. to quantum harmonic oscillators in constant magnetic matrices.  Finally we have focused on the specific example of the Brownian gyrator in dimension $N=2$ to compute the large deviations properties of the entropy production of its stochastic trajectories and to construct the corresponding conditioned processes having a given value of the entropy production per unit time.


\appendix

\section{ Reminder on the classical and quantum mechanics in an electromagnetic potential $(V,\vec A)$  }

\label{app_electromagnetic}

In this Appendix, we recall the standard definitions for the classical and quantum mechanics
for a particle of unit mass and unit charge 
submitted to the scalar potential $V$ and to the vector potential  $ \vec A $.
Here the 'true time' will be denoted by $\theta $ to emphasize the difference with the main text based
on the Euclidean time $t$ obtained by the Wick rotation
\begin{eqnarray}
t= i \theta
\label{Euclideantime}
\end{eqnarray}

\subsection{ Lagrangian perspective }

\subsubsection{ Classical mechanics via the Lagrangian perspective  }

The Lagrangian $L \left (\vec q(\theta), \dot {\vec q}(\theta) \right) $ is 
a function of the position $\vec q (\theta)$ 
and of the velocity $\dot {\vec q}(\theta)=\frac{d \vec q (\theta)}{d\theta} $
that is quadratic with respect to the velocity $\dot {\vec q}(\theta) $
\begin{eqnarray}
L \left (\vec q(\theta), \dot {\vec q}(\theta) \right) = 
\frac{1}{2}\dot {\vec q  }{\ }^2(\theta) 
+\dot {\vec q}(\theta) .\vec A (\vec q (\theta)) 
  - V (\vec q(\theta))
\label{lagrangiantheta}
\end{eqnarray}
while the dependence with respect to the position $ {\vec q}(\theta) $ is contained  
in the vector potential $ \vec A (\vec q (\theta)) $
and in the scalar potential $V (\vec q(\theta)) $.
The classical Lagrange equations of motion 
\begin{eqnarray}
 \frac{d}{d\theta} \left(\frac{\partial L (\vec q(\theta), \dot {\vec q} (\theta) ) } {\partial \dot {\vec q} (\theta)} \right) 
 = \frac{\partial L (\vec q (\theta), \dot {\vec q}  (\theta)) } {\partial \vec q (\theta)} 
\label{lagrangemotiontheta}
\end{eqnarray}
then determine the acceleration $\ddot {\vec q} (\theta)$ in terms of the position 
$\vec q(\theta) $ and of the velocity $\dot {\vec q} (\theta)$.

\subsubsection{ Quantum mechanics via the Feynman path-integral involving the Lagrangian }

In the Feynman perspective of quantum mechanics \cite{feynman},
the amplitude $\psi( \vec q, \Theta \vert \vec q_0, 0 ) $ to go from the initial position $\vec q_0$ at time $\theta=0$
to the final position $\vec q$ at time $\Theta$
can be written as an integral over the paths $\vec q(0 \leq \theta \leq \Theta)$
of a complex phase factor that can be computed from the Lagrangian of Eq. \ref{lagrangiantheta}
\begin{eqnarray}
\psi( \vec q, \Theta \vert \vec q_0, 0 )  && = \int_{\vec q(0)=\vec q_0}^{\vec q(\Theta)=\vec q} {\cal D}(\vec q(\theta) )
e^{\displaystyle  i \int_0^{\Theta} d\theta L (\vec q(\theta), \dot {\vec q}(\theta) ) }
\label{feynman}
\end{eqnarray}

\subsubsection{ Feynman path-integral in the Euclidean time  $t= i \theta$}

The translation into the Euclidean time $t= i \theta$ and $T=i \Theta$ of Eq. \ref{Euclideantime}
involves the change of variables
\begin{eqnarray}
 \vec q(\theta) && = \vec x(t= i \theta) 
 \nonumber \\
 \dot {\vec q}(\theta)  = \frac{d  {\vec q}(\theta)  }{d\theta}  && = i  \frac{d  {\vec x}(t)  }{dt} = i \dot {\vec x}(t)
\label{qxEuclidean}
\end{eqnarray}
So the Feynman amplitude of Eq. \ref{feynman} becomes in Euclidean time
\begin{eqnarray}
\psi_{Euclidean}( \vec x, T \vert \vec x_0, 0 ) = \psi( \vec q, \Theta \vert \vec q_0, 0 )  
= \int_{\vec x(0)=\vec x_0}^{\vec x(T)=\vec x} {\cal D}(\vec x(t) )
e^{\displaystyle  - \int_0^{T}  d t {\cal L} (\vec x(t), \dot {\vec x}(t) ) }
\label{feynmanEuclidean}
\end{eqnarray}
where the Euclidean Lagrangian $ {\cal L} (\vec x(t), \dot {\vec x}(t) ) $ 
is obtained from the initial Lagrangian $ L (\vec q(\theta), \dot {\vec q}(\theta) ) $ of Eq. \ref{lagrangiantheta}
using the change of variables of Eq. \ref{qxEuclidean}
\begin{eqnarray}
 {\cal L} (\vec x(t), \dot {\vec x}(t) ) && = - L (\vec q(\theta), \dot {\vec q}(\theta) )
 = - \frac{1}{2}\dot {\vec q}{\ }^2(\theta) 
- \dot {\vec q}(\theta) .\vec A (\vec q (\theta)) 
  + V (\vec q(\theta))
  \nonumber \\
&& =   \frac{1}{2}\dot {\vec x}^2(t) 
-  i \dot {\vec x}(t) .\vec A ({\vec x}(t)) 
  + V ({\vec x}(t))
\label{lagrangianEuclidean}
\end{eqnarray}
The classical Lagrange equations of motion of Eq. \ref{lagrangemotiontheta} translate into
the similar form in the Euclidean time  $t= i \theta$
\begin{eqnarray}
 \frac{d}{dt} \left(\frac{\partial {\cal L} (\vec x(t), \dot {\vec x}(t) ) } {\partial \dot {\vec x} (t)} \right) 
 = \frac{\partial {\cal L} (\vec x(t), \dot {\vec x}(t) ) } {\partial \vec x (t)} 
\label{lagrangemotionEuclidean}
\end{eqnarray}
since these classical equations of motion correspond to the optimization of the action of Eq. \ref{feynman}
and \ref{feynmanEuclidean}.

\subsubsection{ Euclidean Lagrangian  
when the vector potential is purely imaginary $\vec A ({\vec x}) = -i \vec A^{new} ({\vec x})$  }

The real Lagrangian $L \left (\vec q(\theta), \dot {\vec q}(\theta) \right)$ for real time $\theta$ of Eq. \ref{lagrangiantheta}
has become the complex Euclidean Lagrangian  ${\cal L} (\vec x(t), \dot {\vec x}(t) ) $ for Euclidean time $t$
of Eq. \ref{lagrangianEuclidean}. However if 
the vector potential is purely imaginary 
\begin{eqnarray}
 \vec A ({\vec x}) = -i \vec A^{new} ({\vec x}) \ \ \ {\rm with } \ \ \vec A^{new} ({\vec x}) \ \ {\rm real }
\label{Aimaginary}
\end{eqnarray}
then the Euclidean Lagrangian  ${\cal L} (\vec x(t), \dot {\vec x}(t) ) $ of Eq. \ref{lagrangianEuclidean} becomes real
\begin{eqnarray}
 {\cal L} (\vec x(t), \dot {\vec x}(t) )  =   \frac{1}{2}\dot {\vec x}^2(t) 
-   \dot {\vec x}(t) .\vec A^{new} ({\vec x}(t)) 
  + V ({\vec x}(t))
\label{lagrangianEuclideanimaginary}
\end{eqnarray}
This is the Lagrangian that appears in Eq. \ref{lagrangian} of the main text , where $\vec A^{new} ({\vec x}) $ 
is denoted by $ \vec A ({\vec x}) $ to simplify the notation.


\subsection{ Hamiltonian perspective }

\subsubsection{ Classical mechanics via the Hamiltonian perspective}

In the presence of a vector potential $\vec A (\vec q(\theta)) $, the classical momentum $\vec \pi(\theta) $
defined as the derivative of the Lagrangian $L (\vec q(\theta), \dot {\vec q} (\theta) ) $ of Eq. \ref{lagrangiantheta}
with respect to the velocity $\dot {\vec q}(\theta) $ involves both the velocity $\dot {\vec q}(\theta) $ 
and the vector potential $\vec A (\vec q(\theta)) $
\begin{eqnarray}
\vec \pi(\theta)   \equiv \frac{\partial L (\vec q(\theta), \dot {\vec q} (\theta) ) } {\partial \dot {\vec q} (\theta)}
=  \dot {\vec q}(\theta) + \vec A (\vec q(\theta))
\label{momentum}
\end{eqnarray} 
The classical Hamiltonian $H_{Classical} (\vec q(\theta), \vec \pi(\theta) ) $ 
is a function of the position $ \vec q(\theta)$ and the momentum $\vec \pi(\theta) $
that can be obtained from the Lagrangian $L (\vec q(\theta), \dot {\vec q} (\theta)) $
via the following Legendre transform,
where the velocity $\dot {\vec q}(\theta) $ has to be rewritten in terms of the momentum $\vec \pi(\theta)   $ using Eq. \ref{momentum}
\begin{eqnarray}
H_{Classical} (\vec q(\theta), \vec \pi(\theta) ) && \equiv  
 \dot {\vec q} (\theta). \vec \pi (\theta) - L (\vec q(\theta), \dot {\vec q} (\theta)) 
\nonumber \\
&& = 
\left( \vec \pi (\theta)-  \vec A (\vec q(\theta) ) \right) . \vec \pi (\theta)
- \frac{1}{2}\left( \vec \pi (\theta)-  \vec A (\vec q(\theta))  \right)^2 
- \left( \vec \pi (\theta)-  \vec A (\vec q(\theta) ) \right) .\vec A (\vec q (\theta)) 
  + V (\vec q(\theta))
\nonumber \\
&& =\frac{1}{2} \left(\vec \pi (\theta) -  \vec A (\vec q(\theta)) \right)^2  + V (\vec q(\theta))
\label{hamiltoniantheta}
\end{eqnarray}
In the Hamiltonian perspective, 
the classical equations of motion are first-order differential equations for the position ${\vec q} (\theta) $
and the momentum ${\vec \pi} (\theta) $
\begin{eqnarray}
\dot {\vec q} (\theta) && =  \frac{\partial H_{Classical} (\vec q(\theta), \vec \pi (\theta)) } {\partial \ {\vec \pi}(\theta)} 
= \vec \pi(\theta)   - \vec A (\vec q(\theta))
\nonumber \\
\dot {\vec \pi} (\theta)&& = - \frac{\partial H_{Classical} (\vec q(\theta), \vec \pi (\theta)) } {\partial \ {\vec q}(\theta)} 
\label{hamiltonmotion}
\end{eqnarray}


\subsubsection{ Quantum mechanics via the Schr\"odinger equation involving the Hamiltonian operator}

In the Schr\"odinger perspective of quantum mechanics,
the amplitude $\psi( \vec q, \Theta \vert \vec q_0, 0 )$
satisfies the Schr\"odinger equation
\begin{eqnarray}
i  \frac{ \partial \psi( \vec q, \Theta \vert \vec q_0, 0 ) }{\partial \Theta} =H_{Quantum}   \psi( \vec q, \Theta \vert \vec q_0, 0 )
\label{schrodinger}
\end{eqnarray}
where the Hamiltonian operator $H_{Quantum}$ in the position basis
can be obtained from the classical Hamiltonian $H_{Classical} (\vec q, \vec \pi )$ of Eq. \ref{hamiltoniantheta}
via the replacement of the classical momentum $\vec \pi$ by the Hermitian differential operator $ -i \frac{\partial}{\partial \vec q} $
\begin{eqnarray}
\vec \pi  \to  -i \frac{\partial}{\partial \vec q}
\label{momentumoperator}
\end{eqnarray} 
with the result 
\begin{eqnarray}
H_{Quantum} = H_{Quantum}^{\dagger}  
&&  = \frac{1}{2} \left( -i \frac{\partial}{\partial \vec q}  -  \vec A (\vec q) \right)^2  + V (\vec q) 
\nonumber \\
&&   = - \frac{1}{2}  \frac{\partial^2}{\partial \vec q^2} + i \vec A (\vec q). \frac{\partial}{\partial \vec q}
+ \frac{i}{2}  \left( \frac{\partial }{\partial \vec q} . \vec A (\vec q)\right) + \frac{1}{2} \vec A^2 (\vec q)   + V (\vec q)
\label{hamiltonianquantum}
\end{eqnarray}

\subsubsection{ Schr\"odinger equation in Euclidean time }

The translation of the Schr\"odinger Eq. \ref{schrodinger}
into the Euclidean time $T=i \Theta$ of Eq. \ref{Euclideantime} reads
\begin{eqnarray}
-\frac{ \partial \psi(\vec x,T \vert \vec x_0,0)}{\partial T} = {\cal H}_{Quantum}   \psi(\vec x,T \vert \vec x_0,0)
\label{schrodingerE}
\end{eqnarray}
where the Euclidean quantum Hamiltonian ${\cal H}_{Quantum}  $ is the same as the quantum Hamiltonian of Eq. \ref{hamiltonianquantum} up to the notation replacement $\vec q \to \vec x$
\begin{eqnarray}
{\cal H}_{Quantum}   = {\cal H}_{Quantum}^{\dagger}  && 
  = \frac{1}{2} \left( -i \frac{\partial}{\partial \vec x}  -  \vec A (\vec x) \right)^2  + V (\vec x) 
\nonumber \\
&&   = - \frac{1}{2}  \frac{\partial^2}{\partial \vec x^2} + i \vec A (\vec x). \frac{\partial}{\partial \vec x}
+ \frac{i}{2}   \left( \frac{\partial }{\partial \vec x} . \vec A (\vec x)\right)  + \frac{1}{2} \vec A^2 (\vec x)   + V (\vec x)
\label{hamiltonianquantumEuclidean}
\end{eqnarray}

\subsubsection{ Euclidean quantum Hamiltonian  
when the vector potential is purely imaginary $\vec A ({\vec x}) = -i \vec A^{new} ({\vec x})$  }

The Euclidean quantum Hamiltonian ${\cal H}_{Quantum}  $ of Eq. \ref{hamiltonianquantumEuclidean}
is Hermitian but complex-valued.
 However, when the vector potential is purely imaginary $\vec A ({\vec x}) = -i \vec A^{new} ({\vec x})$, as already considered in Eq. \ref{Aimaginary}, then the Euclidean quantum Hamiltonian ${\cal H}_{Quantum}  $ of Eq. \ref{hamiltonianquantumEuclidean} 
 becomes real-valued
 \begin{eqnarray}
{\cal H}_{Quantum}    && 
  = \frac{1}{2} \left( -i \frac{\partial}{\partial \vec x}  + i  \vec A^{new} (\vec x) \right)^2  + V (\vec x) 
  = -  \frac{1}{2} \left(  \frac{\partial}{\partial \vec x}  -  \vec A^{new} (\vec x) \right)^2  + V (\vec x) 
\nonumber \\
&&   = - \frac{1}{2}  \frac{\partial^2}{\partial \vec x^2} +\vec A^{new} ({\vec x}). \frac{\partial}{\partial \vec x}
+ \frac{1}{2}   \left( \frac{\partial }{\partial \vec x} . \vec A^{new} ({\vec x})\right)  - \frac{1}{2} \left( \vec A^{new} (\vec x) \right)^2  + V (\vec x)
\label{hamiltonianquantumEuclideanAimaginary}
\end{eqnarray}
but non-Hermitian ${\cal H}_{Quantum}   \ne {\cal H}_{Quantum}^{\dagger}$ since its adjoint reads 
  \begin{eqnarray}
 {\cal H}_{Quantum}^{\dagger}  && 
  = \frac{1}{2} \left( -i \frac{\partial}{\partial \vec x}  -i  \vec A^{new} (\vec x) \right)^2  + V (\vec x) 
   = -  \frac{1}{2} \left(  \frac{\partial}{\partial \vec x}  +  \vec A^{new} (\vec x) \right)^2  + V (\vec x) 
\nonumber \\
&&   =  - \frac{1}{2}  \frac{\partial^2}{\partial \vec x^2} - \vec A^{new} ({\vec x}). \frac{\partial}{\partial \vec x}
- \frac{1}{2}   \left( \frac{\partial }{\partial \vec x} . \vec A^{new} ({\vec x})\right)  - \frac{1}{2} \left( \vec A^{new} (\vec x) \right)^2  + V (\vec x)
\label{hamiltonianquantumEuclideanAimaginaryadjoint}
\end{eqnarray}
The non-Hermitian quantum Hamiltonian of Eq. \ref{hamiltonianquantumEuclideanAimaginary} 
plays a major role in the present paper,
since it appears naturally with other notations for $\vec A^{new} ({\vec x}) $ 
both in Eqs \ref{FPhamiltonianpBrown} \ref {FPhamiltonianpBrownNonHermitian} of the Introduction, and in Eq. \ref{FPhamiltonianQuantum} of the main text,
while it has also been previously considered for other purposes \cite{Poland_Malliavin,Poland_Burgers}.
More generally, non-Hermitian physics is nowadays very relevant for many interesting applications 
in various fields
(see the review \cite{ReviewNonHermitian} and references therein).


\section{ Application to diffusions with forces $ F_n( \vec X) $ and uniform diffusion matrix ${\bold D}$ }

\label{app_rescaling}

In this Appendix, we mention how the analysis described in the main text
can be applied to diffusion processes $\vec X(t)$ in dimension $N$
involving a space-dependent force $ \vec F( \vec X) $ and a space-independent diffusion matrix ${\bold D}$.

\subsection{ Initial diffusion process $\vec X(t)$ : forces $ F_n( \vec X) $ and space-independent diffusion matrix ${\bold D}$}

Let us consider the diffusion process $\vec X(t)$,  
where the $N$ components $X_n(t)$ for $n=1,..,N$ follow the Langevin 
stochastic differential system involving the $N$ independent Wiener processes $w_n(t)$ 
\begin{eqnarray}
dX_n(t) =  F_n( \vec X (t) ) \ dt +   \sum_{m=1}^N G_{nm} dw_m(t)
\label{langevinX}
\end{eqnarray}
So the parameters are the $N$ arbitrary space-dependent force-components $F_n( \vec X)  $ 
and the space-independent matrix ${\bold G}=[G_{nm}]$ of size $N\times N$.
The symmetric diffusion matrix ${\bold D}={\bold D}^T $ can be obtained from the matrix ${\bold G}$ and its transpose ${\bold G}^T$
\begin{eqnarray}
{\bold D} = \frac{1}{2} {\bold G} {\bold G}^T
\label{diffusionmatrix}
\end{eqnarray}


\subsection{ Rescaled process $\vec x(t)$ : forces $ f_n( \vec x) $ and trivial diffusion matrix 
${\bold d}= \frac{1}{2} {\bold 1}$} 

We will assume that the matrix ${\bold G}=[G_{nm}]$ is invertible 
to make the following change of variables
from the initial variables $X_n(t)$ to the new variables $x_m(t)$ via
\begin{eqnarray}
X_n(t) &&=   \sum_{m=1}^N G_{nm} x_m(t)
\nonumber \\
x_m(t) &&=   \sum_{n=1}^N ({\bold G}^{-1})_{m n} X_n(t)
\label{bigXsmallx}
\end{eqnarray}
This matrix transformation allows to obtain the simpler Langevin system for the new variables $x_m(t)$
\begin{eqnarray}
dx_m(t) &&=   \sum_{n=1}^N ({\bold G}^{-1})_{m n} dX_n(t)
= \sum_{n=1}^N ({\bold G}^{-1})_{m n}
\left[  F_n( {\bold G} \vec x (t) ) \ dt +   \sum_{l=1}^N G_{nl} dw_l(t) \right]
\nonumber \\
&& \equiv f_m(\vec x (t) ) \ dt + dw_m(t)
\label{langevinx}
\end{eqnarray}
where the new diffusion matrix is simply proportional to the identity ${\bold d}= \frac{1}{2} {\bold 1}$ 
with the diffusion coefficient $\frac{1}{2} $,
while the new $N$ space-dependent force-components $f_m(\vec x ) $ read
\begin{eqnarray}
f_m(\vec x ) && \equiv    \sum_{n=1}^N ({\bold G}^{-1})_{m n} F_n( {\bold G} \vec x  ) 
\label{smallforce}
\end{eqnarray}
So the analysis described in the main text can be applied to the rescaled process $\vec x(t)$
satisfying the Langevin system of Eq. \ref{langevinx}
that coincides with Eq. \ref{langevin}.
This rescaling procedure will be useful in section \ref{sec_Gyr} concerning the Brownian gyrator.



 \section{ Ornstein-Uhlenbeck processes : Explicit solutions via the diagonalization of the matrix  ${\bold \Gamma} $    } 
 
 \label{App_explicitvisdiagoGamma}
 
 In practice, when one wishes to obtain explicit results for the Ornstein-Uhlenbeck processes
 considered in sections \ref{sec_OU} \ref{sec_OUadditive} \ref{sec_Gyr} of the main text,
  one needs to introduce the spectral decomposition of the matrix ${\bold \Gamma} $ \cite{Thouless,CGetJML},
 as recalled in the present Appendix.
 
 \subsection{ Spectral decomposition of the matrix $ {\bold \Gamma}$ }
  
 The spectral decomposition of the real matrix ${\bold \Gamma} $
 involves its $N$ eigenvalues $\gamma_{\alpha}$ 
\begin{eqnarray}
 {\bold \Gamma}  = \sum_{\alpha=1}^N \gamma_{\alpha} \vert \gamma_{\alpha}^R \rangle \langle \gamma_{\alpha}^L \vert
 \label{diagoGamma}
\end{eqnarray} 
where the corresponding right eigenvectors $\vert \gamma_{\alpha}^R \rangle $ and left eigenvectors $\langle \gamma_{\alpha}^L \vert$
\begin{eqnarray}
 {\bold \Gamma} \vert \gamma_{\alpha}^R \rangle && =  \gamma_{\alpha} \vert \gamma_{\alpha}^R \rangle 
 \nonumber \\
\langle \gamma_{\alpha}^L \vert  {\bold \Gamma}  &&=  \gamma_{\alpha}  \langle \gamma_{\alpha}^L \vert
 \label{diagoGammaeigen}
\end{eqnarray} 
satisfy the orthonormalization and closure properties 
 \begin{eqnarray}
\langle \gamma_{\alpha}^L \vert   \gamma_{\beta}^R \rangle && = \delta_{\alpha,\beta}
 \nonumber \\
 \sum_{\alpha=1}^N  \vert \gamma_{\alpha}^L \rangle \langle \gamma_{\alpha}^R \vert && =\mathbf{1}
 \label{orthonor}
\end{eqnarray}
The corresponding spectral decomposition for its transpose ${\bold \Gamma}^T $ that coincide with its adjoint ${\bold \Gamma}^{\dagger}  $ reads
\begin{eqnarray}
 {\bold \Gamma}^T =  {\bold \Gamma}^{\dagger}  = \sum_{\alpha=1}^N \gamma_{\alpha}^* \vert \gamma_{\alpha}^L \rangle \langle \gamma_{\alpha}^R \vert
 \label{diagoGammadagger}
\end{eqnarray} 

 
 \subsection{ Explicit solution for the average values $\vert \vec \mu(t) \rangle $ }
 
Using the spectral decomposition of Eq. \ref{diagoGamma},
the ket $\vert \vec \mu(t) \rangle $ of average values of Eq. \ref{langevinmumatrix} become
\begin{eqnarray}
\vert \vec \mu(t) \rangle = e^{ - {\bold \Gamma} t } \vert \vec y \rangle
  =  \sum_{\alpha=1}^N e^{-\gamma_{\alpha} t} \vert \gamma_{\alpha}^R \rangle \langle \gamma_{\alpha}^L   \vert \vec y \rangle 
  \label{langevinmumatrixdiag}
\end{eqnarray}
So the average values $\mu_n(t)=\overline{x_n(t)}$ 
 are linear combinations of simple exponential behaviors $ e^{-\gamma_{\alpha} t}  $ with respect to the time $t$
that involve the $N$ eigenvalues $\gamma_{\alpha}$
  \begin{eqnarray}
 \mu_n(t) = \langle n \vert  \vec \mu(t) \rangle =
  \sum_{\alpha=1}^N  \langle n \vert \gamma_{\alpha}^R \rangle e^{-\gamma_{\alpha} t}  \langle \gamma_{\alpha}^L   \vert \vec y \rangle    
 \label{muncl}
\end{eqnarray}


  \subsection{ Explicit solution for the connected correlation matrix $ {\bold  C}(t)$ }

Similarly, the connected correlation matrix $ {\bold  C}(t)$ of Eq. \ref{corresol} 
can be rewritten using the spectral decompositions of Eqs \ref{diagoGamma} and \ref{diagoGammadagger}
 \begin{eqnarray}
{\bold  C}(t) &&   =  \int_0^t d\tau e^{ - {\bold \Gamma} \tau }
   e^{ - {\bold \Gamma}^T \tau } 
 = \int_0^t d\tau 
 \left(\sum_{{\alpha}=1}^N e^{-\gamma_{\alpha} \tau} \vert \gamma_{\alpha}^R 
 \rangle \langle \gamma_{\alpha}^L \vert \right)
 \left(  \sum_{{\beta}=1}^N e^{- \gamma_{\beta}^* \tau} \vert \gamma_{\beta}^L \rangle \langle \gamma_{\beta}^R \vert \right)
   \nonumber \\
   &&    
    \equiv  \sum_{{\alpha}=1}^N   \sum_{{\beta}=1}^N  \vert \gamma_{\alpha}^R \rangle
C_{{\alpha}{\beta}}^{RR}(t)    \langle \gamma_{\beta}^R \vert    
 \label{corresoldiag}
\end{eqnarray}
where the matrix elements $C_{{\alpha}{\beta}}^{RR} (t) $ 
 display simple exponential behaviors with respect to the time $t$
\begin{eqnarray}
C_{{\alpha}{\beta}}^{RR} (t) 
 =   \left( \int_0^t d\tau  e^{-(\gamma_{\alpha} + \gamma_{\beta}^*) \tau} \right) \langle \gamma_{\alpha}^L \vert \gamma_{\beta}^L \rangle
 =   \frac{1 -  e^{-(\gamma_{\alpha} + \gamma_{\beta}^*) t} }{ \gamma_{\alpha} + \gamma_{\beta}^*} 
 \langle \gamma_{\alpha}^L \vert \gamma_{\beta}^L \rangle
  \label{CjlR}
\end{eqnarray}
with their limits for $t \to +\infty$
 \begin{eqnarray}
C_{{\alpha}{\beta}}^{RR}  = C_{{\alpha}{\beta}}^{RR} (t=+\infty) = 
 \frac{\langle \gamma_{\alpha}^L \vert \gamma_{\beta}^L \rangle }{ \gamma_{\alpha} + \gamma_{\beta}^*}  
 \label{CjlRinfty}
\end{eqnarray}

Using the matrix ${\bold K}$
 \begin{eqnarray}
 K_{n \alpha} \equiv \langle n \vert \gamma_{\alpha}^R \rangle
 \label{Knalpha}
\end{eqnarray}
the matrix element $C_{nm}(t) $ of Eq. \ref{mucorredef}
can be then obtained via
 \begin{eqnarray}
C_{nm}(t) &&   
 =  \sum_{{\alpha}=1}^N   \sum_{{\beta}=1}^N  \langle n \vert \gamma_{\alpha}^R \rangle
C_{{\alpha}{\beta}}^{RR}(t)    \langle \gamma_{\beta}^R \vert    m \rangle
= ({\bold K } {\bold C^{RR}(t) } {\bold K }^{\dagger} )_{nm}
 \label{cnmfromdiag}
\end{eqnarray}
In particular, the determinant of the matrix ${\bold C}(t)=[C_{nm}(t)]$ can be obtained from the product
\begin{eqnarray}
\det( {\bold C}(t) )&&   
 = \det ({\bold K } ) \det({\bold C^{RR}(t) } ) \det({\bold K }^{\dagger})
 = \det({\bold C^{RR}(t) } ) \det({\bold K }^{\dagger} {\bold K })
 \label{detct}
\end{eqnarray}
where the matrix ${\bold K }^{\dagger} {\bold K }$
 \begin{eqnarray}
({\bold K }^{\dagger} {\bold K })_{\beta \alpha} = \sum_{n=1}^L K^{\dagger}_{\beta n}  K_{n \alpha} 
= \sum_n \langle \gamma_{\beta}^R \vert n \rangle \langle n \vert \gamma_{\alpha}^R \rangle
=  \langle\gamma_{\beta}^R   \vert \gamma_{\alpha}^R \rangle
 \label{kdaggerk}
\end{eqnarray}
involves the scalar products of the right eigenvectors.


  \subsection{ Explicit solution for the inverse ${\bold \Phi}(t) =  [{\bold C}(t) ]^{-1}$ }

The matrix ${\bold \Phi}(t)  = [{\bold C}(t) ]^{-1} $ of Eq. \ref{PhiCinverset}
will  
have simple matrix elements $\Phi_{{\alpha}{\beta}}^{LL} (t) $ in the basis 
\begin{eqnarray}
 {\bold \Phi} (t)=   \sum_{{\alpha}=1}^N   \sum_{{\beta}=1}^N  \vert \gamma_{\alpha}^L \rangle
\Phi_{{\alpha}{\beta}}^{LL}(t)
    \langle \gamma_{\beta}^L \vert   
\label{PhibasisLL}
\end{eqnarray}
Indeed, using Eq. \ref{orthonor}  the conditions for the matrix inversion
\begin{eqnarray}
{\bold 1} && = {\bold \Phi} (t) {\bold C}(t)  
=  \left(  \sum_{{\alpha}=1}^N   \sum_{{\beta}=1}^N  \vert \gamma_{\alpha}^L \rangle
\Phi_{{\alpha}{\beta}}^{LL}(t)    \langle \gamma_{\beta}^L \vert   \right)
\left(  \sum_{{\alpha'}=1}^N   \sum_{{\beta'}=1}^N  \vert \gamma_{\alpha'}^R \rangle
C_{{\alpha'}{\beta'}}^{RR}(t)    \langle \gamma_{\beta'}^R \vert    \right)
        =     \sum_{{\alpha}=1}^N   \sum_{{\beta'}=1}^N    \vert \gamma_{\alpha}^L \rangle
\left(\sum_{{\beta}=1}^N
\Phi_{{\alpha}{\beta}}^{LL}(t)       
C_{{\beta}{\beta'}}^{RR}(t) \right)   \langle \gamma_{\beta'}^R \vert    
    \nonumber \\
{\bold 1} && = {\bold C}(t)  {\bold \Phi} (t) 
=  \left( \sum_{{\alpha'}=1}^N   \sum_{{\beta'}=1}^N  \vert \gamma_{\alpha'}^R \rangle
C_{{\alpha'}{\beta'}}^{RR}(t)    \langle \gamma_{\beta'}^R \vert     \right)
 \left( \sum_{{\alpha}=1}^N   \sum_{{\beta}=1}^N  \vert \gamma_{\alpha}^L \rangle
\Phi_{{\alpha}{\beta}}^{LL}(t)    \langle \gamma_{\beta}^L \vert     \right)
=    \sum_{{\alpha'}=1}^N  \sum_{{\beta}=1}^N
   \vert \gamma_{\alpha'}^R \rangle
   \left( \sum_{{\alpha}=1}^N  C_{{\alpha'}{\alpha}}^{RR}(t)   
\Phi_{{\alpha}{\beta}}^{LL}(t)  \right)  \langle \gamma_{\beta}^L \vert   
 \nonumber \\  
\label{PhiCinverseteq}
\end{eqnarray}
yield that the coefficients should satisfy
\begin{eqnarray} 
\sum_{{\beta}=1}^N
\Phi_{{\alpha}{\beta}}^{LL}(t)       
C_{{\beta}{\beta'}}^{RR}(t) && = \delta_{\alpha \beta'}
      \nonumber \\ 
   \sum_{{\alpha}=1}^N  C_{{\alpha'}{\alpha}}^{RR}(t)   
\Phi_{{\alpha}{\beta}}^{LL}(t)   && = \delta_{\alpha' \beta}
\label{PhiCinversetcoefs}
\end{eqnarray}

The matrix element $\Phi_{nm}(t) $ in the initial basis can be then reconstructed via
\begin{eqnarray}
 {\bold \Phi}_{nm} (t)=   \sum_{{\alpha}=1}^N   \sum_{{\beta}=1}^N  
 \langle n \vert \gamma_{\alpha}^L \rangle
\Phi_{{\alpha}{\beta}}^{LL}(t)
    \langle \gamma_{\beta}^L \vert   m \rangle
\label{PhibasisLLcl}
\end{eqnarray}


\subsection{ Solving the perturbative expansion of subsection \ref{subsec_perturbation}     } 

\label{subsec_upsilonper}

Similarly, the spectral decomposition of the matrix ${\bold \Gamma}$ of Eqs \ref{diagoGamma} and \ref{diagoGammaeigen}
is useful to solve the perturbative expansion
discussed in subsection \ref{subsec_perturbation} of the main text.
For instance,
the projection of Eq. \ref{eqk1} onto the right eigenvectors
on both sides $\langle \gamma_{\alpha}^R \vert ..  \vert \gamma_{\beta}^R \rangle$  
 \begin{eqnarray}
\langle \gamma_{\alpha}^R \vert {\bold W}^{[\cal O]}  \vert \gamma_{\beta}^R \rangle
&& = \langle \gamma_{\alpha}^R \vert   {\bold \Gamma}^T \left( {\mathring {\bold \Upsilon} }^{(1)}  + \frac{{\bold B}^{[\cal O]}  }{2}     \right) \vert \gamma_{\beta}^R \rangle
 +\langle \gamma_{\alpha}^R \vert \left( {\mathring {\bold \Upsilon} }^{(1)}- \frac{{\bold B}^{[\cal O]} }{2}      \right){\bold \Gamma}    \vert \gamma_{\beta}^R \rangle  
 \nonumber \\
 && =\gamma_{\alpha}^* \langle \gamma_{\alpha}^R \vert  \left( {\mathring {\bold \Upsilon} }^{(1)}  + \frac{{\bold B}^{[\cal O]}  }{2}     \right) \vert \gamma_{\beta}^R \rangle
 +\langle \gamma_{\alpha}^R \vert \left( {\mathring {\bold \Upsilon} }^{(1)}- \frac{{\bold B}^{[\cal O]} }{2}      \right)
  \vert \gamma_{\beta}^R \rangle  \gamma_{\beta}
  \nonumber \\
 && =(\gamma_{\alpha}^* +\gamma_{\beta} ) \langle \gamma_{\alpha}^R \vert   {\mathring {\bold \Upsilon} }^{(1)} 
  \vert \gamma_{\beta}^R \rangle 
 +(\gamma_{\alpha}^* - \gamma_{\beta} ) 
  \langle \gamma_{\alpha}^R \vert   \frac{{\bold B}^{[\cal O]}  }{2}    \vert \gamma_{\beta}^R \rangle
 \label{eqk1proj}
\end{eqnarray}
allows to compute the matrix elements of the first-order-correction matrix ${\mathring {\bold \Upsilon} }^{(1)} $ via
 \begin{eqnarray}
 \langle \gamma_{\alpha}^R \vert   {\mathring {\bold \Upsilon} }^{(1)} 
  \vert \gamma_{\beta}^R \rangle 
  = \frac{ \langle \gamma_{\alpha}^R \vert {\bold W}^{[\cal O]}  \vert \gamma_{\beta}^R \rangle
 -(\gamma_{\alpha}^* - \gamma_{\beta} ) 
  \langle \gamma_{\alpha}^R \vert   \frac{{\bold B}^{[\cal O]}  }{2}    \vert \gamma_{\beta}^R \rangle }
  {\gamma_{\alpha}^* +\gamma_{\beta} } 
 \label{ups1solu}
\end{eqnarray}


 \subsection{ Application to the Brownian gyrator in dimension $N=2$   }
 
In this subsection, we apply the general formalism described above
to the Brownian gyrator model of section \ref{sec_Gyr}.

 \subsubsection{ Spectral decomposition of the matrix $\Gamma$   }
 
 The spectral decomposition of Eq. \ref{diagoGamma} reads for the matrix of Eq. \ref{gammamatrixGYR}
     \begin{eqnarray}
 {\bold \Gamma }  = \begin{pmatrix} 
1  & u \rho \\
\frac{u}{\rho} & 1
 \end{pmatrix}  = \gamma_+ \vert \gamma_+^R \rangle \langle \gamma_+^L \vert  + \gamma_- \vert \gamma_-^R \rangle \langle \gamma_-^L \vert 
 \label{gammamatrixGYRdiag}
\end{eqnarray}
where the two eigenvalues $\gamma_{\pm} $ involve the parameter $u$
\begin{eqnarray} 
 \gamma_{\pm} = 1 \pm u
 \label{twoeigen}
\end{eqnarray}
The corresponding right eigenvectors
\begin{eqnarray} 
 \vert \gamma_{\pm}^R \rangle
 =  \frac{1}{\sqrt 2}
 \begin{pmatrix} 
1 \\  
\pm   \frac{1}{\rho }  
  \end{pmatrix}
   \label{tworight}
\end{eqnarray}
and left eigenvectors
\begin{eqnarray} 
 \langle \gamma_{\pm}^L \vert
 = \frac{1}{\sqrt 2} \begin{pmatrix} 
1 &  \pm \rho  
  \end{pmatrix}
   \label{twoleft}
\end{eqnarray}
involve the parameter $\rho$ and
satisfy the orthonormalization relations
\begin{eqnarray} 
 \langle \gamma_{\epsilon}^L \vert \gamma_{\epsilon'}^R \rangle
 = \frac{1}{ 2} \begin{pmatrix} 
1 &  \epsilon \rho  
  \end{pmatrix}
 \begin{pmatrix} 
1 \\  
\epsilon'   \frac{1}{\rho }  
  \end{pmatrix}  
  =  \frac{1}{ 2}  (1+\epsilon \epsilon' )  =  \delta_{\epsilon, \epsilon'} 
       \label{twoortho}
\end{eqnarray}
while other scalar products that will be useful below read
\begin{eqnarray} 
 \langle \gamma_{\epsilon}^L \vert \gamma_{\epsilon'}^L \rangle
&&  = \frac{1}{ 2} \begin{pmatrix} 
1 &  \epsilon \rho  
  \end{pmatrix}
 \begin{pmatrix} 
1 \\  
\epsilon'   \rho   
  \end{pmatrix}  
  =  \frac{1+\epsilon \epsilon' \rho^2}{ 2}    
  \nonumber \\
 \langle \gamma_{\epsilon}^R \vert \gamma_{\epsilon'}^R \rangle
&&  = \frac{1}{ 2}   \begin{pmatrix} 
1 &   \frac{\epsilon}{\rho}  
  \end{pmatrix}
 \begin{pmatrix} 
1 \\  
\frac{\epsilon'}{\rho}  
  \end{pmatrix} =    \frac{1+ \frac{ \epsilon \epsilon' }{\rho^2}}{ 2}    
       \label{twoscalar}
\end{eqnarray}


\subsubsection{ Explicit solution for the average value $\vert \vec \mu(t) \rangle $ }

The average value of \ref{langevinmumatrixdiag}
\begin{eqnarray}
\vert \vec \mu(t) \rangle  
= \overline{\vert \vec x(t) \rangle }
  =  e^{ - {\bold \Gamma} t } \vert \vec y \rangle
  =  e^{- \gamma_+ t} \vert \gamma_+^R \rangle \langle \gamma_+^L \vert  \vec y \rangle
  + e^{- \gamma_- t} \vert \gamma_-^R \rangle \langle \gamma_-^L \vert \vec y \rangle
  = \mu_+^R (t)\vert \gamma_+^R \rangle +\mu_-^R (t)\vert \gamma_-^R \rangle 
  \label{langevinmumatrixGYR}
\end{eqnarray}
involves the coefficients
\begin{eqnarray}
\mu_+^R (t) && = e^{- \gamma_+ t}\langle \gamma_+^L \vert  \vec y \rangle 
=e^{- (1+u) t} \left( \frac{y_1+\rho y_2}{\sqrt 2} \right)
\nonumber \\
\mu_-^R (t) && = e^{- \gamma_- t}\langle \gamma_-^L \vert  \vec y \rangle 
=e^{- (1-u) t} \left( \frac{y_1- \rho y_2}{\sqrt 2} \right)
  \label{mujRgyr}
\end{eqnarray}

The average values $\mu_n(t)=\overline{x_n(t)}$ 
 are then obtained via the linear combination of Eq. \ref{muncl}
 using the right eigenvectors of Eq. \ref{tworight}
 \begin{eqnarray}
 \mu_1(t) && = \langle 1 \vert  \vec \mu(t) \rangle 
 =   \langle 1 \vert  \gamma_+^R  \rangle \mu_+^R (t)
 + \langle 1 \vert  \gamma_-^R  \rangle \mu_-^R (t) 
 = \frac{\mu_+^R (t)+\mu_-^R (t)}{\sqrt 2} =  e^{ -t} \cosh(t u) y_1 - e^{ -t} \sinh(t u) y_2
 \nonumber \\
 \mu_2(t) && = \langle 2 \vert  \vec \mu(t) \rangle 
 =   \langle 2 \vert  \gamma_+^R  \rangle \mu_+^R (t)
 + \langle 2 \vert  \gamma_-^R  \rangle \mu_-^R (t) 
 =   \frac{\mu_+^R (t)-\mu_-^R (t)}{ \rho \sqrt 2} = -\frac{e^{ -t}}{\rho} \sinh(t u) y_1 + e^{ -t} \cosh(t u) y_2
 \label{munclGYR}
\end{eqnarray}
in accordance with \cite{cerasoli} where this result is obtained thanks to standard Laplace transform techniques.

  \subsubsection{ Explicit solution for the connected correlation matrix $ {\bold  C}(t)$ }

The correlation matrix $ {\bold  C}(t)$ of Eq. \ref{corresol} can be expanded via Eq. \ref{corresoldiag}
 \begin{eqnarray}
{\bold  C}(t) 
&&    =  \vert \gamma_+^R \rangle C_{++}^{RR} (t)  \langle \gamma_+^R \vert   
  + \vert \gamma_-^R \rangle C_{--}^{RR} (t)  \langle \gamma_-^R \vert  
  + \vert \gamma_+^R \rangle C_{+-}^{RR} (t)  \langle \gamma_-^R \vert  
  + \vert \gamma_-^R \rangle C_{-+}^{RR} (t)  \langle \gamma_+^R \vert 
 \label{corresoldiaggyr}
\end{eqnarray}
with the components of Eq. \ref{CjlR} using the left eigenvectors of Eq. \ref{twoleft}
\begin{eqnarray}
C_{++}^{RR} (t) &&=  \frac{1 -  e^{- 2 \gamma_+ t } }{ 2 \gamma_+}  \langle \gamma_+^L \vert \gamma_+^L \rangle
=\frac{ (1 -  e^{- 2 (1+u) t } ) ( 1+ \rho^2) }{ 4 (1+u)}  
\nonumber \\
C_{--}^{RR} (t) &&=  \frac{1 -  e^{- 2 \gamma_- t } }{ 2 \gamma_-}  \langle \gamma_-^L \vert \gamma_-^L \rangle
=\frac{(1 -  e^{- 2 (1-u) t } ) (1+ \rho^2 ) }{ 4 (1-u)} 
\nonumber \\
C_{+-}^{RR} (t)  && =
\frac{1 -  e^{-(\gamma_+ + \gamma_-) t} }{ \gamma_+ + \gamma_-}   \langle \gamma_+^L \vert \gamma_-^L \rangle
=\frac{(1 -  e^{-2 t}) ( 1- \rho^2)  }{ 4}
  \nonumber \\
 C_{-+}^{RR} (t) && =  \frac{1 -  e^{-(\gamma_+ + \gamma_-) t} }{ \gamma_+ + \gamma_-}   \langle \gamma_+^L \vert \gamma_-^L \rangle
  =\frac{(1 -  e^{-2 t}) ( 1- \rho^2)  }{ 4}     = C_{+-}^{RR} (t) 
  \label{CjlRGyr}
\end{eqnarray}

The determinant of this symmetric matrix ${\bold C}^{RR}$ 
\begin{eqnarray}
D(t)&&  \equiv \det({\bold C^{RR}(t) } ) = C_{++}^{RR} (t) C_{--}^{RR} (t) - [C_{+-}^{RR}(t)]^2 
\nonumber \\
&& = \frac{ (1 -  e^{- 2 (1+u) t } ) (1 -  e^{- 2 (1-u) t } )( 1+ \rho^2)^2 }{ 16 (1-u^2)}  
-\frac{(1 -  e^{-2 t})^2 ( 1- \rho^2)^2  }{ 16}    
\label{detCRRgyr}
\end{eqnarray}
and the determinant of the matrix of Eq. \ref{kdaggerk} of 
 the scalar products of the right eigenvectors of Eq. \ref{twoscalar}
\begin{eqnarray} 
\det \left(
 \langle \gamma_{\epsilon}^R \vert \gamma_{\epsilon'}^R \rangle \right) 
   = \langle \gamma_{+}^R \vert \gamma_{+}^R \rangle
\langle \gamma_{-}^R \vert \gamma_{-}^R \rangle
- \langle \gamma_{+}^R \vert \gamma_{-}^R \rangle
\langle \gamma_{-}^R \vert \gamma_{+}^R \rangle
 =  \left(  \frac{1+ \frac{ 1 }{\rho^2}}{ 2}   \right)^2-   \left(  \frac{1- \frac{ 1 }{\rho^2}}{ 2}   \right)^2
 =  \frac{ 1 }{\rho^2}
       \label{detscalarright}
\end{eqnarray}
allows to compute the determinant of the matrix ${\bold C}(t)$ via Eq. \ref{detct}
\begin{eqnarray}
\det( {\bold C}(t) )   
 = \det({\bold C^{RR}(t) } ) \det \left(
 \langle \gamma_{\epsilon}^R \vert \gamma_{\epsilon'}^R \rangle \right) =  \frac{ D(t) }{\rho^2}
 \label{detctgyr}
\end{eqnarray}


  \subsubsection{ Explicit solution for the inverse ${\bold \Phi}(t) = [{\bold  C(t)}]^{-1}$ }

The elements of inverse matrix of Eq. \ref{PhibasisLL}
\begin{eqnarray}
 {\bold \Phi} (t)=[{\bold  C(t)}]^{-1}  =
  \vert \gamma_+^L \rangle \Phi_{++}^{LL} (t)  \langle \gamma_+^L \vert   
  + \vert \gamma_-^L \rangle \Phi_{--}^{LL} (t)  \langle \gamma_-^L \vert  
  + \vert \gamma_+^L \rangle \Phi_{+-}^{LL} (t)  \langle \gamma_-^L \vert  
  + \vert \gamma_-^L \rangle \Phi_{-+}^{LL} (t)  \langle \gamma_+^L \vert 
 \label{PhiCinversetGYR}
\end{eqnarray}
 satisfing Eq. \ref{PhiCinverseteq} read in terms of the notation $D(t)$ of Eq. \ref{detCRRgyr}
\begin{eqnarray}
\Phi_{++}^{LL} (t) && = \frac{C_{--}^{RR} (t)}{D(t)}
= \frac{(1 -  e^{- 2 (1-u) t } ) (1+ \rho^2 ) }{ 4 (1-u) D(t) } 
\nonumber \\
\Phi_{--}^{LL} (t) && = \frac{C_{++}^{RR} (t)}{D(t)}
= \frac{ (1 -  e^{- 2 (1+u) t } ) ( 1+ \rho^2) }{ 4 (1+u) D(t) }  
\nonumber \\
\Phi_{+-}^{LL} (t) = \Phi_{-+}^{LL} (t) = \Phi_{off}^{LL} (t)&& = - \frac{C_{+-}^{RR} (t)}{D(t)} 
= - \frac{(1 -  e^{-2 t}) ( 1- \rho^2)  }{ 4 D(t) }  
\label{PhiGyr}
\end{eqnarray}

The matrix element $\Phi_{nm}(t) $ in the initial basis can be then reconstructed via Eq. \ref{PhibasisLLcl}
\begin{eqnarray}
 {\bold \Phi}_{11} (t) && =   \sum_{\epsilon=\pm}   \sum_{\epsilon'=\pm}  
 \langle 1 \vert \gamma_{\epsilon}^L \rangle
\Phi_{\epsilon,\epsilon'}^{LL}(t)
    \langle \gamma_{\epsilon'}^L \vert   1 \rangle 
    = \frac{\Phi_{++}^{LL}(t) +\Phi_{--}^{LL}(t) }{2}  
 +   \Phi_{+-}^{LL}(t) 
 \nonumber \\
 {\bold \Phi}_{22} (t) && =   \sum_{\epsilon=\pm}   \sum_{\epsilon'=\pm}  
 \langle 2 \vert \gamma_{\epsilon}^L \rangle
\Phi_{\epsilon,\epsilon'}^{LL}(t)
    \langle \gamma_{\epsilon'}^L \vert   2 \rangle 
    = \rho^2 \left[  \frac{\Phi_{++}^{LL}(t) + \Phi_{--}^{LL}(t)}{2}   
  -   \Phi_{+-}^{LL}(t)  \right]
   \nonumber \\
 {\bold \Phi}_{12} (t) && =   \sum_{\epsilon=\pm}   \sum_{\epsilon'=\pm}  
 \langle 1 \vert \gamma_{\epsilon}^L \rangle
\Phi_{\epsilon,\epsilon'}^{LL}(t)
    \langle \gamma_{\epsilon'}^L \vert   2 \rangle 
    = \rho \left[  \frac{\Phi_{++}^{LL}(t) - \Phi_{--}^{LL}(t) }{2}  
 \right] 
\label{Phibasisinigyr}
\end{eqnarray}


\subsubsection{ Gaussian propagator $P ( \vec x , t \vert \vec y,0) $  }

The Gaussian propagator of Eq. \ref{gaussinverse} 
\begin{eqnarray}
 P ( \vec x , t \vert \vec y,0)
&& = \frac{1}{ 2 \pi  \sqrt{ \det({\bold C}(t) ) } } 
 e^{ \displaystyle - \Phi_{11}(t)\frac{\left( x_1-\mu_1(t) \right)^2}{2}    
- \Phi_{22}(t)\frac{ \left( x_2-\mu_2(t) \right)^2}{2}   
- \Phi_{12}(t)  \left( x_1-\mu_1(t) \right) \left( x_2-\mu_2(t) \right)
 } 
\label{gaussinversegyr}
\end{eqnarray}
can be then obtained from the average values $\mu_n(t)=\overline{x_n(t)}$ of Eq. \ref{munclGYR},
the matrix $\Phi_{nm}(t) $  of Eq. \ref{Phibasisinigyr}
and the determinant of Eq. \ref{detctgyr}.


\subsubsection{ Gaussian steady state $ P^*(\vec x)$ in the parameter region $  u \in ]-1,+1[$  }

In the parameter region $  u \in ]-1,+1[$,
the average values $\mu_n(t)=\overline{x_n(t)}$ of Eq. \ref{munclGYR} converge towards zero for $t \to +\infty$,
so the propagator of Eq. \ref{gaussinversegyr} will converge towards the steady state
\begin{eqnarray}
 P^* ( \vec x )
&& = \frac{1}{ 2 \pi  \sqrt{ \det({\bold C} ) } } 
 e^{ \displaystyle - \Phi_{11}\frac{x_1^2}{2}    
- \Phi_{22}\frac{ x_2^2}{2}   
- \Phi_{12}  x_1 x_2
 } 
\label{steadygyr}
\end{eqnarray}
The correlation matrix elements of Eq. \ref{CjlRGyr}
  converge toward the finite values
\begin{eqnarray}
C_{++}^{RR}  &&
= \frac{1+ \rho^2}{ 4 (1+u) }   
\nonumber \\
C_{--}^{RR} &&
= \frac{1+ \rho^2}{ 4 (1-u)}     
\nonumber \\
C_{+-}^{RR}
 && =
 \frac{1- \rho^2}{ 4 }  
  \label{CjlRGyrinfty}
\end{eqnarray}
The corresponding limit for the determinant of Eq. \ref{detCRRgyr}
\begin{eqnarray}
D  \equiv \det({\bold C^{RR} } ) = C_{++}^{RR}  C_{--}^{RR} - [C_{+-}^{RR}]^2 
=  \frac{(1+ \rho^2)^2}{ 16 (1-u^2)}  -  \frac{(1- \rho^2)^2}{ 16 }  
 = \frac{  \rho^2 +\left( \frac{ u  (1- \rho^2) }{2} \right)^2 }{ 4 (1-u^2)}  
\label{detCRRgyrtinfty}
\end{eqnarray}
allows us to compute the matrix elements of Eq. \ref{PhiGyr}
\begin{eqnarray}
\Phi_{++}^{LL}  && = \frac{C_{--}^{RR} }{D}
= \frac{(1+ \rho^2)  (1+u)}{    \left[ \rho^2 +\left( \frac{ u  (1- \rho^2) }{2} \right)^2\right]}     
\nonumber \\
\Phi_{--}^{LL}  && = \frac{C_{++}^{RR} }{D}
= \frac{(1+ \rho^2)  (1-u)}{   \left[ \rho^2 +\left( \frac{ u  (1- \rho^2) }{2} \right)^2\right]}   
\nonumber \\
\Phi_{+-}^{LL} && = - \frac{C_{+-}^{RR} }{D} 
= -  \frac{(1- \rho^2)  (1-u^2)}{  \left[ \rho^2 +\left( \frac{ u  (1- \rho^2) }{2} \right)^2\right]}  
\label{PhiGyrinfty}
\end{eqnarray}
Finally Eq. \ref{Phibasisinigyr} yields the matrix elements $\Phi_{nm}$ in the initial basis
\begin{eqnarray}
  \Phi_{11}  && 
    = \frac{\Phi_{++}^{LL} +\Phi_{--}^{LL} }{2}  
 +   \Phi_{+-}^{LL} 
 = \frac{  2 \rho^2+ (1- \rho^2) u^2
 }{    \left[ \rho^2 +\left( \frac{ u  (1- \rho^2) }{2} \right)^2\right]} 
 \nonumber \\
  \Phi_{22}  && 
    = \rho^2 \left[  \frac{\Phi_{++}^{LL} + \Phi_{--}^{LL}}{2}   
  -   \Phi_{+-}^{LL}  \right]
  = \frac{  \rho^2 \left[ 2- (1- \rho^2) u^2 \right]
 }{    \left[ \rho^2 +\left( \frac{ u  (1- \rho^2) }{2} \right)^2\right]} 
   \nonumber \\
  \Phi_{12}  && 
    = \rho \left[  \frac{\Phi_{++}^{LL} - \Phi_{--}^{LL} }{2}  
 \right] =   \frac{ \rho (1+ \rho^2) u
 }{    \left[ \rho^2 +\left( \frac{ u  (1- \rho^2) }{2} \right)^2\right]} 
\label{Phibasisinigyrtinfty}
\end{eqnarray}


\section{ Quantum propagators associated to quadratic Lagrangians }

\label{app_quadratic}

In this appendix, we focus on the non-Hermitian electromagnetic quantum mechanics for the following potentials:

(i) when the scalar potential $V (\vec x ) $ is quadratic
and parametrized by some constant $V_0$ and some symmetric matrix $W_{nm}=W_{mn}$
\begin{eqnarray}
V(\vec x)  = V_0+\sum_{n=1}^N   \sum_{n=1}^N  \frac{W_{nm}}{2} x_n  x_m
\label{scalarpotquadratic}
\end{eqnarray}

(ii) when the vector potential $  \vec A (\vec x ) $ is linear and
parametrized by some matrix $\Lambda_{nm}$
 \begin{eqnarray}
  A_n (\vec x )  = - \sum_{m=1}^N  \Lambda_{nm} x_m
 \label{vectorpotlinear}
\end{eqnarray}
corresponding to the constant antisymmetric magnetic matrix 
\begin{eqnarray}
 B_{nm}     \equiv \partial_n A_m (\vec x ) -  \partial_m A_n (\vec x ) 
  =\Lambda_{nm}-\Lambda_{mn}
 \label{magneticNcstquad}
\end{eqnarray}


 \subsection{ Path-integral involving a quadratic Lagrangian : propagator in terms on the classical action $S_{cl}(\vec x,t \vert \vec y ,0) $ }

The goal is to compute the Euclidean non-Hermitian quantum propagator
\begin{eqnarray}
 \psi(\vec x,t \vert \vec y,0) 
\equiv \int_{\vec x(\tau=0)=\vec y}^{\vec x(\tau=t)=\vec x} {\cal D}   \vec x(\tau)  
 e^{ - \displaystyle \int_0^t d\tau
{\cal L} (\vec x(\tau), \dot {\vec x}(\tau) )
 }
 \label{geneapp}
\end{eqnarray}
where the path-integral involves the quadratic Lagrangian associated to the electromagnetic potential of Eqs \ref{scalarpotquadratic} \ref{vectorpotlinear}
\begin{eqnarray}
{\cal L} (\vec x(\tau), \dot {\vec x}(\tau) )
&& =   \frac{1}{2}  \dot {\vec x}^2  (\tau) 
 - \dot {\vec x } (\tau) . \vec A( \vec x(\tau))
 + V(\vec x)
  \nonumber \\
&& = \frac{1}{2}  \sum_{n=1}^N \dot  x_n^2  (\tau) 
 + \sum_{n=1}^N  \sum_{m=1}^N \dot  x_n  (\tau)    \Lambda_{nm} x_m  (\tau)
 + \sum_{n=1}^N   \sum_{m=1}^N  \frac{W_{nm}}{2} x_n  (\tau) x_m (\tau) +V_0
\label{lagrangianApp}
\end{eqnarray}
 As discussed in textbooks on path-integrals in chapters concerning quadratic Lagrangians, 
 the path-integral of Eq. \ref{geneapp} 
can be explicitly computed in terms of the action $S_{cl}(\vec x,t \vert \vec y ,0) $
of the classical trajectory $\vec x(\tau)$ satisfying
the boundary conditions $\vec x(\tau=0)= \vec y$ and $\vec x(\tau=t)= \vec x$
\begin{eqnarray}
\psi(\vec x,t \vert \vec y,0)
= \sqrt{  \frac{ \det \left( - \frac{ \partial^2 S_{cl}(\vec x,t \vert \vec y, 0)}{\partial_{x_n} \partial_{y_m}}\right)}{(2 \pi)^N}}  e^{ - S_{cl}(\vec x,t \vert \vec y, 0)  } 
\label{pathintegralclass}
\end{eqnarray}
where the prefactor that takes into account the integration over the Gaussian fluctuations 
around the classical trajectory involves the famous Van Vleck determinant of 1928 \cite{Van_Vleck}.
The quantum propagator of Eq. \ref{pathintegralclass} which is exact for Gaussian Lagrangians as recalled above,
 is of course also very much used within the semi-classical approximation for quantum mechanics,
 where it corresponds to the leading saddle-point evaluation of the path-integral.
 
As a final remark, let us stress that besides the formula of Eq. \ref{pathintegralclass} that will be used in the present paper,
there are of course many other methods that have been developed to compute Gaussian functional integrals,
as discussed in textbooks on path-integrals. Let us mention in particular the Fourier-basis method that has  
been used to compute the large deviations of the entropy production for Ornstein-Uhlenbeck processes 
in dimension $N$ \cite{entropyProd_OUdimN} and of the heat in the Brownian gyrator in dimension $N=2$ \cite{Gyr_vanWijland} via an integral over the frequency of the logarithm of the ratio of two determinants.


 \subsection{ Computation of the classical action $S_{cl}(\vec x,t \vert \vec y ,0) $  via the Lagrangian perspective } 

\label{subsec_lagrangian}

 \subsubsection{ Lagrangian equations of motion for the classical trajectory } 

\label{subsec_lagrangianmotion}

As recalled in Eq. \ref{lagrangemotionEuclidean},
the Lagrange equations of motion for the classical trajectory $\vec x(\tau)$ read
\begin{eqnarray}
0 &&= \frac{d}{d\tau} \left(\frac{\partial {\cal L} ({\vec x}(\tau), \dot {\vec x} (\tau) ) } {\partial  \dot  x_n (\tau)}\right) 
- \frac{\partial {\cal L} ({\vec x}(\tau), \dot {\vec x}  (\tau)) } {\partial  x_n (\tau)} 
\label{lagrangeFPmotion}
\end{eqnarray}
The derivatives of the quadratic Lagrangian of Eq. \ref{lagrangianApp}
with respect to the velocities $\dot x_n(\tau) $ represent the classical momenta $p_n(\tau) $
\begin{eqnarray}
p_n(\tau) \equiv \frac{\partial {\cal L} (\vec x(\tau), \dot {\vec x}(\tau) ) } {\partial \dot x_n(\tau)} 
= \dot x_n(\tau) + \sum_{m=1}^N  \Lambda_{nm} x_m  (\tau)
\label{lagrangianFPderidot}
\end{eqnarray}
while the derivatives of the Lagrangian of Eq. \ref{lagrangianApp} with respect to the positions $ x_n(\tau) $
read
\begin{eqnarray}
\frac{\partial {\cal L} (\vec x(\tau), \dot {\vec x}(\tau) ) } {\partial  x_n(\tau)} 
= 
    \sum_{m=1}^N \dot  x_m  (\tau)    \Lambda_{mn} 
 +    \sum_{m=1}^N  W_{nm} x_m (\tau) 
\label{lagrangianderi}
\end{eqnarray}
These derivatives of Eqs \ref{lagrangianFPderidot}
and \ref{lagrangianderi}
can be plugged into Eq. \ref{lagrangeFPmotion}
to obtain the Lagrangian equations of motion as second order differential equations for the positions $x_n(\tau)$
\begin{eqnarray}
0 && =   \frac{d}{d\tau} \left(   \dot x_n(\tau)   + \sum_{m=1}^N  \Lambda_{nm} x_m  (\tau) \right)
-     \sum_{m=1}^N \dot  x_m  (\tau)    \Lambda_{mn} 
 -    \sum_{m=1}^N  W_{nm} x_m (\tau) 
\nonumber \\
&& =    \ddot x_n(\tau) + \sum_{m=1}^N \left(  \Lambda_{nm} - \Lambda_{mn} \right) \dot  x_m(\tau)      -  \sum_{m=1}^N  W_{nm}  x_m (\tau)
\nonumber \\
&& =    \ddot x_n(\tau) + \sum_{m=1}^N B_{nm} \dot  x_m(\tau)      -  \sum_{m=1}^N  W_{nm}  x_m (\tau)
\label{lagrangeFPmotionlinear}
\end{eqnarray}
The second term involving the magnetic matrix $B_{nm}$ of Eq. \ref{magneticNcstquad} 
and the velocities $\dot  x_m(\tau) $ represents the generalization of the Lorentz force in arbitrary dimension $N$
and does not depend upon the gauge choice for the vector potential.
The linear system of Eq. \ref{lagrangeFPmotionlinear}
can be rewritten in matrix form as
\begin{eqnarray}
0  && =      \vert \ddot {\vec x}(\tau) \rangle
+   {\bold B}  \vert \dot {\vec x}(\tau)\rangle   
  - {\bold W} \vert  {\vec x}(\tau)\rangle 
  \nonumber \\
  && =  \left( {\bold 1} \frac{d^2}{d\tau^2}  
+   {\bold B}  \frac{d}{d\tau}    
  - {\bold W} \right) \vert  {\vec x}(\tau)\rangle   
  \equiv {\bold M}(\tau) \vert  {\vec x}(\tau)\rangle  
\label{classicalmotionmatrix}
\end{eqnarray}
where we have introduced the second-order differential operator ${\bold M}(\tau) $
that involves the constant matrices $ {\bold B} $ and $ {\bold W}$
\begin{eqnarray}
{\bold M}(\tau) \equiv  {\bold 1} \frac{d^2}{d\tau^2}  
+   {\bold B}  \frac{d}{d\tau}    
  - {\bold W} 
\label{dynmt}
\end{eqnarray}
The standard method to solve this dynamics is to consider the expansion into eigenmodes
as described in the next subsection.


 \subsubsection{ Solving the Lagrangian classical equations of motion via an expansion into eigenmodes } 

Eigenmodes with a simple exponential time-dependence
\begin{eqnarray}
 \vert  {\vec x}(\tau)\rangle  = e^{\omega \tau}  \vert  {\vec z}_{\omega} \rangle
\label{eigendyn}
\end{eqnarray}
will satisfy the dynamical system of Eq. \ref{classicalmotionmatrix}
\begin{eqnarray}
0  =     \left( \omega^2  {\bold 1}
+ \omega  {\bold B}    - {\bold W} \right) \vert  {\vec z}_{\omega} \rangle
\equiv {\bold M}_{\omega} \vert  {\vec z}_{\omega} \rangle
\label{classicalmotioneigen}
\end{eqnarray}
if $\vert  {\vec z}_{\omega} \rangle $ is a right eigenvector associated to the eigenvalue zero 
for the matrix ${\bold M}_{\omega}  $ of parameter $\omega$
\begin{eqnarray}
{\bold M}_{\omega}  =    \omega^2  {\bold 1}+ \omega  {\bold B}    - {\bold W} 
\label{matrixMomega}
\end{eqnarray}
Therefore, the possible values for $\omega$ can be found from the condition of vanishing determinant
for the matrix ${\bold M}_{\omega}  $
\begin{eqnarray}
0 = \det ({\bold M}_{\omega}  ) = \det \left(\omega^2  {\bold 1}+ \omega  {\bold B}    - {\bold W}  \right)
\label{matrixMomegadet}
\end{eqnarray}
This polynomial equation of degree $(2N)$ will have $(2N)$ solutions $\omega_i$ with $i=1,..,2N$, 
and the corresponding eigenvectors $\vert  {\vec z}_{\omega_i} \rangle$
will satisfy Eq. \ref{classicalmotioneigen}
\begin{eqnarray}
0  =    {\bold M}_{\omega_i} \vert  {\vec z}_{\omega_i} \rangle
\label{classicalmotioneigeni}
\end{eqnarray}
The general solution of the equation of motion \ref{classicalmotionmatrix}
is then obtained as a linear combination of these $(2N)$ eigenmodes
\begin{eqnarray}
\vert  {\vec x}(\tau)\rangle  =    \sum_{i=1}^{2N} \eta_i e^{\omega_i \tau} \vert  {\vec z}_{\omega_i} \rangle
\label{classicalmotioneigeniCL}
\end{eqnarray}
where the $(2N)$ constants $\eta_i$ have to be determined by the boundary conditions at $\tau=0$ and at $\tau=t$
\begin{eqnarray}
\vert  \vec y \rangle = \vert  \vec x (\tau=0)\rangle &&  =    \sum_{i=1}^{2N} \eta_i  \vert  {\vec z}_{\omega_i} \rangle
\nonumber \\
\vert  \vec x \rangle = \vert  \vec x (\tau=t)\rangle &&  =    \sum_{i=1}^{2N} \eta_i e^{\omega_i t} \vert  {\vec z}_{\omega_i} \rangle
\label{classicalmotioneigeniCLBC}
\end{eqnarray}

For the classical action below, one will need the velocity obtained from Eq. \ref{classicalmotioneigeniCL}
\begin{eqnarray}
\vert  \dot {\vec x}(\tau)\rangle  =    \sum_{i=1}^{2N} \eta_i \omega_i e^{\omega_i \tau} \vert  {\vec z}_{\omega_i} \rangle
\label{classicalmotioneigeniCLvelo}
\end{eqnarray}
at the initial time $\tau=0$ and at the final time $\tau=t$.


 \subsubsection{ Evaluation of the action $S_{cl}(\vec x,t \vert \vec y, 0) $ of the classical trajectory } 

For this classical trajectory $ {\vec x}(\tau)$, one needs to evaluate the corresponding 
action $S_{cl}(\vec x,t \vert \vec y, 0) $
that involves the Lagrangian of Eq. \ref{lagrangianApp}  
\begin{eqnarray}
S_{cl}(\vec x,t \vert \vec y, 0) && =  \int_0^t d \tau {\cal L} (\vec x(\tau), \dot {\vec x}(\tau)
\nonumber \\
&&  =  \int_0^t d \tau 
  \left[ 
  \frac{1}{2}  \sum_{n=1}^N \dot  x_n^2  (\tau) 
 + \sum_{n=1}^N  \sum_{m=1}^N \dot  x_n  (\tau)    \Lambda_{nm} x_m  (\tau)
 + \sum_{n=1}^N   \sum_{m=1}^N  \frac{W_{nm}}{2} x_n  (\tau) x_m (\tau) +V_0
  \right]
\label{actiondef}
\end{eqnarray}
The first term corresponding to the kinetic energy can be rewritten via an integration by parts that allows 
to use the equation of motion of Eq. \ref{lagrangeFPmotionlinear}
to replace the accelerations $\ddot x_n (\tau) $
to obtain
\begin{eqnarray}
  \int_0^t d \tau    \dot  x_n^2  (\tau) 
&& =   \left[ x_n (\tau) \dot x_n (\tau) \right]_{\tau=0}^{\tau=t}
 - \int_0^t d \tau x_n (\tau) \ddot x_n (\tau)  
  \nonumber \\
&& = 
\left[ x_n (\tau) \dot x_n (\tau) \right]_{\tau=0}^{\tau=t} 
+  \int_0^t d \tau x_n (\tau) 
\bigg( 
 \sum_{m=1}^N B_{nm} \dot  x_m(\tau)    
  -  \sum_{m=1}^N  W_{nm}  x_m(\tau))
\bigg)   
\label{actionintegparts}
\end{eqnarray}
Plugging this expression into the action of Eq. \ref{actiondef}
yields
\begin{eqnarray}
S_{cl}(\vec x,t \vert \vec y, 0) &&   =  \frac{1}{2} \sum_{n=1}^N 
\left( \left[ x_n (\tau) \dot x_n (\tau) \right]_{\tau=0}^{\tau=t} 
+  \int_0^t d \tau x_n (\tau) 
\bigg( 
 \sum_{m=1}^N B_{nm} \dot  x_m(\tau)    
  -  \sum_{m=1}^N  W_{nm}  x_m(\tau))
\bigg)   \right)
  \nonumber \\
&&+ \int_0^t d \tau 
  \left[ 
   \sum_{n=1}^N  \sum_{m=1}^N \dot  x_n  (\tau)    \Lambda_{nm} x_m  (\tau)
 + \sum_{n=1}^N   \sum_{m=1}^N  \frac{W_{nm}}{2} x_n  (\tau) x_m (\tau) +V_0
  \right]
     \nonumber \\
&&  =  \sum_{n=1}^N    \frac{ x_n (t) \dot x_n (t) - x_n (0) \dot x_n (0)  }{ 2 }
+ \sum_{n=1}^N \sum_{m=1}^N \int_0^t d \tau  \dot  x_n(\tau)
\bigg(  \Lambda_{nm} - \frac{B_{nm}}{2}  
           \bigg)x_m (\tau)
+ t V_0
\label{actioncalcul}
\end{eqnarray}
While the classical trajectory $\vec x(\tau) $ solution of Eq. \ref{lagrangeFPmotionlinear} is gauge-invariant,
the value of the classical action $S_{cl}(\vec x,t \vert \vec y, 0) $ depends on the gauge
via the second term that involves the matrix $\Lambda_{mn}$ parametrizing the vector potential of Eq. \ref{vectorpotlinear} and not only the magnetic matrix of Eq. \ref{magneticNcstquad}.
In terms of the Coulomb vector potential $A^{Coulomb}_n  (\vec x ) $ discussed in Eq. \ref{coulomb} 
of the main text and the corresponding gauge transformation of Eq. \ref{gaugetransfolinearcoulomb} involving the function 
\begin{eqnarray}
\nu^{Coulomb}(\vec x) &&  
 =  \langle \vec x \vert \bigg(  \frac{ 
 {\bold \Lambda}+{\bold \Lambda}^T  }{4} \bigg)  \vert  \vec x \rangle
 \label{muquadraticcoulombLambda}
\end{eqnarray}
the second term of Eq. \ref{actioncalcul}
can be rewritten as
\begin{eqnarray}
&& \sum_{n=1}^N \sum_{m=1}^N \int_0^t d \tau  \dot  x_n(\tau)
\bigg( \Lambda_{nm} - \frac{B_{nm}}{2}  
            \bigg)x_m (\tau)
 = \sum_{n=1}^N  \int_0^t d \tau  \dot  x_n(\tau)
\bigg( A_n^{Coulomb} ( {\vec x}(\tau)) -  A_n ( {\vec x}(\tau))  \bigg)
\nonumber \\
&& =   \int_0^t d \tau \sum_{n=1}^N \dot  x_n(\tau) \partial_n \nu^{Coulomb}(\vec x(\tau))
=  \int_0^t d \tau \partial_{\tau} \nu^{Coulomb}(\vec x(\tau)) = \left[\nu^{Coulomb}(\vec x(\tau)) \right]_{\tau=0}^{\tau=t}
\nonumber \\
&&= \nu^{Coulomb}(\vec x(t)) - \nu^{Coulomb}(\vec x(0))
\label{actioncalculsecond}
\end{eqnarray}
Plugging this result into Eq. \ref{actioncalcul}
yields the final result for the classical action 
\begin{eqnarray}
S_{cl}(\vec x,t \vert \vec y, 0) &&   = 
    \frac{ \vec x (t) . \dot {\vec x }(t) - \vec x (0) . \dot {\vec x} (0)  }{ 2 }
+ \nu^{Coulomb}(\vec x(t)) - \nu^{Coulomb}(\vec x(0))
+ t V_0
     \nonumber \\
&&  =    \frac{ \vec x  . \dot {\vec x }(t) - \vec y . \dot {\vec x} (0)  }{ 2 }
+ \nu^{Coulomb}(\vec x) - \nu^{Coulomb}(\vec y)
+ t V_0
 \label{actionres}
\end{eqnarray}
where one still needs to compute the initial velocity $\dot {\vec x} (0) $ and the final velocity $\dot {\vec x }(t) $
in terms of the initial position $\vec x (\tau=0)=\vec y$
and final position $\vec x (\tau=t)=\vec x$ from the solution of Eqs 
\ref{classicalmotioneigeniCLBC} and 
\ref{classicalmotioneigeniCLvelo}.


 \subsubsection{ Example in dimension $N=2$  } 

\label{subsec_lagrangianN2}
 
To be more concrete, let us now show how the general framework described above
can be applied in dimension $N=2$.
 The $2 \times 2$ matrix ${\bold M}_{\omega}  $ of parameter $\omega$ of Eq. \ref{matrixMomega}
 read in terms of the magnetic field $B=B_{12}=-B_{21}$ and the symmetric matrix ${\bold W}$
\begin{eqnarray}
{\bold M}_{\omega}  =    \omega^2  {\bold 1}+ \omega  {\bold B}    - {\bold W} 
=   \begin{pmatrix} 
\omega^2 - W_{11} & \omega B- W_{12} \\
-\omega B- W_{12} & \omega^2- W_{22}
 \end{pmatrix}  
\label{matrixMomegaN2}
\end{eqnarray}
The possible values for $\omega$ are the solutions of Eq. \ref{matrixMomegadet}  
\begin{eqnarray}
0 && = \det ({\bold M}_{\omega}  ) 
= \left[ \omega^2 - W_{11}\right] \left[ \omega^2- W_{22}\right]
+ \left[ \omega B + W_{12} \right] \left[ \omega B- W_{12} \right] 
\nonumber \\
&& = \omega^4 + \omega^2 \left[ B^2- (W_{11}+W_{22} )   \right] 
+ (W_{11} W_{22} - W_{12}^2) 
\label{matrixMomegadetN2}
\end{eqnarray}
The two solutions $\omega_{\pm}^2 $ of this second-order equation for $\omega^2$ read
\begin{eqnarray}
\omega_{\pm}^2  = \frac{ (W_{11}+W_{22} )-B^2 \pm \sqrt{ \Delta} }{2}
\label{freqN2}
\end{eqnarray}
in terms of the discriminant
\begin{eqnarray}
 \Delta =  \left[ B^2- (W_{11}+W_{22} )   \right] ^2 -4 (W_{11} W_{22} - W_{12}^2) 
 = B^4 - 2 B^2 (W_{11}+W_{22} ) + (W_{11}-W_{22} )^2 + 4 W_{12}^2
\label{deltaN2}
\end{eqnarray}
So one obtains that the four roots of Eq. \ref{matrixMomegadetN2}
are of the form $\omega_i=(\pm \omega_+;\pm \omega_-)$ with
\begin{eqnarray}
\omega_{\pm}  = \sqrt{\omega_{\pm}^2} = \sqrt{ \frac{ (W_{11}+W_{22} )-B^2 \pm \sqrt{ \Delta} }{2} }
\label{freqN2sqrt}
\end{eqnarray}
For later comparisons with other methods, it is useful to mention their product and their sum
\begin{eqnarray}
\omega_+  \omega_- && =  \sqrt{ \frac{ \left[ (W_{11}+W_{22} )-B^2\right]^2  - \Delta }{4} } 
= \sqrt{ W_{11} W_{22} - W_{12}^2} = \sqrt{ \det(\bold W) } 
\nonumber \\
\omega_+  + \omega_- && = \sqrt{ (\omega_+  + \omega_-  )^2} 
=  \sqrt{ (\omega_+^2  + \omega_-^2  ) + 2 \omega_+ \omega_-}
=  \sqrt{ (W_{11}+W_{22} -B^2 ) + 2\sqrt{ \det(\bold W) }  }
\label{freqN2sqrtsumprod}
\end{eqnarray}

The general solution of the equation of motion \ref{classicalmotionmatrix}
can be written as the linear combination of Eq. \ref{classicalmotioneigeniCL}
\begin{eqnarray}
\vert  {\vec x}(\tau)\rangle  
=    \eta_{++} e^{\omega_+ \tau} \vert  {\vec z}^{[+\omega_+]} \rangle
+  \eta_{-+} e^{- \omega_+ \tau} \vert  {\vec z}^{[-\omega_+]} \rangle
+  \eta_{+-} e^{\omega_- \tau} \vert  {\vec z}^{[+\omega_-]} \rangle
+  \eta_{--} e^{- \omega_- \tau} \vert  {\vec z}^{[-\omega_-]} \rangle
\label{classicalmotioneigeniCLGyr}
\end{eqnarray}
with the following notations :

(i)  the four eigeneigenvectors $\vert  {\vec z}^{\omega_i} \rangle$ 
should be computed from Eq. \ref{classicalmotioneigeni}
\begin{eqnarray}
0  =    {\bold M}_{\omega_i} \vert  {\vec z}^{[\omega_i]} \rangle
=   \begin{pmatrix} 
\omega_i^2 - W_{11} & \omega_i B- W_{12} \\
-\omega_i B- W_{12} & \omega_i^2- W_{22}
 \end{pmatrix}   
   \begin{pmatrix} 
z_1^{[\omega_i]}   \\
 z_2^{[\omega_i]}
 \end{pmatrix}
\label{classicalmotioneigeniGyr}
\end{eqnarray}
so that one can choose
\begin{eqnarray}
z_1^{[\omega_i]}   && =\omega_i^2- W_{22}
\nonumber \\
 z_2^{[\omega_i]} && = \omega_i B + W_{12}
\label{classicalmotioneigeniGyrchoice}
\end{eqnarray}

(ii) The four constants $\eta_{\pm,\pm}$ have to be determined by the $4 \times 4$ linear system
fixing the boundary conditions at $\tau=0$ and at $\tau=t$ of Eq. \ref{classicalmotioneigeniCLBC}
\begin{eqnarray}
y_1 && =x_1(\tau=0)  = \eta_{++}  z_1^{[+\omega_+]} 
+  \eta_{-+}  z_1^{[-\omega_+]} 
+  \eta_{+-}  z_1^{[+\omega_-]} 
+  \eta_{--}    z_1^{[-\omega_-]}  
\nonumber \\
&& =(\omega_+^2- W_{22} )  (\eta_{++}  +  \eta_{-+} )
+(\omega_-^2- W_{22} ) (\eta_{+-}  +  \eta_{--} ) 
\nonumber \\
 y_2 && =x_2(\tau=0)= \eta_{++}  z_2^{[+\omega_+]} 
+  \eta_{-+}  z_2^{[-\omega_+]} 
+  \eta_{+-}  z_2^{[+\omega_-]} 
+  \eta_{--}    z_2^{[-\omega_-]}  
\nonumber \\
&&=  \left[ B\omega_+ +W_{12}\right]\eta_{++}
+   \left[ -B\omega_+ +W_{12}\right]\eta_{-+}
+   \left[ B\omega_- +W_{12}\right]\eta_{+-}
+      \left[ -B\omega_- +W_{12}\right]\eta_{--}
\nonumber \\
x_1 &&=x_1(\tau=t)  = \eta_{++} e^{\omega_+ t} z_1^{[+\omega_+]} 
+  \eta_{-+} e^{-\omega_+ t} z_1^{[-\omega_+]} 
+  \eta_{+-} e^{\omega_- t} z_1^{[+\omega_-]} 
+  \eta_{--} e^{-\omega_- t}   z_1^{[-\omega_-]}  
\nonumber \\
&&= (\omega_+^2- W_{22} )  (e^{\omega_+ t}\eta_{++}  + e^{-\omega_+ t} \eta_{-+} )
+(\omega_-^2- W_{22} ) (e^{\omega_- t}\eta_{+-}  + e^{-\omega_- t} \eta_{-+} )
\nonumber \\
 x_2 &&=x_2(\tau=t) = \eta_{++} e^{\omega_+ t} z_2^{[+\omega_+]} 
+  \eta_{-+} e^{-\omega_+ t} z_2^{[-\omega_+]} 
+  \eta_{+-} e^{\omega_- t} z_2^{[+\omega_-]} 
+  \eta_{--} e^{-\omega_- t}   z_2^{[-\omega_-]}  
\nonumber \\
&&= \left[ B\omega_+ +W_{12}\right]e^{\omega_+ t}\eta_{++}
+   \left[ -B\omega_+ +W_{12}\right]e^{-\omega_+ t}\eta_{-+}
+   \left[ B\omega_- +W_{12}\right]e^{\omega_- t}\eta_{+-}
+      \left[ -B\omega_- +W_{12}\right]e^{-\omega_- t}\eta_{--}
\nonumber \\
\label{classicalmotioneigeniCLBCGyr}
\end{eqnarray}

The initial and final velocities $\dot {\vec x}(\tau) $ of Eq. \ref{classicalmotioneigeniCLvelo}
at $\tau=0$ and $\tau=t$
\begin{small}
\begin{eqnarray}
\dot x_1(0)  &&  =(\omega_+^2- W_{22} ) \omega_+ (\eta_{++}  -  \eta_{-+} )
+(\omega_-^2- W_{22} ) \omega_-(\eta_{+-}  -  \eta_{--} ) 
\nonumber \\
\dot x_2(0)
&&=  \left[ B\omega_+ +W_{12}\right] \omega_+\eta_{++}
+   \left[ B\omega_+ - W_{12}\right] \omega_+\eta_{-+}
+   \left[ B\omega_- +W_{12}\right]\omega_-\eta_{+-}
+      \left[ B\omega_- - W_{12}\right]\omega_-\eta_{--}
\nonumber \\
\dot x_1(t)  
&&= (\omega_+^2- W_{22} ) \omega_+ (e^{\omega_+ t}\eta_{++}  - e^{-\omega_+ t} \eta_{-+} )
+(\omega_-^2- W_{22} )\omega_- (e^{\omega_- t}\eta_{+-}  - e^{-\omega_- t} \eta_{-+} )
\nonumber \\
 \dot x_2(t) 
&&= \left[ B\omega_+ +W_{12}\right]\omega_+e^{\omega_+ t}\eta_{++}
+   \left[ B\omega_+ -W_{12}\right]\omega_+e^{-\omega_+ t}\eta_{-+}
+   \left[ B\omega_- +W_{12}\right]\omega_-e^{\omega_- t}\eta_{+-}
+      \left[ B\omega_- -W_{12}\right]\omega_-e^{-\omega_- t}\eta_{--}
\nonumber \\
\label{classicalmotioneigeniCLvelogyr}
\end{eqnarray}
\end{small}
and the function $\nu^{Coulomb}(\vec x) $ of Eq. \ref{muquadraticcoulombLambda}
\begin{eqnarray}
\nu^{Coulomb}(\vec x) &&  
 =  \langle \vec x \vert \bigg(  \frac{ 
 {\bold \Lambda}+{\bold \Lambda}^T  }{4} \bigg)  \vert \vec x \rangle
 =    \frac{   \Lambda_{11} }{2} x_1^2  
  +  \frac{   \Lambda_{22}}{2} x_2^2 
   +  \frac{   \Lambda_{12}+\Lambda_{21} }{2} x_1 x_2 
 \label{muquadraticcoulombGyr}
\end{eqnarray}
allows to compute
the action of Eq. \ref{actionres}
  \begin{eqnarray}
S_{cl}(x_1,x_2,t \vert y_1,y_2, 0) &&  
  =    \frac{ x_1  . \dot x_1(t)+ x_2  . \dot x_2(t) - y_1  . \dot x_1(0) - y_2  . \dot x_2(0)  }{ 2 }
+ \nu^{Coulomb}(\vec x) - \nu^{Coulomb}(\vec y)
+ t V_0
 \label{actionresGyr}
\end{eqnarray}

So even in dimension $N=2$, this method is somewhat heavy and
it is thus much simpler to obtain the classical action via another method that will be described below.


 \subsection{ Computation of the classical action $S_{cl}(\vec x,t \vert \vec y ,0) $  via the alternative Hamiltonian perspective } 
 
 \label{subsec_hamiltonian}

  \subsubsection{ Hamilton's equations of motion for the classical trajectory  } 

Instead of the Lagrangian second-order equation of motion for the position $\vec x(\tau)$
discussed in subsection \ref{subsec_lagrangianmotion} of the present appendix,
one can use the Hamiltonian perspective where one writes first-order differential equations
for the positions $x_n(t)$ and the momenta $p_n(t)$ of Eq. \ref{lagrangianFPderidot}.
This definition of Eq. \ref{lagrangianFPderidot} for the momenta 
already gives the first-order dynamics for the positions $x_n(t)$
\begin{eqnarray}
\dot x_n(\tau) =p_n(\tau) - \sum_{m=1}^N  \Lambda_{nm} x_m  (\tau)
\label{Hamilton1}
\end{eqnarray}
Then Eqs \ref{lagrangeFPmotion} and \ref{lagrangianderi}
give the first-order dynamics for the momenta $p_n(t)$
after the elimination of the velocities $\dot x_n(\tau) $ via Eq. \ref{Hamilton1}
\begin{eqnarray}
      \dot p_n(\tau)  && =
     \sum_{m=1}^N \dot  x_m  (\tau)    \Lambda_{mn} 
 +    \sum_{m=1}^N  W_{nm} x_m (\tau) 
\nonumber \\
&& =      \sum_{m=1}^N   p_m  (\tau)    \Lambda_{mn} 
  +    \sum_{l=1}^N  \left(  W_{nl} -  \sum_{m=1}^N    \Lambda_{ml}     \Lambda_{mn}  \right)x_l (\tau) 
 \nonumber \\
&& =        \sum_{m=1}^N  ({\bold \Lambda}^T)_{nm}   p_m  (\tau)  
  +    \sum_{l=1}^N  \left(  W_{nl} -   ({\bold \Lambda}^T {\bold \Lambda})_{nl}     \right)x_l (\tau) 
\label{Hamilton2}
\end{eqnarray}

The dynamics Eqs \ref{Hamilton1}
and \ref{Hamilton2}
can be summarized in the matrix form as
\begin{eqnarray}
  \vert \dot {\vec x}(\tau)\rangle   && =   \vert  {\vec p} (\tau)\rangle  - {\bold \Lambda}  \vert  {\vec x}(\tau)\rangle
  \nonumber \\
  \vert \dot {\vec p} (\tau)\rangle    && =   {\bold \Lambda}^T\vert  {\vec p} (\tau)\rangle
  + \left(  {\bold W} -   {\bold \Lambda}^T {\bold \Lambda}    \right) \vert  {\vec x}(\tau)\rangle
\label{Hclassicalmotion}
\end{eqnarray}
or equivalently with a ket of size $(2N)$ containing both the $N$ positions $x_n(\tau)$ and the $N$ momenta $p_n(\tau)$
\begin{eqnarray}
 \frac{d}{d \tau}   \begin{pmatrix} 
\vert  {\vec x}(\tau)\rangle   \\
\vert  {\vec p} (\tau)\rangle 
 \end{pmatrix} =  {\cal M} 
  \begin{pmatrix} 
\vert  {\vec x}(\tau)\rangle   \\
\vert  {\vec p} (\tau)\rangle 
 \end{pmatrix}    
\label{Hclassicalmotionmatrix}
\end{eqnarray}
where the matrix ${\cal M }$ of size $(2N)\times(2N)$
displays the following structure in terms of four blocks of size $N \times N$
  \begin{eqnarray}
{\cal M }  = \begin{pmatrix} 
- {\bold \Lambda}   & {\bold 1}  \\
\left(  {\bold W} -   {\bold \Lambda}^T {\bold \Lambda}    \right)&   {\bold \Lambda}^T
 \end{pmatrix}    
 \label{HamiltonMmatrix}
\end{eqnarray}


 \subsubsection{ Solving the Hamiltonian classical equations of motion via the spectral decomposition of the matrix ${\cal M }  $ } 
 
 The technical advantage of the Hamilton first-order dynamics of Eq. \ref{Hclassicalmotionmatrix}
 is that the solution can be written in matrix form in terms of the initial condition
\begin{eqnarray}
  \begin{pmatrix} 
\vert  {\vec x}(\tau)\rangle   \\
\vert  {\vec p} (\tau)\rangle 
 \end{pmatrix} = e^{ {\cal M} \tau }
  \begin{pmatrix} 
\vert  {\vec x}(0)\rangle   \\
\vert  {\vec p} (0)\rangle 
 \end{pmatrix}    
\label{Hclassicalmotionmatrixsol}
\end{eqnarray}

It is thus useful to introduce the spectral decomposition of the matrix ${\cal M} $ of Eq. \ref{HamiltonMmatrix}
using block notations
  \begin{eqnarray}
{\cal M }  = \sum_{i=1}^{2N} \omega_i    
\begin{pmatrix} 
\vert {\vec z}^R_{\omega_i} \rangle   \\
\vert  {\vec \pi}^R_{\omega_i}\rangle 
 \end{pmatrix}
 \begin{pmatrix} 
\langle {\vec z}^L_{\omega_i} \vert   &
\langle  {\vec \pi}^L_{\omega_i}\vert 
 \end{pmatrix}
 = \sum_{i=1}^{2N} \omega_i    
\begin{pmatrix} 
\vert {\vec z}^R_{\omega_i} \rangle \langle {\vec z}^L_{\omega_i} \vert & \vert {\vec z}^R_{\omega_i} \rangle \langle  {\vec \pi}^L_{\omega_i}\vert \\
\vert  {\vec \pi}^R_{\omega_i}\rangle \langle {\vec z}^L_{\omega_i} \vert& \vert  {\vec \pi}^R_{\omega_i}\rangle\langle  {\vec \pi}^L_{\omega_i}\vert 
 \end{pmatrix}
 \label{HamiltonMmatrixdiag}
\end{eqnarray}
in terms of its $(2N)$ eigenvalues $\omega_i$ 
and the corresponding right eigenvectors and left eigenvectors 
\begin{eqnarray}
{\cal M }     
\begin{pmatrix} 
\vert {\vec z}^R_{\omega_i} \rangle   \\
\vert  {\vec \pi}^R_{\omega_i}\rangle 
 \end{pmatrix} && =  \omega_i    
\begin{pmatrix} 
\vert {\vec z}^R_{\omega_i} \rangle   \\
\vert  {\vec \pi}^R_{\omega_i}\rangle 
 \end{pmatrix}
 \nonumber \\
 \begin{pmatrix} 
\langle {\vec z}^L_{\omega_i} \vert   &
\langle  {\vec \pi}^L_{\omega_i}\vert 
 \end{pmatrix} {\cal M }     &&=  \omega_i
  \begin{pmatrix} 
\langle {\vec z}^L_{\omega_i} \vert   &
\langle  {\vec \pi}^L_{\omega_i}\vert 
 \end{pmatrix}
 \label{diagoMeigen}
\end{eqnarray} 
satisfying the standard orthonormalization and closure properties.
 
 Plugging the spectral decomposition of Eq. \ref{HamiltonMmatrixdiag} into the solution of Eq. \ref{Hclassicalmotionmatrixsol}
 yields the expansion into eigenmodes with simple exponential time-dependence 
 \begin{eqnarray}
  \begin{pmatrix} 
\vert  {\vec x}(\tau)\rangle   \\
\vert  {\vec p} (\tau)\rangle 
 \end{pmatrix} = 
 \sum_{i=1}^{2N} e^{\omega_i   \tau } 
\begin{pmatrix} 
\vert {\vec z}^R_{\omega_i} \rangle   \\
\vert  {\vec \pi}^R_{\omega_i}\rangle 
 \end{pmatrix}
 \begin{pmatrix} 
\langle {\vec z}^L_{\omega_i} \vert   &
\langle  {\vec \pi}^L_{\omega_i}\vert 
 \end{pmatrix}
  \begin{pmatrix} 
\vert  {\vec x}(0)\rangle   \\
\vert  {\vec p} (0)\rangle 
 \end{pmatrix}    
 =  \sum_{i=1}^{2N} \eta_i e^{\omega_i \tau}  
  \begin{pmatrix} 
\vert {\vec z}^R_{\omega_i} \rangle   \\
\vert  {\vec \pi}^R_{\omega_i}\rangle 
 \end{pmatrix}
\label{Hclassicalmotionmatrixsoleigen}
\end{eqnarray}
with the coefficients
\begin{eqnarray}
 \eta_i =\langle {\vec z}^L_{\omega_i} \vert  {\vec x}(0)\rangle  +\langle  {\vec \pi}^L_{\omega_i}\vert  {\vec p} (0)\rangle 
\label{etaiHamilton}
\end{eqnarray}

Let us now mention the correspondence with the expansion of Eq. \ref{classicalmotioneigeniCL}
within the Lagrangian perspective. 
The eigenvalue Eq. \ref{diagoMeigen} for the right eigenvector 
can be decomposed into the two following equations for the blocks using Eq. \ref{HamiltonMmatrix}
\begin{eqnarray}
0 && = ( \omega_i {\bold 1} + {\bold \Lambda})   \vert {\vec z}^R_{\omega_i} \rangle - \vert  {\vec \pi}^R_{\omega_i}\rangle 
  \nonumber \\
0 && =( \omega_i {\bold 1} - {\bold \Lambda}^T) \vert  {\vec \pi}^R_{\omega_i}\rangle 
-  \left(  {\bold W} -   {\bold \Lambda}^T {\bold \Lambda}    \right) \vert {\vec z}^R_{\omega_i} \rangle
  \label{diagoMeigenblock}
\end{eqnarray} 
As a consequence, one can apply $( \omega_i {\bold 1} - {\bold \Lambda}^T) $ to the first 
equation and use the second equation to obtain a closed equation for $\vert {\vec z}^R_{\omega_i} \rangle $
\begin{eqnarray}
0 && =( \omega_i {\bold 1} - {\bold \Lambda}^T) ( \omega_i {\bold 1} + {\bold \Lambda})   \vert {\vec z}^R_{\omega_i} \rangle - ( \omega_i {\bold 1} - {\bold \Lambda}^T)\vert  {\vec \pi}^R_{\omega_i}\rangle 
  \nonumber \\
 && =( \omega_i {\bold 1} - {\bold \Lambda}^T) ( \omega_i {\bold 1} + {\bold \Lambda})   \vert {\vec z}^R_{\omega_i} \rangle
 - \left(  {\bold W} -   {\bold \Lambda}^T {\bold \Lambda}    \right) \vert {\vec z}^R_{\omega_i} \rangle
   \nonumber \\
 && = \left[ \omega_i^2 {\bold 1} 
 + \omega_i ( {\bold \Lambda}- {\bold \Lambda})^T) -  {\bold W} \right] 
    \vert {\vec z}^R_{\omega_i} \rangle
     \nonumber \\
 && = \left[ \omega_i^2 {\bold 1} 
 + \omega_i  {\bold B} -  {\bold W} \right] 
    \vert {\vec z}^R_{\omega_i} \rangle =   {\bold M}_{\omega_i} \vert  {\vec z}^R_{\omega_i} \rangle
  \label{diagoMeigenblockzonly}
\end{eqnarray} 
which coincides with Eqs \ref{matrixMomega} and \ref{classicalmotioneigeni}
of the Lagrangian perspective, as it should for consistency.


 \subsubsection{ Evaluation of the classical action $S_{cl}(\vec x,t \vert \vec y, 0) $ in the Hamilton perspective }

To translate the action of Eq. \ref{actionres} into Hamilton's variables, one just needs to 
use Eq. \ref{Hamilton1} to replace the 
initial velocity $\dot {\vec x} (0) $ and the final velocity $\dot {\vec x }(t) $
in terms of the initial and final positions and momenta
\begin{eqnarray}
\dot x_n(0) =p_n(0) - \sum_{m=1}^N  \Lambda_{nm} x_m  (0) = p_n(0) - \sum_{m=1}^N  \Lambda_{nm} y_m  
\nonumber \\
\dot x_n(t) =p_n(t) - \sum_{m=1}^N  \Lambda_{nm} x_m  (t) = p_n(t) - \sum_{m=1}^N  \Lambda_{nm} x_m 
\label{Hamilton1inifin}
\end{eqnarray}
to obtain
\begin{eqnarray}
S_{cl}(\vec x,t \vert \vec y, 0) &&    
  =    \frac{ \vec x  .  {\vec p }(t) - \vec y .  {\vec p} (0)  }{ 2 }
+ \xi(\vec x) - \xi(\vec y)
+ t V_0
 \label{actionresHcalcul}
\end{eqnarray}
where the new function $\xi(\vec x) $ 
is found to vanish when using the symmetrization of the first part and the explicit form of 
$\nu^{Coulomb}(\vec x) $ in Eq. \ref{muquadraticcoulomb}
\begin{eqnarray}
\xi(\vec x) \equiv  - \sum_{n=1}^N  \sum_{m=1}^N x_n  \frac{ \Lambda_{nm} }{2} x_m+ \nu^{Coulomb}(\vec x) 
= - \sum_{n=1}^N \sum_{m=1}^N  \frac{\Lambda_{nm}+\Lambda_{mn}}{4} x_n x_m
 + \nu^{Coulomb}(\vec x)=0
 \label{nudef}
\end{eqnarray}

So the classical action of Eq. \ref{actionresHcalcul}
reduces to
\begin{eqnarray}
S_{cl}(\vec x,t \vert \vec y, 0) &&    
  =    \frac{\langle \vec x   \vert  {\vec p} (t)\rangle - \langle \vec y   \vert  {\vec p} (0)\rangle  }{ 2 }
+ t V_0
 \label{actionresHres}
\end{eqnarray}
where one still needs to compute the initial momentum $ {\vec p} (0) $ and the final momentum $ {\vec p }(t) $
in terms of the initial position $\vec x (0)=\vec y$
and final position $\vec x (t)=\vec x$ from the solution of Eq \ref{Hclassicalmotionmatrix}.


 \subsubsection{ Taking advantage of the gauge freedom to choose a new gauge ${\mathring {\bold \Lambda} } $ that simplifies the Hamiltonian dynamics } 
 
 \label{subsec_ring}

Hamilton's equations of motion involve the momenta $p_n(\tau)$ as variables
and thus depend on the gauge choice for the vector potential via the matrix ${\bold \Lambda}$,
in contrast to the Lagrangian equations of motion of Eq. \ref{classicalmotionmatrix} that only involved the magnetic matrix.
As a consequence, a natural question is whether it can be helpful technically to choose another gauge 
\begin{eqnarray}
  {\mathring A}_n (\vec x )  = -  \sum_{m=1}^N {\mathring {\bold \Lambda} }_{nm} x_m 
   \label{newgaugelinearups}
\end{eqnarray} 
parametrized by the new matrix ${\mathring {\bold \Lambda} } $ instead of $\bold \Lambda$, 
with the same antisymmetric part fixed by the magnetic matrix ${\bold B} =-{\bold B}^T$ 
  \begin{eqnarray}
{\mathring {\bold \Lambda} }    - {\mathring {\bold \Lambda} }^T={\bold B} = {\bold \Lambda}   
  -  {\bold \Lambda}^T
 \label{upsulontildeBtilde}
\end{eqnarray}
but whose symmetric part could be chosen
to simplify the corresponding matrix ${\mathring {\cal M}}$ of Eq. \ref{HamiltonMmatrix}
  \begin{eqnarray}
{\mathring {\cal M}} = \begin{pmatrix} 
- {\mathring {\bold \Lambda} }   & {\bold 1}  \\
\left(  {\bold W} -   {\mathring {\bold \Lambda} }^T {\mathring {\bold \Lambda} }    \right)&  
 {\mathring {\bold \Lambda} }^T
 \end{pmatrix}    
 \label{HamiltonMmatrixtilde}
\end{eqnarray}
by making its left-lower block vanish
 \begin{eqnarray}
    {\mathring {\bold \Lambda} }^T {\mathring {\bold \Lambda} }  =  {\bold W} 
   \label{Leftlower}
\end{eqnarray} 
Indeed, as described in subsection \ref{subsec_PropagatorClassicalAction} of the main text
concerning the matrix ${\cal M}$ of Eq. \ref{HamiltonMmatrixSimple}
whose left-lower block vanishes, the Hamiltonian dynamics is then very simple.
The huge simplification produced by the vanishing of left-lower block
in the matrix of Eq. \ref{HamiltonMmatrixtilde} 
  \begin{eqnarray}
{\mathring {\cal M}} = \begin{pmatrix} 
- {\mathring {\bold \Lambda} }   & {\bold 1}  \\
0 &  
 {\mathring {\bold \Lambda} }^T
 \end{pmatrix}    
 \label{HamiltonMmatrixtildeSimple}
\end{eqnarray}
can also be seen by considering
its $(2N)$ eigenvalues $\omega_i$ that are the solutions of the characteristic polynomial of $ {\mathring {\cal M}}$
that can now be factorized with respect to its two diagonal blocks
  \begin{eqnarray}
0=\det ( \omega {\bold 1}_{2N} - {\mathring {\cal M}} )= 
\det( \omega {\bold 1} +{\mathring {\bold \Lambda} } )  \times
 \det( \omega {\bold 1} -{\mathring {\bold \Lambda} }^T )   
 \label{HamiltonMmatrixtildeSimpledet}
\end{eqnarray}
Hence, the $(2N)$ eigenvalues $\omega_i$ of $ {\mathring {\cal M}}$
are directly related to the $N$ eigenvalues ${\mathring {\bold \lambda} }_i  $ of the matrix ${\mathring {\bold \Lambda} } $ : 

(i) the vanishing of the first determinant
gives the first $N$ roots $\omega_i=- {\mathring {\bold \lambda} }_i$ for $i=1,..,N$.

(ii)  the vanishing of the second determinant
gives the other $N$ roots $\omega_{N+i}= {\mathring {\bold \lambda} }_i^*$ for $i=1,..,N$.

Since the condition of Eq. \ref{Leftlower}
 corresponds to $\frac{N(N+1)}{2} $ quadratic equations for the $\frac{N(N+1)}{2} $ matrix elements of
the symmetric part $( {\mathring {\bold \Lambda} }    + {\mathring {\bold \Lambda} }^T)$
that should be computed in terms of the $\frac{N(N-1)}{2} $ matrix elements of the 
antisymmetric magnetic matrix ${\bold B} $ and the $\frac{N(N+1)}{2} $ matrix elements
of the given symmetric matrix ${\bold W} $, the solution of Eq. \ref{Leftlower} is not unique.
Among the various solutions ${\mathring {\bold \Lambda} } $ satisfying Eq. \ref{Leftlower}, 
it will be convenient to choose the solution with good relaxation properties as 
 the matrix ${\bold \Gamma}$ of the main text, 
 i.e. the solution ${\mathring {\bold \Lambda} } $ whose $N$ eigenvalues $ {\mathring {\bold \lambda} }_i $ have positive real parts
 \begin{eqnarray}
 {\rm Re} ( {\mathring {\bold \lambda} }_i ) >0
 \label{conditionRealPartsapp}
\end{eqnarray}

The example of the dimension $N=2$ is described 
 in the next subsection.


 \subsubsection{ Computing the simple gauge ${\mathring {\bold \Lambda} } $ in dimension $N=2$   } 
 
 \label{subsec_ringN2}

Let us now describe how the simple gauge ${\mathring {\bold \Lambda} } $ satisfying 
 Eq. \ref{Leftlower} can be found in dimension $N=2$.
 The real matrix $ {\mathring {\bold \Lambda} } $  can be decomposed on the basis of Pauli matrices 
 $(\sigma_0={\bold 1},\sigma_x,\sigma_y,\sigma_z)$ with four real coefficients $(g_0,g_x,g_y,g_z)$ as follows
\begin{eqnarray}
 {\mathring {\bold \Lambda} } && = g_0 {\bold 1} + g_x  \sigma_x + g_y (i  \sigma_y) + g_z  \sigma_z
 =  \begin{pmatrix} 
g_0+g_z  & g_x+g_y \\
g_x-g_y & g_0-g_z
 \end{pmatrix}   
 \nonumber \\
 {\mathring {\bold \Lambda} }^T && = g_0 {\bold 1} + g_x \sigma_x - g_y (i \sigma_y) + g_z \sigma_z
 =  \begin{pmatrix} 
g_0+g_z  & g_x-g_y \\
g_x+g_y & g_0-g_z
 \end{pmatrix}    
 \label{basePauli}
\end{eqnarray}
The coefficient $g_y$ is fixed by the antisymmetry of Eq. \ref{upsulontildeBtilde}
\begin{eqnarray}
 {\mathring {\bold \Lambda} } -  {\mathring {\bold \Lambda} }^T&& = 2 g_y (i  \sigma_y) 
 =  \begin{pmatrix} 
0  & 2g_y \\
-2g_y & 0
 \end{pmatrix}  
 \label{basePaulianti}
\end{eqnarray}
The three other coefficients $(g_0,g_x,g_z)$ that parametrize the symmetric part
$  ({\mathring {\bold \Lambda} } + {\mathring {\bold \Lambda} }^T)$ should be chosen to satisfy Eq. \ref{Leftlower}.
In the Pauli basis, the identification of the product $ {\mathring {\bold \Lambda} }^T  {\mathring {\bold \Lambda} } $  
\begin{eqnarray}
 {\mathring {\bold \Lambda} }^T  {\mathring {\bold \Lambda} }&&  
 =  \begin{pmatrix} 
g_0+g_z  & g_x-g_y \\
g_x+g_y & g_0-g_z
 \end{pmatrix}   
  \begin{pmatrix} 
g_0+g_z  & g_x+g_y \\
g_x-g_y & g_0-g_z
 \end{pmatrix}   
\nonumber \\
&&  =   \begin{pmatrix} 
(g_0+g_z)^2+ (g_x-g_y)^2  & 2 (g_0 g_x+g_z g_y) \\
2 (g_0 g_x+g_z g_y) & (g_0-g_z)^2+ (g_x+g_y)^2 
 \end{pmatrix} 
 \nonumber \\
&& =
 \left[g_0^2+ g_x^2+g_y^2+ g_z^2 \right]  {\bold 1} 
 + 2 (g_0 g_x+g_z g_y) \sigma_x  
 +  2 (g_0 g_z-g_x g_y)  \sigma_z
 \label{basePauliproduit}
\end{eqnarray}
with the given symmetric matrix
\begin{eqnarray}
  {\bold W}   = \begin{pmatrix} 
W_{11}  & W_{12} \\
W_{12} & W_{22} 
 \end{pmatrix} 
 = w_0  {\bold 1} 
 + w_x \sigma_x  
+ w_z  \sigma_z
 \label{basePauliW}
\end{eqnarray}
which involves the coefficients
\begin{eqnarray}
w_0 && =  \frac{W_{11}  + W_{22}}{2}
\nonumber \\
w_x && = W_{12}
\nonumber \\
w_z && =  \frac{W_{11}  - W_{22}}{2}
 \label{basePauliproduitWcoefs}
\end{eqnarray}
leads to the following three quadratic equations for the three variables $(g_0,g_x,g_z)$
 \begin{eqnarray}
g_0^2+ g_x^2+g_y^2+ g_z^2 && = w_0
 \nonumber \\
g_0 g_x+ g_y g_z&& =  \frac{w_x   }{2}
\nonumber \\
-g_y g_x + g_0 g_z && = \frac{w_z}{2}   
 \label{basePaulieqW}
\end{eqnarray}
The last two equations can be used to write $g_x$ and $g_z$ in terms of $g_0$
  \begin{eqnarray}
 g_x && = \frac{ g_0 w_x   -g_y w_z  }{ 2(g_0^2+g_y^2)}
\nonumber \\
 g_z && = \frac{g_0 w_z + g_y w_x        }{ 2(g_0^2+g_y^2)}
 \label{basePaulieqWxz}
\end{eqnarray}
One can then plug the corresponding value for
\begin{eqnarray}
g_x^2+g_z^2  = \frac{ \left[ g_0 w_x   -g_y w_z  \right]^2+\left[  g_0 w_z + g_y w_x    \right]^2 }{4(g_0^2+g_y^2)^2} 
 =   \frac{  w_x^2  + w_z^2 }{4(g_0^2+g_y^2)}
 \label{gxgzchoicesquare}
\end{eqnarray}
into the first Eq. \ref{basePaulieqW} to obtain the 
following closed equation for the remaining coefficient $g_0$
  \begin{eqnarray}
0 && = (g_0^2+g_y^2)+ \frac{ w_x^2  + w_z^2  }{4(g_0^2+g_y^2)}  - w_0  
 \label{basePaulieqWfirst}
\end{eqnarray}
 After multiplication by $(g_0^2+g_y^2) $, one obtains
 the following second order equation for $(g_0^2+g_y^2)$
\begin{eqnarray}
0  = (g_0^2+g_y^2)^2 - w_0  (g_0^2+g_y^2)
+ \frac{ w_x^2  + w_z^2  }{4} 
 \label{coefidentityfactor2dorder}
\end{eqnarray}
Using Eq. \ref{basePauliproduitWcoefs}, the discriminant is found to coincide with the determinant of
the matrix ${\bold W}$
\begin{eqnarray}
w_0^2 - (w_x^2  + w_z^2 ) = \frac{(W_{11}  + W_{22})^2-(W_{11}  - W_{22})^2}{4}-W_{12}^2
= W_{11}   W_{22}-W_{12}^2 = \det (\bold W) 
 \label{basePauliproduitWcoefsdetW}
\end{eqnarray}
while $w_0 = \frac{W_{11}  + W_{22}}{2}$ represents half the trace of the matrix ${\bold W}$
so that the two solutions of Eq. \ref{coefidentityfactor2dorder} for $g_0^2$ read
  \begin{eqnarray}
g_0^2+g_y^2 
= \frac{  w_0 \pm \sqrt{W_{11}   W_{22}-W_{12}^2} }{2} 
=   \frac{  w_0 \pm \sqrt{\det (\bold W)  } }{2} 
=  \frac{  \frac{ {\rm tr}(\bold W) }{2} \pm \sqrt{\det (\bold W)  } }{2} 
 \label{coefidentityfactor2dordersol}
\end{eqnarray}
The absolute value of the determinant 
of $ {\mathring {\bold \Lambda} }$ is fixed by the condition Eq. \ref{Leftlower}
that gives  
\begin{eqnarray}
 \det (\bold W) = \det( {\mathring {\bold \Lambda} }^T ) \det( {\mathring {\bold \Lambda} } )  = \left( \det( {\mathring {\bold \Lambda} } )\right)^2
 \label{detWdetUps}
\end{eqnarray}
but it is useful to compute the sign corresponding to the two solutions of Eq. \ref{coefidentityfactor2dordersol}
\begin{eqnarray}
\det( {\mathring {\bold \Lambda} } ) && = (g_0^2+g_y^2)-(g_z^2+g_x^2) 
=  \frac{ w_0 \pm \sqrt{\det (\bold W)  } }{2} 
-  \frac{  w_0^2-\det (\bold W)  }{ 2(w_0 \pm \sqrt{\det (\bold W) } ) }
= \frac{ w_0 \pm \sqrt{\det (\bold W)  } }{2} 
-  \frac{  w_0 \mp\sqrt{\det (\bold W) } }{ 2 }
\nonumber \\
&& = \pm \sqrt{\det (\bold W)}
 \label{basePaulidet}
\end{eqnarray}
The two eigenvalues ${\mathring  \lambda_{\pm} }  $ for the matrix $ {\mathring {\bold \Lambda} } $
can be computed from their product and sum given by the trace and the determinant 
of $ {\mathring {\bold \Lambda} }  $ respectively
  \begin{eqnarray}
  {\mathring  \lambda_+ }{\mathring  \lambda_- }&& = \det( {\mathring {\bold \Lambda} } ) =  \pm \sqrt{\det (\bold W)}
\nonumber \\
{\mathring  \lambda_+ }+{\mathring  \lambda_- } && = {\rm tr } ({\mathring {\bold \Lambda} })=2 g_0 
 \label{sumandproduct}
\end{eqnarray}
As explained around Eq. \ref{conditionRealPartsapp}, we wish to choose the matrix $ {\mathring {\bold \Lambda} } $ whose two eigenvalues $ {\mathring {\bold \lambda} }_{\pm} $ have strictly positive real parts
\begin{eqnarray}
 {\rm Re} ( {\mathring {\bold \lambda} }_{\pm} ) >0
 \label{conditionRealPartsN2}
\end{eqnarray}
So we choose the $+$ solution for $g_0^2$ in Eq. \ref{coefidentityfactor2dordersol} corresponding to the positive determinant $\det( {\mathring {\bold \Lambda} } ) = +  \sqrt{\det (\bold W)} $ in Eq. \ref{basePaulidet},
and then we choose the positive square-root for $g_0$
corresponding to the positive trace ${\rm tr } ({\mathring {\bold \Lambda} })=2 g_0>0$ 
  \begin{eqnarray}
g_0 
= \sqrt{  \frac{  \frac{ {\rm tr}(\bold W) }{2} + \sqrt{\det (\bold W)  } }{2} - g_y^2 }
 \label{coefidentityfactor2dordersolchoice}
\end{eqnarray}
The two eigenvalues of $ {\mathring {\bold \Lambda} } $ 
  \begin{eqnarray}
{\mathring  \lambda_{\pm} }  = g_0 \pm \sqrt{ g_0^2- \sqrt{\det (\bold W)}}
 \label{eigenlambdaring}
\end{eqnarray}
are in agreement with $\omega_{\pm}$ found in Eq. \ref{freqN2sqrtsumprod} via the Lagrangian perspective.
Then one can compute the corresponding real values of $(g_x,g_z)$ given by Eq. \ref{basePaulieqWxz}
  \begin{eqnarray}
 g_x && = \frac{ g_0 W_{12}   -g_y \frac{W_{11}  - W_{22}}{2}  }{ \frac{ {\rm tr}(\bold W) }{2} \pm \sqrt{\det (\bold W)  } }
\nonumber \\
 g_z && = \frac{g_0 \frac{W_{11}  - W_{22}}{2} + g_y W_{12}        }{ \frac{ {\rm tr}(\bold W) }{2} \pm \sqrt{\det (\bold W)  } }
 \label{basePaulieqWxzini}
\end{eqnarray}


\section{ Ornstein-Uhlenbeck processes : Diagonalization of the Hamiltonian ${\hat {\cal  H}} $ in the irreversible gauge } 

\label{app_diagoHirrcanonique}

In this Appendix, the quantum Hamiltonian ${\hat {\cal  H}} $ in the irreversible gauge
for Ornstein-Uhlenbeck processes discussed in subsection \ref{subsec_HirrOU} of the main text
is diagonalized in terms of canonical creation and annihilation operators.

\subsection{ Rewriting the quantum Hamiltonian ${\hat {\cal  H}} $ in terms of canonical operators $q_{ \alpha}$ and $q_{ \alpha}^{\dagger}$ }

The commutation relations of Eq. \ref{QnsusycommutPhi} 
yields that it is useful to introduce the spectral decomposition of the real symmetric matrix ${\bold \Phi } $
involving its $N$ real eigenvalues $\varphi_{\alpha}$ and the corresponding orthonormal eigenvectors $\vert \varphi_{\alpha} \rangle $
 \begin{eqnarray}
{\bold \Phi }= \sum_{\alpha=1}^N \varphi_{\alpha} \vert \varphi_{\alpha} \rangle \langle \varphi_{\alpha} \vert
  \label{diagoPhi}
\end{eqnarray}

The new annihilation and creation operators defined by the following linear combinations 
 \begin{eqnarray}
q_{ \alpha} \equiv \frac{1}{ \sqrt{ \varphi_{\alpha}}} \sum_{n=1}^N \langle \varphi_{\alpha} \vert n \rangle Q_n
\nonumber \\
q_{ \alpha}^{\dagger} \equiv  \frac{1}{ \sqrt{\varphi_{\alpha}}} \sum_{n=1}^N Q_n^{\dagger} \langle n \vert \varphi_{\alpha} \rangle 
  \label{qfrombigQ}
\end{eqnarray}
inherit from Eq. \ref{Qnsusycommut0} the vanishing commutators between two annihilation or two creation operators
\begin{eqnarray}
[q_{ \alpha} , q_{\beta} ]  = 0 = [q_{ \alpha}^{\dagger} , q_{\beta}^{\dagger} ]
\label{Qnsusycommut0small}
\end{eqnarray}
The commutators of Eq.  \ref{QnsusycommutPhi} using \ref{diagoPhi}
\begin{eqnarray}
[  Q_n, Q_m^{\dagger}]  =      \Phi_{nm}
= \sum_{\alpha'=1}^N \varphi_{\alpha'} \langle n \vert \varphi_{\alpha'} \rangle \langle \varphi_{\alpha'} \vert m \rangle
\label{QnsusycommutPhidiag}
\end{eqnarray}
yield that the commutators between one new annihilation operator $q_{ \alpha} $ and one new creation operator $q_{\beta}^{\dagger} $
 \begin{eqnarray}
[q_{ \alpha} ,q_{\beta}^{\dagger} ] && = \frac{1}{\sqrt{ \varphi_{\alpha} \varphi_{\beta}} } \sum_{n=1}^N \langle \varphi_{\alpha} \vert n \rangle  \sum_{m=1}^N  \langle m \vert \varphi_{\beta} \rangle 
[Q_n, Q_m^{\dagger}]
\nonumber \\
&& = \frac{1}{\sqrt{ \varphi_{\alpha} \varphi_{\beta}} } 
\sum_{\alpha'=1}^N  \varphi_{\alpha'} 
\left( \sum_{n=1}^N \langle \varphi_{\alpha} \vert n \rangle  \langle n \vert \varphi_{\alpha'} \rangle \right)
\left( \sum_{m=1}^N  
  \langle \varphi_{\alpha'} \vert m \rangle \langle m \vert \varphi_{\beta} \rangle  \right)
  \nonumber \\
&& = \frac{1}{\sqrt{ \varphi_{\alpha} \varphi_{\beta}} } 
\sum_{\alpha'=1}^N  \varphi_{\alpha'} \delta_{\alpha,\alpha'}\delta_{\alpha',\beta} = \delta_{\alpha,\beta}
  \label{smallqcomm}
\end{eqnarray}
are canonical.

Inverting the change of operators of Eq. \ref{qfrombigQ}
 \begin{eqnarray}
Q_n = \sum_{\alpha=1}^N \langle n \vert \varphi_{\alpha} \rangle \sqrt{ \varphi_{\alpha}} q_{ \alpha} 
\nonumber \\
Q_n^{\dagger} = \sum_{\alpha=1}^N \langle \varphi_{\alpha} \vert n \rangle \sqrt{ \varphi_{\alpha}} q^{\dagger}_{ \alpha} 
  \label{bigQfromq}
\end{eqnarray}
one obtains that
the reversible Hamiltonian of Eq. \ref{hamiltonianhateqsusyQOU} reads
\begin{eqnarray}
  {\hat {\cal  H}}_{rev}  && =   \frac{1}{2}   \sum_{n=1}^N 
Q_n^{\dagger}  Q_n 
=  \frac{1}{2}   \sum_{\alpha=1}^N \sqrt{ \varphi_{\alpha}}  q^{\dagger}_{ \alpha} 
\sum_{\beta=1}^N \sqrt{ \varphi_{\beta}} q_{ \beta} 
\left( \sum_{n=1}^N 
 \langle \varphi_{\alpha} \vert n \rangle 
  \langle n \vert \varphi_{\beta} \rangle  \right)
  = \frac{1}{2}   \sum_{\alpha=1}^N \sqrt{ \varphi_{\alpha}}  q^{\dagger}_{ \alpha} 
\sum_{\beta=1}^N \sqrt{ \varphi_{\beta}} q_{ \beta} 
\delta_{\alpha,\beta}
 \nonumber \\
 &&  =  \sum_{\alpha=1}^N    q^{\dagger}_{ \alpha}  \frac{\varphi_{\alpha}}{2}  q_{ \alpha} 
 \label{hrevsmallq}
\end{eqnarray}
while the irreversible Hamiltonian of Eq. \ref{HirrOUomegaq}
becomes
\begin{eqnarray}
  {\hat {\cal  H}}_{irr}  && =    
   \sum_n    \sum_m \Omega_{nm}   Q_n^{\dagger} Q_m 
   =  \sum_{\alpha=1}^N \sqrt{ \varphi_{\alpha}} q^{\dagger}_{ \alpha} 
    \sum_{\beta=1}^N \sqrt{ \varphi_{\beta}} q_{ \beta} 
   \left(   \sum_n    \sum_m   
    \langle \varphi_{\alpha} \vert n \rangle \Omega_{nm} 
 \langle m \vert \varphi_{\beta} \rangle \right)
  \nonumber \\
 && = \sum_{\alpha=1}^N    \sum_{\beta=1}^N q^{\dagger}_{ \alpha} 
    \left( \sqrt{ \varphi_{\alpha}}     \langle \varphi_{\alpha} \vert {\bold \Omega} \vert \varphi_{\beta} \rangle 
   \sqrt{ \varphi_{\beta}} \right)    q_{ \beta} 
 \label{hrirrevsmallq}
\end{eqnarray}

Using Eq. \ref{gammamatrix} and the inverse of Eq. \ref{diagoPhi}
 \begin{eqnarray}
{\bold \Phi }^{-1}= \sum_{\alpha=1}^N \frac{1}{\varphi_{\alpha} } \vert \varphi_{\alpha} \rangle \langle \varphi_{\alpha} \vert
  \label{diagoPhiinv}
\end{eqnarray}
the total Hamiltonian reads in terms of the new canonical operators
\begin{eqnarray}
 {\hat {\cal  H}} && =  {\hat {\cal  H}}_{rev} +  {\hat {\cal  H}}_{irr}   =   
\sum_{\alpha=1}^N    \sum_{\beta=1}^N q^{\dagger}_{ \alpha} 
    \left( \sqrt{ \varphi_{\alpha}}     \langle \varphi_{\alpha} \vert \left( \frac{1}{2} {\bold 1}+{\bold \Omega} \right) \vert \varphi_{\beta} \rangle 
   \sqrt{ \varphi_{\beta}} \right)    q_{ \beta} 
   \nonumber \\
&& =  
\sum_{\alpha=1}^N    \sum_{\beta=1}^N q^{\dagger}_{ \alpha} 
    \left( \sqrt{ \varphi_{\alpha}}     \langle \varphi_{\alpha} \vert \left(  {\bold \Gamma}{\bold \Phi}^{-1} \right) \vert \varphi_{\beta} \rangle 
   \sqrt{ \varphi_{\beta}} \right)    q_{ \beta}  
    \nonumber \\
&& =  
\sum_{\alpha=1}^N    \sum_{\beta=1}^N q^{\dagger}_{ \alpha} 
    \left( \sqrt{ \varphi_{\alpha}}     \langle \varphi_{\alpha} \vert  {\bold \Gamma} \vert \varphi_{\beta} \rangle 
  \frac{1}{ \sqrt{ \varphi_{\beta}} }\right)    q_{ \beta}     
 \label{hfullsmallq}
\end{eqnarray}


\subsection{ Diagonalization of the quantum Hamiltonian ${\hat {\cal  H}} $ to construct its full spectrum }

Plugging the spectral decomposition of ${\bold \Gamma} $ of Eq. \ref{diagoGamma}
into Eq. \ref{hfullsmallq}
\begin{eqnarray}
 {\hat {\cal  H}} &&  =  
\sum_{\alpha=1}^N    \sum_{\beta=1}^N q^{\dagger}_{ \alpha} 
    \left( \sqrt{ \varphi_{\alpha}}     \langle \varphi_{\alpha} \vert  \left(  \sum_{\alpha'=1}^N \gamma_{\alpha'} \vert \gamma_{\alpha'}^R \rangle \langle \gamma_{\alpha'}^L \vert
\right)
         \vert \varphi_{\beta} \rangle 
  \frac{1}{ \sqrt{ \varphi_{\beta}} }\right)    q_{ \beta}     
     \nonumber \\
&& =\sum_{\alpha'=1}^N \gamma_{\alpha'}
\left( \sum_{\alpha=1}^N    
q^{\dagger}_{ \alpha} 
    \sqrt{ \varphi_{\alpha}}     \langle \varphi_{\alpha} \vert   \gamma_{\alpha'}^R \rangle
\right) 
\left( \sum_{\beta=1}^N  \langle \gamma_{\alpha'}^L \vert \varphi_{\beta} \rangle 
  \frac{1}{ \sqrt{ \varphi_{\beta}} }    q_{ \beta}     
\right)
\equiv \sum_{\alpha'=1}^N \gamma_{\alpha'} r^{\dagger}_{ \alpha'} l_{ \alpha'} 
 \label{boldHgamma}
\end{eqnarray}
suggests to introduce the following creation operators $r^{\dagger}_{ \alpha'} $ involving the right eigenvectors $\vert   \gamma_{\alpha'}^R \rangle $ of $\bold \Gamma$
\begin{eqnarray}
r^{\dagger}_{ \alpha'} \equiv \sum_{\alpha=1}^N    
q^{\dagger}_{ \alpha} 
    \sqrt{ \varphi_{\alpha}}     \langle \varphi_{\alpha} \vert   \gamma_{\alpha'}^R \rangle
 \label{rdaggerdef}
\end{eqnarray} 
and the following annihilation operators $l_{ \alpha'} $ involving the left eigenvectors $\langle \gamma_{\alpha'}^L \vert $ of $\bold \Gamma$
\begin{eqnarray}
l_{ \alpha'} \equiv \sum_{\beta=1}^N  \langle \gamma_{\alpha'}^L \vert \varphi_{\beta} \rangle 
  \frac{1}{ \sqrt{ \varphi_{\beta}} }    q_{ \beta}    
 \label{lannihildef}
\end{eqnarray} 
Eq. \ref{Qnsusycommut0small}
 leads to the vanishing commutators between two annihilation or two creation operators
\begin{eqnarray}
[l_{ \alpha'} , l_{\beta'} ]  = 0 = [r_{ \alpha'}^{\dagger} , r_{\beta'}^{\dagger} ]
\label{Qnsusycommutllrr}
\end{eqnarray}
while the canonical commutators of Eq. \ref{smallqcomm}
 lead to the commutators
\begin{eqnarray}
[l_{ \alpha'} , r_{\beta'}^{\dagger} ] && =
\sum_{\beta=1}^N  \langle \gamma_{\alpha'}^L \vert \varphi_{\beta} \rangle 
  \frac{1}{ \sqrt{ \varphi_{\beta}} }    
    \sum_{\alpha=1}^N    
    \sqrt{ \varphi_{\alpha}}     \langle \varphi_{\alpha} \vert   \gamma_{\beta'}^R \rangle
    [ q_{ \beta}, q^{\dagger}_{ \alpha} ] 
    \nonumber \\
&& =
\sum_{\beta=1}^N  \langle \gamma_{\alpha'}^L \vert \varphi_{\beta} \rangle 
  \frac{1}{ \sqrt{ \varphi_{\beta}} }    
    \sum_{\alpha=1}^N    
    \sqrt{ \varphi_{\alpha}}     \langle \varphi_{\alpha} \vert   \gamma_{\beta'}^R \rangle
    \delta_{\alpha,\beta}   
       \nonumber \\
&& =   \sum_{\alpha=1}^N    
  \langle \gamma_{\alpha'}^L \vert \varphi_{\alpha} \rangle 
      \langle \varphi_{\alpha} \vert   \gamma_{\beta'}^R \rangle
      =  \langle \gamma_{\alpha'}^L \vert   \gamma_{\beta'}^R \rangle = \delta_{\alpha',\beta'}
\label{commalrdagger}
\end{eqnarray}
As a consequence, the number operators
\begin{eqnarray}
{\hat n}_{ \alpha} \equiv r^{\dagger}_{ \alpha} l_{ \alpha} 
 \label{Numbersop}
\end{eqnarray} 
which appear in the Hamiltonian of Eq. \ref{boldHgamma} commute
\begin{eqnarray}
[{\hat n}_{ \alpha} ,{\hat n}_{\beta} ]= 0 
 \label{Numbersopcomm}
\end{eqnarray} 
and satisfy
\begin{eqnarray}
[{\hat n}_{ \alpha} ,r^{\dagger}_{\beta} ]=  r^{\dagger}_{ \alpha} l_{ \alpha} r^{\dagger}_{\beta} - r^{\dagger}_{\beta} r^{\dagger}_{ \alpha} l_{ \alpha} = r^{\dagger}_{\beta} \delta_{\alpha,\beta}
 \label{Numbersoprdagger}
\end{eqnarray}

As discussed around Eq. \ref{annihilationboth}, 
the right ground-state ${\hat r}_0(\vec x) = \langle \vec x \vert {\hat r}_0\rangle$ of Eq. \ref{changetowardshateigen}
of the Hamiltonian $  {\hat {\cal  H}}$ for the energy $E=0$ 
is annihilated by all the annihilation operators $Q_n$ (Eq. \ref{Qnsusyannihil}),
so it is also annihilated by all the annihilation operators $q_{\alpha}$ of Eq. \ref{qfrombigQ}
and by all the annihilation operators $l_{\alpha}$ of Eq. \ref{lannihildef}
\begin{eqnarray}
l_{ \alpha} \vert {\hat r}_0\rangle =0    
 \label{lannihildefsurgs}
\end{eqnarray} 
The diagonalization of Eq. \ref{boldHgamma} involving the commuting number operators ${\hat n}_{ \alpha} $
\begin{eqnarray}
 {\hat {\cal  H}} &&  =  \sum_{\alpha=1}^N \gamma_{\alpha} r^{\dagger}_{ \alpha} l_{ \alpha} 
 = \sum_{\alpha=1}^N \gamma_{\alpha}{\hat n}_{ \alpha}
 \label{boldHgamman}
\end{eqnarray}
allows us to construct the full spectrum via the application of creation operators on the ground-state :
the unnormalized state parametrized by the $N$ integers $n_{\alpha}=0,1,... $ for $\alpha=1,..,N$
\begin{eqnarray}
\vert {\hat r}_{\{n_1,..,n_{\alpha},...,n_N\}} \rangle \equiv  \prod_{\alpha=1}^N (r^{\dagger}_{ \alpha})^{n_{\alpha}} \vert {\hat r}_0\rangle     
 \label{excited}
\end{eqnarray} 
is an eigenstate of the $N$ commuting number operators ${\hat n}_{ \alpha} $ with eigenvalues $n_{\alpha} $
\begin{eqnarray}
{\hat n}_{ \alpha}\vert {\hat r}_{\{n_1,..,n_{\alpha},...,n_N\}} \rangle =n_{\alpha} \vert {\hat r}_{\{n_1,..,n_{\alpha},...,n_N\}} \rangle
 \label{excitedeigenn}
\end{eqnarray} 
and is thus an eigenstate of the Hamiltonian ${\hat {\cal  H}} $ 
\begin{eqnarray}
{\hat H}\vert {\hat r}_{\{n_1,..,n_{\alpha},...,n_N\}} \rangle ={\hat E}_{n_1,..,n_{\alpha},...} \vert {\hat r}_{\{n_1,..,n_{\alpha},...,n_N\}} \rangle
 \label{excitedeigenH}
\end{eqnarray} 
of energy
\begin{eqnarray}
 {\hat E}_{n_1,..,n_{\alpha},...}   =   \sum_{\alpha=1}^N \gamma_{\alpha}n_{ \alpha}
 \label{energynalpha}
\end{eqnarray}
So the energy spectrum is simply given by linear combination involving the integer numbers $ n_{ \alpha}$
of the eigenvalues $ \gamma_{\alpha} $ of the matrix ${\bold \Gamma}$.

\end{document}